\documentclass[aps,prl,twocolumn,showpacs,superscriptaddress, floatfix,10pt]{revtex4-2}
\usepackage{graphicx}  
\usepackage{dcolumn}   
\usepackage{bm}        
\usepackage{amssymb}   
\usepackage[caption=false]{subfig}
\usepackage{xcolor}
\usepackage[normalem]{ulem}  
\usepackage{placeins}   

\makeatletter
\newif\ifpreprintoption
\@ifclasswith{revtex4-2}{preprint}{\preprintoptiontrue}{\preprintoptionfalse}
\makeatother

\newcommand{\minerva}{MINERvA}

\newcommand{\pt}{\ensuremath{p_{t}}}
\newcommand{\pz}{\ensuremath{p_{||}}}
\newcommand{\eavail}{\ensuremath{E_{\rm\scriptstyle avail}}}
\newcommand{\sumtp}{\ensuremath{\Sigma T_{p}}}

\newcommand{\qzeroqe}{\ensuremath{q_{0}^{\rm\scriptstyle (QE)}}}

\newcommand{\deborahAdd}[1]{{\color{black} \protect #1}}

\newcommand{\collabAdd}[1]{{\color{black} \protect #1}}

\newlength{\triplet}
\newlength{\doublet}
\newlength{\allBins}
\newlength{\quadruplet}

\begin{document}

\hspace{5.2in} \mbox{FERMILAB-PUB-26-0381-PPD}


\title{Comparisons of triple-differential cross sections for quasielastic-like $\nu_\mu$-hydrocarbon interactions using $\langle E_\nu\rangle \sim$ 3~GeV versus $\sim$ 6~GeV beams in MINERvA}

\newcommand{\Rutgers}{Rutgers, The State University of New Jersey, Piscataway, New Jersey 08854, USA}
\newcommand{\Hampton}{Hampton University, Dept. of Physics, Hampton, VA 23668, USA}
\newcommand{\Dortmund}{Institute of Physics, Dortmund University, 44221, Germany }
\newcommand{\Otterbein}{Department of Physics, Otterbein University, 1 South Grove Street, Westerville, OH, 43081 USA}
\newcommand{\JMU}{James Madison University, Harrisonburg, Virginia 22807, USA}
\newcommand{\Florida}{University of Florida, Department of Physics, Gainesville, FL 32611}
\newcommand{\UCIrvine}{Department of Physics and Astronomy, University of California, Irvine, Irvine, California 92697-4575, USA}
\newcommand{\CBPF}{Centro Brasileiro de Pesquisas F\'{i}sicas, Rua Dr. Xavier Sigaud 150, Urca, Rio de Janeiro, Rio de Janeiro, 22290-180, Brazil}
\newcommand{\PUCP}{Secci\'{o}n F\'{i}sica, Departamento de Ciencias, Pontificia Universidad Cat\'{o}lica del Per\'{u}, Apartado 1761, Lima, Per\'{u}}
\newcommand{\INRM}{Institute for Nuclear Research of the Russian Academy of Sciences, 117312 Moscow, Russia}
\newcommand{\Jlab}{Jefferson Lab, 12000 Jefferson Avenue, Newport News, VA 23606, USA}
\newcommand{\Pittsburgh}{Department of Physics and Astronomy, University of Pittsburgh, Pittsburgh, Pennsylvania 15260, USA}
\newcommand{\Guanajuato}{Campus Le\'{o}n y Campus Guanajuato, Universidad de Guanajuato, Lascurain de Retana No. 5, Colonia Centro, Guanajuato 36000, Guanajuato M\'{e}xico.}
\newcommand{\Athens}{Department of Physics, University of Athens, GR-15771 Athens, Greece}
\newcommand{\Tufts}{Physics Department, Tufts University, Medford, Massachusetts 02155, USA}
\newcommand{\WM}{Department of Physics, William \& Mary, Williamsburg, Virginia 23187, USA}
\newcommand{\FNAL}{Fermi National Accelerator Laboratory, Batavia, Illinois 60510, USA}
\newcommand{\Purdue}{Department of Chemistry and Physics, Purdue University Calumet, Hammond, Indiana 46323, USA}
\newcommand{\MCLA}{Massachusetts College of Liberal Arts, 375 Church Street, North Adams, MA 01247}
\newcommand{\UMD}{Department of Physics, University of Minnesota -- Duluth, Duluth, Minnesota 55812, USA}
\newcommand{\Northwestern}{Northwestern University, Evanston, Illinois 60208}
\newcommand{\UNI}{Facultad de Ciencias, Universidad Nacional de Ingenier\'{i}a, Apartado 31139, Lima, Per\'{u}}
\newcommand{\Rochester}{Department of Physics and Astronomy, University of Rochester, Rochester, New York 14627 USA}
\newcommand{\Austin}{Department of Physics, University of Texas, 1 University Station, Austin, Texas 78712, USA}
\newcommand{\USM}{Departamento de F\'{i}sica, Universidad T\'{e}cnica Federico Santa Mar\'{i}a, Avenida Espa\~{n}a 1680 Casilla 110-V, Valpara\'{i}so, Chile}
\newcommand{\Geneva}{University of Geneva, 1211 Geneva 4, Switzerland}
\newcommand{\Chicago}{Enrico Fermi Institute, University of Chicago, Chicago, IL 60637 USA}
\newcommand{\hired}{}
\newcommand{\OregonState}{Department of Physics, Oregon State University, Corvallis, Oregon 97331, USA}
\newcommand{\oxford}{Oxford University, Department of Physics, Oxford, OX1 3PJ United Kingdom}
\newcommand{\umiss}{University of Mississippi, Oxford, Mississippi 38677, USA}
\newcommand{\upenn}{Department of Physics and Astronomy, University of Pennsylvania, Philadelphia, PA 19104}
\newcommand{\AMU}{AMU Campus, Aligarh, Uttar Pradesh 202001, India}
\newcommand{\wroclaw}{University of Wroclaw, plac Uniwersytecki 1, 50-137 Wroa\l{}aw, Poland}
\newcommand{\Mohali}{Department of Physical Sciences, IISER Mohali, Knowledge City, SAS Nagar, Mohali - 140306, Punjab, India}
\newcommand{\CINVESTAV}{Departamento de Fisica Col. San Pedro Zacatenco, 07360 Mexico, DF, Av. Instituto Politecnico Nacional, Mexico}
\newcommand{\york}{York University, Department of Physics and Astronomy, Toronto, Ontario, M3J 1P3 Canada}
\newcommand{\ND}{Department of Physics, University of Notre Dame, Notre Dame, Indiana 46556, USA}
\newcommand{\ICL}{The Blackett Laboratory,  Imperial College London,  London SW7 2BW, United Kingdom}
\newcommand{\warwick}{Department of Physics, University of Warwick, Coventry, CV4 7AL, UK}
\newcommand{\qmul}{G O Jones Building, Queen Mary University of London, 327 Mile End Road, London E1 4NS, UK}
\newcommand{\LLNL}{Nuclear and Chemical Sciences Division, Lawrence Livermore National Laboratory, Livermore, CA 94550, USA}
\newcommand{\adrianThanks}{Now at Department of Physics, Drexel University, Philadelphia, Pennsylvania 19104, USA}
\newcommand{\lazazuetareyesThanks}{now at Syracuse University, Syracuse, NY 13244, USA}


\author{D.~Ruterbories}                   \affiliation{\Rochester}
\author{S.~Akhter}                        \affiliation{\AMU}
\author{Z.~Ahmad~Dar}                     \affiliation{\WM}  \affiliation{\AMU}
\author{M.~Sajjad~Athar}                  \affiliation{\AMU}
\author{M.~Betancourt}                    \affiliation{\FNAL}
\author{S.~Boyd}                          \affiliation{\warwick}  \affiliation{\Pittsburgh}
\author{H.~da~Motta}                      \affiliation{\CBPF}
\author{J.~Felix}                         \affiliation{\Guanajuato}
\author{L.~Fields}                        \affiliation{\ND}
\author{R.~Fine}
\affiliation{\Rochester}
\author{A.M.~Gago}                        \affiliation{\PUCP}
\author{H.~Gallagher}                     \affiliation{\Tufts}
\author{P.K.Gaur}                         \affiliation{\AMU}
\author{S.M.~Gilligan}                    \affiliation{\OregonState}
\author{R.~Gran}                          \affiliation{\UMD}
\author{E.Granados}                       \affiliation{\Guanajuato}  \affiliation{\Guanajuato}
\author{D.A.~Harris}                      \affiliation{\york}  \affiliation{\FNAL}
\author{A.L.~Hart}                        \affiliation{\qmul}
\author{A.~Klustov\'{a}}                  \affiliation{\ICL}
\author{M.~Kordosky}                      \affiliation{\WM}
\author{D.~Last}                          \affiliation{\Rochester}  \affiliation{\upenn}
\author{Z.~Lin}                           \affiliation{\Rochester}
\author{A.~Lozano}\thanks{\adrianThanks}  \affiliation{\CBPF}
\author{S.~Manly}                         \affiliation{\Rochester}
\author{W.A.~Mann}                        \affiliation{\Tufts}
\author{C.~Mauger}                        \affiliation{\upenn}
\author{K.S.~McFarland}                   \affiliation{\Rochester}
\author{M.~Mehmood}                       \affiliation{\york}
\author{O.~Moreno}                        \affiliation{\WM}  \affiliation{\Guanajuato}
\author{J.G.~Morf\'{i}n}                  \affiliation{\FNAL}
\author{J.K.~Nelson}                      \affiliation{\WM}
\author{C.~Nguyen}                        \affiliation{\Florida}
\author{V.~Paolone}                       \affiliation{\Pittsburgh}
\author{G.N.~Perdue}                      \affiliation{\FNAL}  \affiliation{\Rochester}
\author{C.~Pernas}                        \affiliation{\WM}
\author{M.A.~Ram\'{i}rez}                 \affiliation{\upenn}  \affiliation{\Guanajuato}
\author{R.D.~Ransome}                     \affiliation{\Rutgers}
\author{N.~Roy}                           \affiliation{\york}
\author{H.~Schellman}                     \affiliation{\OregonState}
\author{C.J.~Solano~Salinas}              \affiliation{\UNI}
\author{N.H.~Vaughan}                     \affiliation{\OregonState}
\author{A.V.~Waldron}                     \affiliation{\qmul}  \affiliation{\ICL}
\author{L.~Zazueta}\thanks{\lazazuetareyesThanks}  \affiliation{\WM}

\collaboration{The \minerva\ Collaboration}\ \noaffiliation
\date{\today}

\begin{abstract}
Neutrino charged-current quasielastic-like scattering, a reaction category extensively used in neutrino oscillation measurements, receives contributions from single nucleon knockout processes, multinucleon processes, and inelastic scattering with subsequent rescattering or absorption in the nucleus to produce only nucleons in the final state.   In this article, comparisons are presented of the same measurement in two different wideband neutrino beams:  one beam peaks near 3~GeV with few neutrinos above 6~GeV;  the other peaks near 6~GeV with few neutrinos above 10~GeV.    Comparisons of differential cross sections in muon and proton kinematics for these two exposures probe deviations from free-neutron scattering that arise from the processes involving the nuclear medium, and provide a test of neutrino interaction models used to infer neutrino energies in oscillation experiments.  Discrepancies are observed between the data and predictions that point to overestimates of the final state interactions of both protons and charged pions in quasielastic-like events.

\end{abstract}
\maketitle

\section{Introduction}
A leading contributor to charged-current (CC) neutrino interactions at a few GeV energies is the quasielastic-like reaction in which one or more nucleons is knocked out of the nucleus: 
\begin{equation}
\label{signal-channel}
\nu_{\mu}+\mathcal{A}\rightarrow\mu^{-}+\textrm{nucleons} + \mathcal{A'.}
\label{eqn:qelike-reaction}
\end{equation}
\noindent
Charged-current neutrino-nucleus interactions, even those with apparent two-body quasielastic final states, are altered by a number of poorly understood nuclear effects: The struck nucleons of the initial state are bound and in motion~\cite{Moniz:1971mt,Benhar:1994hw}; short-range multinucleon processes give rise to enhanced reaction rates relative to scattering on free nucleons~\cite{Marteau:1999kt,Martini:2010ex,Nieves:2011pp,Gran:2013kda,Fiorentini:2013ezn,Rodrigues:2015hik}, and hadrons produced in the parent $\nu_{\mu}$ interactions with nucleons undergo intranuclear final-state interactions (FSI) within the target nuclei.  

While the reaction $\nu_{\mu}+\mathcal{A}\rightarrow\mu^{-}+ p + \mathcal{A'}$, wherein nearly all final-state energy is visible, is thought to be the main contributor to the quasielastic-like channel (Eq.~\ref{eqn:qelike-reaction}) a significant number of events may have energy deposited in undetected neutrons or light nuclear fragments.  Final states of the latter kind complicate the task of inferring neutrino energy from samples of quasielastic-like events.

This paper reports measurements of the quasielastic-like channel in which the final-state muon transverse (\pt) and longitudinal (\pz) momenta are measured in each event simultaneously with the total \lq\lq available\rq\rq, or calorimetrically visible, recoil energy (\eavail) used in previous analyses of data from MINERvA~\cite{Rodrigues:2015hik,MINERvA:2018nab,MINERvA:2021wjs,MINERvA:2022mnw}.  Since the signal requires final state muon plus nucleons only, \eavail\ is  the sum of the kinetic energies of all protons, denoted \sumtp.  

For the reaction on a free nucleon, $\nu_\mu n\to mu^- p$, which is free of nuclear effects, a measurement of either \pt and \pz, or of $E_\mu$ and \sumtp\, would uniquely specify the neutrino energy. 
Therefore the cross section in one bin of \pt, \pz, and \sumtp\ should not depend on the distribution of neutrino energies incident on the detector without such nuclear effects.  The extent to which the cross sections from these two different fluxes differ for each (\pt, \pz, \sumtp )  bin is therefore a measure of how different the smearing is between \lq\lq invisible\rq\rq energy (for example, energy carried by final state neutrons, or from pions being absorbed in the nucleus) and the 
\lq\lq visible\rq\rq energy.  

Neutrino oscillation experiments estimate event-by-event neutrino energy in different ways depending on the detector technology.  In general, the final-state lepton energy is well-measured, however experiments differ in their estimation of hadronic recoil energy to be added to the lepton energy to 
obtain the neutrino energy. 
For experiments whose detectors can measure final hadronic kinetic energies, such as NOvA or DUNE, this recoil energy is simply the sum of the kinetic energies of the protons, or \sumtp~ (since many final state neutrons deposit no visible energy).  For the case of water Cerenkov-based experiments such as T2K and Hyper-K where only final state lepton kinematics are measured, the recoil energy added to the lepton energy is formed by assuming a stationary target neutron in $\nu_\mu + n$(bound)$ \rightarrow \mu^- + p$:
\begin{equation}
    \qzeroqe \equiv\frac{m_p^2-(m_n-E_b)^2-m_\mu^2+2(E_\mu-p_\mu cos\theta_\mu)E_\mu}{2\left[(m_n-E_b)-E_\mu+p_\mu cos\theta_\mu\right]}. \label{eqn:q0qe}
\end{equation}
Here, $m_\mu$, $m_p$, and $m_n$ are the masses of the muon, proton, and neutron, $E_b$ is the average binding energy of $34$~MeV~\cite{Bodek:2018lmc,Katori:2008zz,Moniz:1971mt} for an carbon nucleus, and $E_\mu$, $p_\mu$, and $\theta_\mu$ are the muon energy, momentum and angle with respect to the neutrino beam.
Combined T2K-NOvA analyses~\cite{T2K:2025wet}
rely on interaction models to correctly predict the relationship between these quantities, in particular as a function of neutrino energy.  

For a given muon momentum, either the \qzeroqe\ or the \sumtp\ will specify the incident neutrino energy, assuming no nuclear effects. Comparisons of the cross section for one \qzeroqe\ bin as a function of \sumtp\ for the same muon momentum, but using two different fluxes is another way to explore the extent to which energy is lost to processes that are invisible in a detector.  

The MINERvA experiment can test predictions of energy loss mechanisms because it collected high statistics data samples in neutrino beams whose average energies differ by a over factor of two, using a fine-grained detector that is able to directly measure \pt, \pz\ and \sumtp.   
In this paper, a new quasielastic-like cross section is reported using the same MINERvA detector and methodology as previously used for the cross section reported in Ref.~\cite{MINERvA:2022mnw}, however the analyzed data comes from a neutrino flux with half the peak energy of the beam used for the previous result.  This paper also re-extracts the cross sections reported in Ref.~\cite{MINERvA:2022mnw}, but only uses the kinematic regions and binning that is appropriate for the lower energy beam.
Comparisons that incorporate the correlations between the two measurements are then provided.  
Such comparisons illuminate the dependence on neutrino energy of the mismodeling of the cross section in specific regions of \sumtp\ and \pt\  observed in Ref.~\cite{MINERvA:2022mnw}.  Moreover they confirm the conclusion of that previous work, namely that the overestimate of final state interactions observed appears to be independent of $\pz$ and are therefore likely to be independent of neutrino energy.

\begin{figure}[tp]
    \centering
    \includegraphics[width=0.9\linewidth]{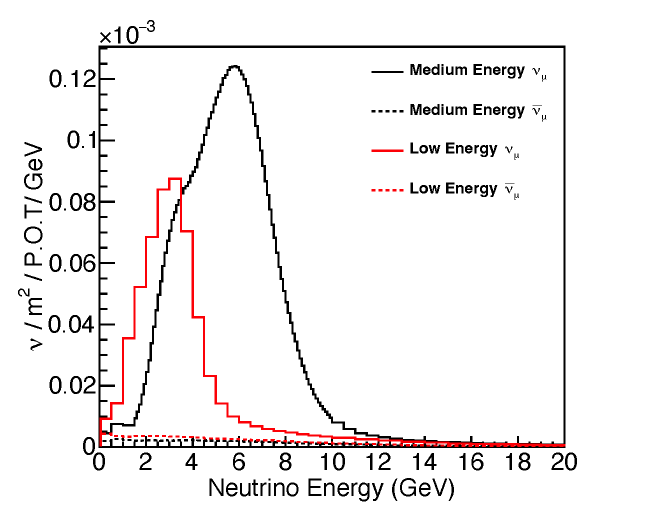}
    \caption{{Medium and Low Energy fluxes in the neutrino focused mode at \minerva, per proton on target (P.O.T.). In addition to the $\nu_\mu$ flux, the $\bar{\nu_\mu}$ contamination in each beam is shown.}}
    \label{fig:fluxes}
\end{figure}

\section{Experimental Setup}
The neutrino beams used in these analyses were both produced by 120~GeV protons from Fermilab's Main Injector striking graphite targets, producing pions and kaons which were then focused in a magnetic two-horn system and sent to a 675~m decay pipe, but with two important differences that change the resulting neutrino energy distribution.  In the Low Energy beam the upstream edge of the target was located 45~cm upstream of the first horn, while in the Medium Energy beam the target was 119~cm upstream of the first horn.  In addition, the separation between the two horns was 10~m in the Low Energy configuration but was changed to 23~m in the Medium Energy configuration~\cite{Adamson:2015dkw}.  The resulting neutrino fluxes, shown in Fig.~\ref{fig:fluxes}, were calculated using a GEANT4 simulation with input from hadronic interaction data~\cite{Aliaga:2016oaz}.  The fluxes were also constrained in both cases using neutrino-electron scattering in each beam~\cite{Park:2015eqa}\cite{MINERvA:2022vmb}.  The neutrino-mode Low Energy (Medium Energy) beam comes from a total of $3.34\times 10^{20}$  $(10.61\times 10^{20})$ protons on target collected between 2010-2012 (2013-2017).  
The neutrino interactions analyzed occurred in the 5.3~ton central plastic scintillator region of the MINERvA detector~\cite{Aliaga:2013uqz} and the muons that originated in these interactions were charge- and momentum-analyzed in the MINOS Near Detector which is made of magnetized steel and scintillator planes~\cite{Michael:2008bc}.  

Because the Low Energy data used here was taken roughly four years before the Medium energy data it is important to take into account the time-dependence of the detector response.  The time dependence of the scintillator response was measured using through-going muons and then simulated accordingly using GEANT4~\cite{Agostinelli:2002hh} version 4.9.4p2 with the QGSP\_BERT physics list throughout the entire data-taking period. 
 The calibrations are described in Refs.~\cite{Aliaga:2013uqz,MINERvA:2021mpk}.  
In addition, the instantaneous neutrino event rate between the two beams differed by up to a factor of 10 because of the higher neutrino intensity per proton on target in the medium energy beam, and the accelerator complex's improvement in instantaneous protons on target delivery between 2010 and 2017.  
The time-dependent accidental activity arising from the neutrino intensity was simulated by overlaying hits from the data in both MINERvA and MINOS for the relevant time the data were taken.  The detector response to protons relative to muons was measured in scaled-down version of the MINERvA detector in a test beam~\cite{Aliaga:2015aqe} and used for both data-taking periods.  

An analysis done in the Medium Energy beam of events with low hadronic energy resulted in a shift of the MINOS muon energy scale by 3.6$\pm$1.0\%~\cite{MINERvA:2021mpk}, a shift of 1.8 times the {\it a priori} uncertainty on that quantity.  That shift results in a correlation between the flux uncertainty and the muon energy scale uncertainty.  Since the same detector was used in the Low Energy neutrino beam that same muon energy scale shift was applied, but because the Low Energy data lacks the precision to repeat the analysis described above, the uncertainty on the muon energy scale in the Low Energy data is kept at the original 2\%.  These different uncertainties in scale between the Low Energy and Medium Energy data are taken to be completely correlated.  

The reference signal and background models for this analysis are based on a modified version of the GENIE~\cite{Andreopoulos:2009rq} v2.8.4 for the Low Energy cross section.  The original cross sections reported in Ref.~\cite{MINERvA:2022mnw} used the GENIE v2.12.6 event generator as a base model.  The modifications and options used in the GENIE v2.8.4 generator for the Low Energy analysis make that model nearly equivalent to the GENIE v2.12.6 model used for the signal processes, with the primary difference coming from small changes in the pion production model.  Details of the specific models used here can be found in ~\cite{MINERvA:2022mnw}, but the key model choices are that quasielastic interactions are modeled using the Llewellyn Smith formalism~\cite{LlewellynSmith:1971zm} with BBBA05 vector form factors~\cite{Bradford:2006yz} coming from fits to electron scattering data and an axial-vector form factor based on a z-expansion fit to neutrino data on deuterium~\citep{Meyer:2016oeg}. The initial state nucleons are modeled as a relativistic Fermi gas~\cite{Smith:1972xh} and with a Bodek-Ritchie high momentum tail~\cite{Bodek:1981wr}.  Multinucleon quasielastic-like interactions, referred to here as 2p2h, are simulated by the Valencia model described in Refs.~\cite{Nieves:2011pp,Gran:2013kda,Schwehr:2016pvn}.  Intranuclear final-state interactions of produced hadrons are modeled using the INTRANUKE-hA package~\cite{Dytman:2007zz}.  There was an error in GENIE's treatment of Elastic FSI~\cite{Harewood:2019rzy} which has a negligible effect in these variables; this bug was fixed for the Medium Energy prediction but not for the Low Energy prediction.  

To account for an observed excess in specific regions of three-momentum transfer and \sumtp~in the Low Energy neutrino charged current data~\cite{Rodrigues:2015hik}, the multinucleon cross section in the simulation is increased based on fits to those data.  MINERvA also developed changes to its pion production model based on measurements in its medium~\cite{MINERvA:2022djk} and low~\cite{MINERvA:2019kfr} energy beams.  In events with identified pions, MINERvA sees a suppression of pion production strength at low momentum transfer squared, which would then have an impact on the prediction for quasielastic-like events that come from pion production. Furthermore, based on fits to $\nu_{\mu}$-hydrogen data~\cite{Rodrigues:2016xjj}, the non-resonant CC pion production is decreased by \deborahAdd{57\%}, 
the overall baryon-resonance pion production is increased by 15\%, and $M_{A}^{RES}$~is set to 0.94~GeV.  
Finally, a Random Phase Approximation correction is 
applied to quasielastic scattering to account for nuclear screening of the weak charge~\cite{Nieves:2004wx,Gran:2017psn}.
These two modifications are identical in the Low Energy and Medium Energy models. 

\section{Analysis Procedure}

The analysis procedure for both neutrino fluxes is identical and is summarized here.  More details are provided in \cite{MINERvA:2022mnw}.  Because the potential to mis-model the background contributions is large, there are signal and background-dominated samples that are collected in parallel in the data, and then a fit is done to the background-dominated samples to add an additional constraint to the predicted background levels in the signal sample.  Both the 
 candidate quasielastic-like interactions and the background samples require a negatively-charged muon track that starts in the fiducial volume of MINERvA and is identified in MINOS.  All other tracked particles originating from the beginning of the muon track must have $dE/dx$ consistent with a proton.  
Signal and background samples 
are formed by counting the number of Michel electron candidates within $600$~mm long, $600$~mm diameter cylinders centered on the neutrino vertex and on endpoints of tracked particles, and by counting isolated clusters constructed from two-dimensional clusters with at least 1~MeV visible energy. The former identify $\pi^+$, and the latter identify photons from $\pi^0$ decays. Clusters with an energy less than 10~MeV per hit are assumed to
be caused by neutrons producing low energy protons and are not used. 

The primary backgrounds come from events that contain either a $\pi^+$\, a $\pi^0$\, or both.  
In each beam configuration, four exclusive samples are assembled using the criteria of $0$ or $\ge 1$ Michel electrons, and $\le 1$ or $\ge 2$ isolated clusters.    Signal events are required to have no Michel electron candidates, and $\le 1$ isolated clusters.  The sample dominated by single charged pion events requires one Michel electron and $\le 1$ isolated clusters per event.  The sample dominated by single $\pi^0$ events requires $\ge 2$ isolated clusters, and no Michel electrons in the event.  The sample dominated by multipion events requires both a Michel electron and $\ge 2$ isolated clusters.  
These data samples in each beam configuration are analyzed using a simultaneous joint fit in each bin of \pt\ and \sumtp, while integrating over $1.5< \pz (GeV/c)<4.5$.  A single background scale factor is determined for each of the three background categories, and an additional scale factor is returned for the signal sample.  The fit minimizes a $\chi^2$ over the four scale factors using a singular value decomposition (SVD) technique which drops singular values with condition number $<10^{-3}$ to avoid numerical instability.  The fit does not allow negative scale factors for any component. 
\begin{figure}[h]
    \centering
    \includegraphics[width=\linewidth]{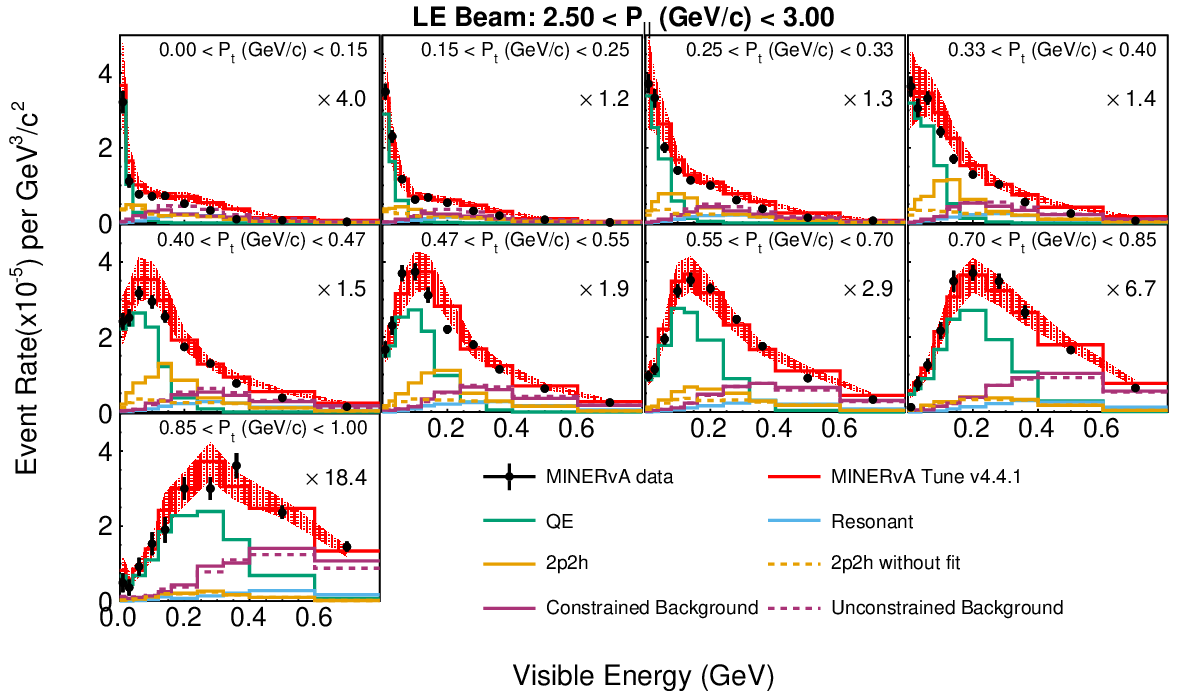}
    \includegraphics[width=\linewidth]{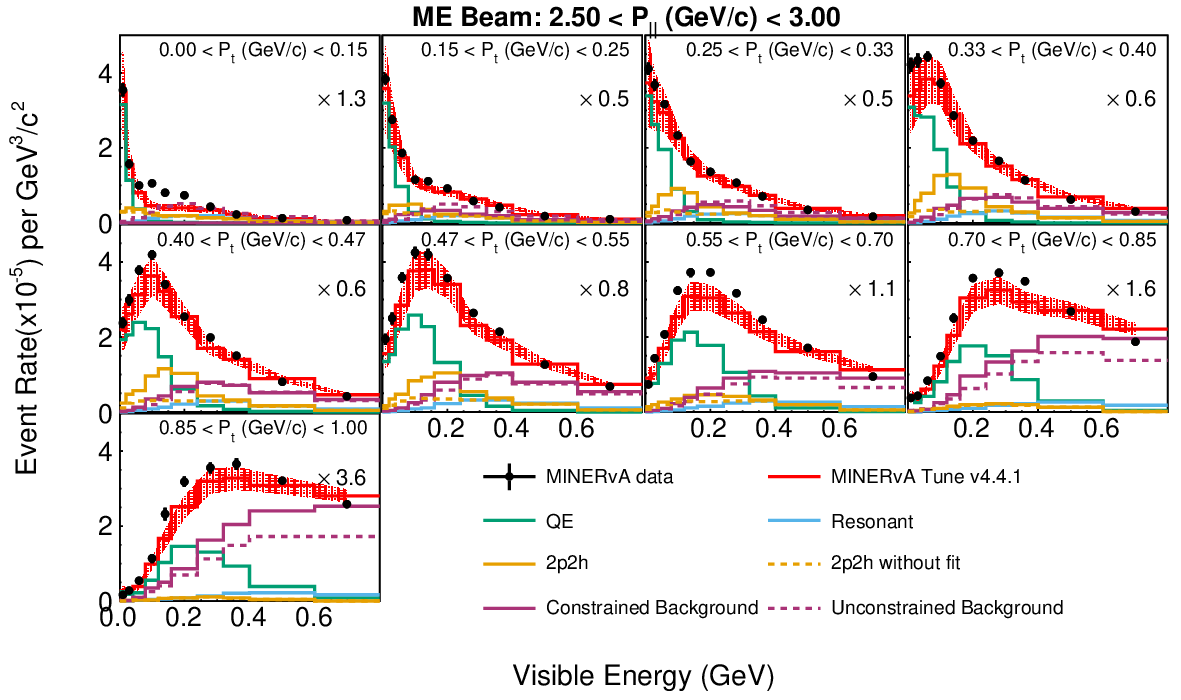}
    \caption{Event distributions in data and prediction after the background fits described in the text, for both the Low (top) and Medium (bottom) energy data, for one \pz\ bin as a function of visible energy and \pt.  The labels QE, 2p2h, and Resonant refer to the contributions from quasielastic-like processes, while the Constrained (unconstrained) Background line is the prediction for the non-quasielastic-like events, based on (before) the fits to alternate data samples described in the text.  The contribution labeled 2p2h without fit is what the 2p2h prediction would have been without the modification based on the Low Energy analysis described in Ref.~~\cite{Rodrigues:2015hik}. The shaded band represents the total uncertainty in the predicted number of events.}
    \label{fig:bckgd}
\end{figure}

Figure~\ref{fig:bckgd} shows the event distributions in the Low Energy (LE) and Medium Energy (ME) Energy in the data and simulation for one \pz\ bin as a function of \pt\ and the \lq\lq visible\rq\rq hadronic energy.  This figure also shows the predicted quasielastic-like signal channels as well as the backgrounds before and after the fits described above.  This \pz\ bin is shown because it contains high statistics event samples in both beams.  The event distributions for all \pz\ bins can be found in the Supplement.  The MINERvA tune reproduces the data trends observed in both beam configurations, however the model starts underpredicting the data in the region dominated by the quasielastic peak for \pt\ above $0.33~$GeV/c in the ME beam data.

The backgrounds in both beams are most important at high \pt\ and high \sumtp, and are larger in the Medium Energy beam than in the Low Energy beam.  The latter trend reflects the presence of the $\Delta$ invariant mass threshold and the factor of two difference between the peak energies of the two beams.  For both the Low and Medium energy beams, the additional strength of the 2p2h contribution based on the fit to Low Energy inclusive charged current data alone improves the agreement with the simulation.  

Each background-subtracted event rate is unfolded using an iterative technique~\citep{D'Agostini:1994zf} from the RooUnfold framework~\citep{Adye:2011gm} which is regularized by the number of iterations. A regularization of ten iterations was used for both beam configurations.  

Because of the lower average neutrino energy in the Low Energy beam, the results reported here for both beam configurations only extend to a muon longitudinal momentum of 4.5~GeV (the original muon cut in the higher energy beam was 20~GeV).  The muon's transverse momentum bin boundary of 2.5~GeV was maintained.  In addition, the bin boundaries for the muon kinematics have changed with respect to Ref.~~\cite{MINERvA:2022mnw}.   

\section{Systematic Uncertainties}

The systematic uncertainties on these cross-section measurements arise from three different sources:  the flux, neutrino interactions, and the combination of the MINERvA and MINOS detector responses.  These uncertainties are evaluated for each neutrino beam configuration by re-extracting the cross section in each beam after varying each source of uncertainty, so that the correlations between different bins (and the correlations between the low and medium energy beam) can be incorporated.  

The flux uncertainty comes from uncertainty in hadron production and focusing effects, and is constrained in the Low Energy beam at 6\% and in the Medium Energy beam at 3.9\% using neutrino-electron scattering interactions~\cite{Park:2015eqa,MINERvA:2022vmb}. 
In reality, the hadron production uncertainties between the Low Energy and Medium Energy beams are correlated, but for the ratio analysis we make the conservative assumption that they are not.  The neutrino electron scattering constraints are dominated by statistics and are therefore also uncorrelated.  

Neutrino interaction uncertainties are dominated by the modeling uncertainties in background processes, in particular the final-state interaction uncertainties.  Because the uncertainties are evaluated in parallel for the Medium and Low Energy configurations, there is some cancellation of the systematic uncertainties in the ratio measurement.  
Detector uncertainties are dominated by uncertainties in muon reconstruction, which are small, but increase at high \pt\ where the cross section is falling steeply and small changes in the muon energy scale have a large effect on the accepted interactions.  Again the muon reconstruction uncertainty is correlated between the Low and Medium Energy beam configurations.

\begin{figure}[tp]
    \centering
\includegraphics[width=\linewidth]{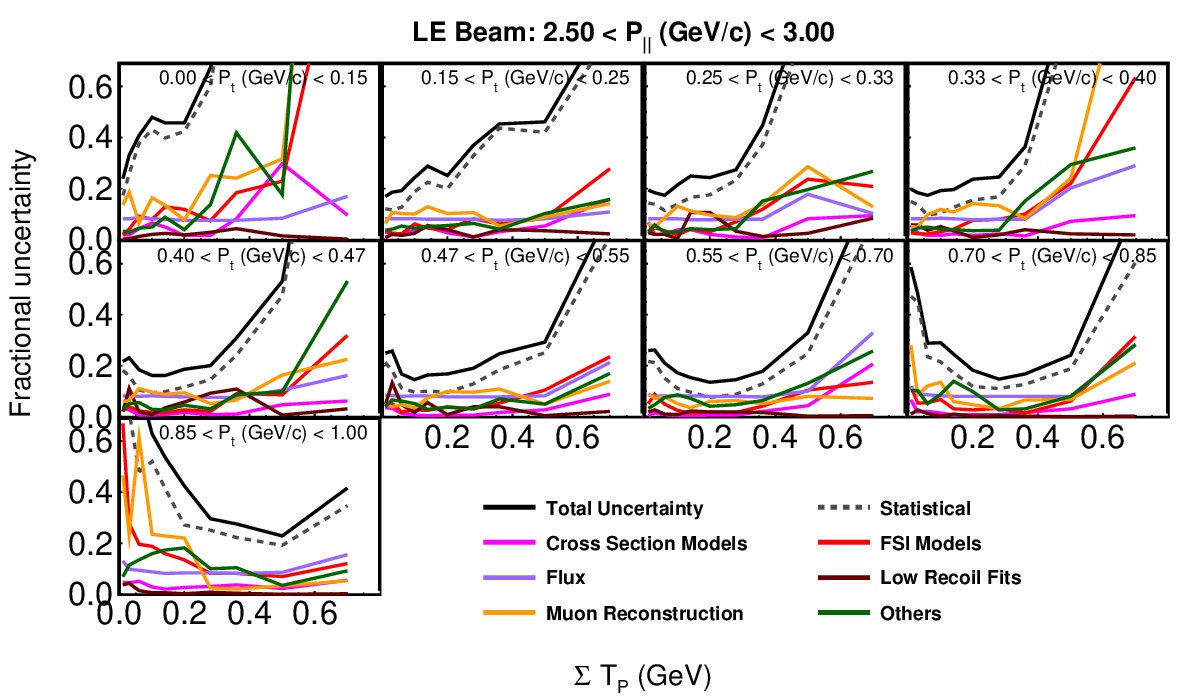}
\includegraphics[width=\linewidth]{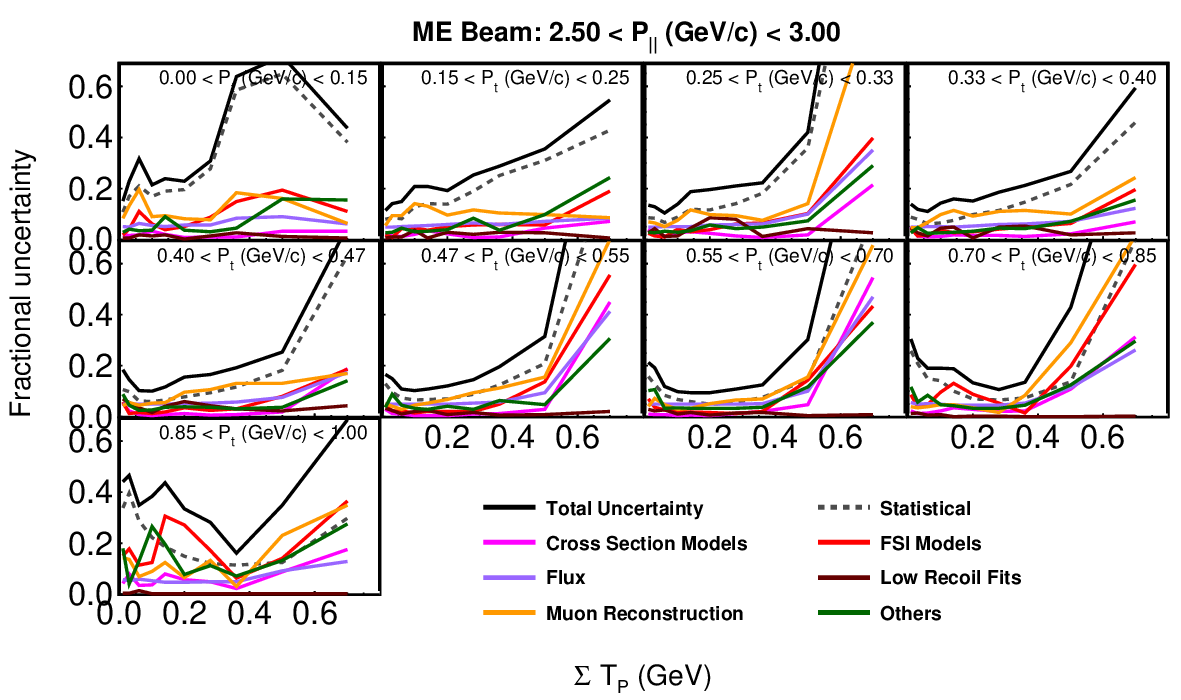}
\vspace{.1cm}

\includegraphics[width=\linewidth,trim={0 0 0 0.950cm},clip]
{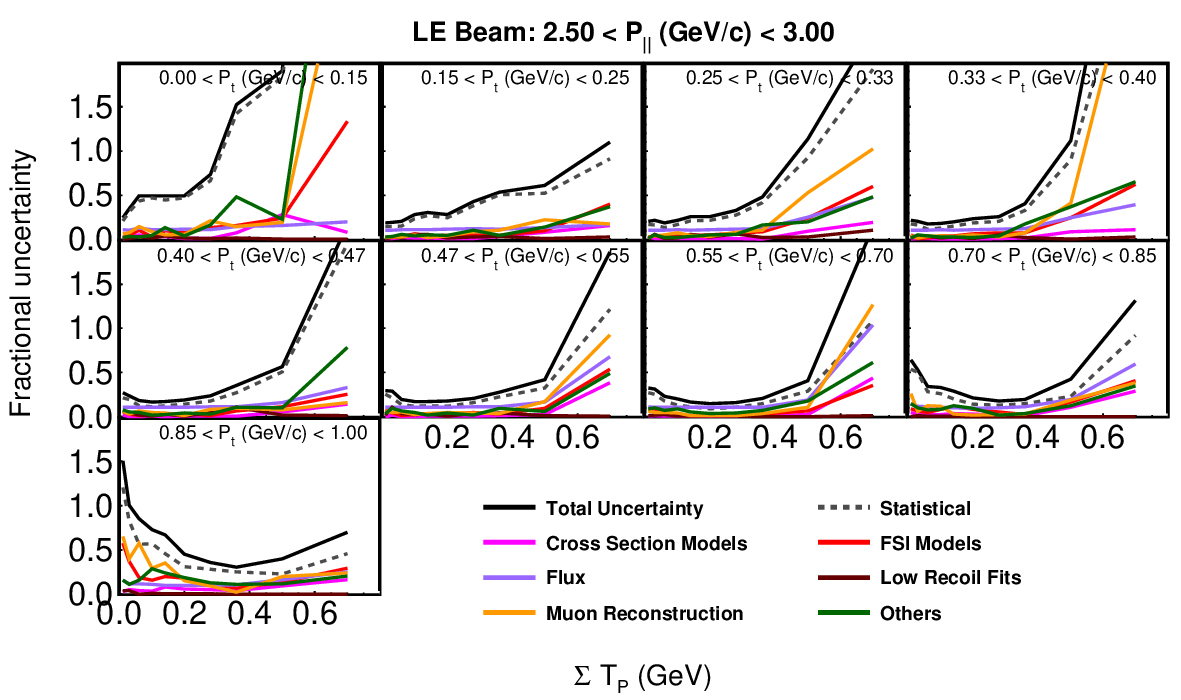}
    \caption{Fractional uncertainty on the cross section in the Low Energy (Top) Medium Energy (Middle) beam configurations as well as the double ratio (Bottom) as a function of the sum of proton kinetic energies, for different muon transverse momenta for muons with longitudinal momenta between 2.5~GeV/c and 3.0~GeV/c.
}
    \label{fig:syst_uncert}
\end{figure}

Although the detector's response to muons is statistically uncorrelated between the Low and Medium energy data, the proton energy scale and the muon energy scale are 100\% correlated between the two configurations since they depend on the same test beam measurements with a smaller version of the MINERvA detector~\cite{Aliaga:2015aqe}.  Similarly, the neutrino interaction uncertainties are also 100\% correlated between the two configurations.  Figure~\ref{fig:syst_uncert} shows the fractional uncertainties on the Low (top) and Medium (middle) Energy beam cross sections, as well as the uncertainty on the ratio of cross sections (bottom) for one \pz\ bin as a function of \sumtp\ and \pt.  At the lowest \pt\ the uncertainties are dominated by the statistics, and in most regions the systematic uncertainty is dominated by the muon uncertainty and the FSI model.  
~\\
\section{MEASUREMENTS AND COMPARISONS}
\subsection{ Triple Differential Cross Sections}

\begin{figure*}[tp]
    \centering
\includegraphics[width=0.9\linewidth]{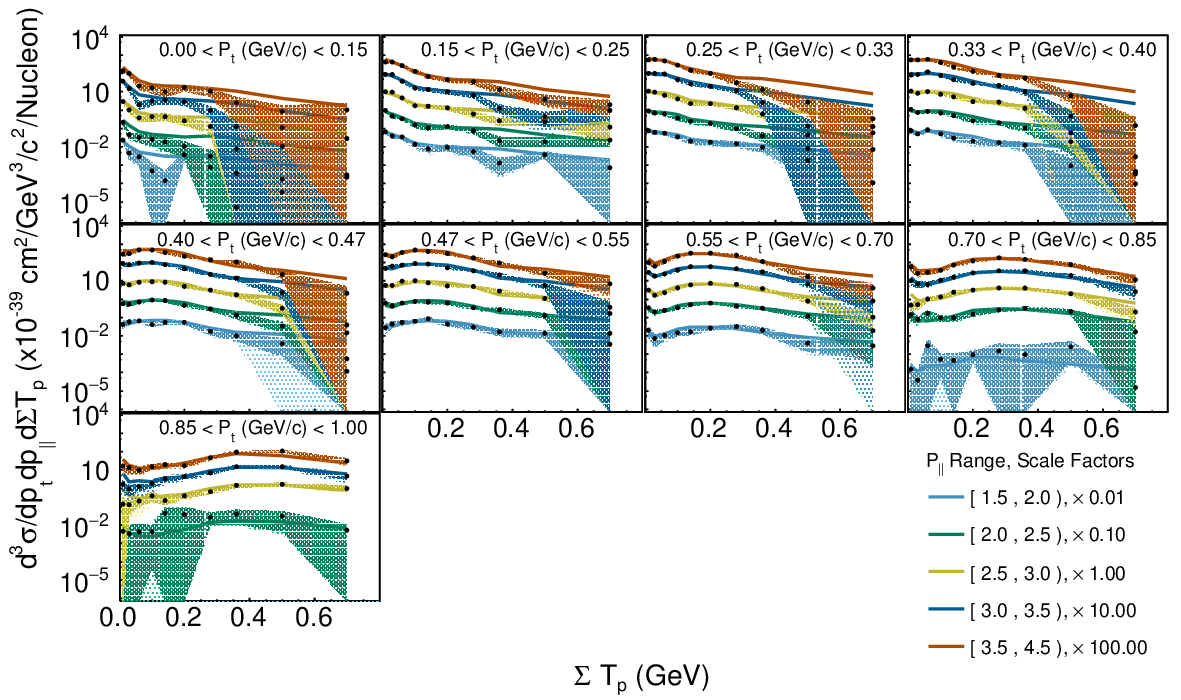}
    \includegraphics[width=0.9\linewidth]{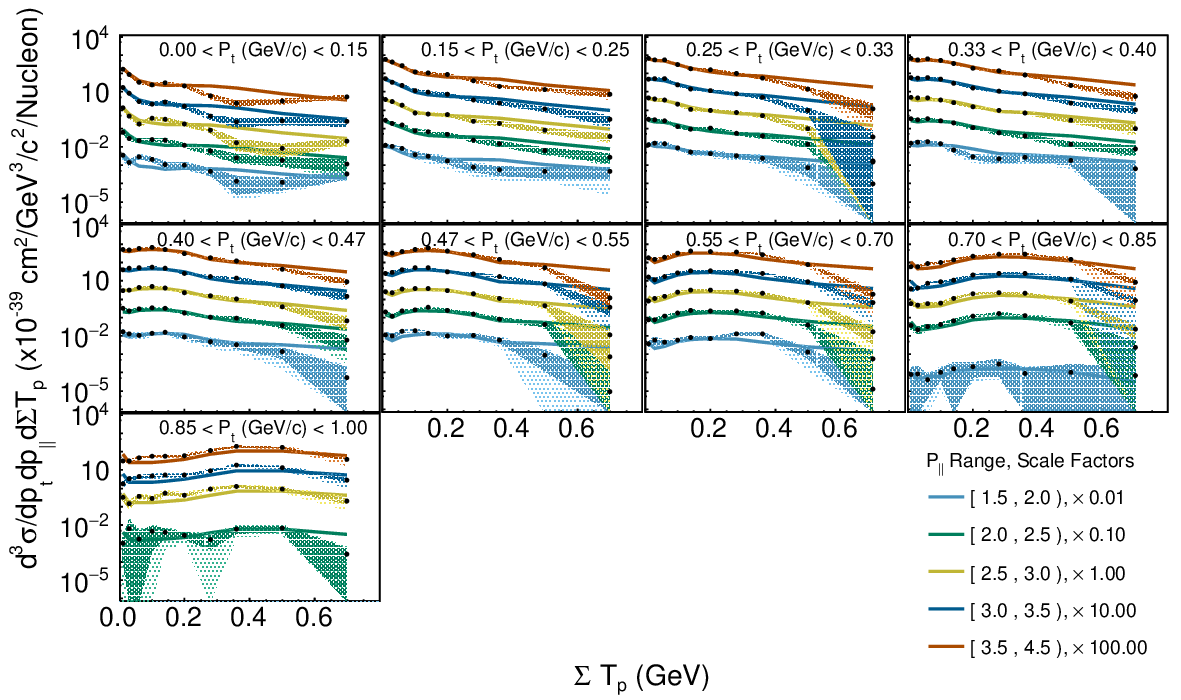}
    \caption{Cross section in the Low (top) and Medium (bottom) Energy beam as a function of the sum of proton kinetic energies, for different muon longitudinal and transverse momenta.  The black points represent the data and the colored lines show the predictions from MINERvA's base model.  The colored bands indicate the uncertainties on the data points, and the color of the band matches the color of the \pz\ bin's base model prediction line (units of \pz\ in the legend are in GeV).  The binning presented here is different than the previously published result~\cite{MINERvA:2022mnw} in order to match the Low Energy binning}
    \label{fig:MEtriplexsec}
\end{figure*}

The Low and Medium Energy cross sections are shown in Fig.~\ref{fig:MEtriplexsec}.  Each panel has a horizontal axis of \sumtp, is a different value of \pt, and has different vertical offsets for each longitudinal momentum bin (\pz ).  In this analysis \pz\ is the variable most closely related to the original neutrino energy, and \pt\ is the variable most closely related to the negative of the four-momentum transfer squared, $Q^2$.  

Although the uncertainties in the Low Energy beam are significantly larger than in the Medium Energy beam, the trends for the falling cross section as a function of \sumtp\ persist in a similar way across different \pz\ bins and across different \pt\ bins in both beams.  There is general agreement between the data and the MINERvA model prediction across both a broad range of phase space and over a wide range of absolute cross sections. 

Discrepancies between data and prediction are visible in these plots at high \sumtp, are most pronounced at low \pt, and can be seen in both data sets.  
Furthermore, the discrepancies are more similar in different \pz\ bins than in different \pt\ bins.  

\begin{figure}[tp]
    \centering
\includegraphics[width=0.95\linewidth]{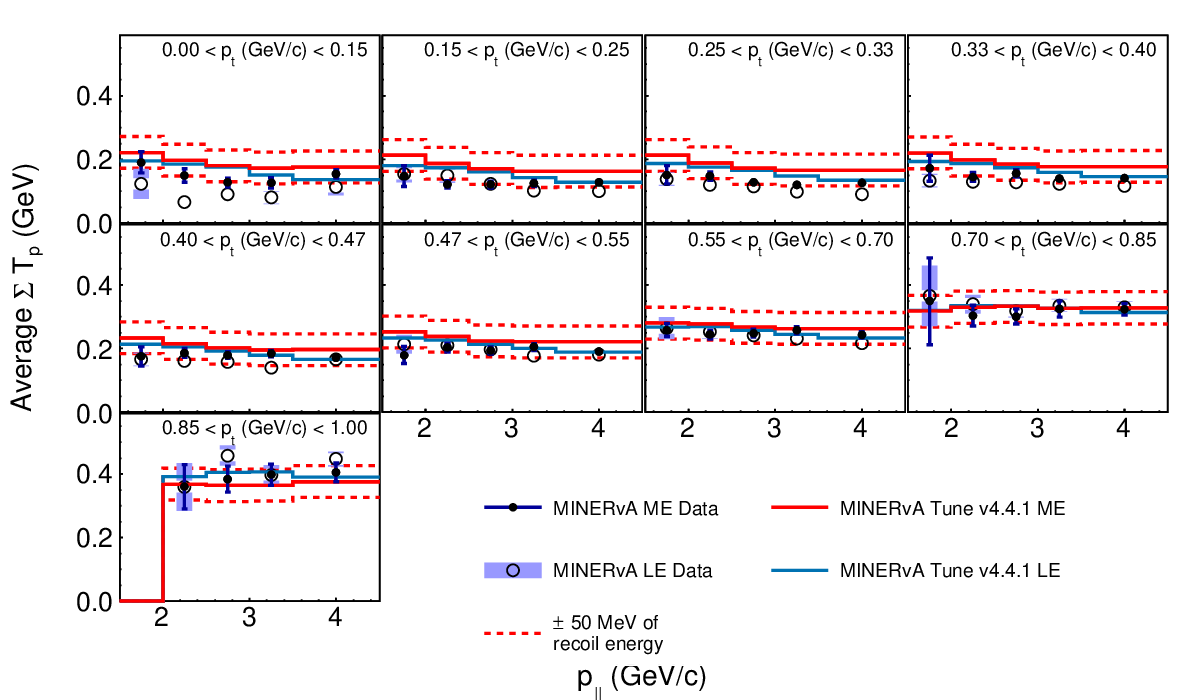}
    \caption{Average recoil for the events after background subtraction as a function of \pz, separated out by different \pt\  regions, for the Low and Medium Energy quasielastic-like data and predictions from the simulation. The Low Energy (Medium) Energy data are in open (filled) circles, and the Low (Medium) Energy prediction is in blue (red) solid lines.  }
    \label{fig:LEMEavgrecoil}
\end{figure}

One way to summarize these results is to plot the average \sumtp\  as a function of \pt\ and \pz ; this distribution is shown in Fig.~\ref{fig:LEMEavgrecoil}. Each panel is a bin of \pt\ and the horizontal axis is \pz. The Low Energy (Medium) Energy data are in open (filled) circles, and the Low (Medium) Energy prediction is in blue (red) solid lines.
The average recoil in the LE (ME) beam is about 75~MeV (50~MeV) lower than the prediction in the lowest \pt\ bins.  As \pt\ increases, however the over-prediction in the simulation diminishes.  For the highest \pt\ bin the average recoil in the LE beam is about 50~MeV higher than the prediction, while for the ME beam the simulation is in agreement with the data.  

The fact that the average \sumtp\ values for a given muon bin in \pt\ and \pz\ is lower in the Low Energy beam than in the Medium Energy beam is because that specific bin in muon kinematics is averaging over a different set of incoming neutrino energies due to the very different fluxes.  The fact that the data and simulation are in disagreement by different amounts in the two beams points to a mis-modeling of the neutrino energy dependence of the sharing between lepton and hadronic energies.  Given that these are quasielastic-like events with a small amount of hadronic activity in the first place, the mismodeling of that activity is substantial.    

\subsection{Comparison in Peak Overlap Region}

To better understand the three-dimensional measurement of the quasielastic-like process, 
it is helpful to consider the cross section in one longitudinal muon momentum bin for both Low and Medium Energy beams, and then compare the measurements from the two beams as a function of \pt\ and \sumtp.  The results for the cross sections and ratios for all bins of \pz\  are provided in the Supplement. 

When evaluating cross-section trends, it is useful to keep in mind that there are three different processes that contribute to quasielastic-like events: 
 true quasielastic interactions on an initial state neutron, 2p2h interactions which occur on a correlated pair of nucleons, and pion production followed by absorption.  The fact that these interactions occur inside a nucleus are responsible for the second two processes.  The nuclear environment also modifies the first and third processes because the initial state nucleon need not be at rest.  Finally, the nuclear environment can modify the particles on their way out of the nucleus, and indeed there are regions dominated by a process where the model predicts that the final state nucleon with the highest momentum is a neutron rather than a proton.  Each of these processes can contribute different amounts in each bin.  

\setlength{\triplet}{1.4\linewidth}
\setlength{\doublet}{1.8\linewidth}
\setlength{\allBins}{0.8\linewidth}
\setlength{\quadruplet}{1.0\linewidth}

\begin{figure*}[tp]
    \centering
    \includegraphics[width=\quadruplet]{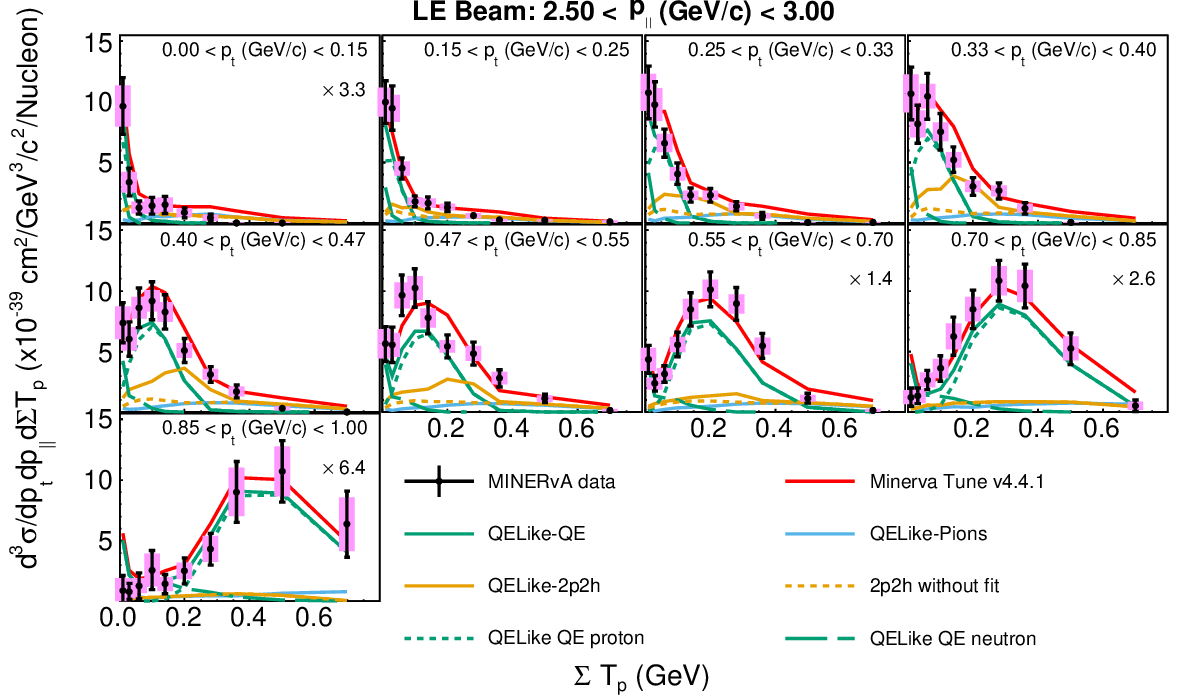}
    \includegraphics[width=\quadruplet]{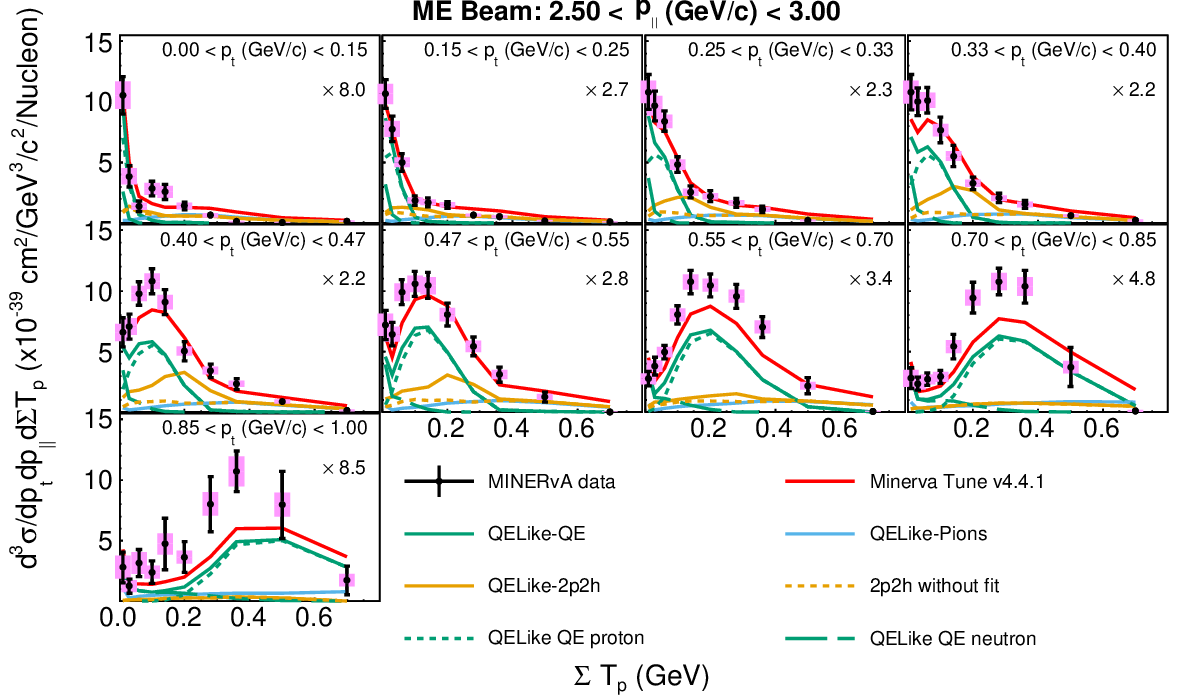}
    \includegraphics[width=\quadruplet]{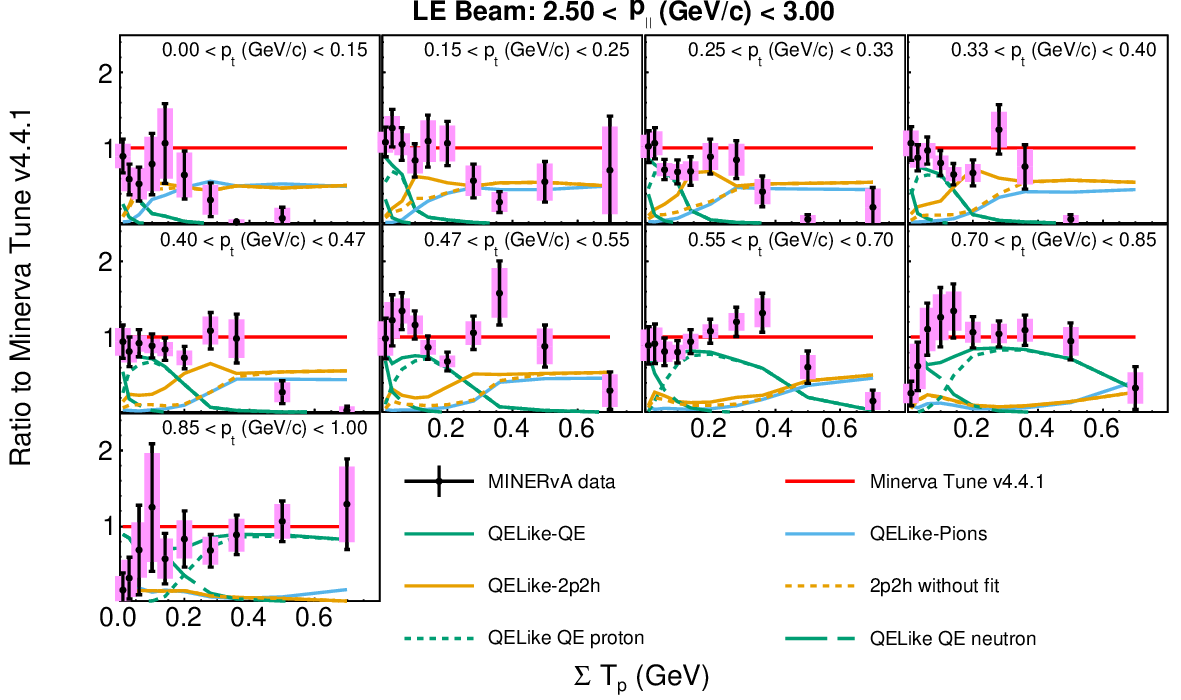}
    \includegraphics[width=\quadruplet]{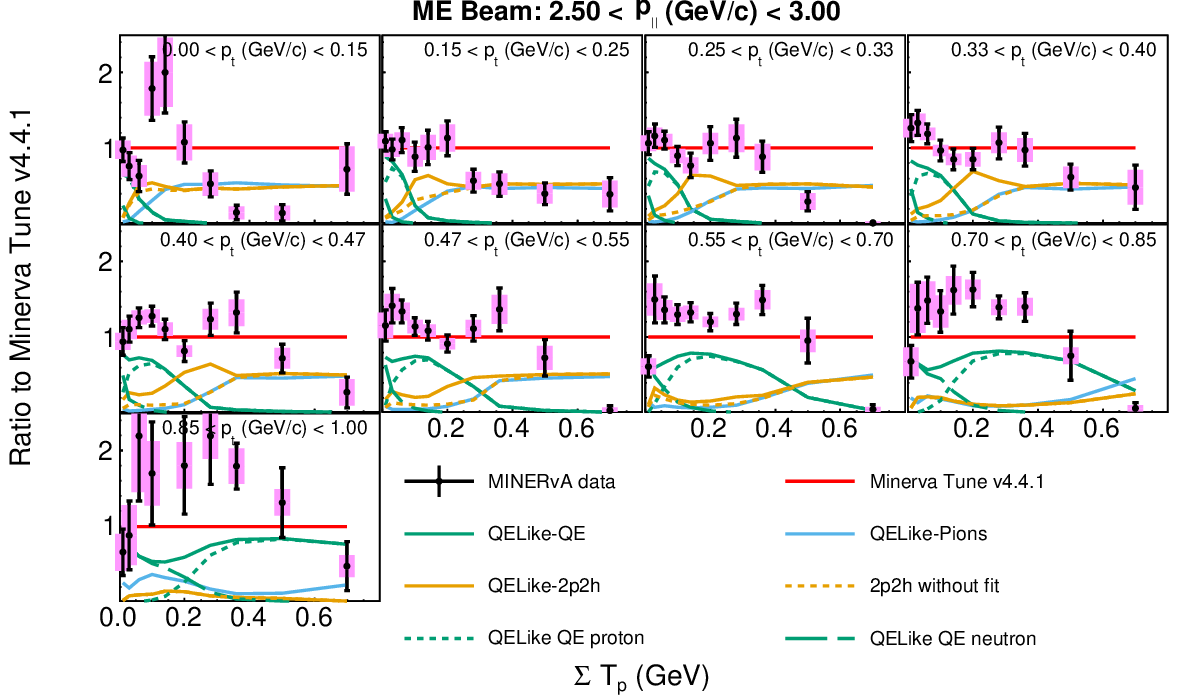}
    \caption{Cross Sections (top) and Ratios between the cross sections and MINERvA's Tune to GENIE (bottom) as a function of \pt\ and \sumtp\ for the bin where \pz\ is between $2.5<\pz\/(GeV/c)<3.0$, for the Low Energy (Left) and Medium Energy (Right) beams.  The shaded boxes (error bars) represent the statistical (total) uncertainty on each point.  
    The QE subsample is labeled by the identity of the highest energy nucleon as \lq\lq QE proton\rq\rq and \lq\lq QE neutron\rq\rq. The \lq\lq 2p2h without fit\rq\rq is the prediction without the 2p2h-enhancement. 
    }
    \label{fig:ptpzsumtp_bin3m}
\end{figure*}

The longitudinal momentum bin with the most precise cross section ratio is the bin from $2.5<\pz (GeV/c) <3.0$.  This bin corresponds to the flux peak of the Low Energy beam, and the rising edge of the flux peak in the Medium Energy beam.  Therefore, the contribution to this visible energy bin will come from higher-energy neutrinos in the Medium Energy than in the Low Energy result.  
The top plots in Fig.~\ref{fig:ptpzsumtp_bin3m} show the cross sections for the Low (left) and Medium (right) beams, along with the predicted contributions from individual quasielastic-like processes.  The bottom plots in Fig.~\ref{fig:ptpzsumtp_bin3m}
show the ratios between the LE (left) and ME (right) data and the prediction, along with the predicted fractions of each of the different quasielastic-like processes.  Because different nuclear effects will affect different regions of the phase space, the discrepancies between the data and the prediction, shown in this way, can provide insight into where the models need modification.  By considering the discrepancies separately from high \sumtp\ to low \sumtp\, different conclusions can be drawn about the underlying model.

Ratio plots in Fig.~\ref{fig:ptpzsumtp_bin3m} show that the largest discrepancies with predictions are at \sumtp\ above 0.4~GeV, which were also seen in Fig.~\ref{fig:MEtriplexsec} across most \pt\ and \pz\ bins.  
At \sumtp\ above 0.3~GeV and low \pt , a region untouched by previous 2p2h modifications, the model over-predicts the cross section in both beams.   In the model, this region is primarily 2p2h and pion production plus absorption events.  
 \collabAdd{The MINERvA experiment previously found evidence that pion production processes are overpredicted at low muon $p_T$, both in the Medium Energy~\cite{MINERvA:2022djk} and Low Energy~\cite{MINERvA:2019kfr} beams, and in the background constraints for this analysis~\cite{MINERvA:2022mnw}.}

 For intermediate \sumtp\ and \pt\, where the modification from the 2p2h fit to the inclusive Low Energy neutrino charged current data~\cite{Rodrigues:2015hik} is the largest, (seen by comparing the \lq\lq 2p2h without fit\rq\rq prediction with the \lq\lq 2p2h\rq\rq  prediction) the model appears to overpredict the data in both beams, especially in the \pt\ region between 0.25~GeV/c and 0.4~GeV/c.  
 The data in the region where this contribution is important lie between where the two predictions would be\collabAdd{, which suggests that the modification is too strong, either at high visible recoil or at low neutrino energies.  It should be noted that the Low Energy Data tune in Ref.~\cite{Rodrigues:2015hik} was based on an inclusive data set whereas the cross sections reported here only concern quasielastic-like events. } 

Careful inspection at \sumtp\ below 0.1~GeV and high \pt\  indicates that there is also an over-prediction 
in both beams for quasielastic events where the highest momentum nucleon is a neutron.  
This discrepancy suggests a need to decrease the effects of proton final state interactions in the simulation to better agree with the data. Such deficiencies in the model for final state interactions could be addressed by tuning to measurements of proton transparency in nuclei~\cite{Ankowski:2014yfa}.

The fact that most discrepancies are similar in both the Medium and Low Energy data points to the fact that the mismodeling is less a function of neutrino energy than a function of momentum transfer.  The absolute cross sections for the low and medium energy data and prediction from MINERvA's base model for all bins of \pz\ can be found in the Supplement. 
The discrepancies in the  
regions listed above persist, within the statistical uncertainties, for nearly all reported bins of \pz.   The one exception is that for the region of \pz\ between 1.5 and 2~GeV, the prediction and data do agree in the bins that are dominated by final state neutrons, although the statistics there are limited.  There is an underprediction of the simulation at intermediate \sumtp\ that is spread over a broad range of \qzeroqe\ or \pt\ in the Medium Energy beam but is less pronounced in the Low Energy beam.
\subsection{Comparisons versus inferred recoil energy}
\begin{figure}[tp]
    \centering
    \includegraphics[width=\quadruplet]{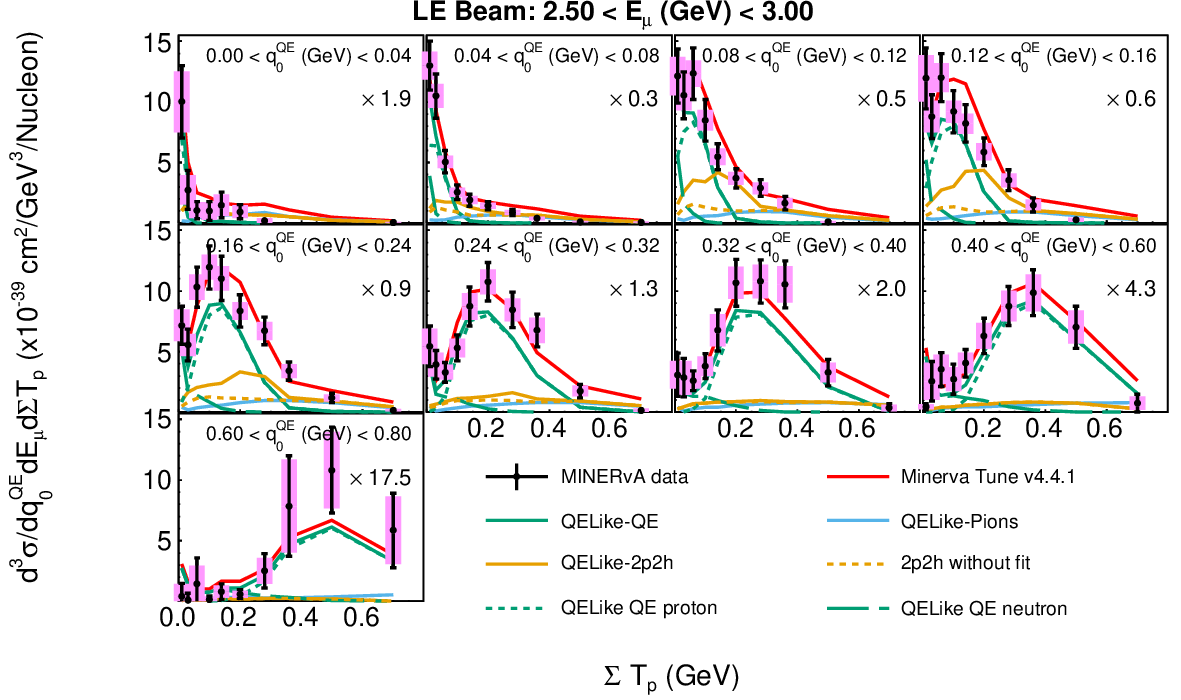}
\includegraphics[width=\quadruplet]{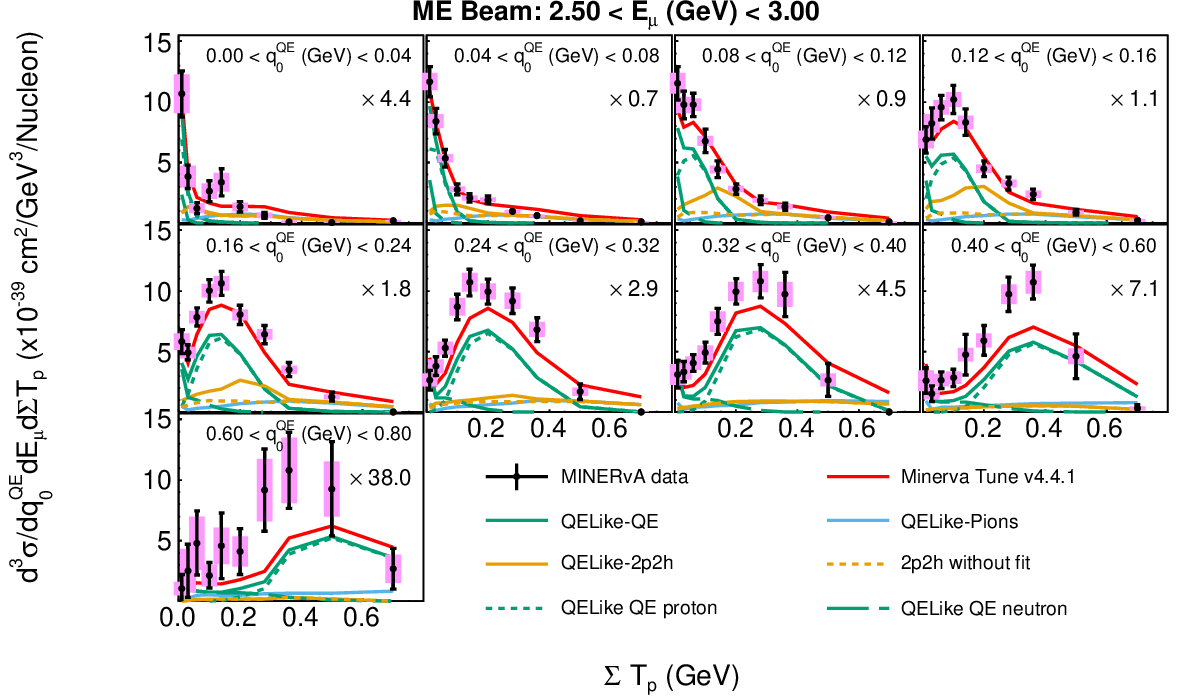}

    \caption{Cross Sections as a function of \qzeroqe\ and \sumtp\ for the bin where $E_\mu$ is between $2.5<E_\mu\/(GeV)<3.0$, for the Low Energy (left) and Medium Energy (right) beams.  The shaded boxes (error bars) represent the statistical (total) uncertainty on each point.  The predicted contributions from different quasielastic-like processes are also shown.  The QE subsample is labeled by the identity of the highest energy nucleon as \lq\lq QE proton\rq\rq and \lq\lq QE neutron\rq\rq . The \lq\lq 2p2h without fit\rq\rq is the prediction without the 2p2h-enhancement.
    }
    \label{fig:q0qesumtp_bin3m}
\end{figure}

Oscillation experiments need to add energy to the measured muon energy to estimate the neutrino energy, as described in the Introduction.  For quasielastic-like events the added energy, in the case of Cerenkov detectors is \qzeroqe, and in the case of calorimetric detectors is \sumtp. The cross sections and ratios reported in the earlier sections can also be expressed in terms of \qzeroqe\ and \sumtp\ for each muon energy bin, thereby providing a more direct comparison of the two different recoil energy (and hence neutrino energy) estimators for oscillation experiments.  The region of maximum overlap is now the $E_\mu$ bin between 2.5 and 3~GeV.  The cross sections in that bin but binned in \qzeroqe\ instead of \pt\ are shown in Fig.~\ref{fig:q0qesumtp_bin3m} for the Low (left) and Medium (right) Energy beams.   As was seen in Fig.~\ref{fig:ptpzsumtp_bin3m}, there is an underprediction of the simulation at intermediate \sumtp\ that is spread over a broad range of \qzeroqe\ or \pt\ in the Medium Energy beam and is less pronounced in the Low Energy beam. 

When the cross sections are binned in these variables, the 2p2h contribution is more isolated in a few \qzeroqe\ bins compared to the cross sections reported as a function of \pt.  Conversely, in this binning the contribution from quasielastic interactions with FSI producing high energy neutrons is less isolated.  The same conclusions about mismodeling can be drawn from these plots, although the other representation allows a better separation of the quasielastic events that undergo FSI to produce neutrons.  The absolute cross sections binned in \qzeroqe\ and \sumtp\ for each \pz\ bin for the Low and Medium Energy beams separately are presented in the Supplement.

Since the effects are similar in the two different representations, the remainder of this paper will focus on the measurements versus \pt\ rather than \qzeroqe.  

\section{Model Comparisons of Absolute Cross Sections}
Given the size of the discrepancies with MINERvA's model, and the fact that the previous modifications to GENIE by MINERvA do not affect the regions where the current discrepancies are largest, it is important to examine how other current model predictions compare with these new results.  

\begin{figure}[tp]
    \centering    
    \includegraphics[width=\quadruplet]{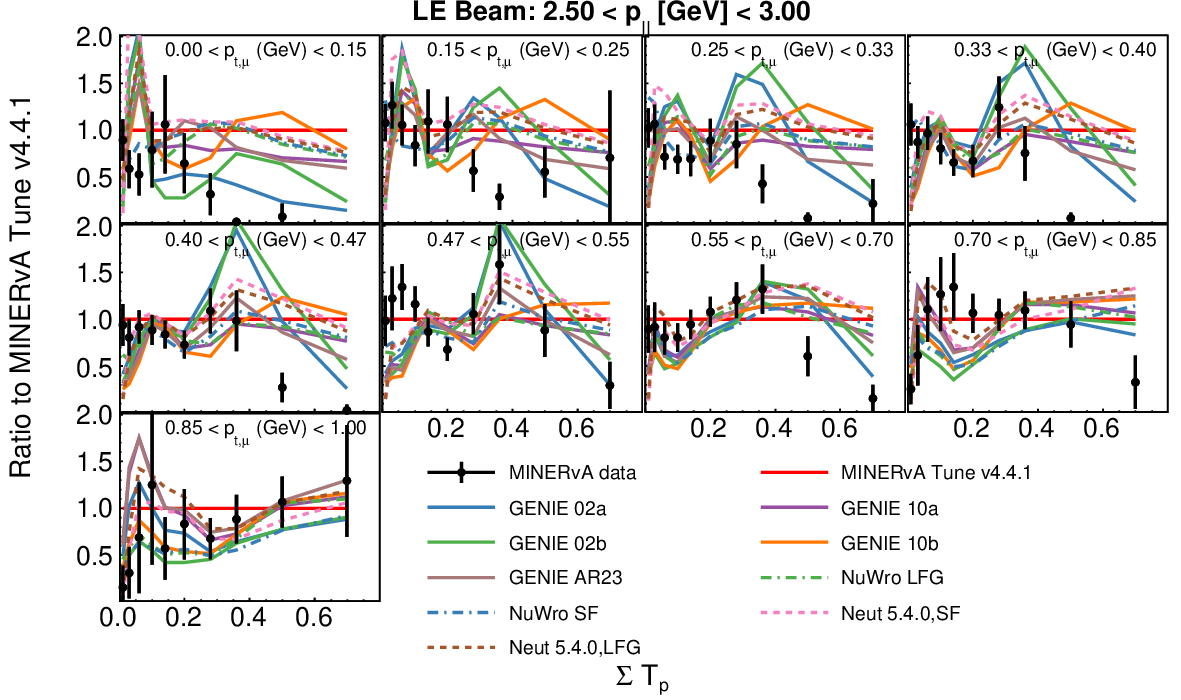}
    \includegraphics[width=\quadruplet]{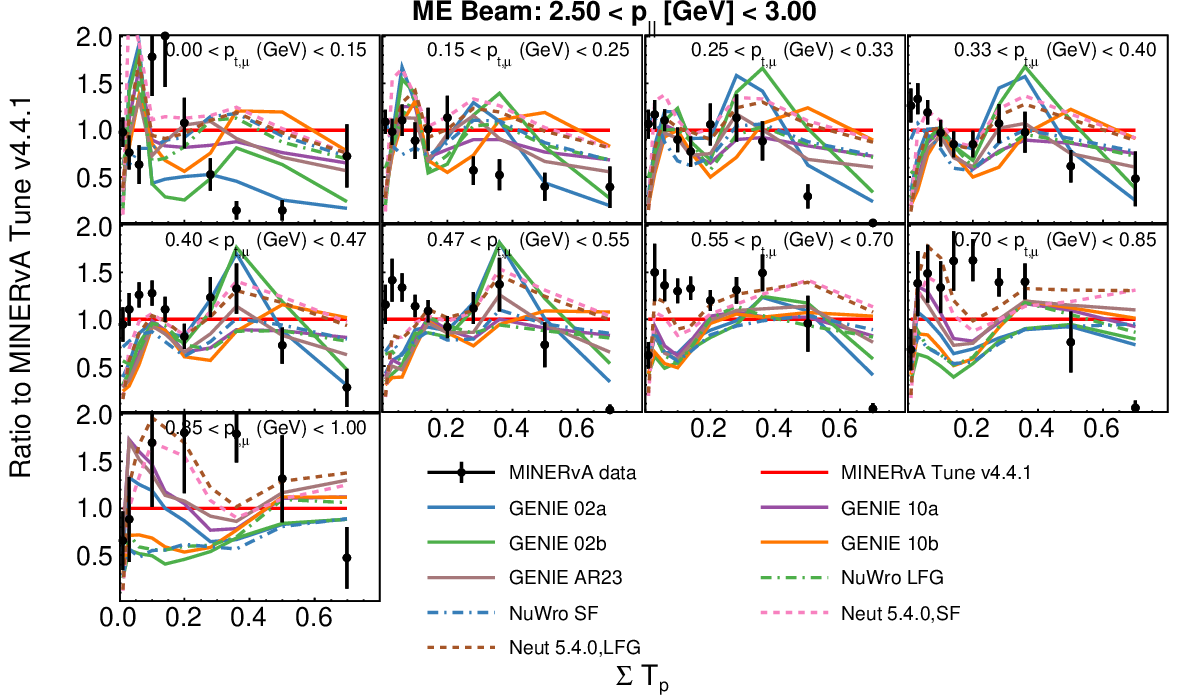}
    \includegraphics[width=\quadruplet]{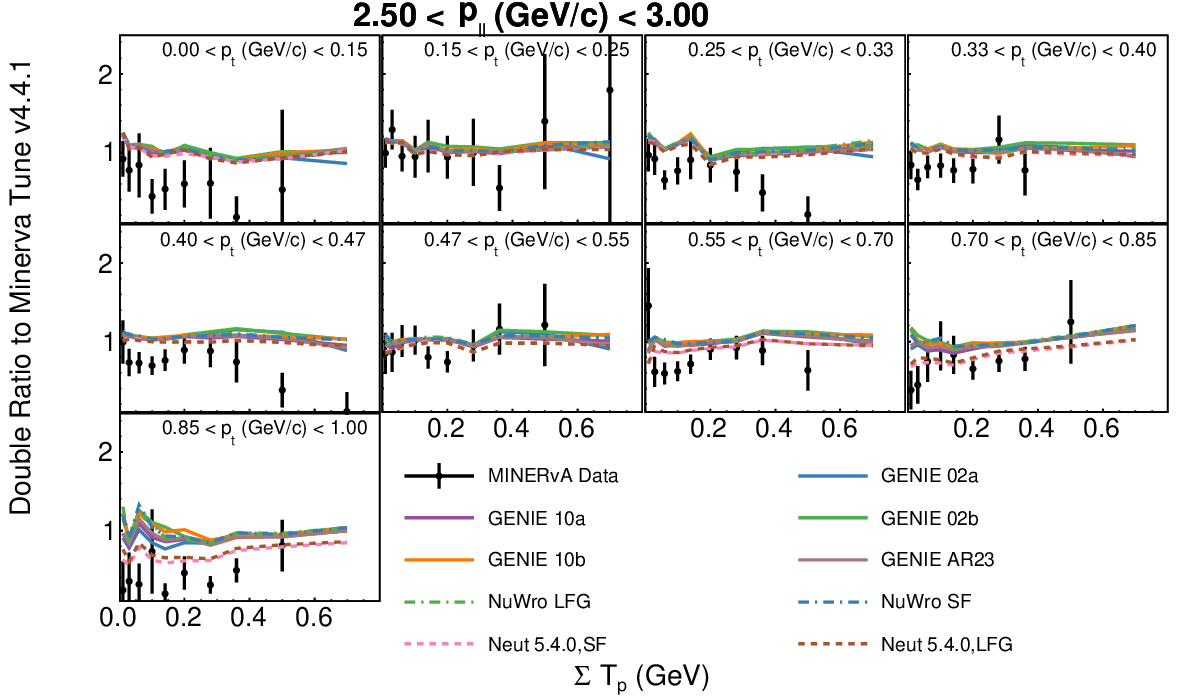}
    \caption{Ratios of both MINERvA data and alternate generator predictions to MINERvA's Tune to GENIE as a function of \pt\ and \sumtp\ for the bin where \pz\ is between 2.5$<$\pz\/(GeV/c)$<$3.0, for the Low Energy (top) and Medium Energy (middle) beams, and for the \lq\lq Double Ratio\rq\rq, (Low Energy to Medium Energy)(bottom).   \collabAdd{For \pt $>$ 0.4GeV/c the double ratios that do not appear on the plots are above two but consistent with unity while the missing double ratios for \pt$<$0.4GeV/c are consistent with zero but again with large uncertainties.}  Final state interaction model choices in GENIE are either \lq\lq a\rq\rq (the effective intranuclear cascade, or \lq\lq hA\rq\rq) or \lq\lq b\rq\rq (the full intranuclear cascade, or \lq\lq hN\rq\rq), and the \lq\lq GENIE 02\rq\rq indicates a Relativistic Fermi Gas while the \lq\lq GENIE 10\rq\rq indicates a Local Fermi Gas model.   
     }
\label{fig:ptpzsumtp_bin3_otherModelsm}
\end{figure}

Figure~\ref{fig:ptpzsumtp_bin3_otherModelsm} repeats the data to prediction ratio plots shown in the bottom panels of Fig.~\ref{fig:ptpzsumtp_bin3m}, but adds curves for the ratio between other generator choices and MINERvA's base model, for both the Low and Medium Energy beams.  The GENIE event generator, used by NOvA, MINOS, DUNE, and the SBN neutrino oscillation experiments, has a series of models to choose from for both the initial nucleon momentum distributions and for final state interactions.  
The NEUT event generator~\cite{Hayato:2009zz}, used by T2K and HyperK, can also switch between different models for the initial state nucleon kinematics.  NuWro is another neutrino event generator which employs a different nuclear shell model and correlated nucleon scattering~\cite{Juszczak:2005zs}.  
Two initial state nucleon models that are in common use by all three generators are the spectral function (SF) model and the Local Fermi Gas (LFG) model.  For the sake of simplicity, only two FSI models in GENIE are compared, and two different initial nucleon distributions are compared for each generator.  The label of \lq\lq AR23\rq\rq\  indicates GENIE with the settings currently used by DUNE.

It can be seen that none of the models predict the strength of the cross section at high \sumtp\ and low \pt\ in either beam.  Moreover, none of the other models describe the low \sumtp\ and intermediate \pt.  
By comparing different model choices in the two beams one can understand what regions of the quasielastic-like cross section are affected by which choices, and how much that choice is a function of neutrino energy.  For example, the region at low \pt\ and high \sumtp\ is affected most in GENIE by switching between two final state interaction treatments.  

Figure~\ref{fig:ptpzsumtp_bin3_otherModelsm} also compares these results with predictions for the ME and LE cross sections for different choices of the quasielastic model and of final state interaction models available in GENIE-3.   In general, changes to the FSI Model (GENIE02a and GENIE10a compared to GENIE 02b and GENIE10b) have much bigger impact in the ME beam compared to the LE beam, and bigger impact at high \pt\ and high \sumtp.  
Changes to the quasielastic interaction and 2p2h have larger effects at low \pt\ and low \sumtp.  None of these models agree with the shape of the cross section at lowest \pt\ although some do better at intermediate \pt\  and \sumtp\ than does the base GENIE-2 model.  
None of the different choices of quasielastic models agree well with the MINERvA data in the regions where the QE contributions are the largest.  Finally, the GENIE3 models all differ from the MINERvA tune in the lowest \sumtp\ bins at most \pt, yet none of them are in better agreement with the data in those regions.  The predictions for other \pz\ bins are presented in the Supplement.  

\subsection{LE/ME Cross Section Ratios}
\begin{figure*}[tp]
    \centering    
    \includegraphics[width=\quadruplet]{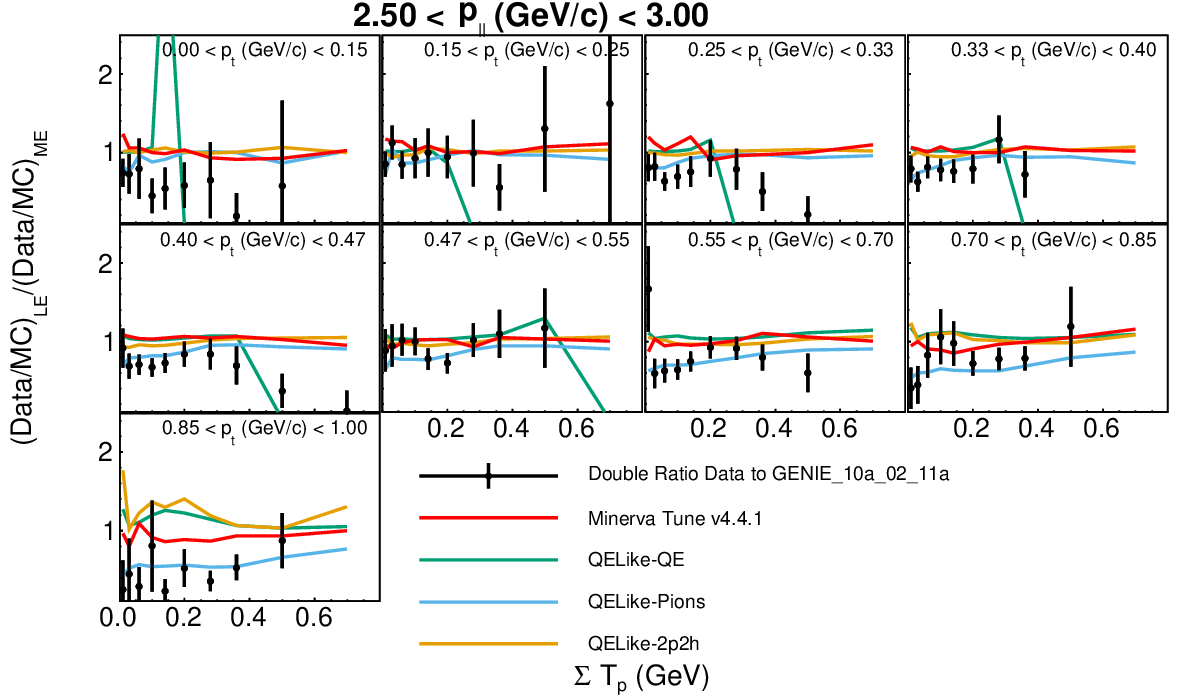}
   \includegraphics[width=\quadruplet]{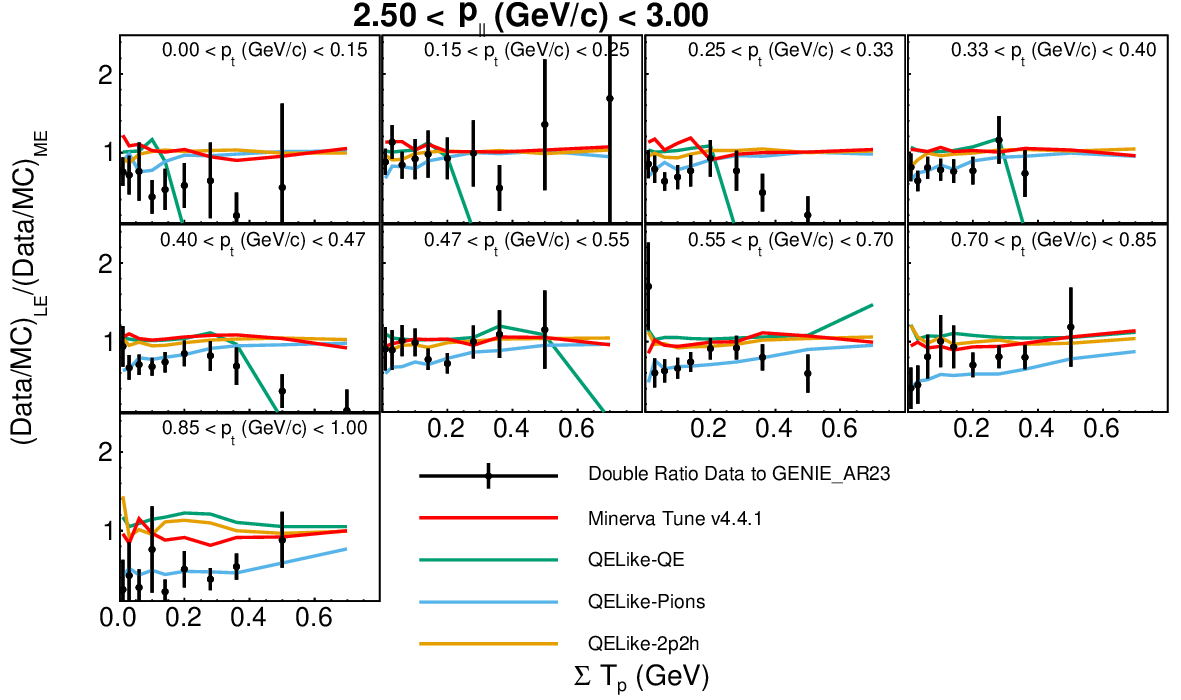}
   \includegraphics[width=\quadruplet]{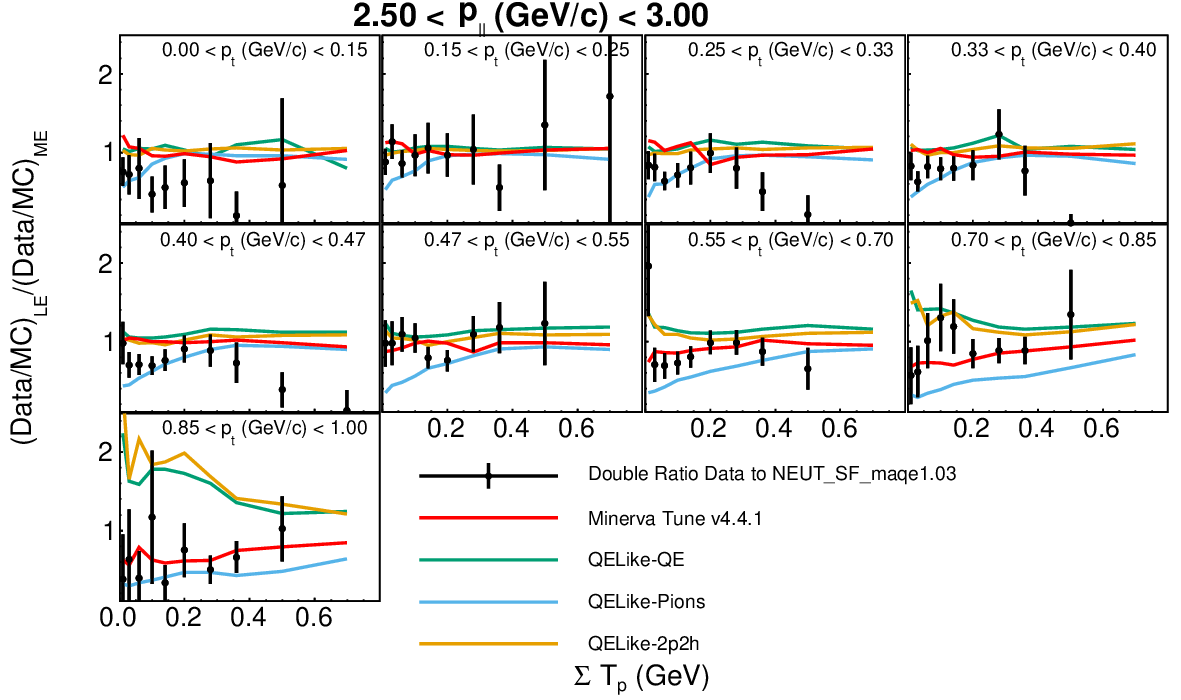}
    \includegraphics[width=\quadruplet]{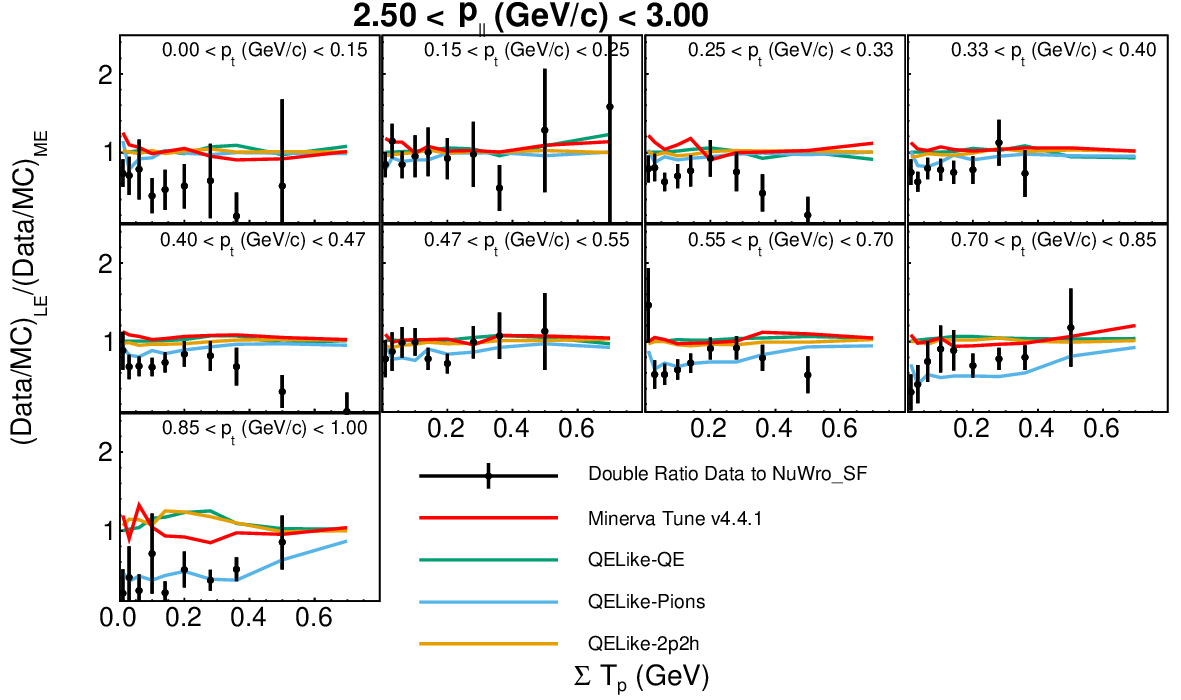}
    \caption{Ratio of LE/ME cross section ratio to the simulation's ratio as a function of \pz,\pt, where the simulation is either GENIE with LFG and hA FSI (top left), GENIE with settings for DUNE (AR23) (top right), NEUT SF (bottom left) or NuWro SF (bottom right).  The ratio for individual quasielastic-like processes in the four different predictions is also shown.  Green is for Quasielastic events, Mustard is for 2p2h events, and blue is for pion production plus absorption events. \collabAdd{The ratios for GENIE at high \sumtp\ and low \pt\  have large fluctuations for the QELike-QE contribution because GENIE predicts very few true quasielastic events in that region.} 
}
\label{fig:ptpzsumtp_bin3_doubleRatio_mmodelsm}
\end{figure*}

The advantage of doing a simultaneous measurement of these cross sections with two different fluxes but using the same detector is that ratios of the cross sections provide some cancellation of the systematic uncertainties, thereby further constraining the underlying models.  
Figure~\ref{fig:ptpzsumtp_bin3_doubleRatio_mmodelsm} shows the Low to Medium energy cross section ratios in data divided by the predicted ratio for different models for one \pz\ bin.  The models shown include GENIE, NEUT, and NuWro event generators.  This figure also shows what that ratio should be for the different individual contributions to quasielastic-like events.  For all generators, the pion production-plus-absorption channel drives this ratio to below unity, while for all but the NEUT generator at the highest \pt, the cross section ratio for the other processes are quite close to unity. 
The different generators do vary in how much pion production-plus-absorption they predict.  Although the absolute cross sections of these different generators, shown in Fig.~\ref{fig:ptpzsumtp_bin3_otherModelsm} vary at the $\pm50\%$ level, they still all predict that the ratio of LE to ME cross sections are close to unity to within a few per cent.  

For both the quasielastic process and the 2p2h process, the ratio is close to one because in both cases the final \lq\lq visible\rq\rq energy is predicted to be close to the incoming neutrino energy, and so the fact that the peak energy in the two fluxes differ by a factor of two is not predicted to have an impact.  For pion production followed by absorption, the ratio is not unity, and in particular since pion production is more important at higher neutrino energies for the same visible energy, for a ratio of LE/ME beams, that ratio would be below unity.  

The fact that the double ratio in the data is also below unity suggests that the energy dependence of pion absorption is not accurately modeled across a broad range of \pt\ and \sumtp\, and in particular that more quasielastic-like events that involve pion absorption are present in the ME data (or less in the LE data) than in the prediction.
The ratios in Fig.~\ref{fig:ptpzsumtp_bin3_doubleRatio_mmodelsm}  show that the effects of pion production followed by absorption is quite different for GENIE compared to NEUT, and in particular, pion absorption in NEUT modifies this LE to ME ratio more than in GENIE.  
Though increasing the pion production-plus-absorption strength would improve the agreement between the LE$/$ME ratio prediction and the measurement, this channel is predicted to be only a small fraction of the total cross section at high \pt\, so in that region a large increase in strength combined with a decrease of the other QE-like processes would be needed for the prediction to match the data.  Another implication of these results is that the relationship between incoming and visible energy is not well-modeled even for the other QE-like processes.
\begin{figure*}[tp]
    \centering    
    \includegraphics[width=\quadruplet]
{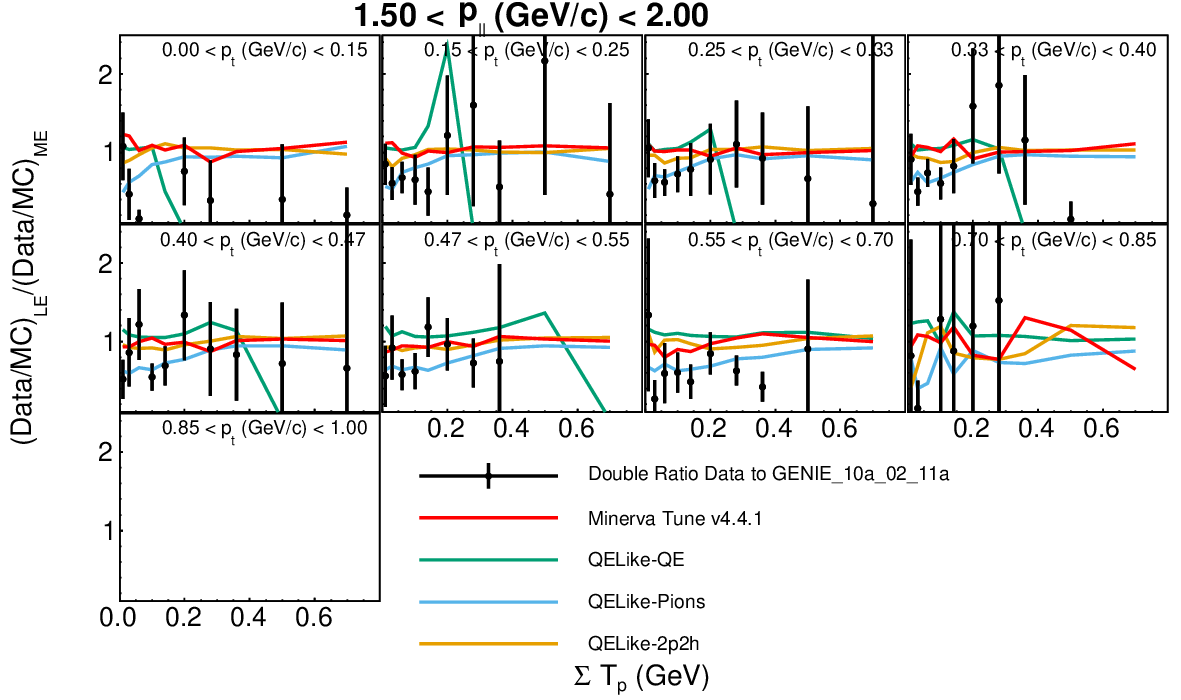}
\includegraphics[width=\quadruplet]
{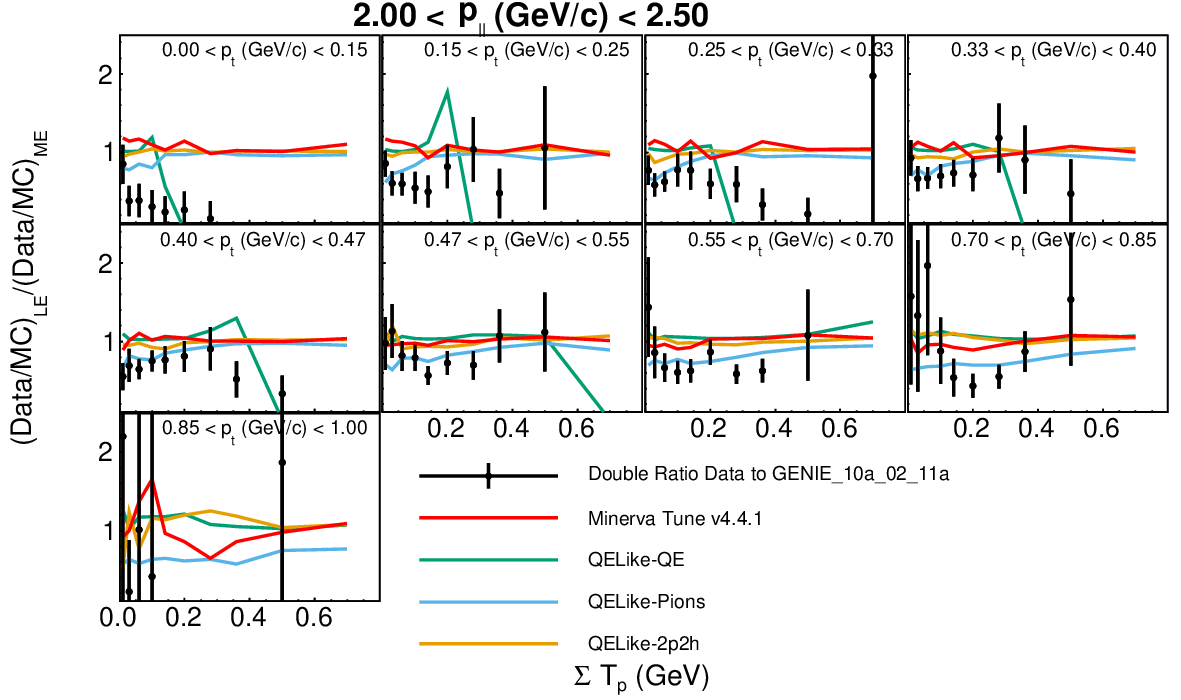}
\includegraphics[width=\quadruplet]{NewPlots/DoubleRatio_ExternalModels_Data/GENIE_10a/LE_ME_RATIO_3.eps}
\includegraphics[width=\quadruplet]
{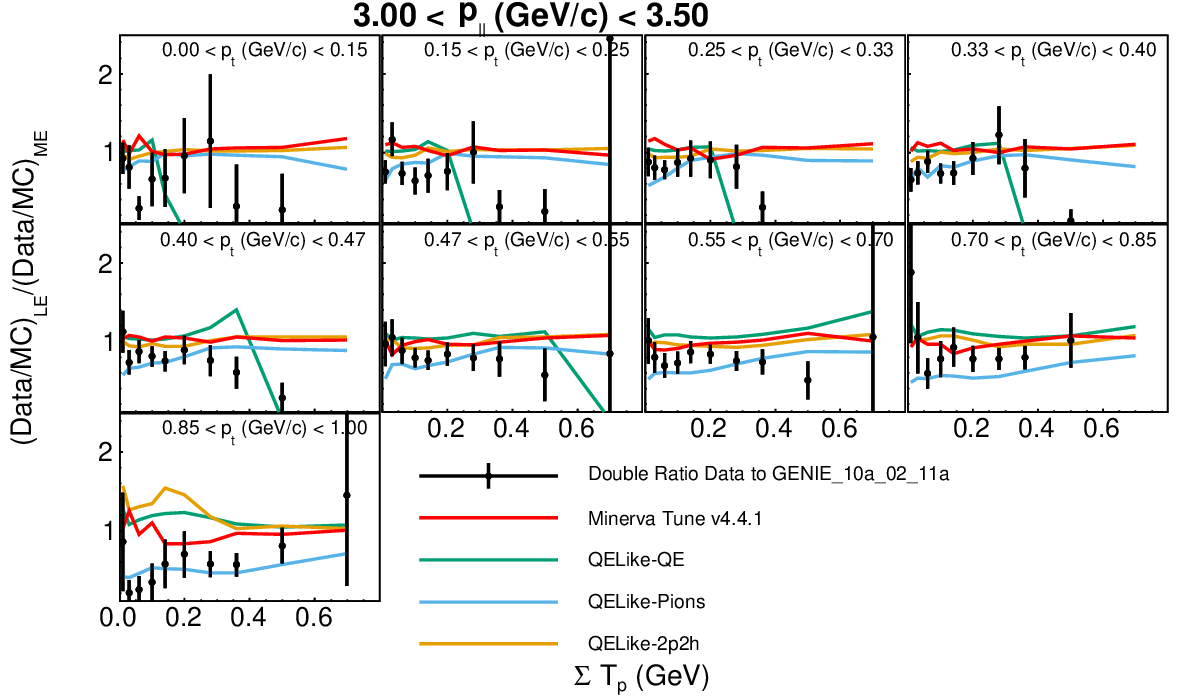}
\includegraphics[width=\quadruplet]
{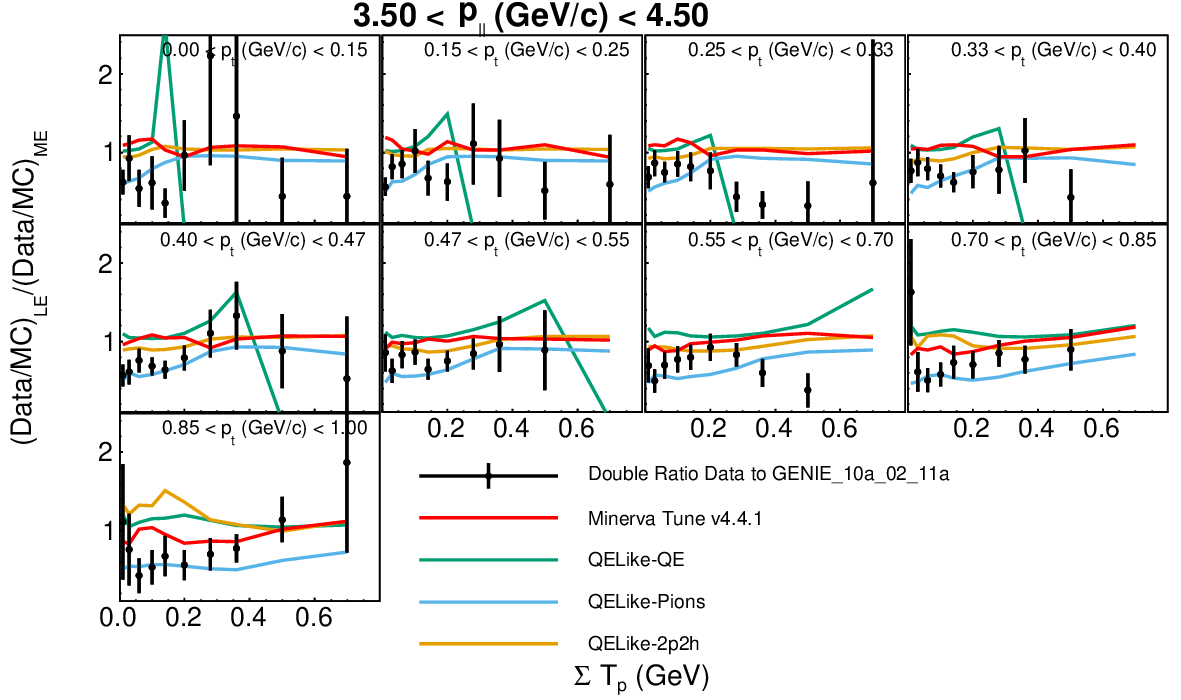}
\caption{Ratio of LE/ME cross section ratio in the data to GENIE3's ratio as a function of \pz,\pt, for several different values of \pz.  The ratio for indiviual quasielastic-like processes is also shown.  Green is for Quasielastic events, Mustard is for 2p2h events, and blue is for pion production plus absorption events.\collabAdd{The ratios at high \sumtp\ and low \pt\  have large fluctuations for the QELike-QE contribution because GENIE Predicts very few true quasielastic events in that region.}}
\label{fig:plong_dependence}
\end{figure*}

Figure~\ref{fig:plong_dependence} shows the LE/ME double ratio for the data and the MINERvA tune prediction along with the different components, where each set of plots is the cross section double ratio for a different \pz\ bin.  The  LE/ME double ratio is consistently below unity for most bins, indicating that for most visible neutrino energies, there is more pion absorption in the ME data compared to the LE data than is currently predicted by the simulation.  

Given the LE/ME ratio reported in the lowest \pz\ bin, it is clear that this region is dominated by events where a pion was produced but then was absorbed in the nucleus.  This is well below the peak neutrino energy of both the Low and Medium Energy beams, yet this is precisely the region where oscillation experiments may be looking for a second oscillation maximum.  Understanding the feed-down from higher energy neutrino events is therefore important. The ratio of LE/ME cross sections does approach unity as \pz\ gets closer to the peak neutrino energy beams, but does stay mostly below unity.  

The double ratio exhibits the largest disagreement at both low \pt\ and high \sumtp\  for all \pz, and also at the highest \pt\ values for all \pz, which is also where the effects of pion absorption are expected to be the largest.  Although the total quasielastic-like cross section in these regions is small, these plots indicate that the pion production contribution may be overestimated by almost 100\% in the ME beam compared to the LE beam.  

\section{Conclusions}

The quasielastic-like cross section measured in the three kinematic variables, \pt, \sumtp, and \pz\ has been measured in MINERvA's Low Energy data, thereby complementing the experiment's previous measurement using Medium Energy data.  
\deborahAdd{As was seen at higher neutrino energies}, there is a large over-prediction of cross section strength in the \sumtp\ region above 200-300~MeV, which is most significant \deborahAdd{in the Low Energy Beam at \pt\ below 400~MeV$/$c and in the Medium Energy Beam at \pt\ below 330~MeV$/$c}, and which is absent above \pt\ of 700~MeV$/$c.  In addition, there is also an over-prediction in both beams of the FSI strength that causes protons to knock out neutrons in the quasielastic interaction, in particular above \pt\ of 700~MeV.  Similar discrepencies are also observed when the data is expressed in the related kinematic variables that are used by oscillation experiments to estimate neutrino energy, namely \qzeroqe, \sumtp, and $E_\mu$.  

To facilitate the separation of processes that lead to the discrepancies seen in the absolute cross sections, this paper has reported the cross section ratio between two beams of peak energies that vary by a factor of two.  The discrepancies between the double ratio of the MINERvA data and several modern neutrino interaction models are largely a function of \sumtp\ and \pt, and are approximately consistent across different bins of \pz.  This suggests a problem with the modeling of the momentum transfer to the nucleus rather than a modeling of the neutrino energy dependence of that momentum transfer.   
The discrepancies between both data sets and the predictions are larger in the regions dominated by pion production plus absorption than in the region where 2p2h is a significant contributor.

The ability to model the visible energy distribution of neutrino interactions across a broad range of incoming neutrino energy spectra is of utmost importance to neutrino oscillation experiments.  The high statistics MINERvA data in Medium and Low Energy beams on the same detector 
provide a benchmark for current and future neutrino event generators.  
  
\begin{acknowledgments}

This document was prepared by members of the MINERvA collaboration using the resources of the Fermi National Accelerator Laboratory (Fermilab), a U.S. Department of Energy, Office of Science, Office of High Energy Physics HEP User Facility. Fermilab is managed by Fermi Forward Discovery Group, LLC, acting under Contract No. 89243024CSC000002.  
These resources included support for the MINERvA construction project, and support
for construction also
was granted by the United States National Science Foundation under
Award No. PHY-0619727 and by the University of Rochester. Support for
participating scientists was provided by NSF and DOE (USA); by CAPES
and CNPq (Brazil); by CoNaCyT (Mexico); by ANID PIA / APOYO AFB180002, CONICYT PIA ACT1413, and Fondecyt 3170845 and 11130133 (Chile); 
by CONCYTEC (Consejo Nacional de Ciencia, Tecnolog\'ia e Innovaci\'on Tecnol\'ogica), DGI-PUCP (Direcci\'on de Gesti\'on de la Investigaci\'on  - Pontificia Universidad Cat\'olica del Peru), and VRI-UNI (Vice-Rectorate for Research of National University of Engineering) (Peru); NCN Opus Grant No. 2016/21/B/ST2/01092 (Poland); by Science and Technology Facilities Council (UK); by EU Horizon 2020 Marie Skłodowska-Curie Action; by a Cottrell Postdoctoral Fellowship from the Research Corporation for Scientific Advancement; by an Imperial College London President's PhD Scholarship.  We thank the MINOS Collaboration for use of its near detector data. Finally, we thank the staff of
Fermilab for support of the beam line, the detector, and computing infrastructure.
\end{acknowledgments}
\bibliography{main}{}

\clearpage
\onecolumngrid
\pagebreak
\section{Supplemental Material}

The article describes the measurement and provides cross sections, cross section ratios, and model comparisons for one bin of longitudinal muon momentum or muon energy, and the supplement provides these same plots but for all muon momentum bins.  \lq\lq Double Ratio\rq\rq means the LE/ME cross section in the data divided by the LE/ME cross section in the simulation.  

The supplement is organized as follows:  
\begin{itemize}
\item Cross Sections for low and medium energy beams along with the cross section ratios, where individual qe-like components are shown, as a function of \pt,\sumtp, for different \pz\ bins. 
\item Cross Sections for low and medium energy beams along with the cross section ratios, where individual qe-like components are shown, as a function of \qzeroqe,\sumtp, for different $E_\mu$ bins. 
\item Cross Section Double Ratios using GENIE and using NEUT as a function of \pt,\sumtp, for different \pz\ bins.
\item Model Comparisons for the individual LE and ME cross sections along with the cross section ratios.
\item Cross section ratios to MINERvA's Tune, showing the effect on each ratio for different modifications to GENIE
\item Cross Section LE/ME Ratios for different models and each of their components
\item Event Distributions for data and prediction after background fits as function of \pt,\sumtp, for all of the \pz\ bins.  
\item Ratio of Event distributions to the MINERvA Tune after background fits, for the data, the simulation, showing the resulting background fraction and predicted signal process fractions \pt,\sumtp, for different \pz\ bins.
\item Double cross section ratios for different models individually, along with the prediction of the double ratio for each of the signal processes separately.  Shown for a few models, but plots are available for all of the following models: 
\end{itemize}

\clearpage
\FloatBarrier

\setcounter{figure}{109}

\begin{figure*}[tp]
    \centering
    \includegraphics[width=\triplet]{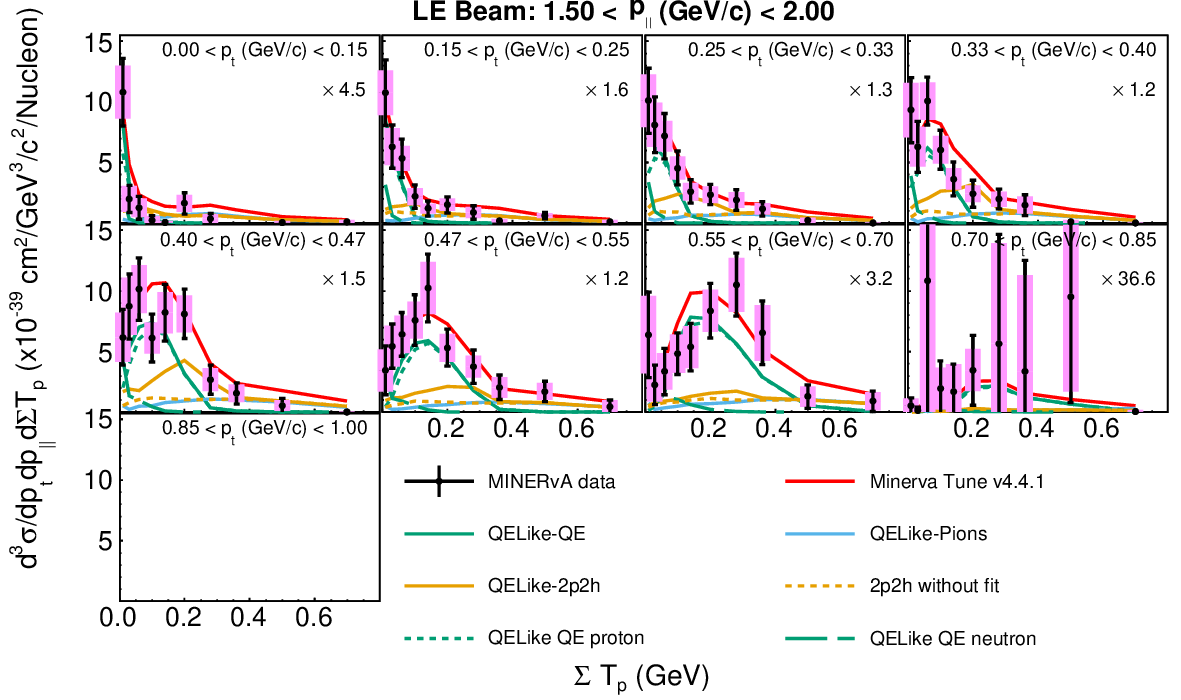}
    \includegraphics[width=\triplet]{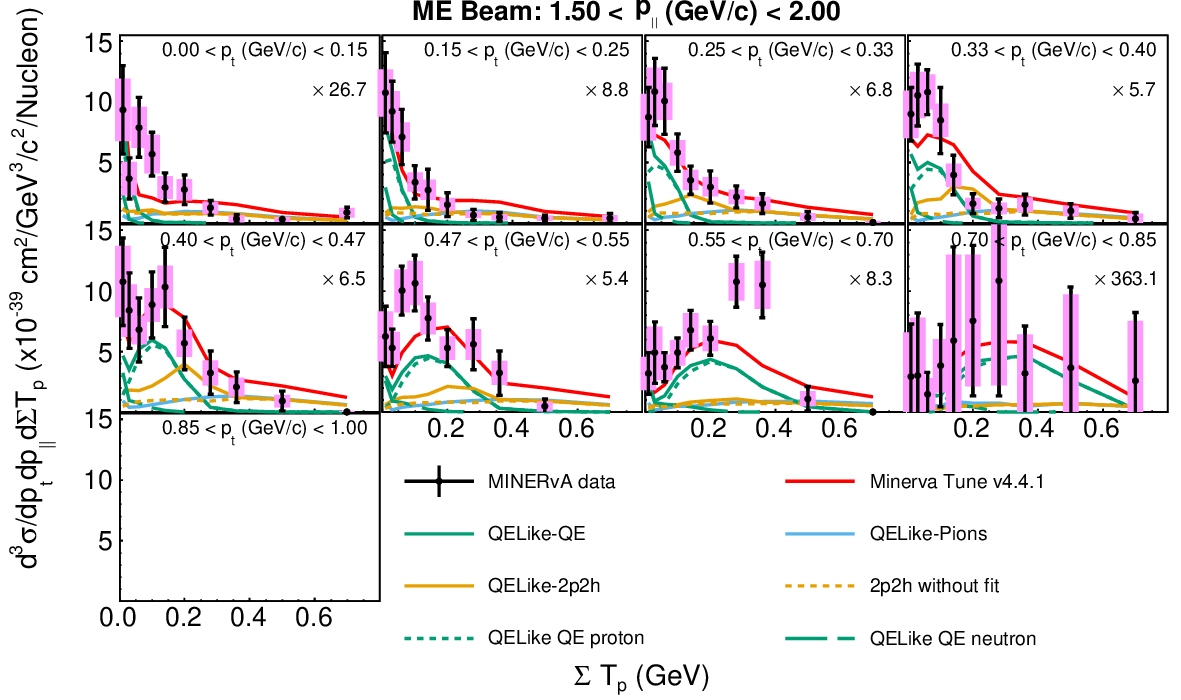}
    \includegraphics[width=\triplet]{NewPlots/DoubleRatio_ExternalModels_Data/GENIE_10a//LE_ME_Ratio_1.eps}
    \caption{Cross Sections as a function of \pt\ and \sumtp\ for the bin where \pz\ is between $1.5<p_z/(GeV/c)<2.0$, for the Low Energy (top) and Medium Energy (middle) beams.  The predicted contributions from different quasielastic-like processes are also shown. The Low Energy/Medium Energy Cross Section ratio for the data, the total prediction, and predictions for individual channels are shown in the bottom plot.}     \label{fig:ptpzsumtp_bin1}
\end{figure*}
\begin{figure*}[tp]
    \centering
    \includegraphics[width=\triplet]{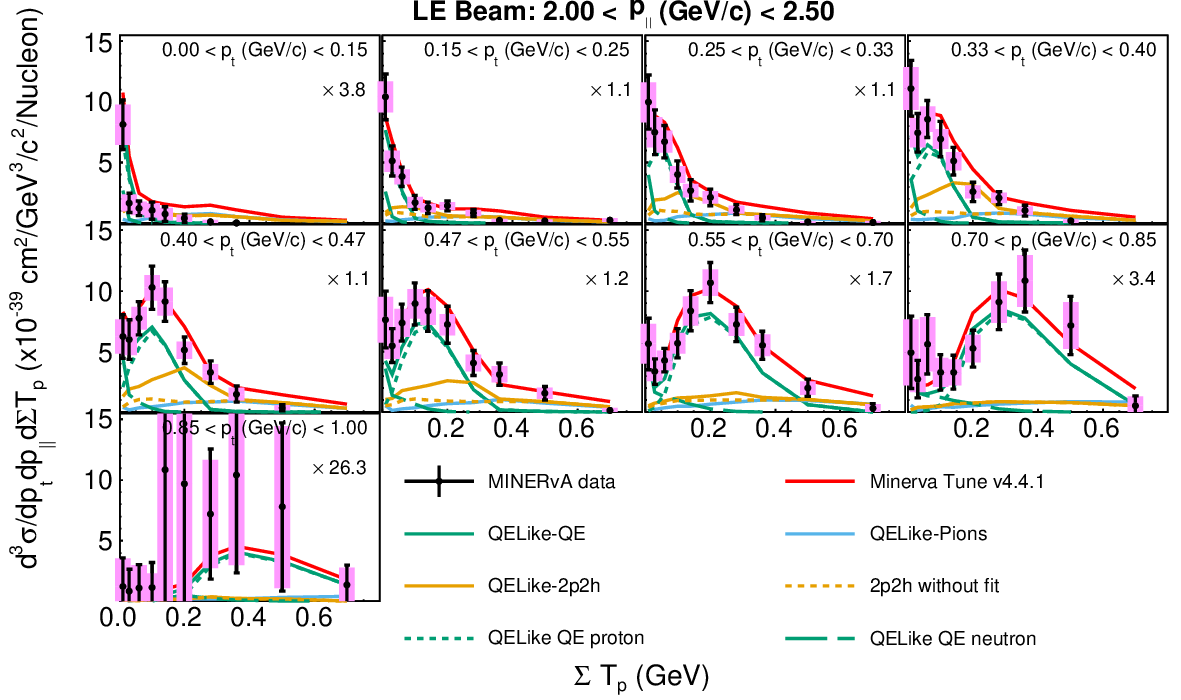}
    \includegraphics[width=\triplet]{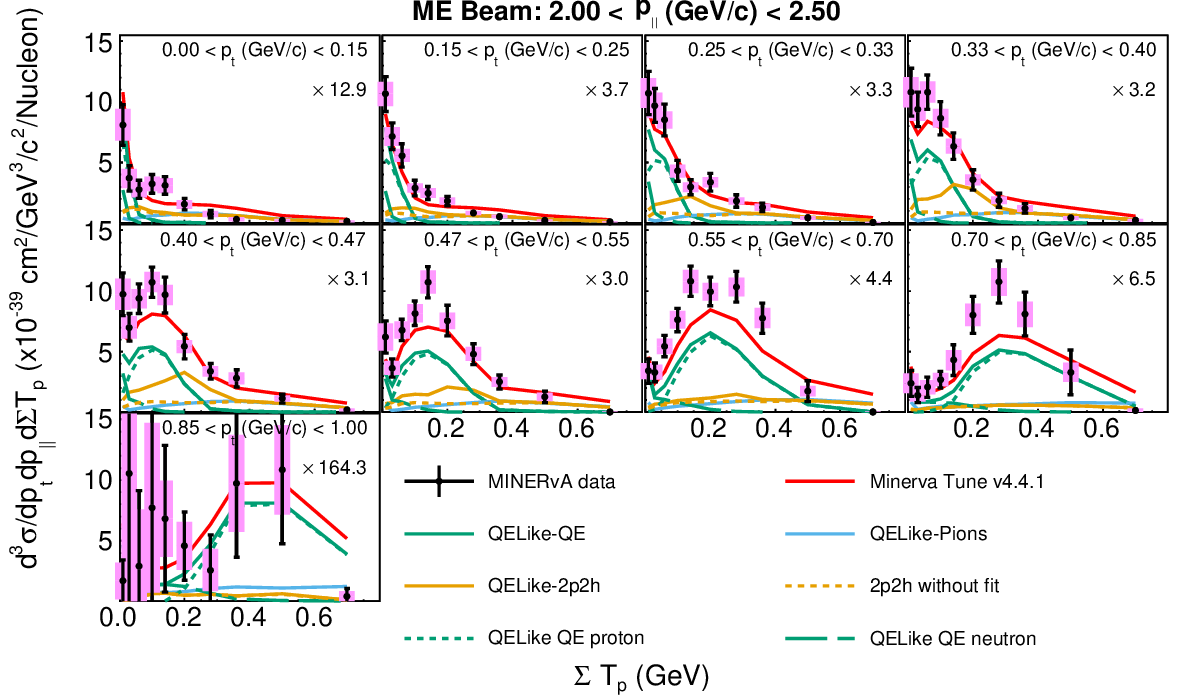}
    \includegraphics[width=\triplet]{NewPlots/DoubleRatio_ExternalModels_Data/GENIE_10a//LE_ME_Ratio_2.eps}
    \caption{Cross Sections as a function of \pt\ and \sumtp\ for the bin where \pz\ is between $2.0<p_z/(GeV/c)<2.5$, for the Low Energy (top) and Medium Energy (middle) beams.  The predicted contributions from different quasielastic-like processes are also shown. The Low Energy/Medium Energy Cross Section ratio for the data, the total prediction, and predictions for individual channels are shown in the bottom plot.}
    \label{fig:ptpzsumtp_bin2}
\end{figure*}
\begin{figure*}[tp]
    \centering
    \includegraphics[width=\triplet]{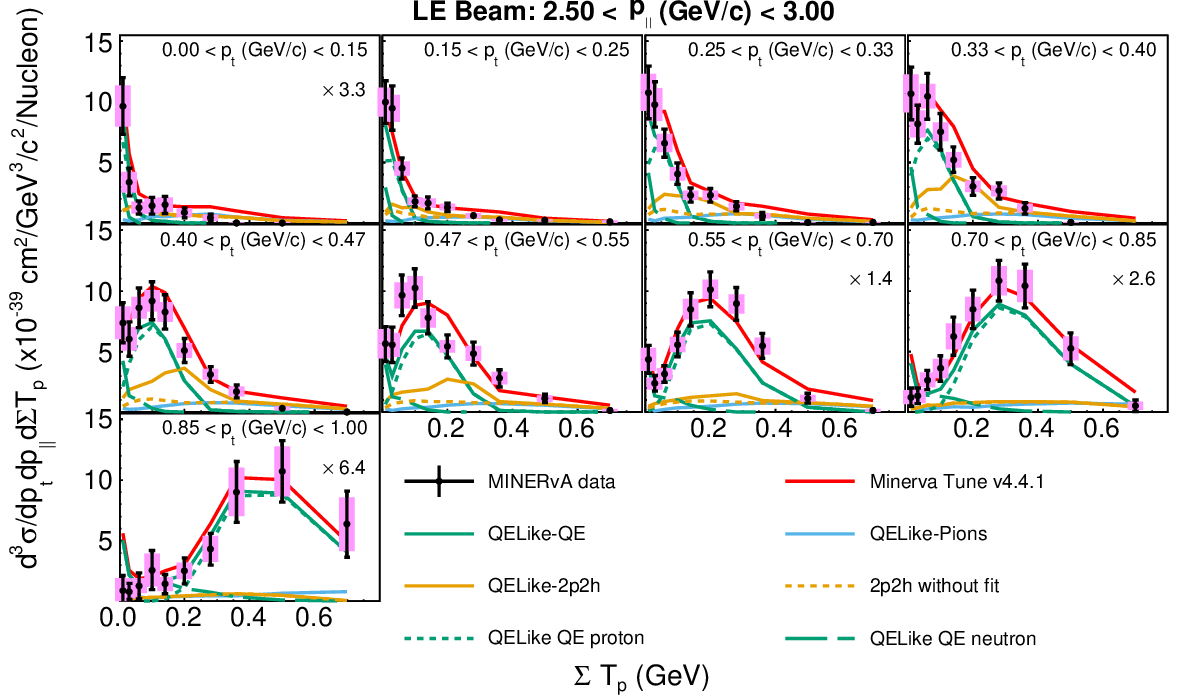}
    \includegraphics[width=\triplet]{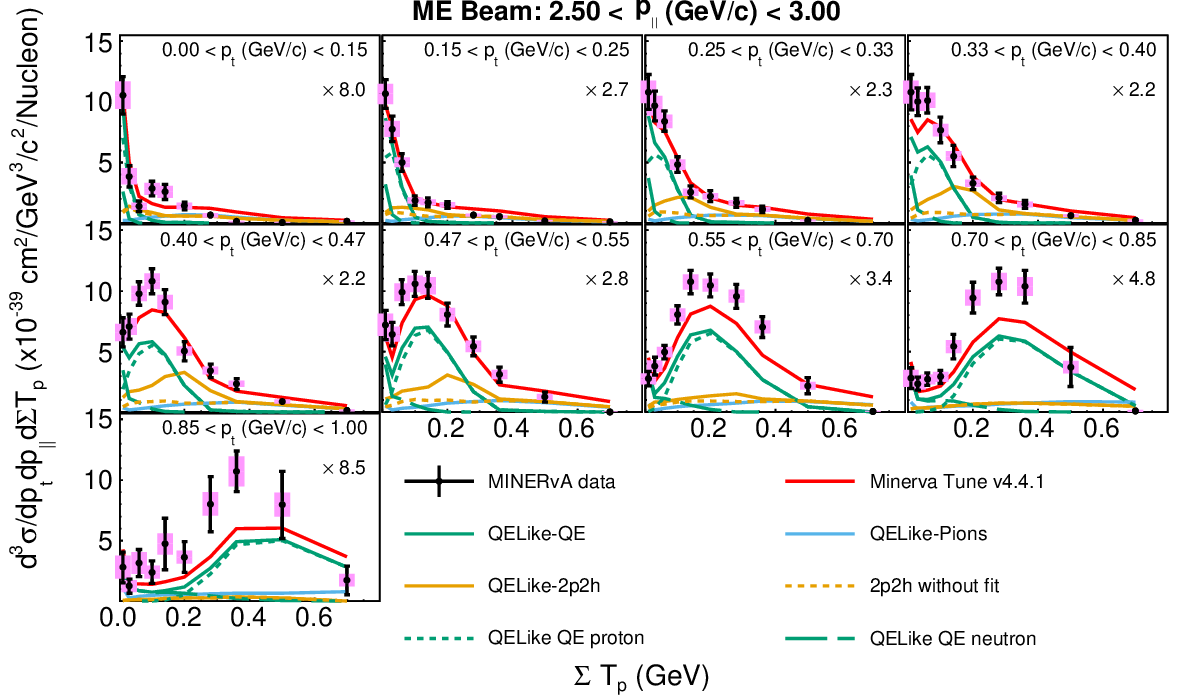}
    \includegraphics[width=\triplet]{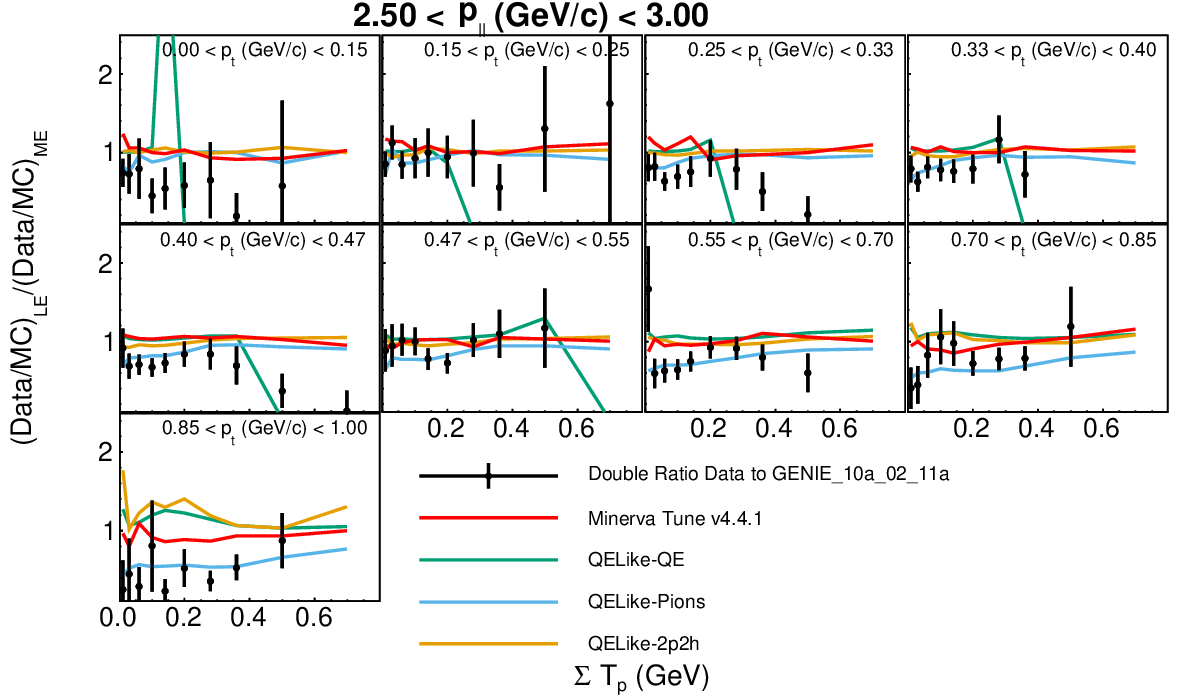}
    \caption{Cross Sections as a function of \pt\ and \sumtp\ for the bin where \pz\ is between $2.5<p_z/(GeV/c)<3.0$, for the Low Energy (top) and Medium Energy (middle) beams.  The predicted contributions from different quasielastic-like processes are also shown. The Low Energy/Medium Energy Cross Section ratio for the data, the total prediction, and predictions for individual channels are shown in the bottom plot.}
    \label{fig:ptpzsumtp_bin3}
\end{figure*}
\begin{figure*}[tp]
    \centering
    \includegraphics[width=\triplet]{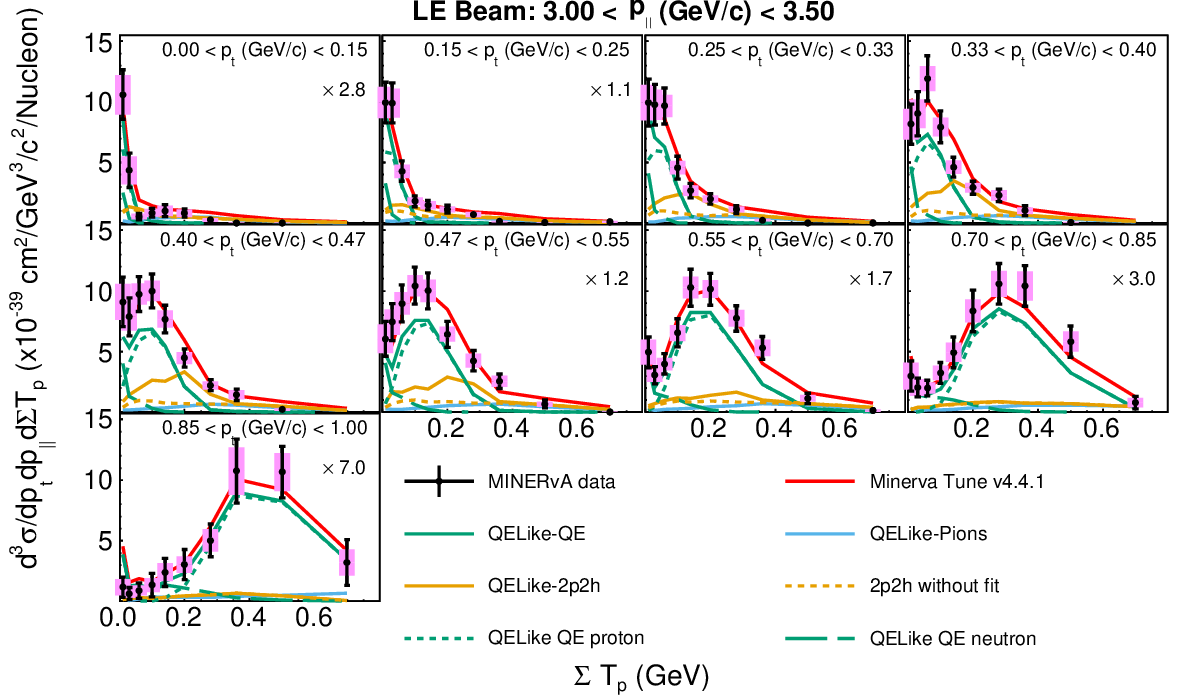}
    \includegraphics[width=\triplet]{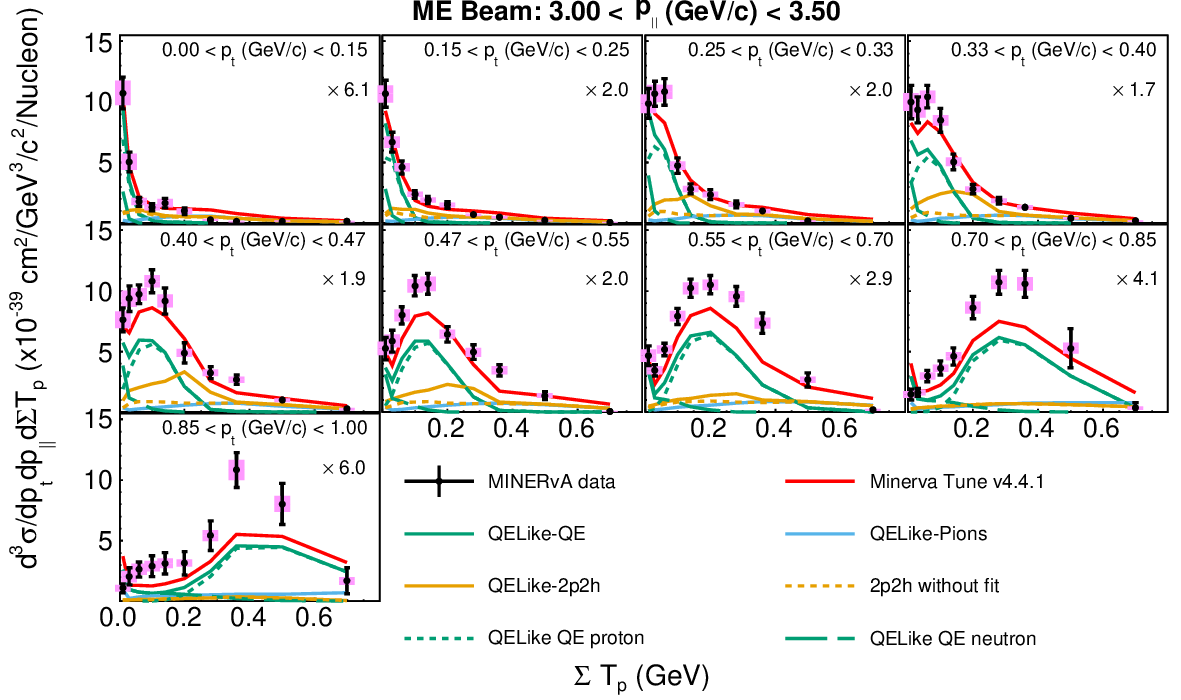}
    \includegraphics[width=\triplet]{NewPlots/DoubleRatio_ExternalModels_Data/GENIE_10a//LE_ME_Ratio_4.eps}
    \caption{Cross Sections as a function of \pt\ and \sumtp\ for the bin where \pz\ is between $3.0<p_z/(GeV/c)<3.5$, for the Low Energy (top) and Medium Energy (middle) beams.  The predicted contributions from different quasielastic-like processes are also shown. The Low Energy/Medium Energy Cross Section ratio for the data, the total prediction, and predictions for individual channels are shown in the bottom plot.}
    \label{fig:ptpzsumtp_bin4}
\end{figure*}\begin{figure*}[tp]
    \centering
    \includegraphics[width=\triplet]{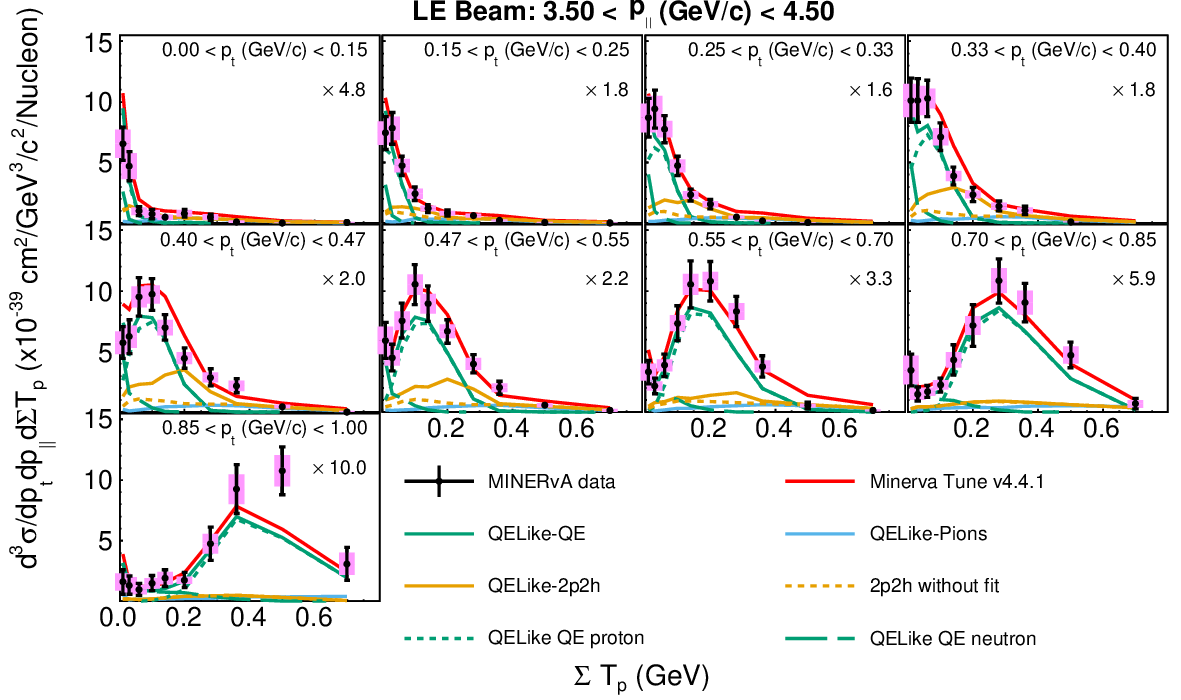}
    \includegraphics[width=\triplet]{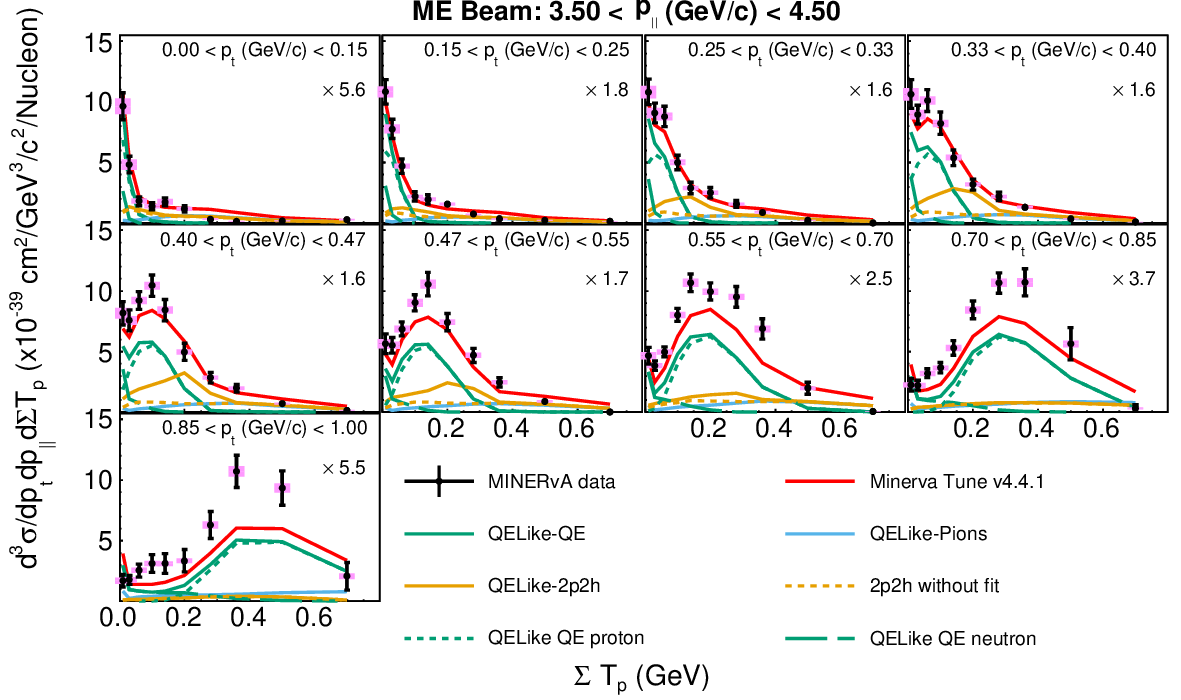}
    \includegraphics[width=\triplet]{NewPlots/DoubleRatio_ExternalModels_Data/GENIE_10a/LE_ME_Ratio_5.eps}
    \caption{Cross Sections as a function of \pt\ and \sumtp\ for the bin where \pz\ is between $3.5<p_z/(GeV/c)<4.5$, for the Low Energy (top) and Medium Energy (middle) beams.  The predicted contributions from different quasielastic-like processes are also shown. The Low Energy/Medium Energy Cross Section ratio for the data, the total prediction, and predictions for individual channels are shown in the bottom plot.}
    \label{fig:ptpzsumtp_bin5}
\end{figure*}


\begin{figure*}[tp]
    \centering
    \includegraphics[width=\triplet]{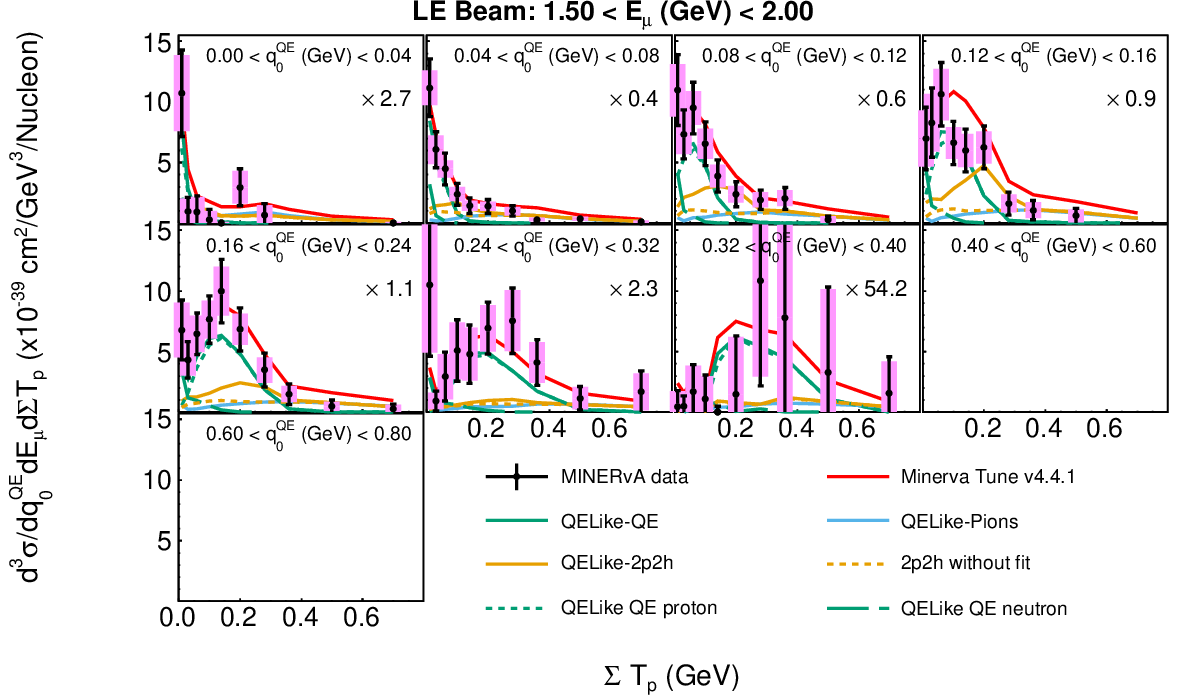}
    \includegraphics[width=\triplet]{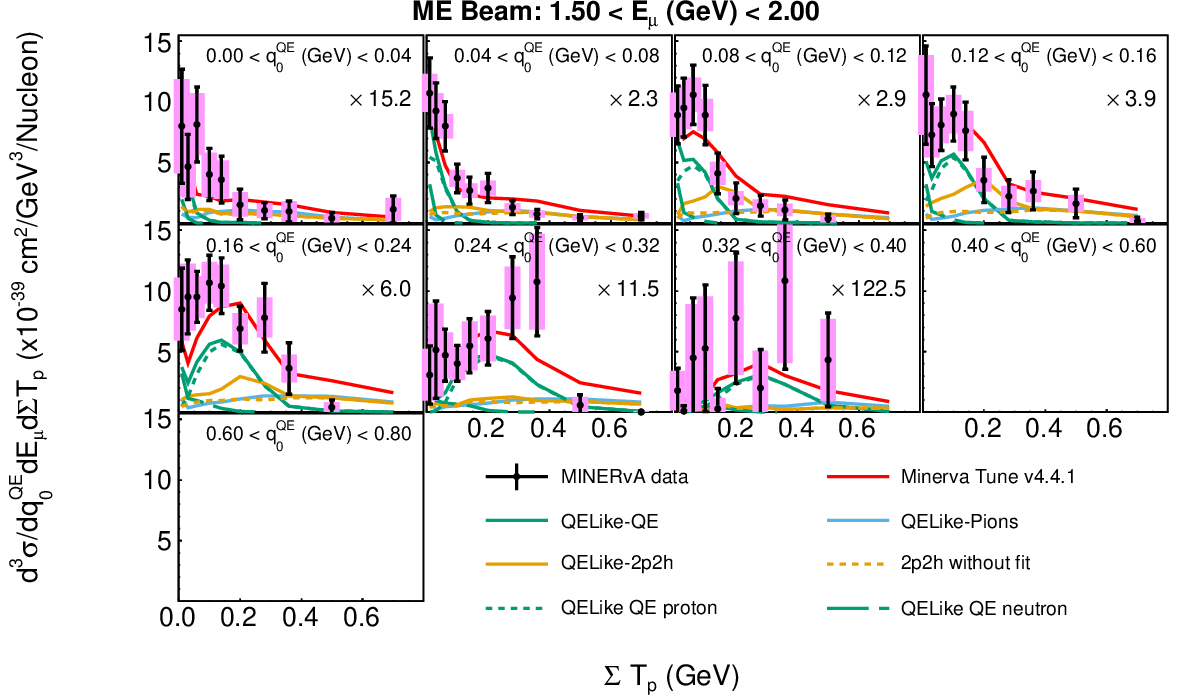}
    \includegraphics[width=\triplet]{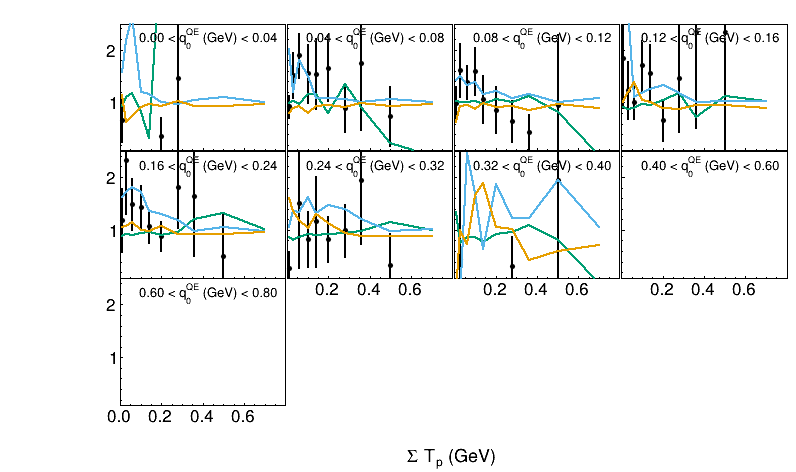}
    \caption{Cross Sections as a function of \qzeroqe\ and \sumtp\ for the bin where $E_\mu$ is between $1.5<p_z/(GeV/c)<2.0$, for the Low Energy (top) and Medium Energy (middle) beams.  The predicted contributions from different quasielastic-like processes are also shown. The Low Energy/Medium Energy Cross Section ratio for the data, the total prediction, and predictions for individual channels are shown in the bottom plot.
    }
    \label{fig:q0qesumtp_bin1}
\end{figure*}
\begin{figure*}[tp]
    \centering
    \includegraphics[width=\triplet]{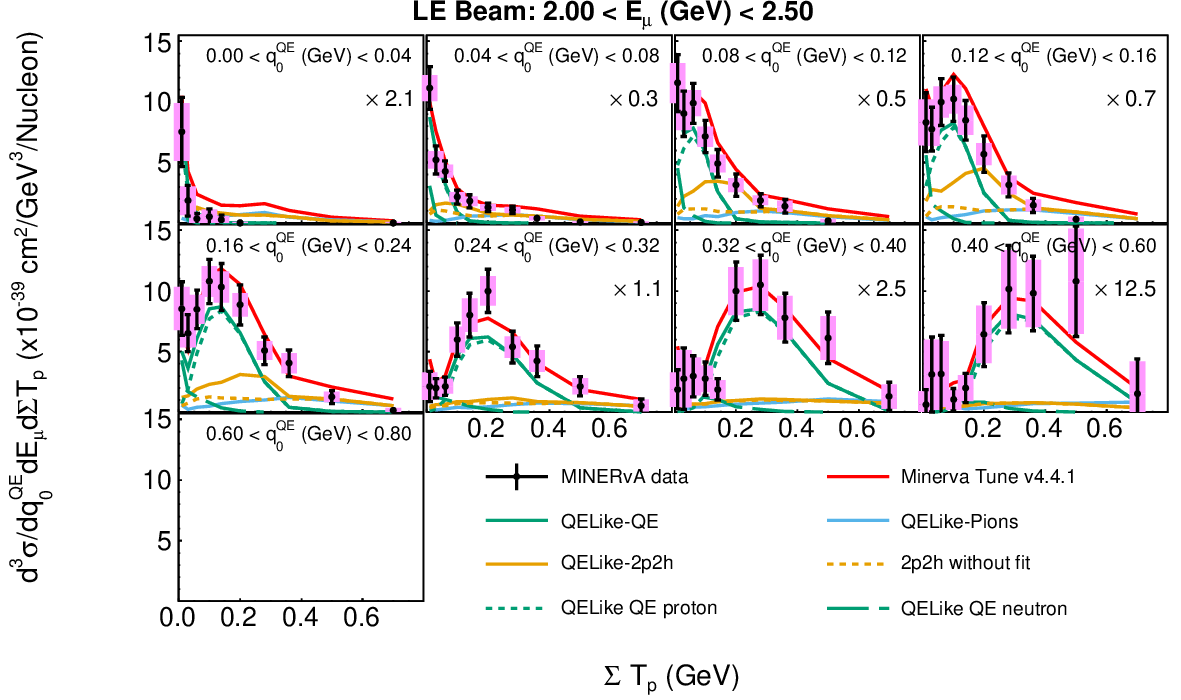}
    \includegraphics[width=\triplet]{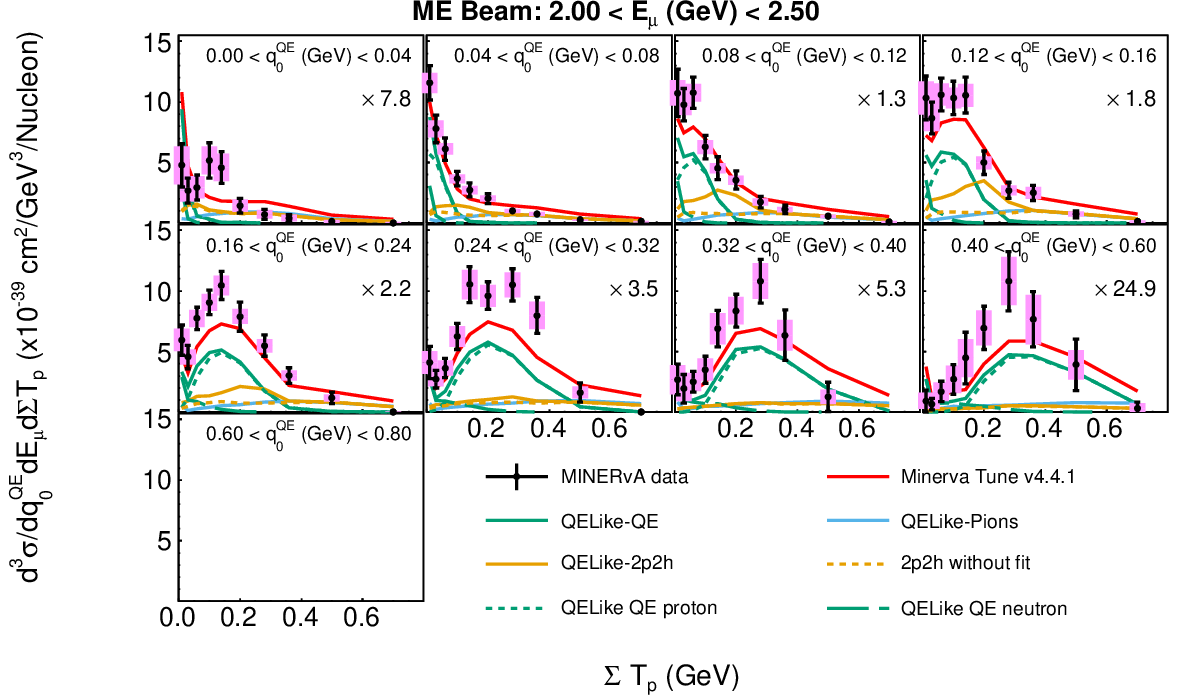}
    \includegraphics[width=\triplet]{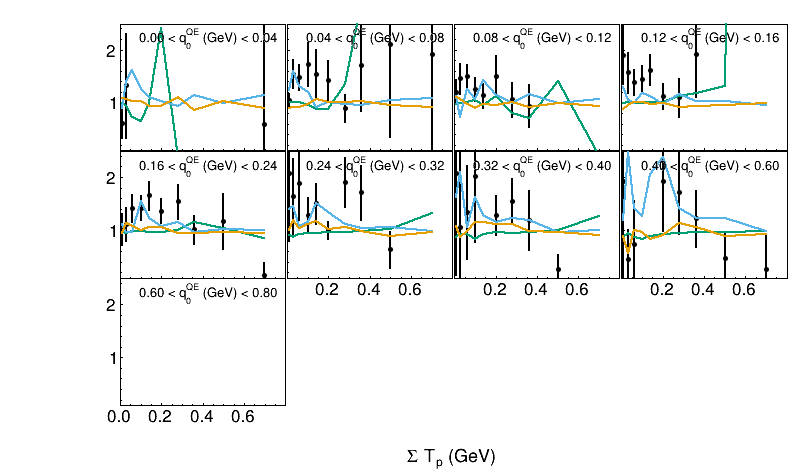}
    \caption{Cross Sections as a function of \qzeroqe\ and \sumtp\ for the bin where $E_\mu$ is between $2.0<p_z/(GeV/c)<2.5$, for the Low Energy (top) and Medium Energy (middle) beams.  The predicted contributions from different quasielastic-like processes are also shown. The Low Energy/Medium Energy Cross Section ratio for the data, the total prediction, and predictions for individual channels are shown in the bottom plot.}
    \label{fig:q0qesumtp_bin2}
\end{figure*}
\begin{figure*}[tp]
    \centering
    \includegraphics[width=\triplet]{NewPlots/LE/nu-3d-xsec-comps-enuproxyE_no2p2Tune-0_resfsi-0_qefis-1_resisi-0_2p2htunes-0_ratio-0-pz-multiplier_bin_3.eps}
    \includegraphics[width=\triplet]{NewPlots/ME/nu-3d-xsec-comps-enuproxyE_no2p2Tune-0_resfsi-0_qefis-1_resisi-0_2p2htunes-0_ratio-0-pz-multiplier_bin_3.eps}
    \includegraphics[width=\triplet]{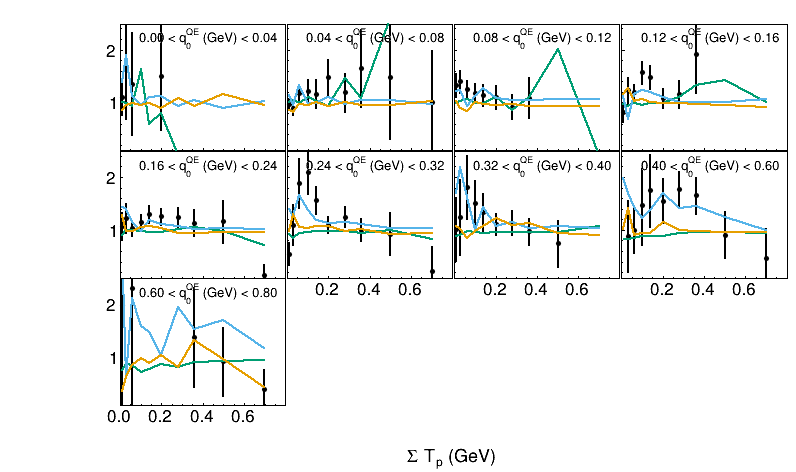}
    \caption{Cross Sections as a function of \qzeroqe\ and \sumtp\ for the bin where $E_\mu$ is between $2.5<p_z/(GeV/c)<3.0$, for the Low Energy (top) and Medium Energy (middle) beams.  The predicted contributions from different quasielastic-like processes are also shown. The Low Energy/Medium Energy Cross Section ratio for the data, the total prediction, and predictions for individual channels are shown in the bottom plot.}
    \label{fig:q0qesumtp_bin3}
\end{figure*}
\begin{figure*}[tp]
    \centering
    \includegraphics[width=\triplet]{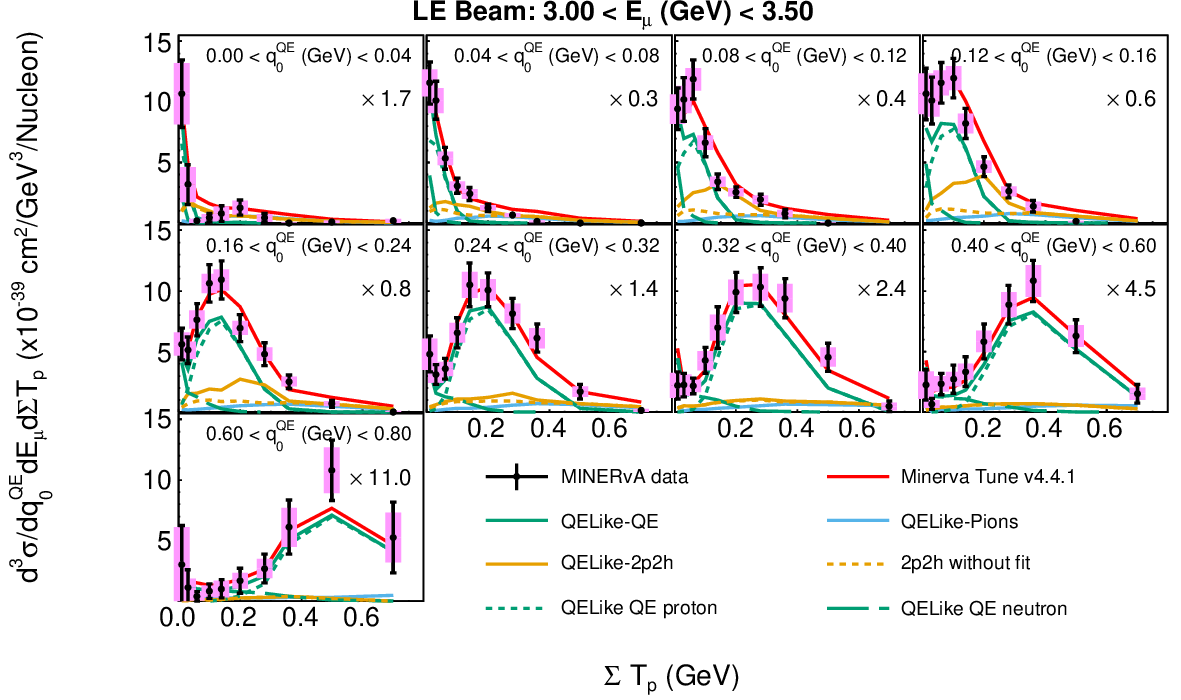}
    \includegraphics[width=\triplet]{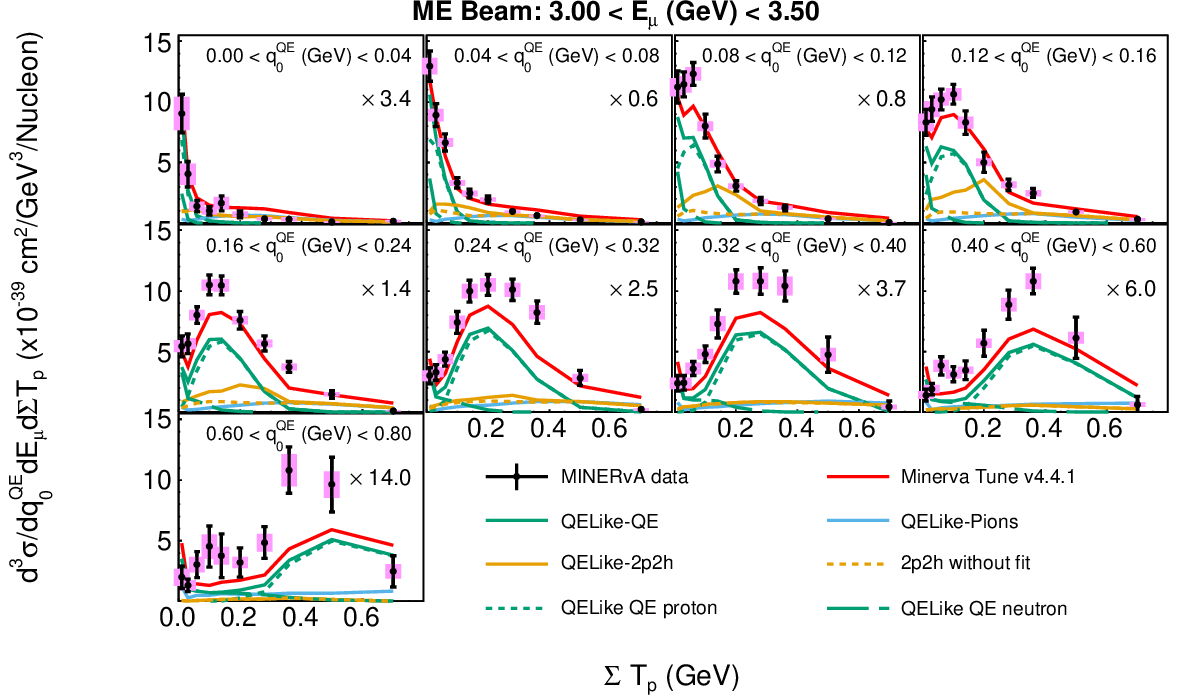}
    \includegraphics[width=\triplet]{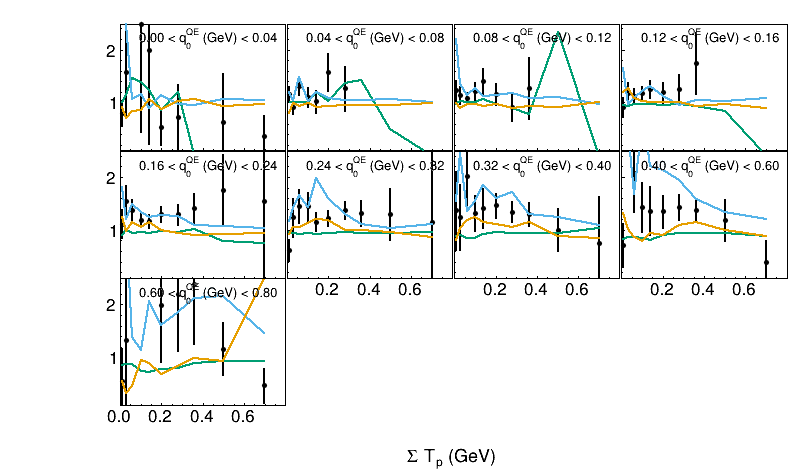}
    \caption{Cross Sections as a function of \qzeroqe\ and \sumtp\ for the bin where $E_\mu$ is between $3.0<p_z/(GeV/c)<3.5$, for the Low Energy (top) and Medium Energy (middle) beams.  The predicted contributions from different quasielastic-like processes are also shown. The Low Energy/Medium Energy Cross Section ratio for the data, the total prediction, and predictions for individual channels are shown in the bottom plot.}
    \label{fig:q0qesumtp_bin4}
\end{figure*}
\begin{figure*}[tp]
    \centering
    \includegraphics[width=\triplet]{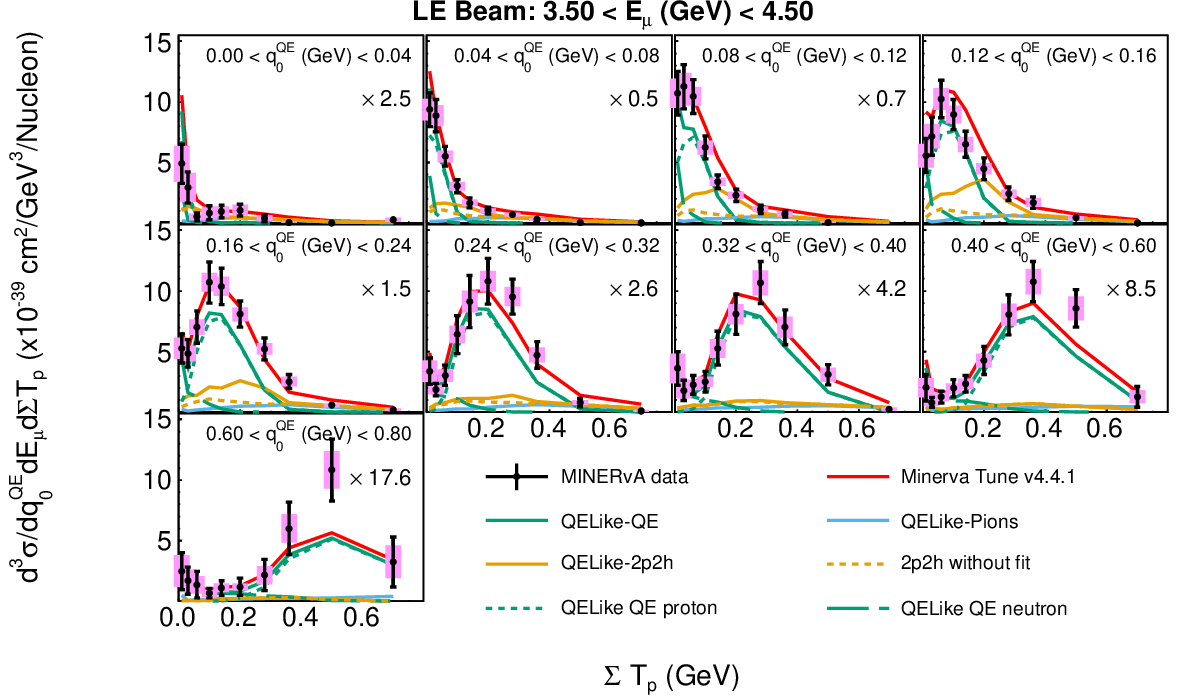}
    \includegraphics[width=\triplet]{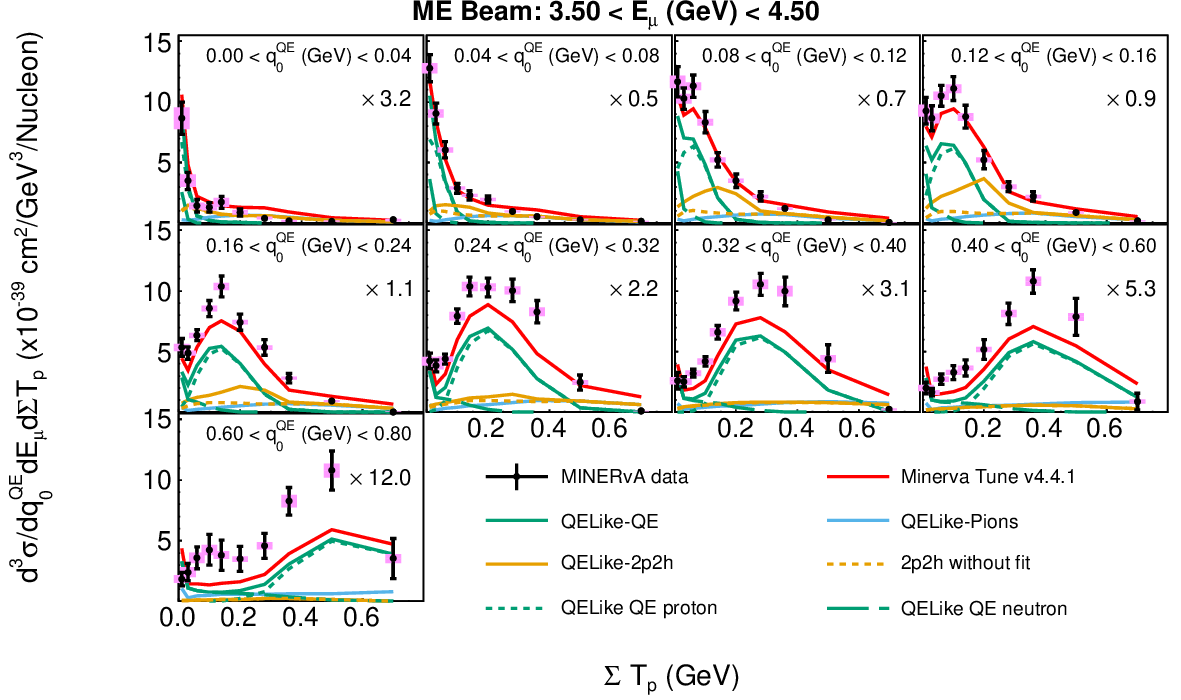}
    \includegraphics[width=\triplet]{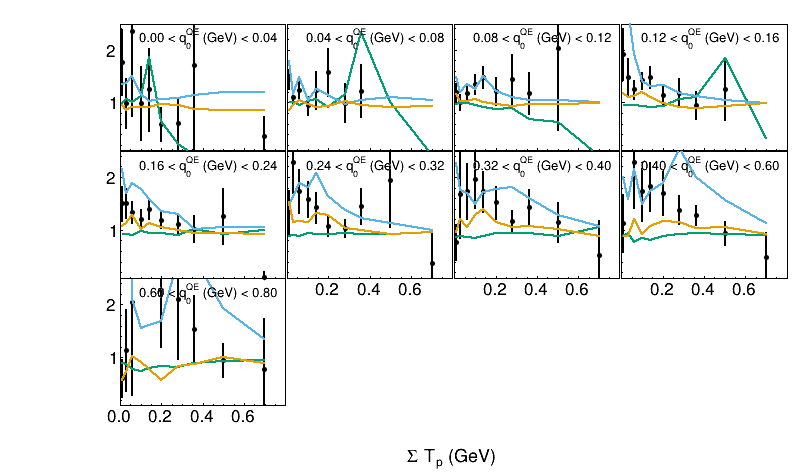}
    \caption{Cross Sections as a function of \qzeroqe\ and \sumtp\ for the bin where $E_\mu$ is between $3.5<p_z/(GeV/c)<4.5$, for the Low Energy (top) and Medium Energy (middle) beams.  The predicted contributions from different quasielastic-like processes are also shown. The Low Energy/Medium Energy Cross Section ratio for the data, the total prediction, and predictions for individual channels are shown in the bottom plot.}
    \label{fig:q0qesumtp_bin5}
\end{figure*}

\begin{figure*}[tp]
    \centering    
    \includegraphics[width=\triplet]{NewPlots/DoubleRatio_ExternalModels_Data/GENIE_10a/LE_ME_RATIO_1.eps}
    \includegraphics[width=\triplet]{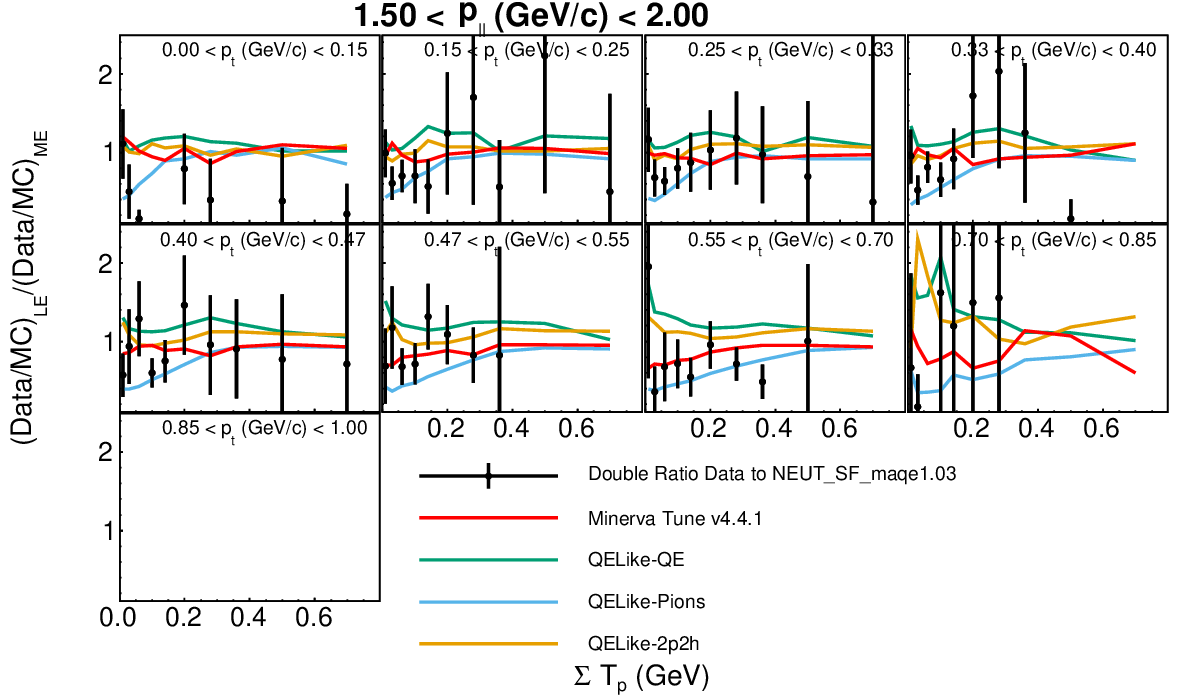}
    \includegraphics[width=\triplet]{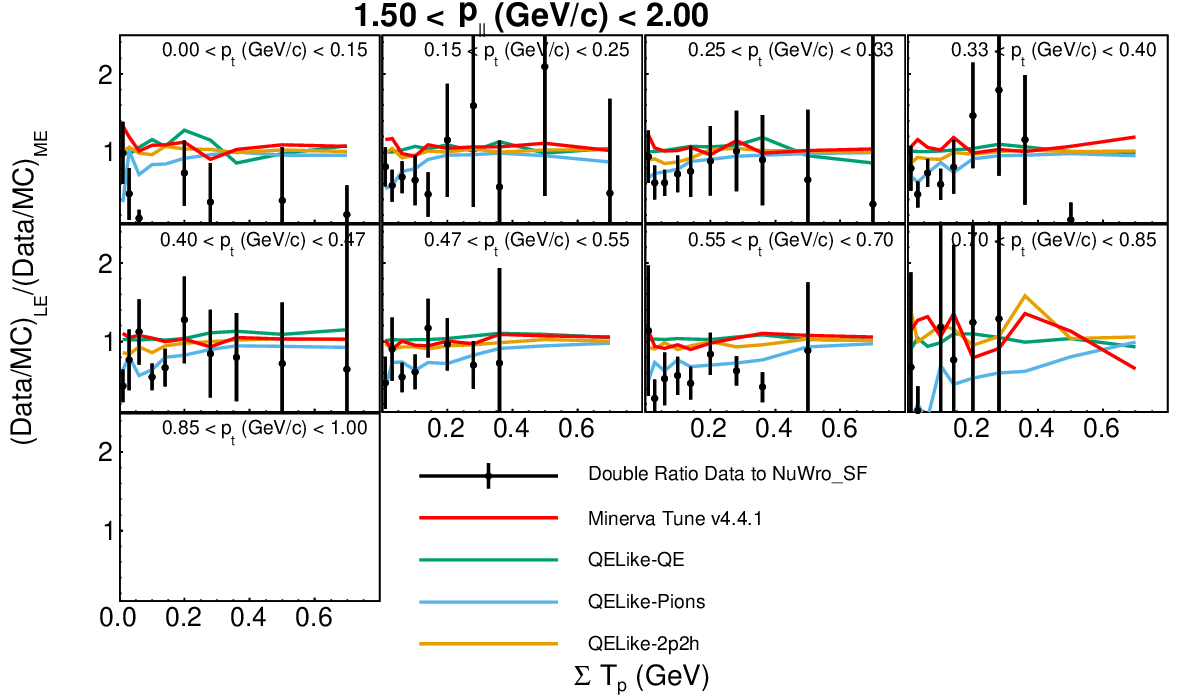}
    \caption{Double ratio of $(data/model)_{LE}/(data/model)_{ME}$.GENIE 10a (top) and NEUT with a spectral function (middle) and NuWro with a spectral function (bottom) are shown.      }
    \label{fig:ptpzsumtp_bin1_doubleRatio_models}
\end{figure*}
\begin{figure*}[tp]
    \centering    
    \includegraphics[width=\triplet]{NewPlots/DoubleRatio_ExternalModels_Data/GENIE_10a//LE_ME_RATIO_2.eps}
    \includegraphics[width=\triplet]{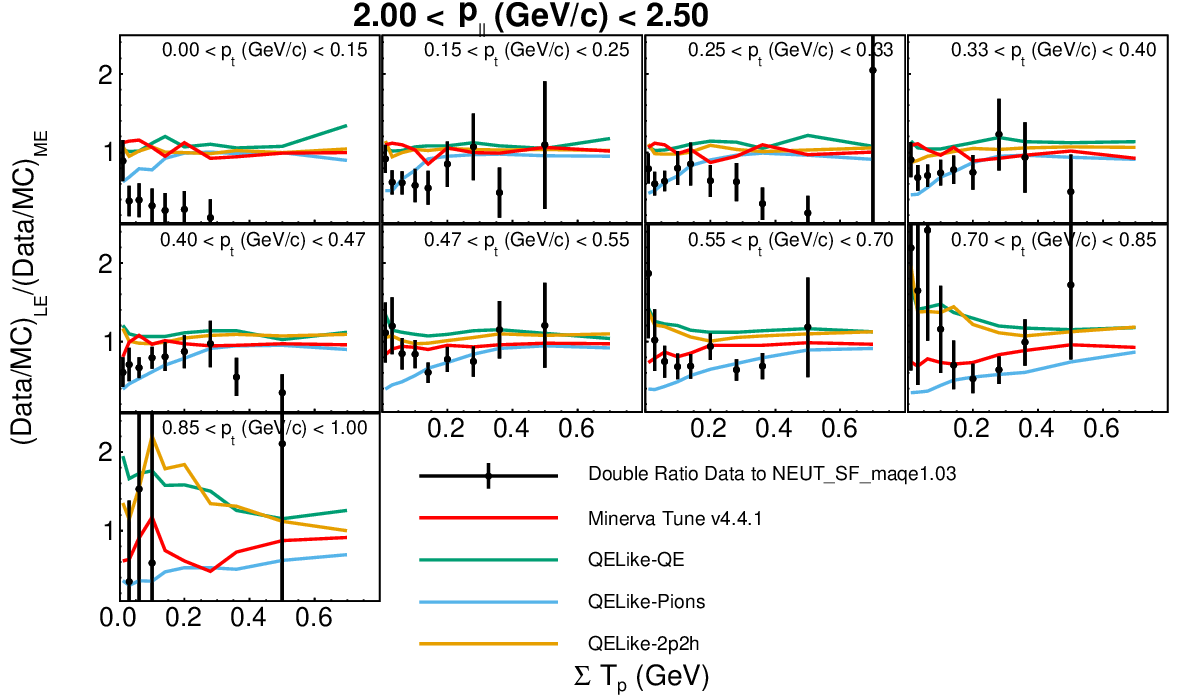}
        \includegraphics[width=\triplet]{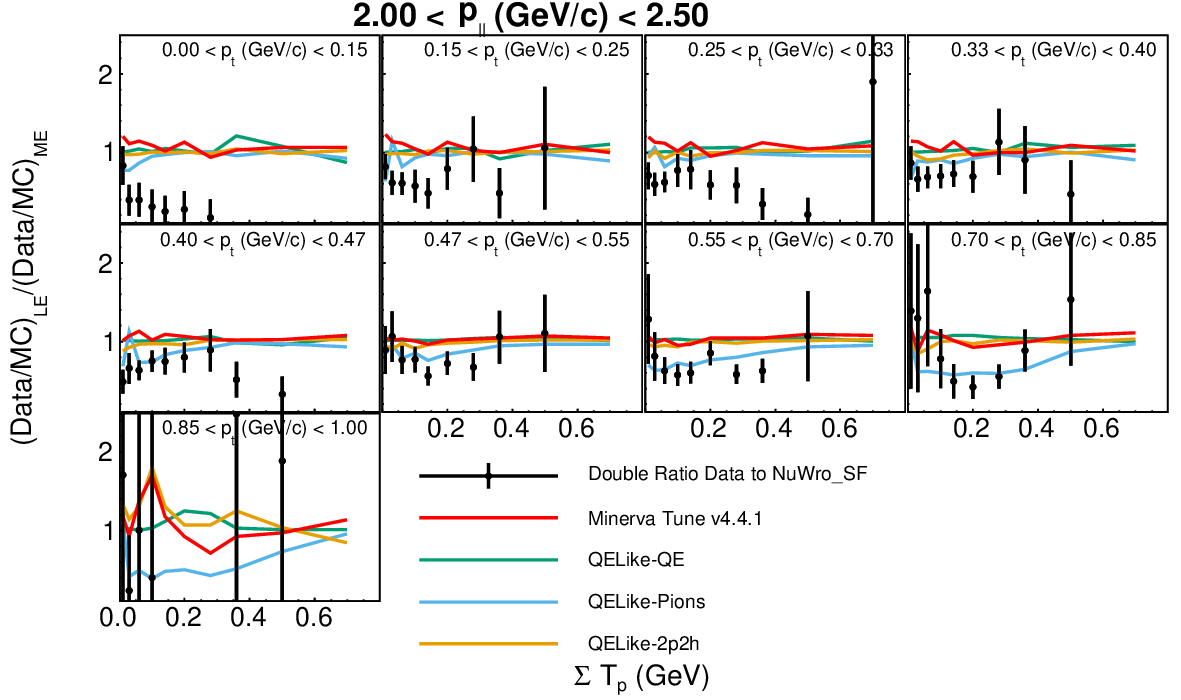}

    \caption{Double ratio of $(data/model)_{LE}/(data/model)_{ME}$.GENIE 10a (top) and NEUT with a spectral function (middle)and NuWro (bottom) with a spectral function (bottom) are shown.}
    \label{fig:ptpzsumtp_bin2_doubleRatio_mmodels}
\end{figure*}
\begin{figure*}[tp]
    \centering    
    \includegraphics[width=\triplet]{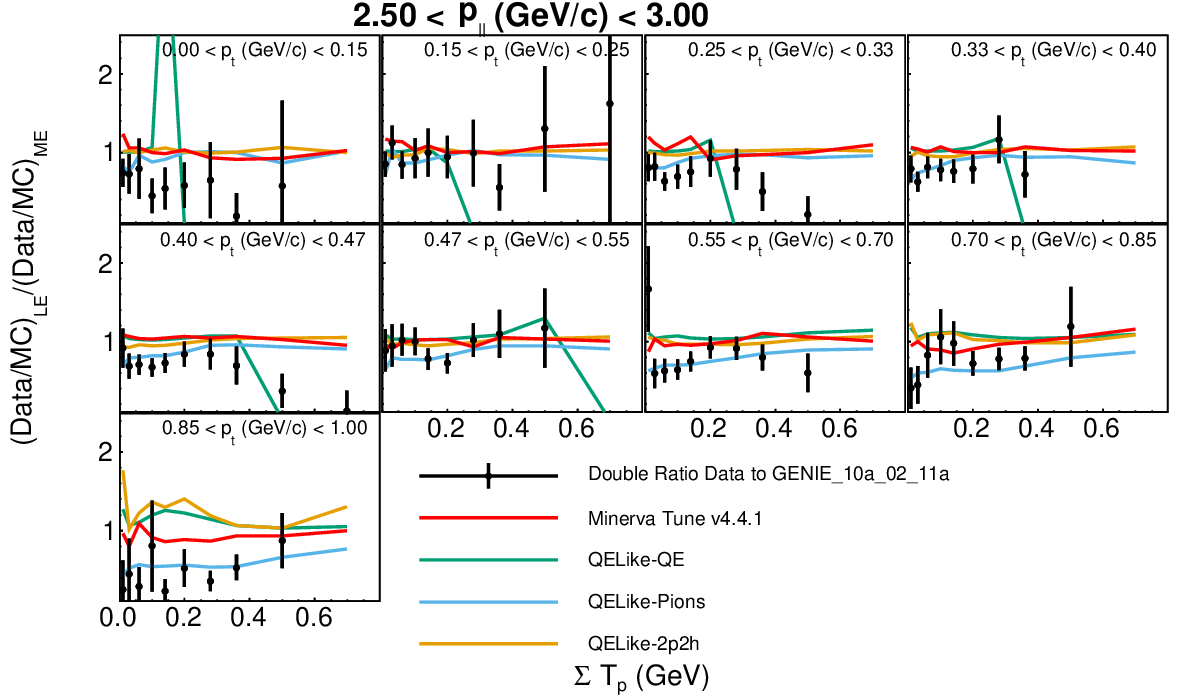}
    \includegraphics[width=\triplet]{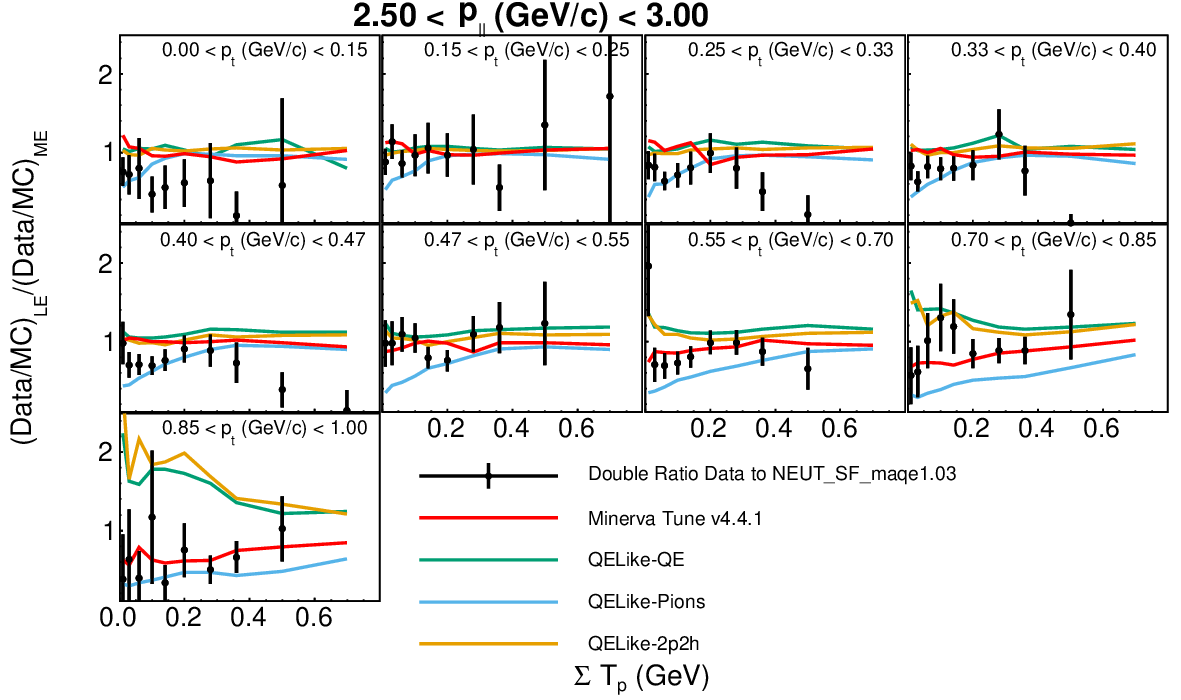}
            \includegraphics[width=\triplet]{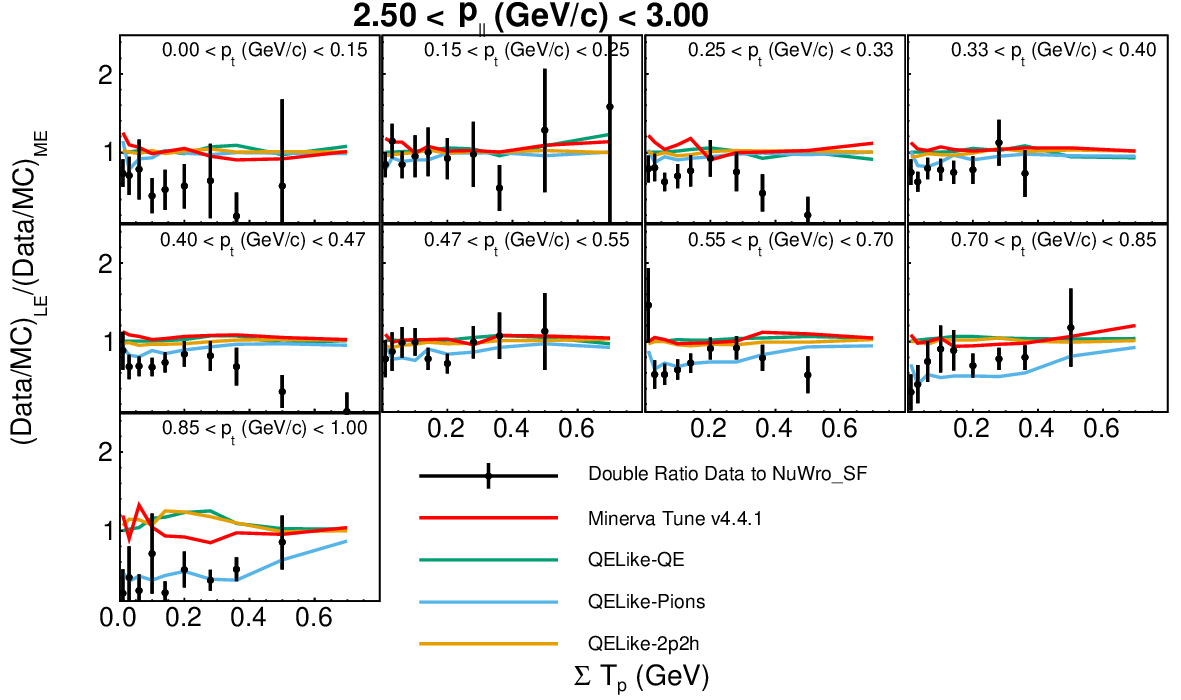}
    \caption{Double ratio of $(data/model)_{LE}/(data/model)_{ME}$.GENIE 10a (top) and NEUT with a spectral function (middle) and NuWro with a spectral function (bottom) are shown}
    \label{fig:ptpzsumtp_bin3_doubleRatio_mmodels}
\end{figure*}
\begin{figure*}[tp]
    \centering    
    \includegraphics[width=\triplet]{NewPlots/DoubleRatio_ExternalModels_Data/GENIE_10a//LE_ME_RATIO_4.eps}
    \includegraphics[width=\triplet]{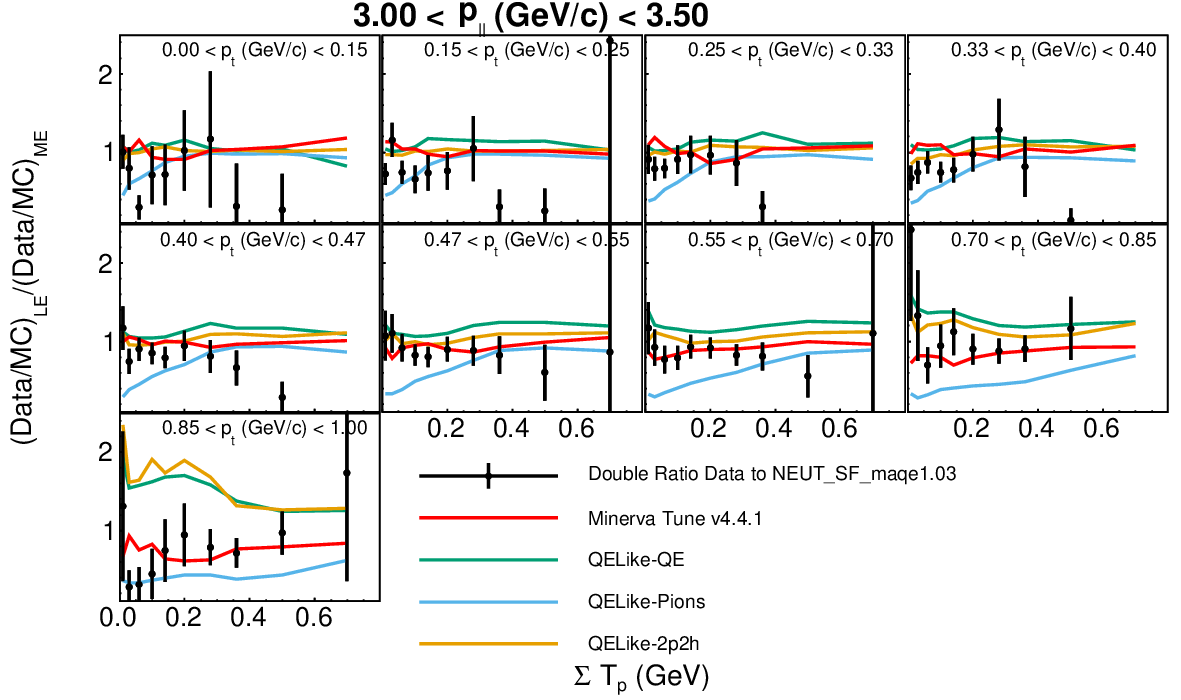}
            \includegraphics[width=\triplet]{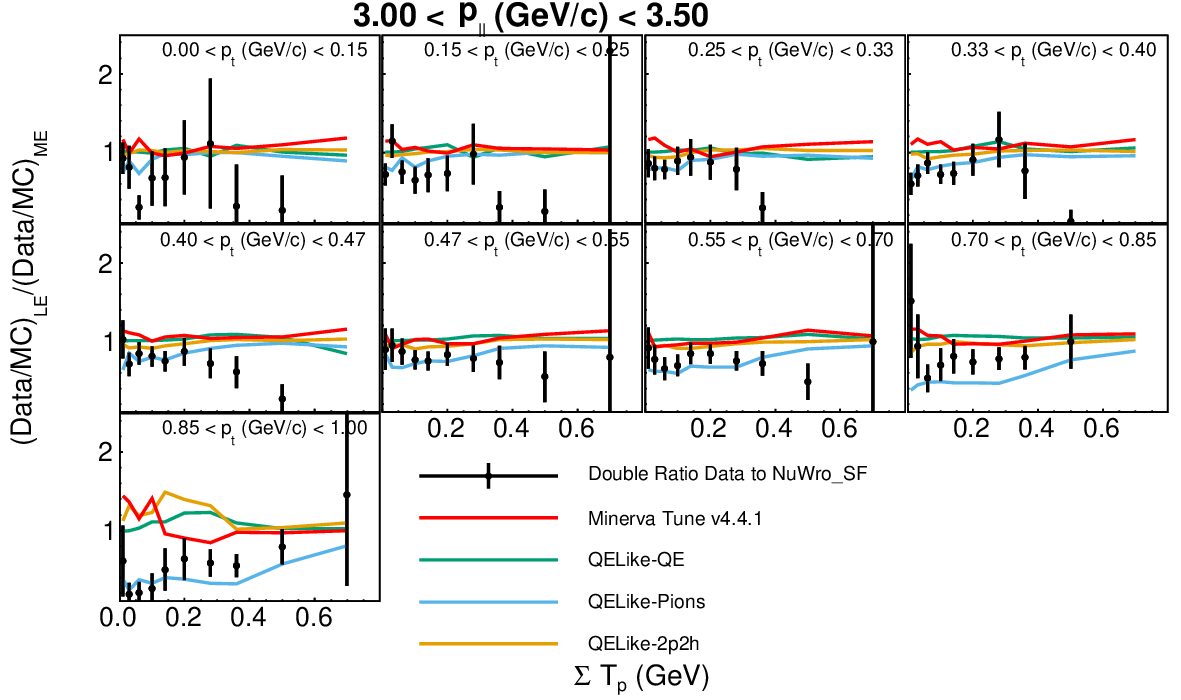}
    \caption{Double ratio of $(data/model)_{LE}/(data/model)_{ME}$.GENIE 10a (top) and NEUT with a spectral function (middle) and NuWro with a spectral function (bottom) are shown.}
    \label{fig:ptpzsumtp_bin4_doubleRatio_models}
\end{figure*}
\begin{figure*}[tp]
    \centering    
    \includegraphics[width=\triplet]{NewPlots/DoubleRatio_ExternalModels_Data/GENIE_10a//LE_ME_RATIO_5.eps}
    \includegraphics[width=\triplet]{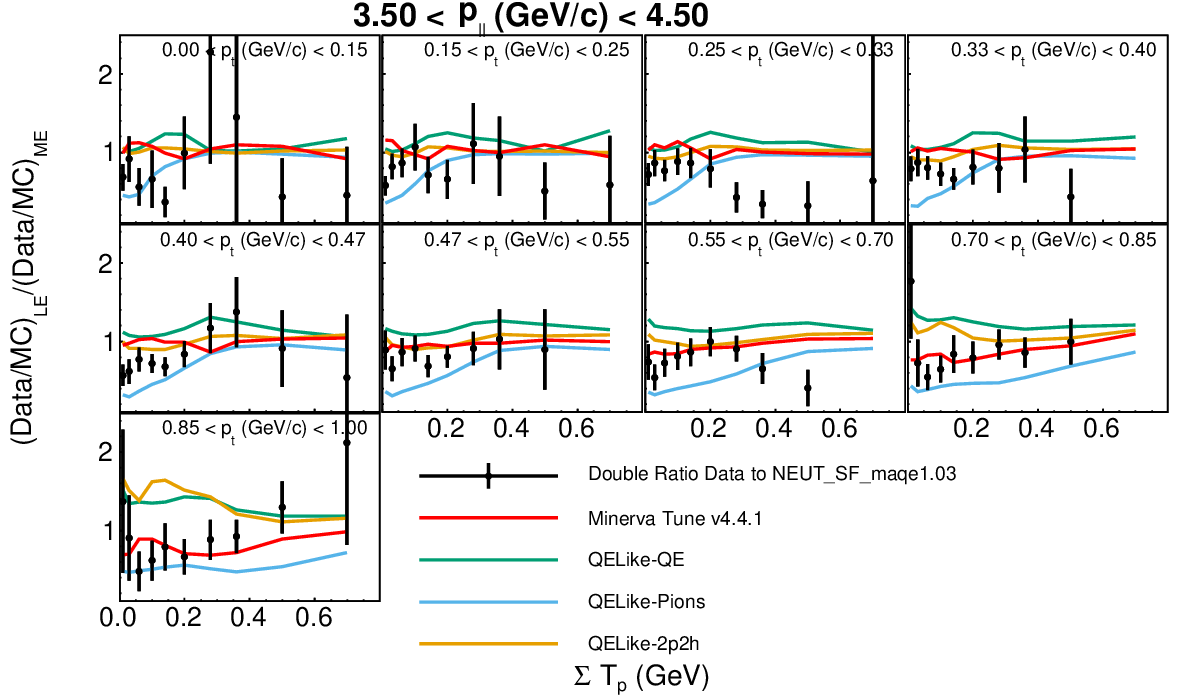}
            \includegraphics[width=\triplet]{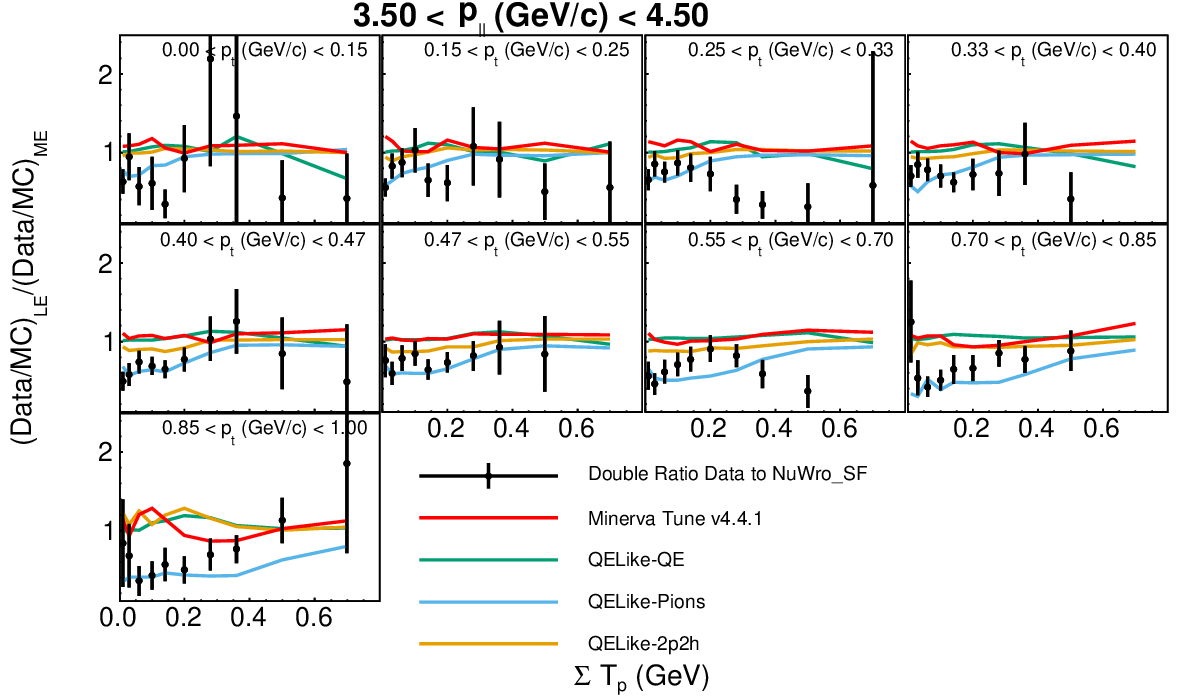}
    \caption{Double ratio of $(data/model)_{LE}/(data/model)_{ME}$.GENIE 10a (top) and NEUT with a spectral function (middle) and NuWro with a spectral function (bottom) are shown.}
    \label{fig:ptpzsumtp_bin5_doubleRatio_models}
\end{figure*}


\begin{figure*}[tp]
    \centering    
\includegraphics[width=\triplet]{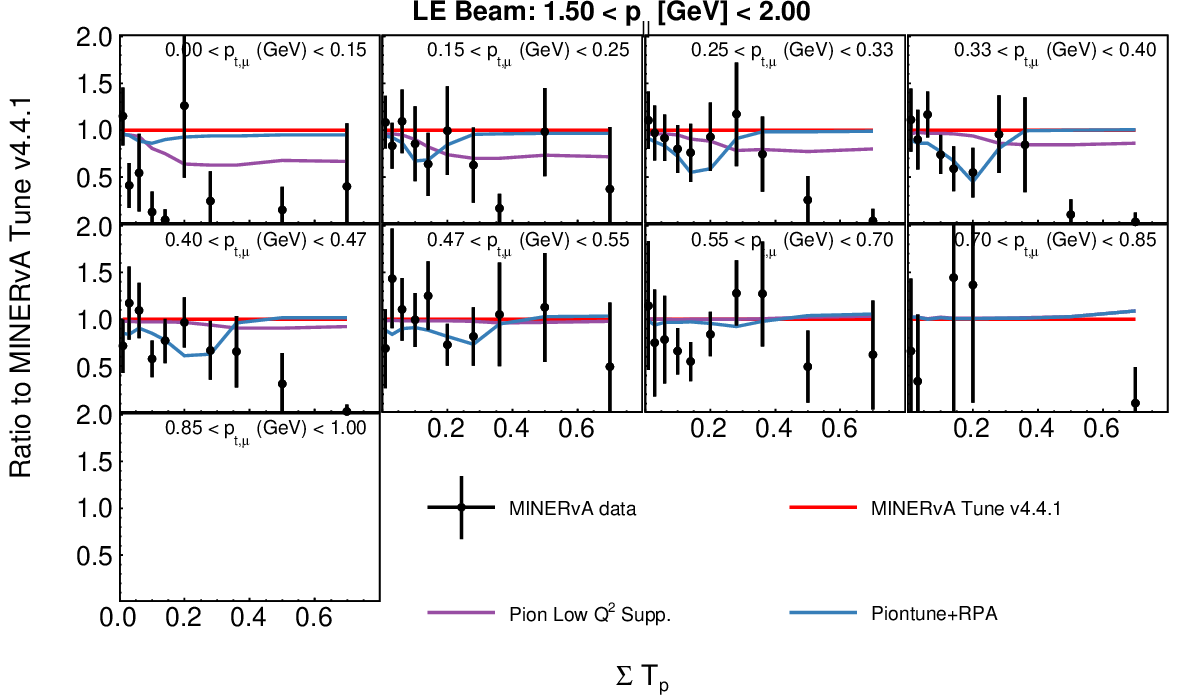}
\includegraphics[width=\triplet]{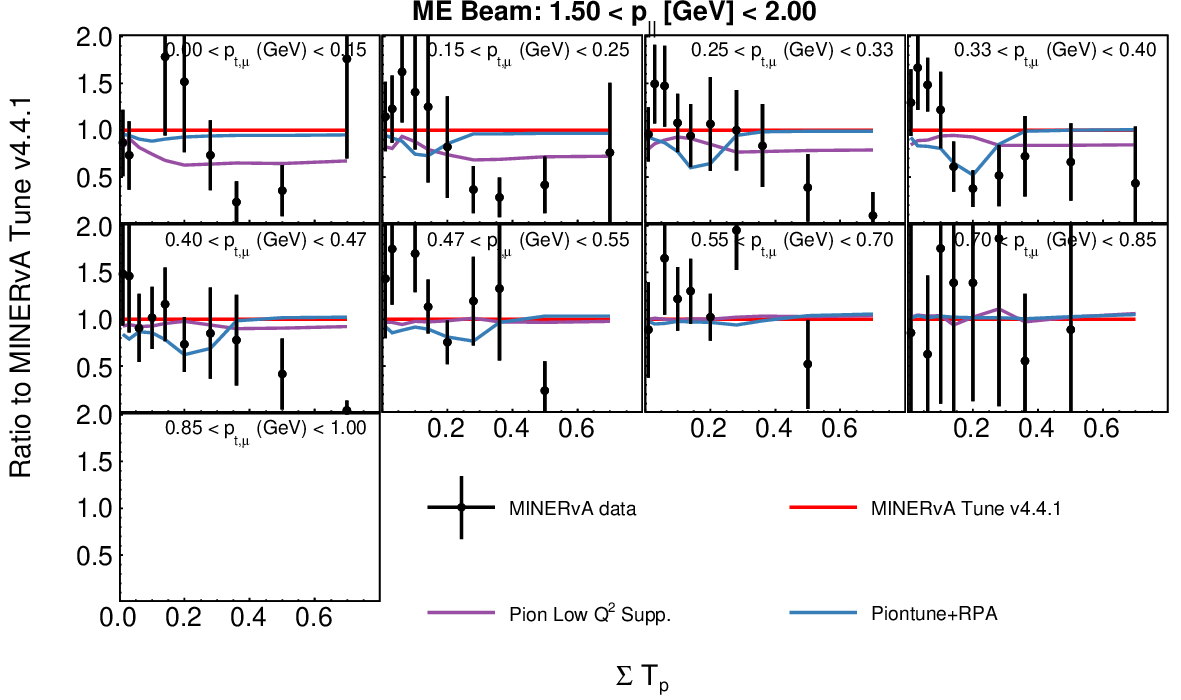}
    \caption{Data to simulation ratio as a function of \pt\ and \sumtp\ for \pz\ between $1.5~GeV/c$ and $2.0~GeV/c$, showing the effect of different modifications to GENIE for the MINERvA tune: LE (top), ME (bottom) 
    }
\label{fig:ptpzsumtp_bin1_MINERvAmodels}
\end{figure*}
\begin{figure*}[tp]
    \centering    
    \includegraphics[width=\doublet]{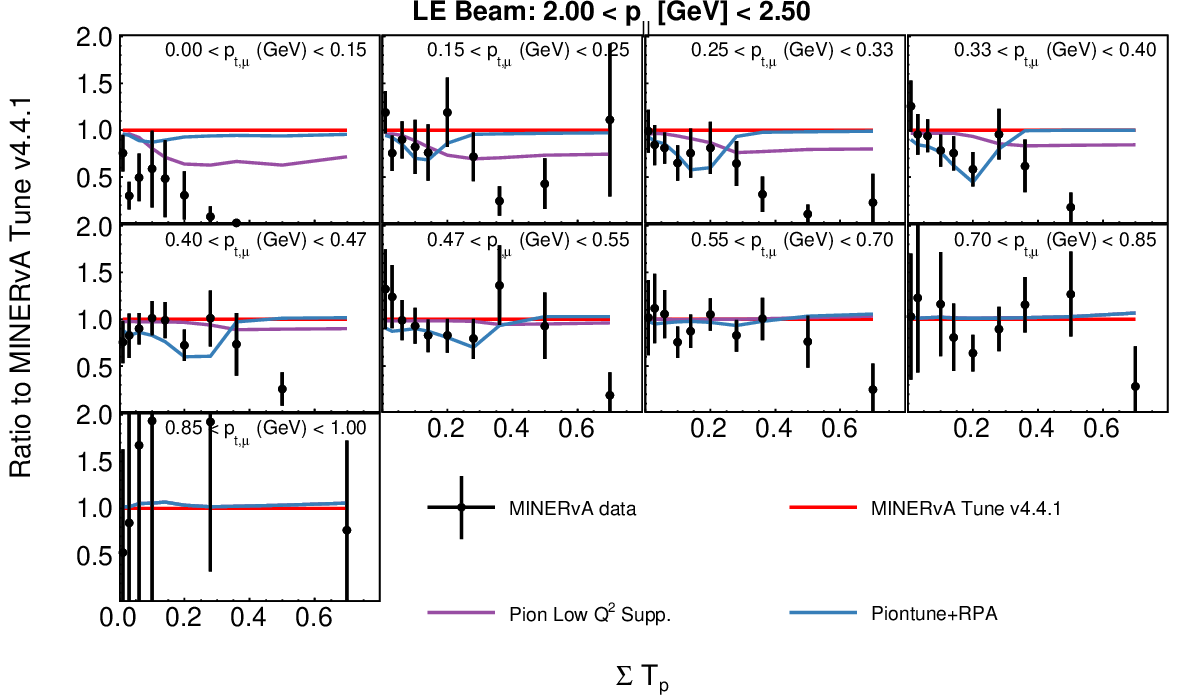}
    \includegraphics[width=\doublet]{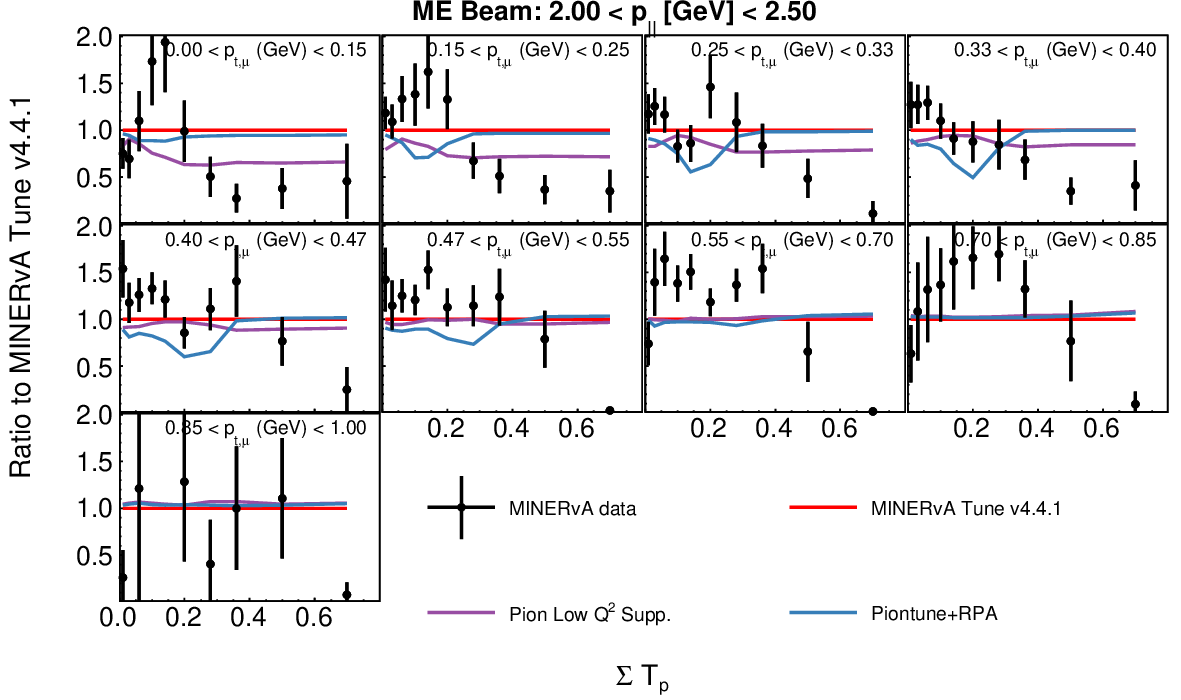}
    \caption{Data to simulation ratio as a function of \pt\ and \sumtp\ for \pz\ between $2.0~GeV/c$ and $2.5~GeV/c$, showing the effect of different modifications to GENIE for the MINERvA tune: LE (top), ME (bottom) }
    \label{fig:ptpzsumtp_bin2_MINERvAmodels}
\end{figure*}
\begin{figure*}[tp]
    \centering    
    \includegraphics[width=\doublet]{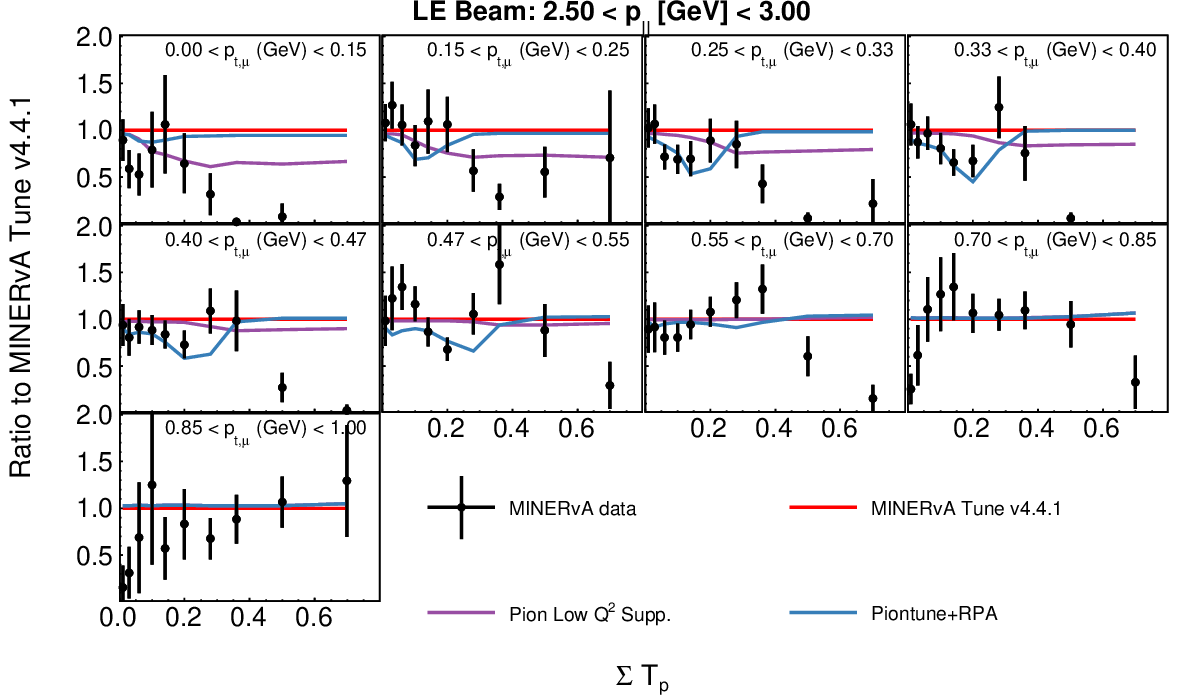}
    \includegraphics[width=\doublet]{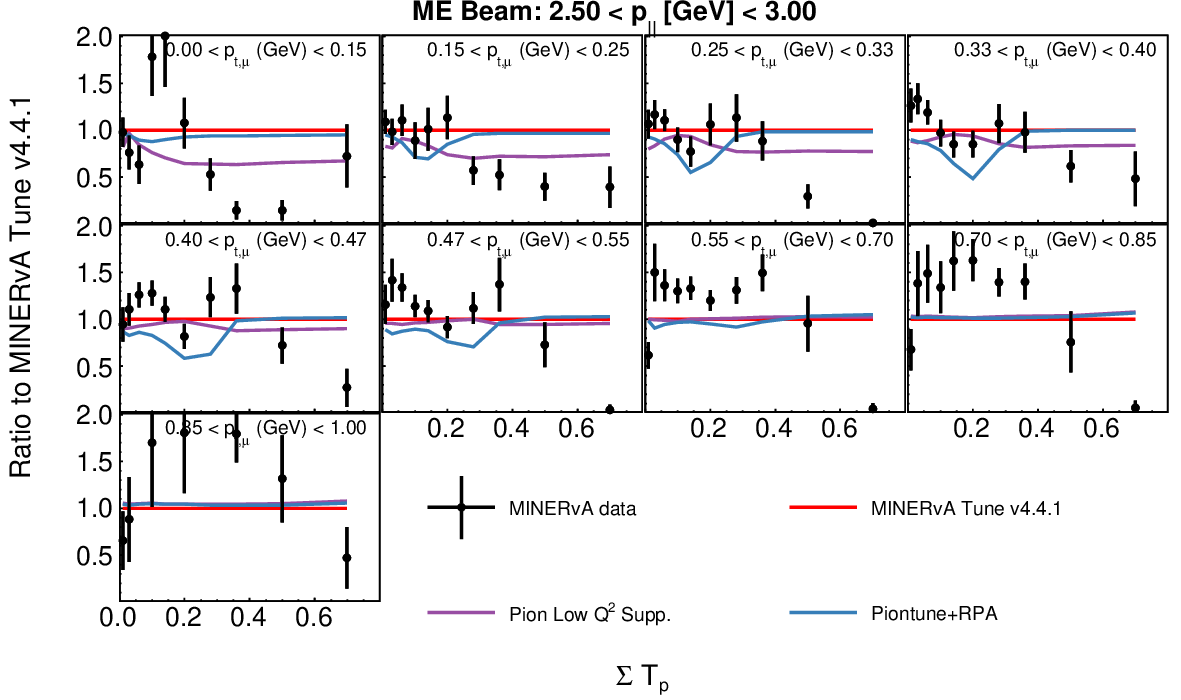}
    \caption{Data to simulation ratio as a function of \pt\ and \sumtp\ for \pz\ between $2.5~GeV/c$ and $3.0~GeV/c$, showing the effect of different modifications to GENIE for the MINERvA tune: LE (top), ME (bottom) }
    \label{fig:ptpzsumtp_bin3_MINERvAmodels}
\end{figure*}
\begin{figure*}[tp]
    \centering    
    \includegraphics[width=\doublet]{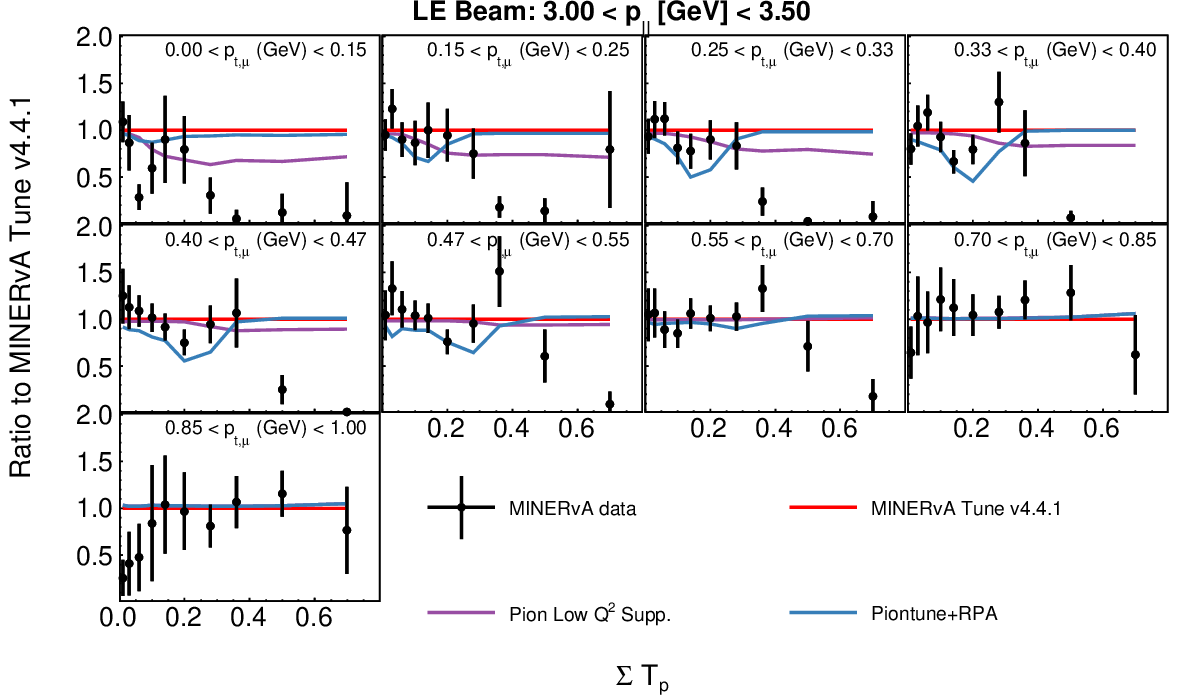}
    \includegraphics[width=\doublet]{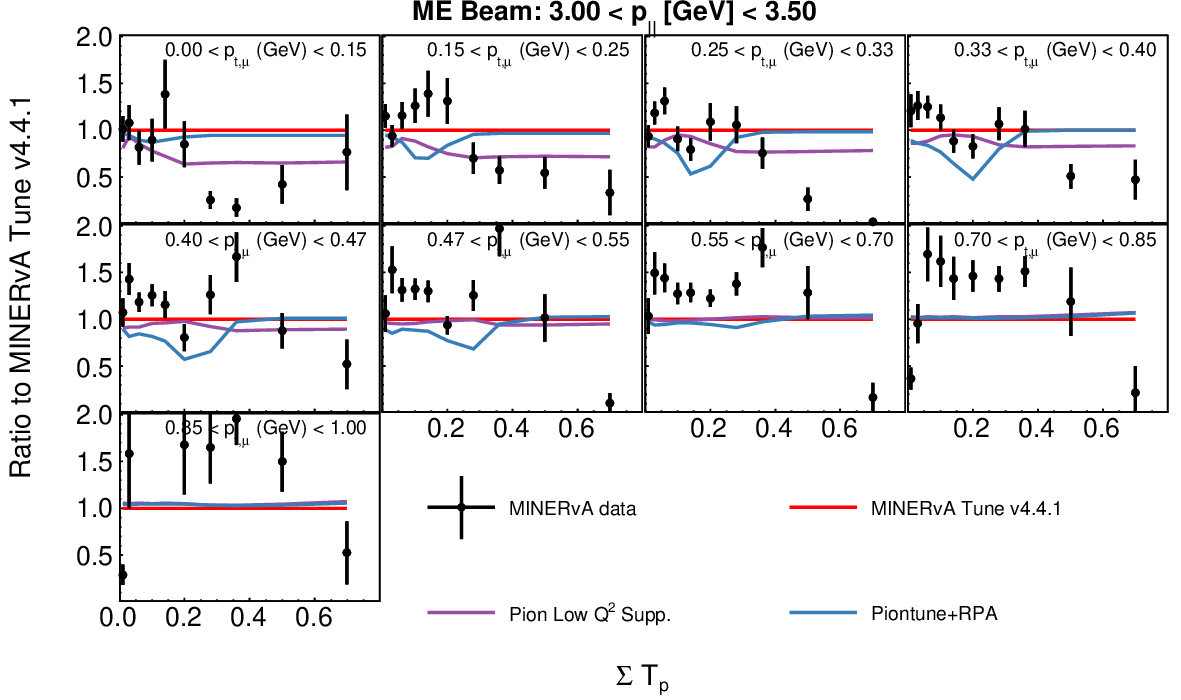}
    \caption{Data to simulation ratio as a function of \pt\ and \sumtp\ for \pz\ between $3.0~GeV/c$ and $3.5~GeV/c$, showing the effect of different modifications to GENIE for the MINERvA tune: LE (top), ME (bottom) }
    \label{fig:ptpzsumtp_bin4_MINERvAmodels}
\end{figure*}
\begin{figure*}[tp]
    \centering    
    \includegraphics[width=\doublet]{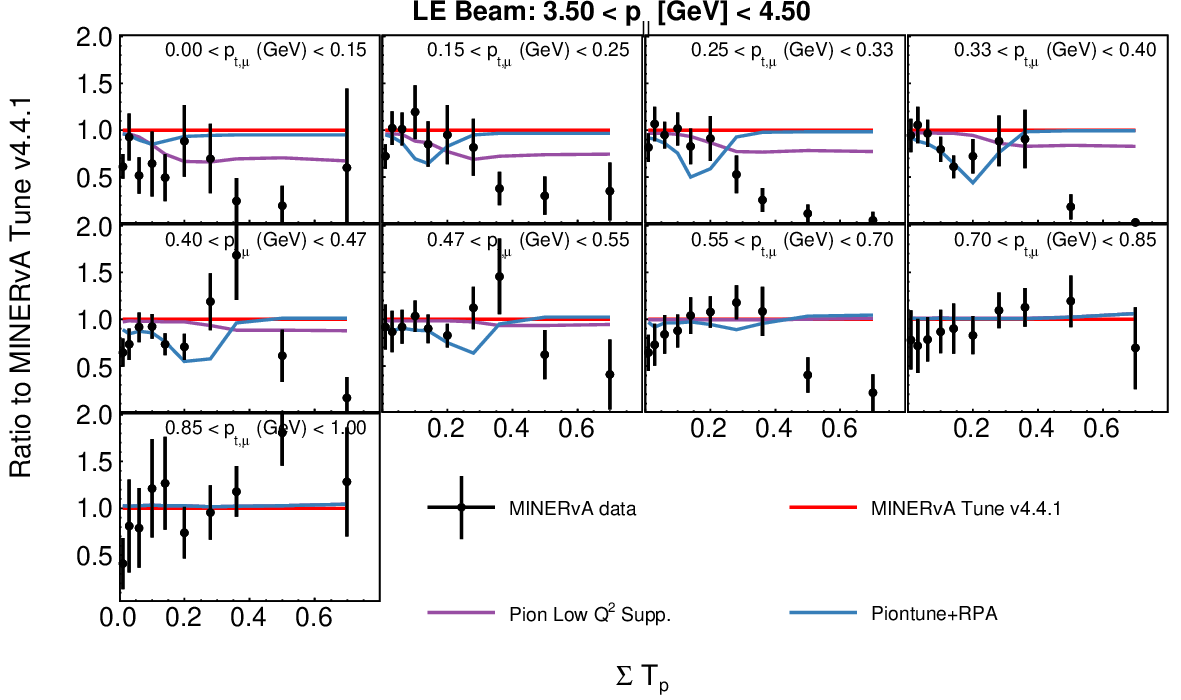}
    \includegraphics[width=\doublet]{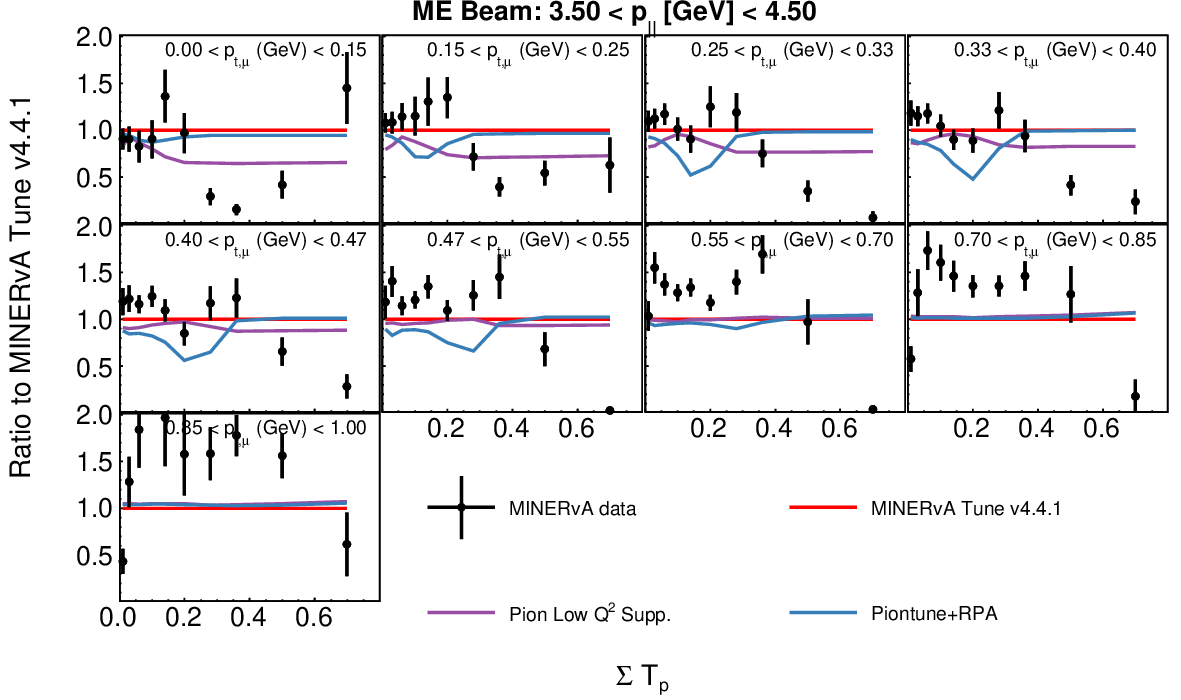}
    \caption{Data to simulation ratio as a function of \pt\ and \sumtp\ for \pz\ between $3.5~GeV/c$ and $4.5~GeV/c$, showing the effect of different modifications to GENIE for the MINERvA tune: LE (top), ME (bottom) }
    \label{fig:ptpzsumtp_bin5_MINERvAmodels}
\end{figure*}
%
%
\begin{figure*}[tp]
    \centering    
    \includegraphics[width=\triplet]{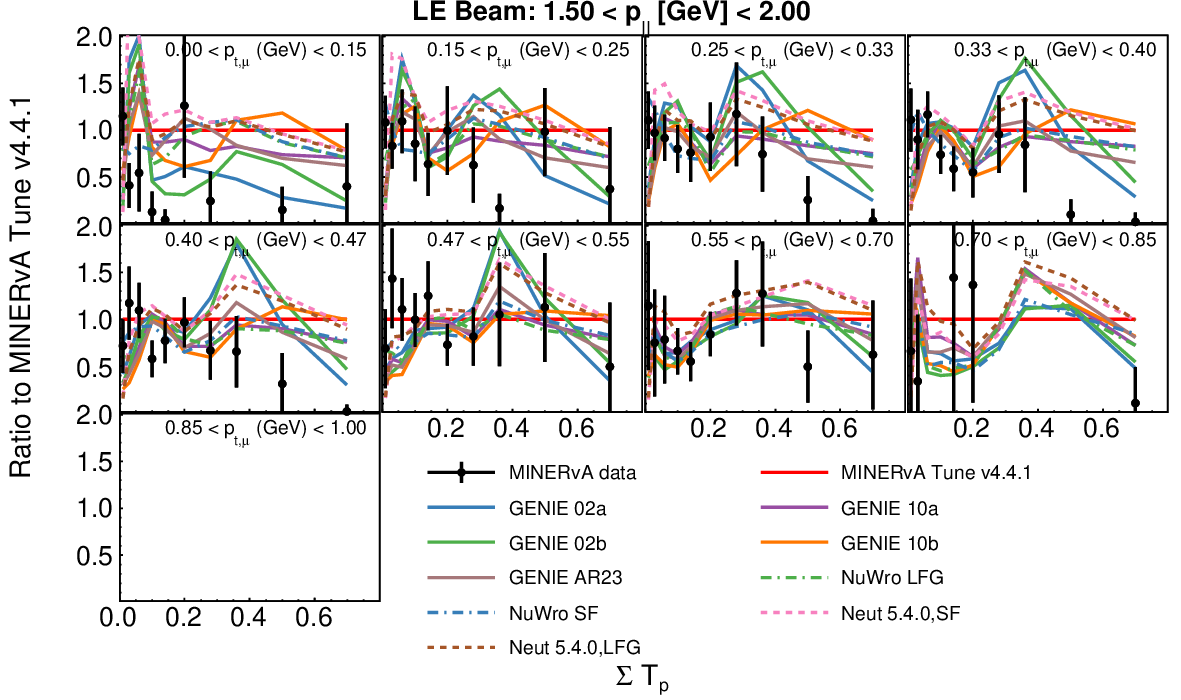}
    \includegraphics[width=\triplet]{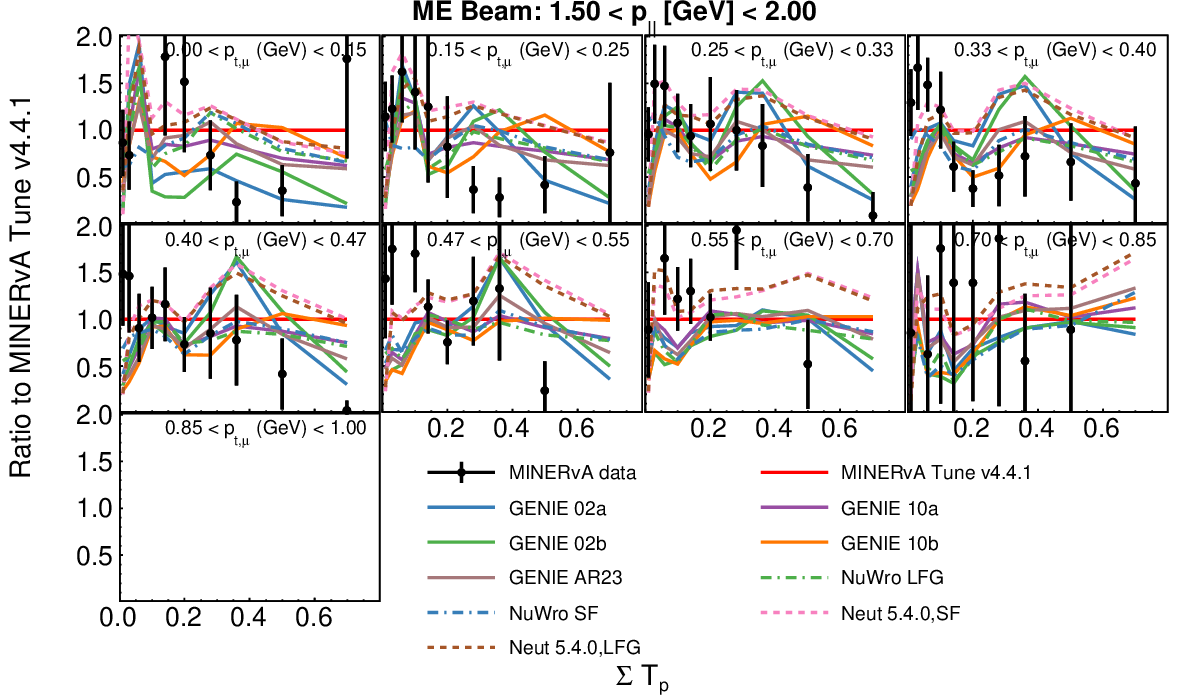}         \includegraphics[width=\triplet]{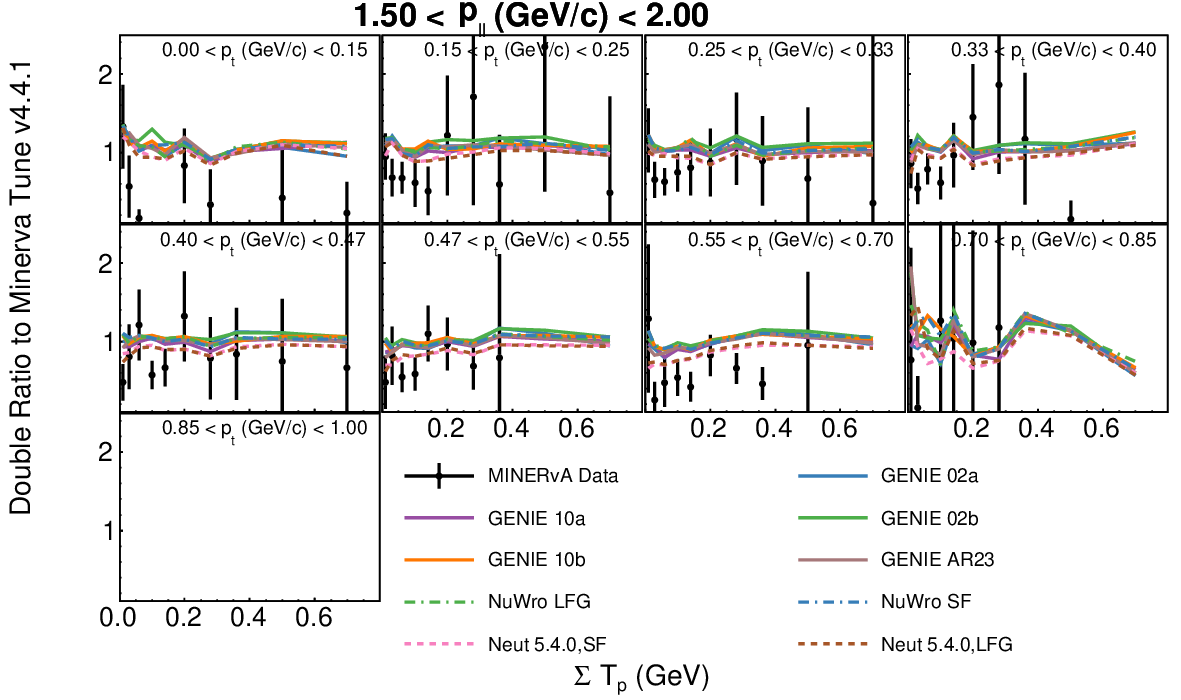}
    \caption{Top (Low Energy), Middle (Medium Energy): Measured cross sections divided by the MINERvA tune, and other generator choices divided by the MINERvA tune as a function of \pt,\sumtp, for \pz\ in bin 1, and Bottom (Low to Medium Energy) ratio divided by the MINERvA tune ratio, and other generator choices divided by the MINERvA tune ratio.}
    \label{fig:ptpzsumtp_bin1_otherModels}
\end{figure*}
\begin{figure*}[tp]
    \centering    
    \includegraphics[width=\triplet]{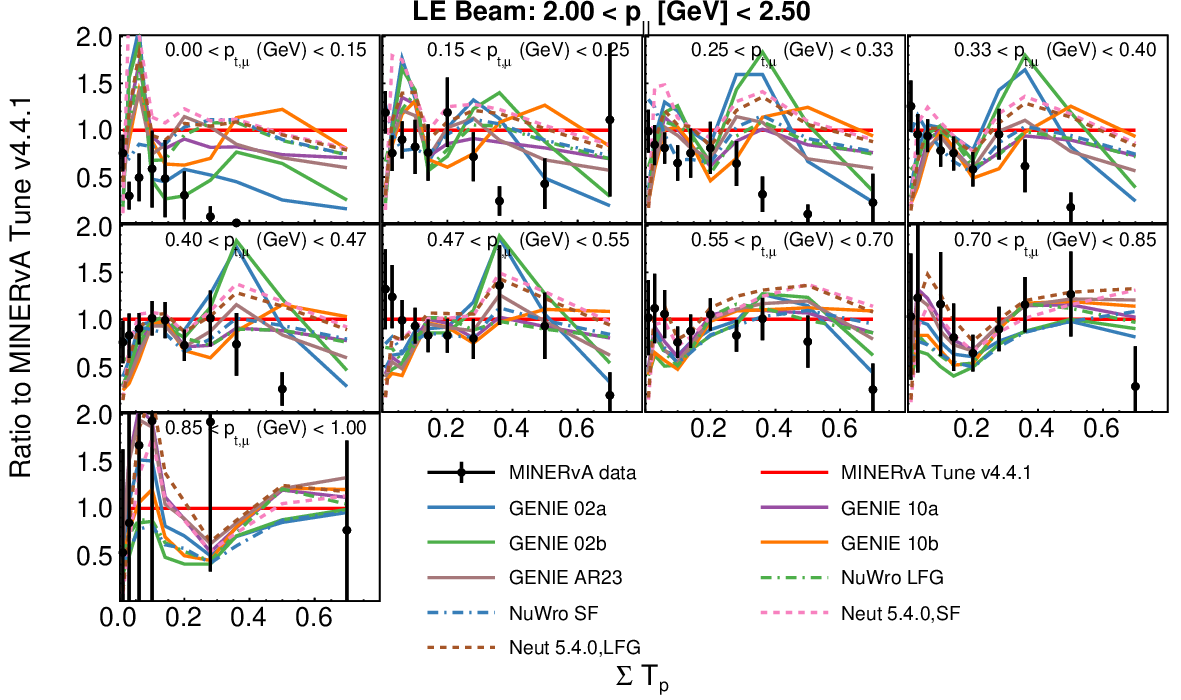}
    \includegraphics[width=\triplet]{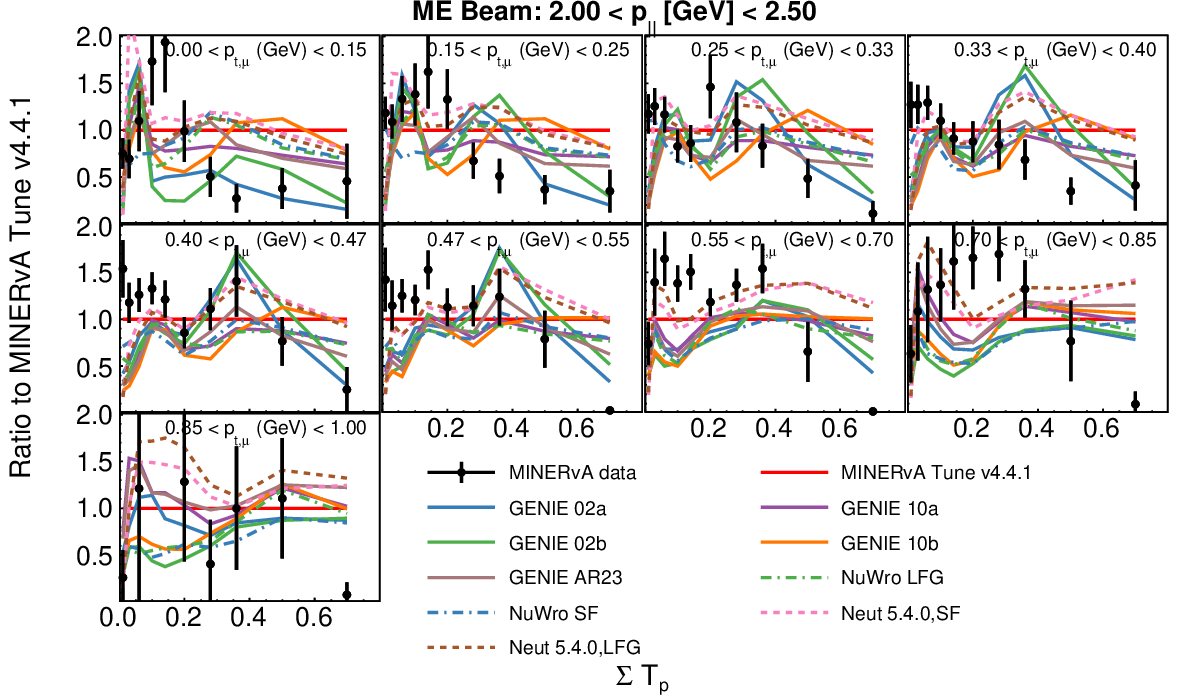}
    \includegraphics[width=\triplet]{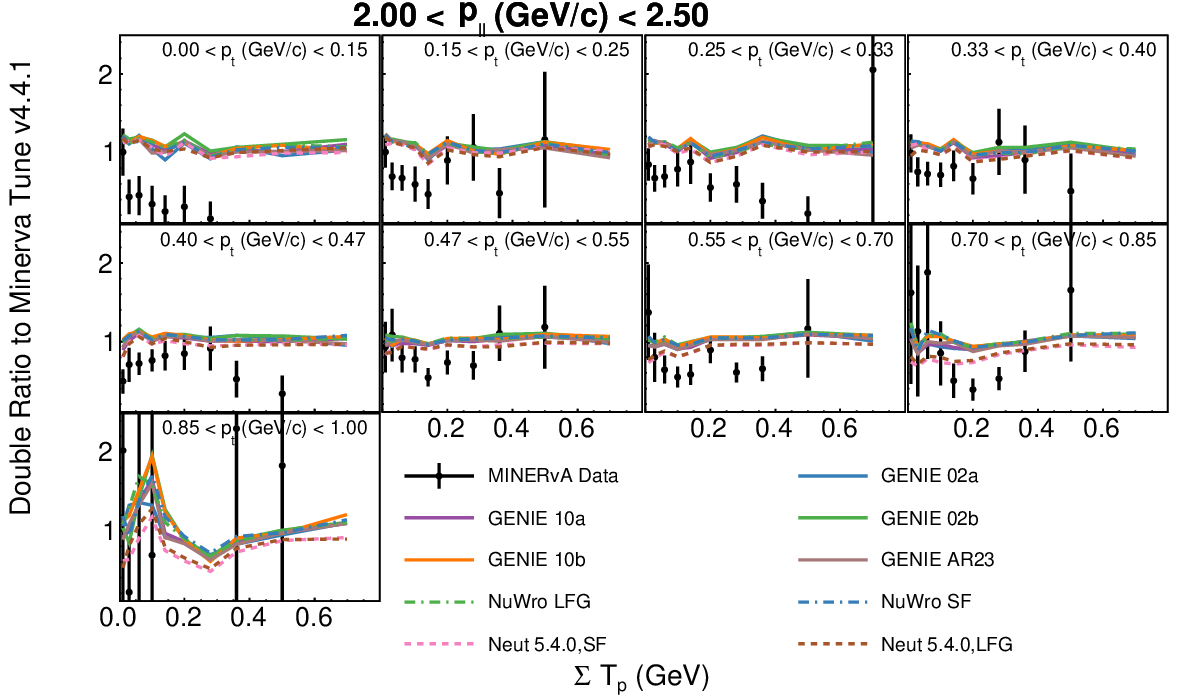}
    \caption{Top (Low Energy), Middle (Medium Energy): Measured cross sections divided by the MINERvA tune, and other generator choices divided by the MINERvA tune as a function of \pt,\sumtp, for \pz\ in bin 2, and Bottom (Low to Medium Energy) ratio divided by the MINERvA tune ratio, and other generator choices divided by the MINERvA tune ratio.}
\label{fig:ptpzsumtp_bin2_otherModels}
\end{figure*}
\begin{figure*}[tp]
    \centering    
    \includegraphics[width=\triplet]{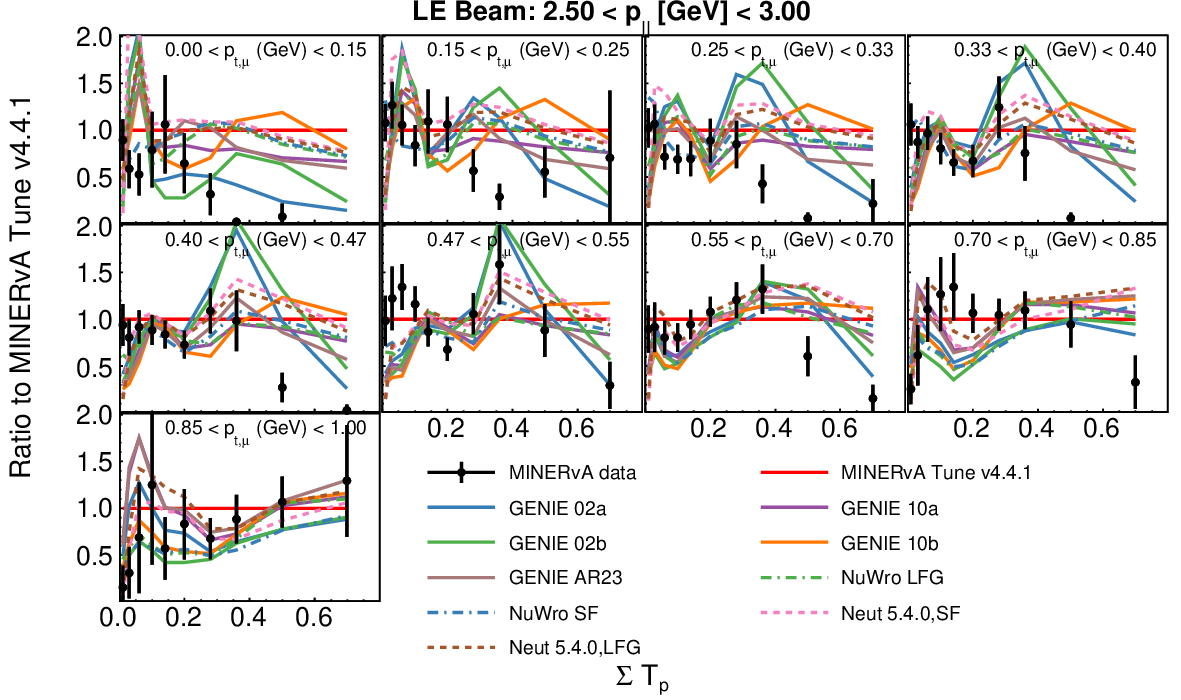}
    \includegraphics[width=\triplet]{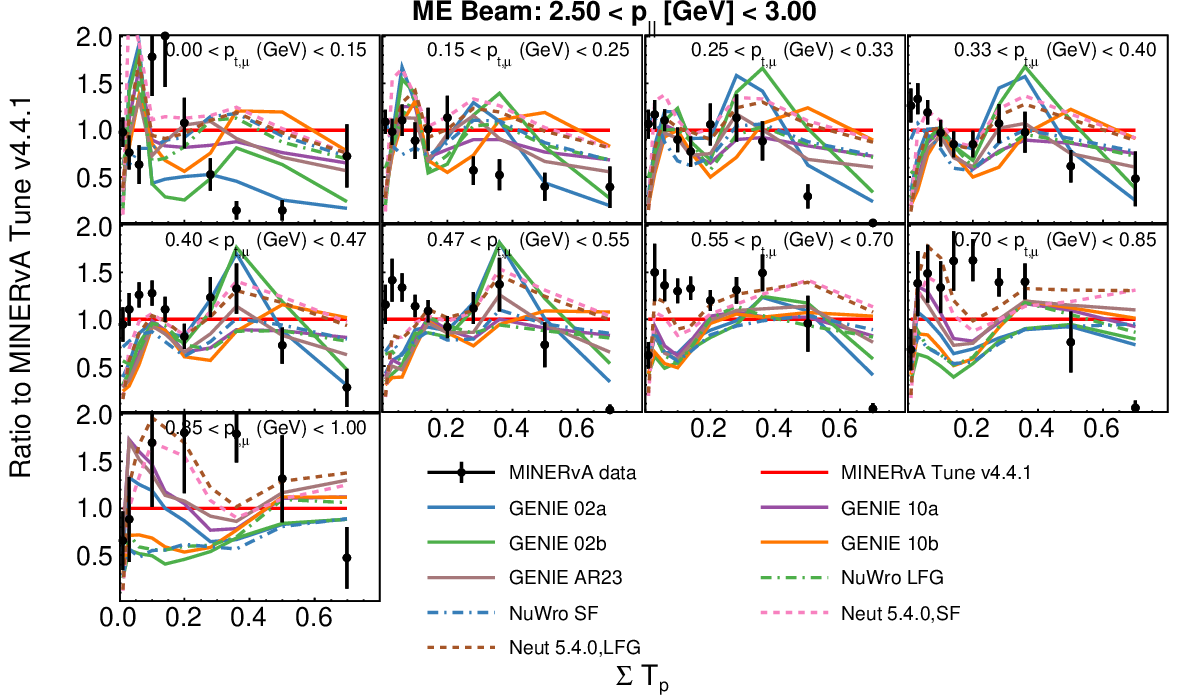}
        \includegraphics[width=\triplet]{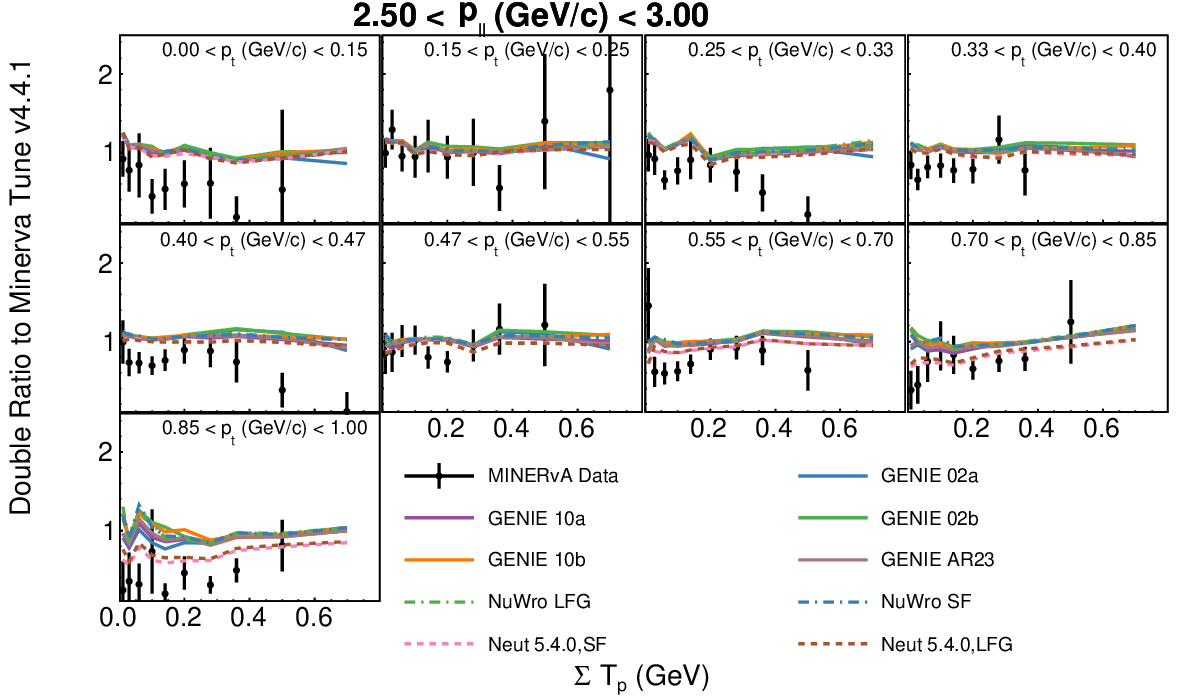}    
        \caption{Top (Low Energy), Middle (Medium Energy): Measured cross sections divided by the MINERvA tune, and other generator choices divided by the MINERvA tune as a function of \pt,\sumtp, for \pz\ in bin 3, and Bottom (LE/ME) ratio divided by the MINERvA tune ratio, and other generator choices divided by the MINERvA tune ratio.}
\label{fig:fig:ptpzsumtp_bin3_otherModels}
\end{figure*}
\begin{figure*}[tp]
    \centering    
    \includegraphics[width=\triplet]{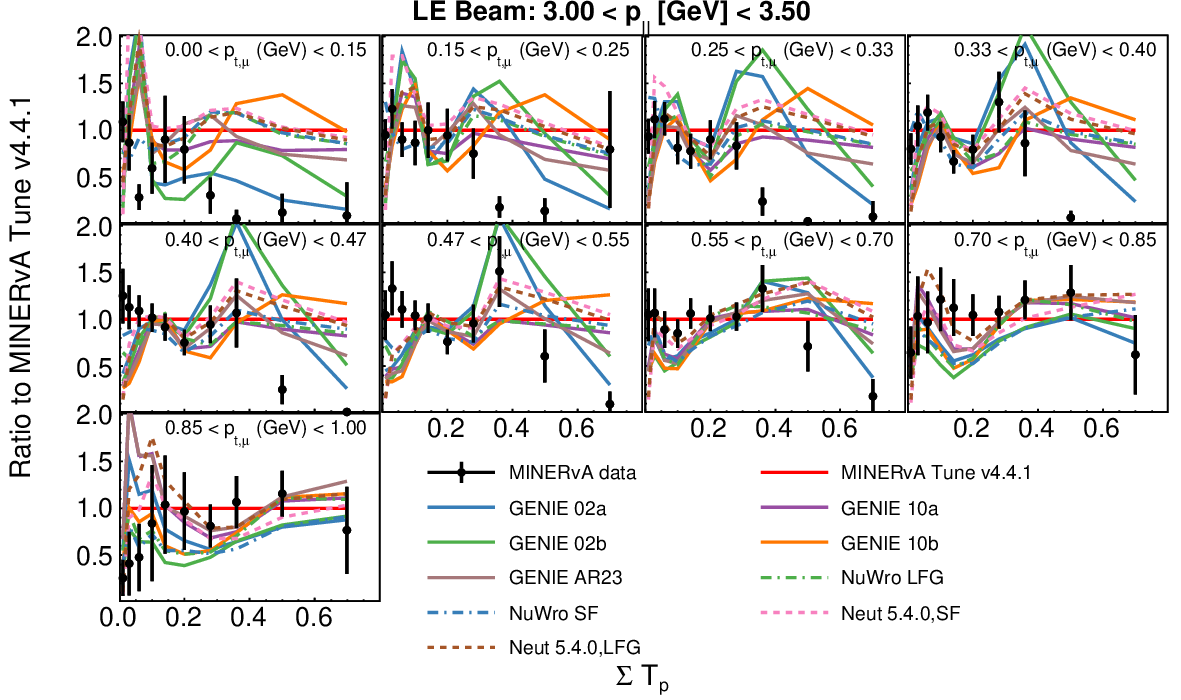}
    \includegraphics[width=\triplet]{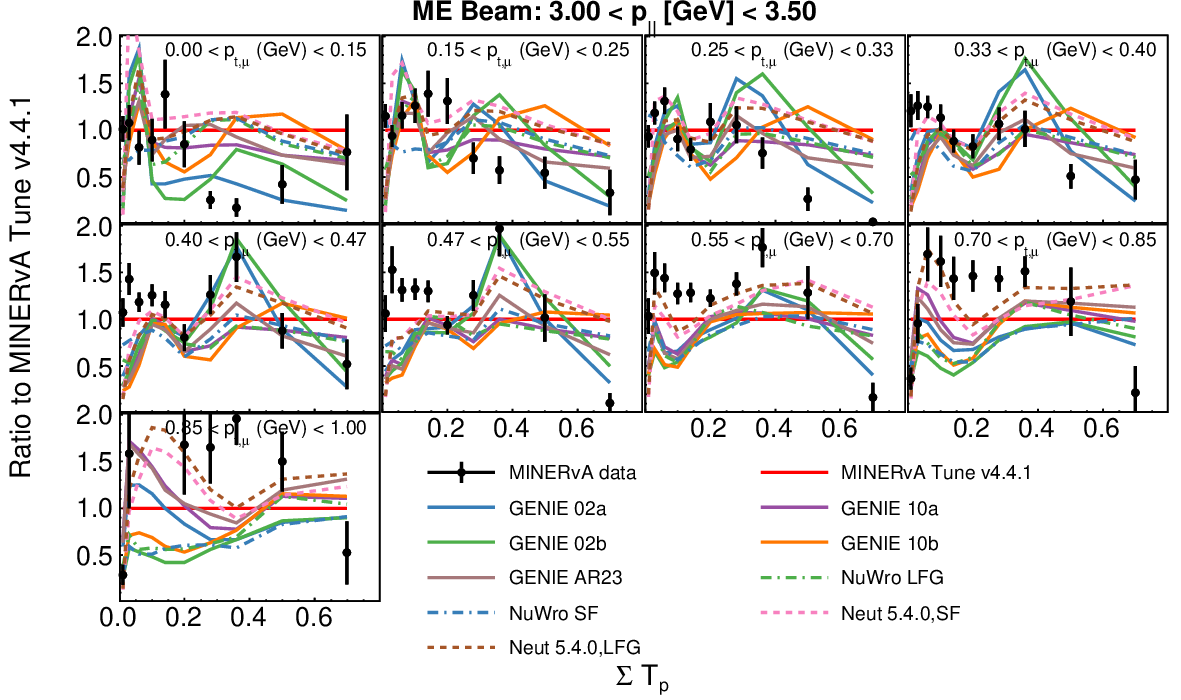}
        \includegraphics[width=\triplet]{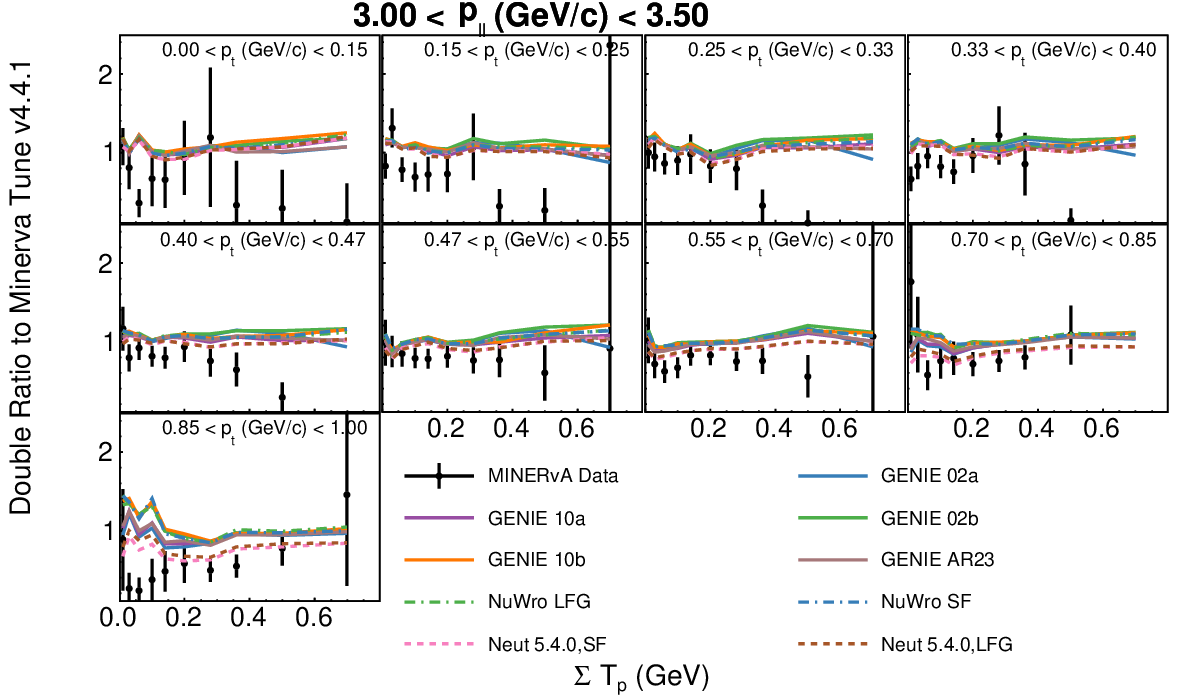}
    \caption{Top (Low Energy), Middle (Medium Energy): Measured cross sections divided by the MINERvA tune, and other generator choices divided by the MINERvA tune as a function of \pt,\sumtp, for \pz\ in bin 4, and Bottom (Low to Medium Energy) ratio divided by the MINERvA tune ratio, and other generator choices divided by the MINERvA tune ratio.}
    \label{fig:fig:ptpzsumtp_bin4_otherModels}
\end{figure*}
\begin{figure*}[tp]
    \centering    
    \includegraphics[width=\triplet]{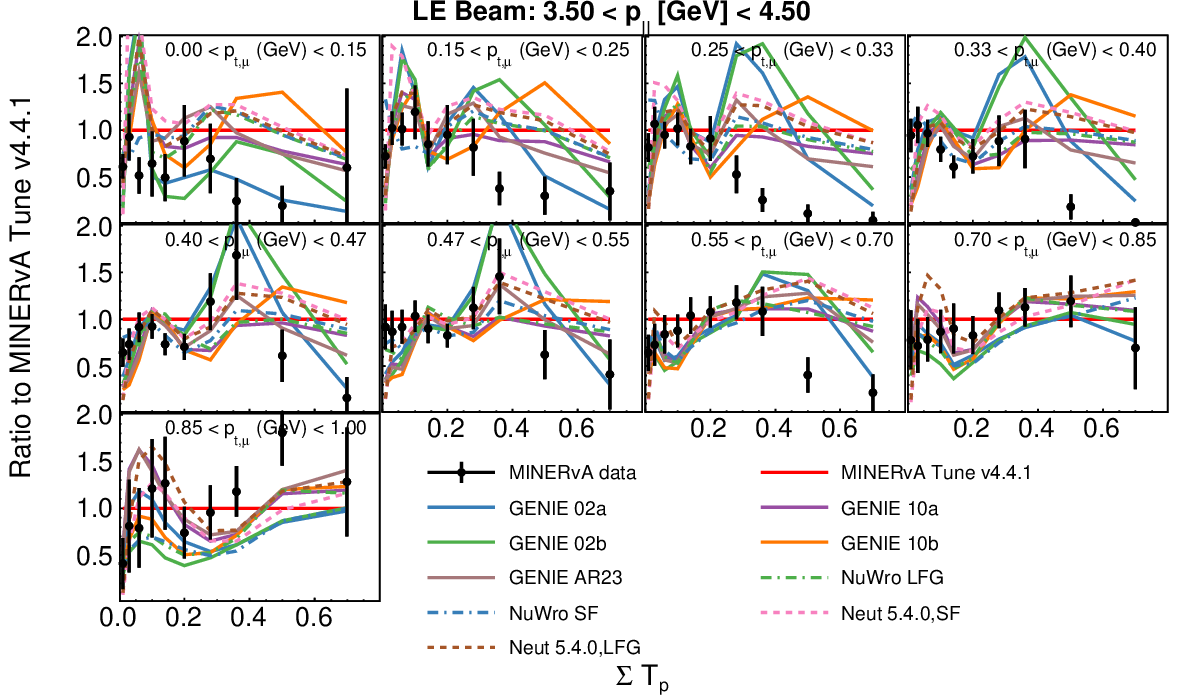}
    \includegraphics[width=\triplet]{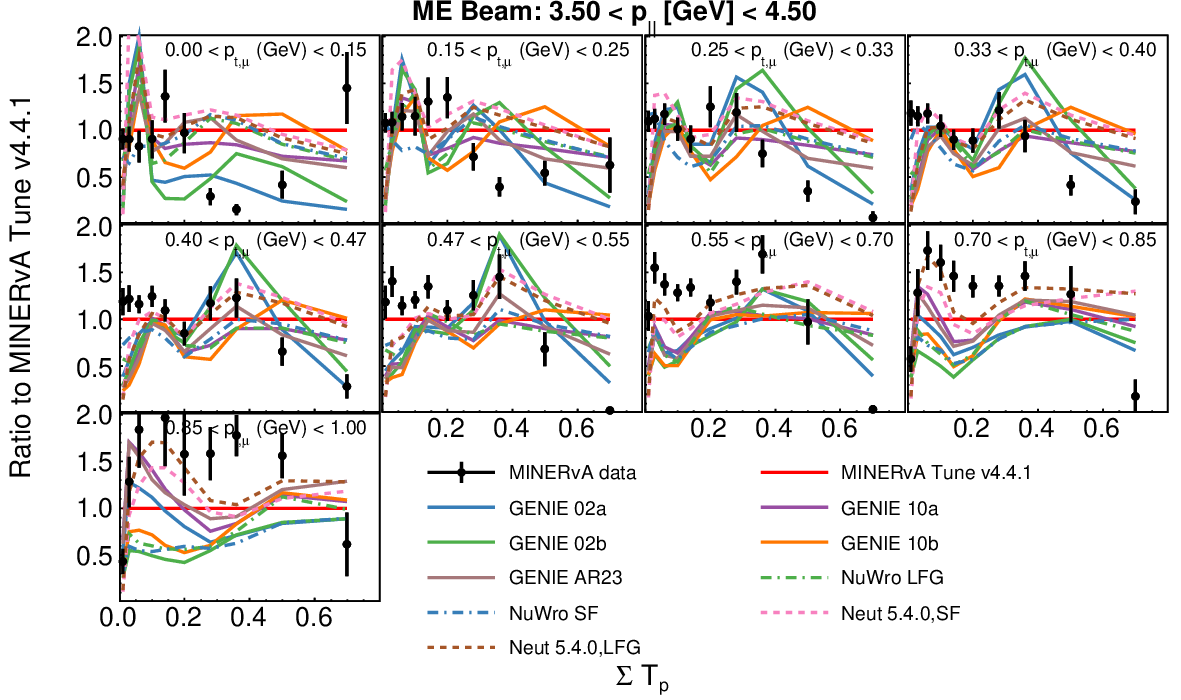}
        \includegraphics[width=\triplet]{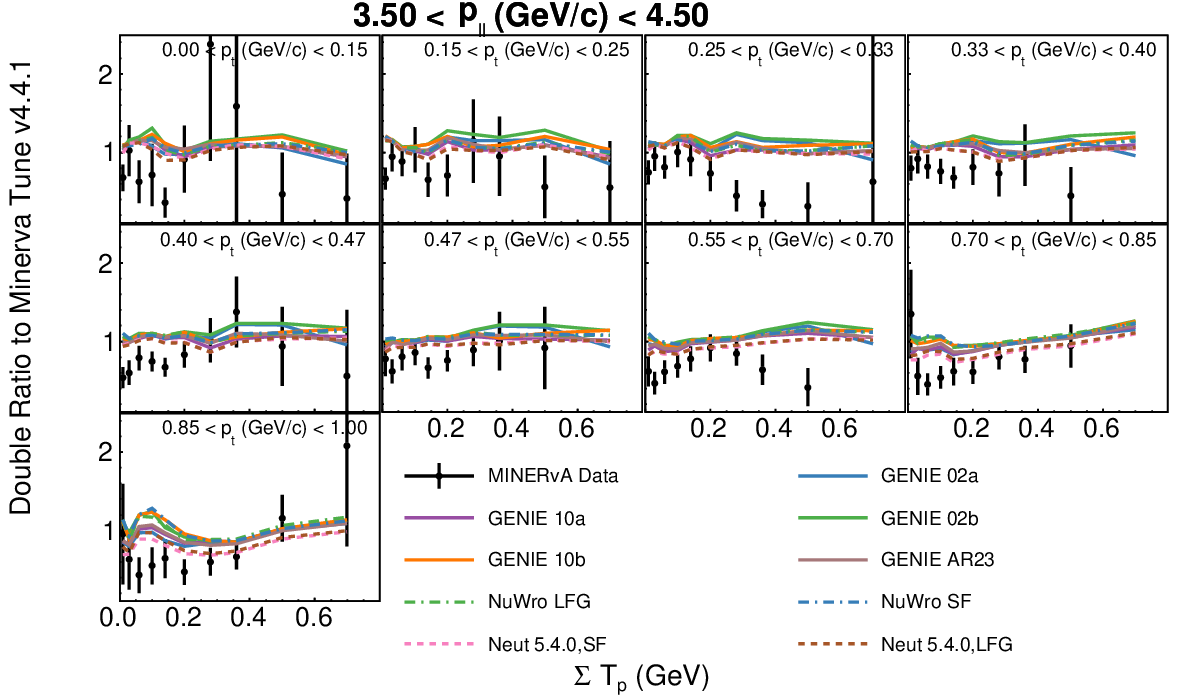}
    \caption{Top (Low Energy), Middle (Medium Energy): Measured cross sections divided by the MINERvA tune, and other generator choices divided by the MINERvA tune as a function of \pt,\sumtp, for \pz\ in bin 5, and Bottom (Low to Medium Energy) ratio divided by the MINERvA tune ratio, and other generator choices divided by the MINERvA tune ratio.}
\label{fig:ptpzsumtp_bin5_otherModels}
\end{figure*}

\FloatBarrier
\newpage 

\subsection{Model Component and Energy Dependence}
All plots are (model/GENIE10a)LE / (model/GENIE10a)ME. GENIE10a was chosen as the basis model as that is what is used for a number of data double ratio plots in the triple panel section of the paper, aka LE (top) ME (middle) double ratio (bottom)

\subsection{GENIE 02a}
\begin{figure*}[tp]
    \centering    
    \includegraphics[width=\triplet]{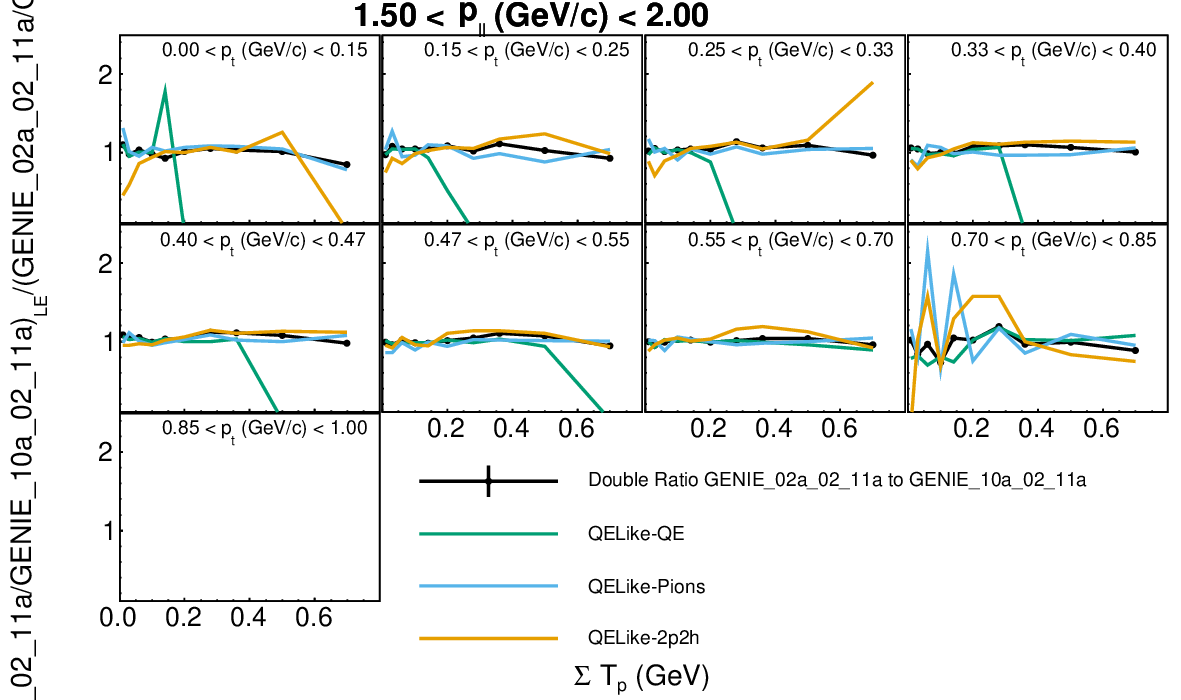}
    \includegraphics[width=\triplet]{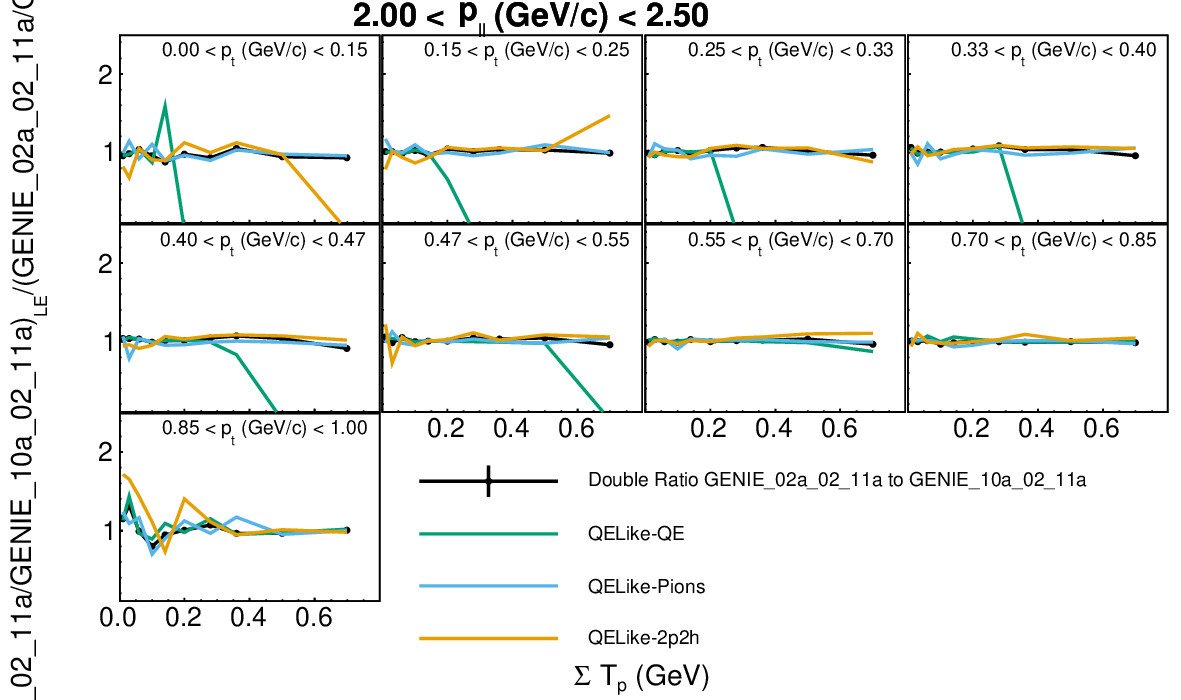}
        \includegraphics[width=\triplet]{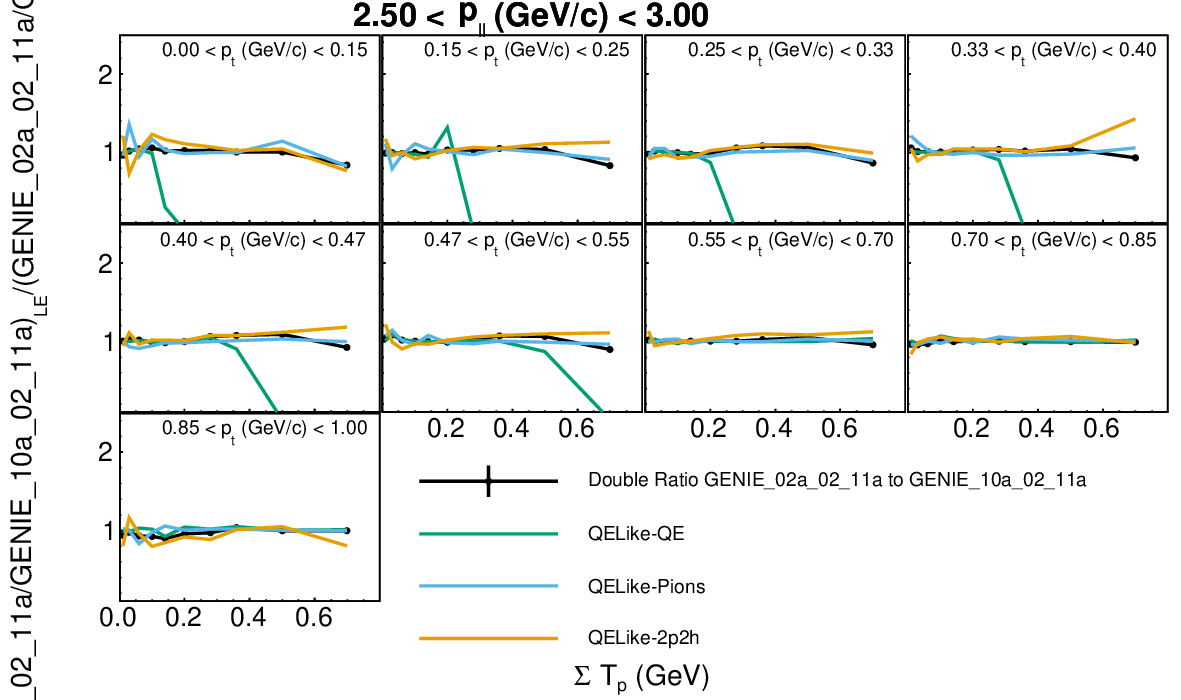}
    \caption{Pz bins 1 (top) and 2(middle) and 3 (bottom) comparing GENIE 02a against GENIE 10a.  The double ratios are also shown for individual sub-processes contributing to the signal, quasielastic (QE), multinucleon knockout (2p2h), and inelastic production of new particles (Pions).}
    \label{fig:ptpzsumtp_bin123_otherModels_G2a}
\end{figure*}

\begin{figure*}[tp]
    \centering    
    \includegraphics[width=\triplet]{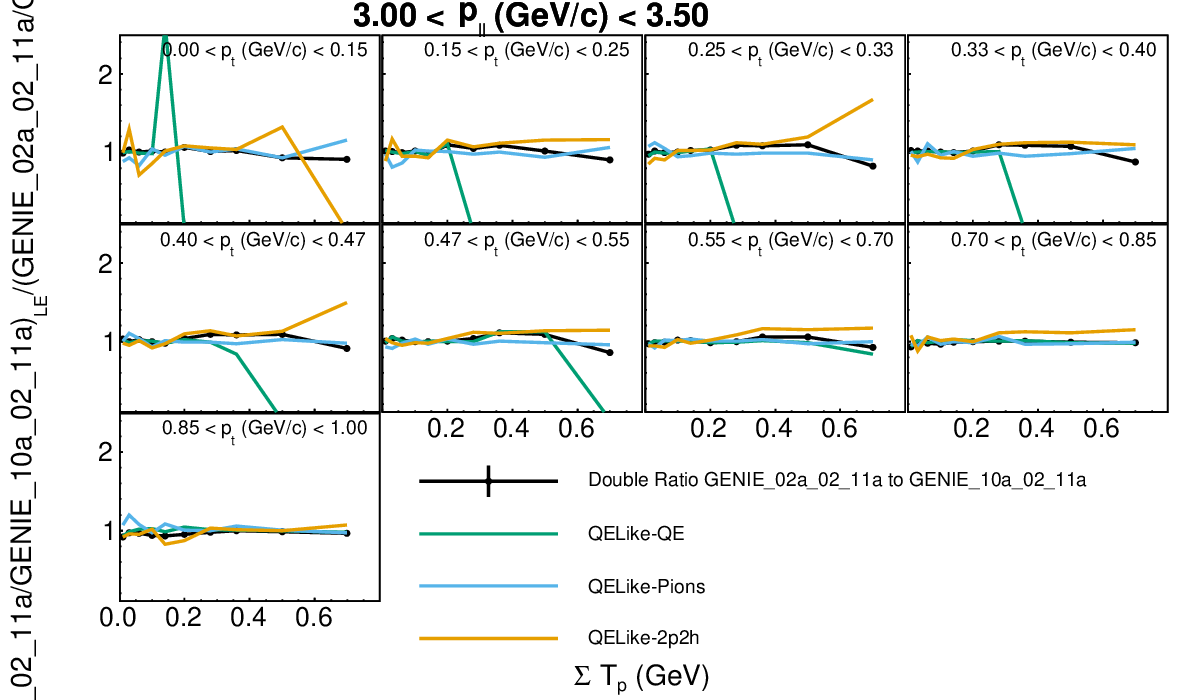}
    \includegraphics[width=\triplet]{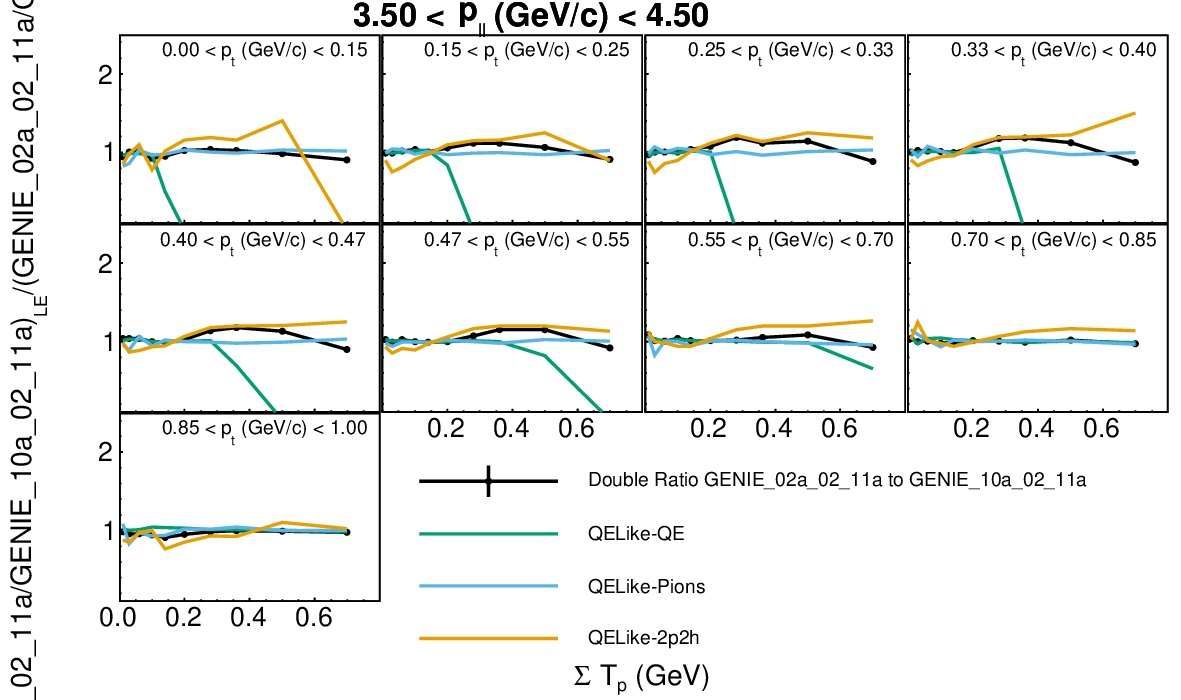}
    \caption{Pz bins 4 (top) and 5(bottom) comparing GENIE 02a against GENIE 10a.  The double ratios are also shown for individual sub-processes contributing to the signal, quasielastic (QE), multinucleon knockout (2p2h), and inelastic production of new particles (Pions).}
    \label{fig:ptpzsumtp_bin45_otherModels_G2a}
\end{figure*}

\subsection{GENIE 02b}
\begin{figure*}[tp]
    \centering    
    \includegraphics[width=\triplet]{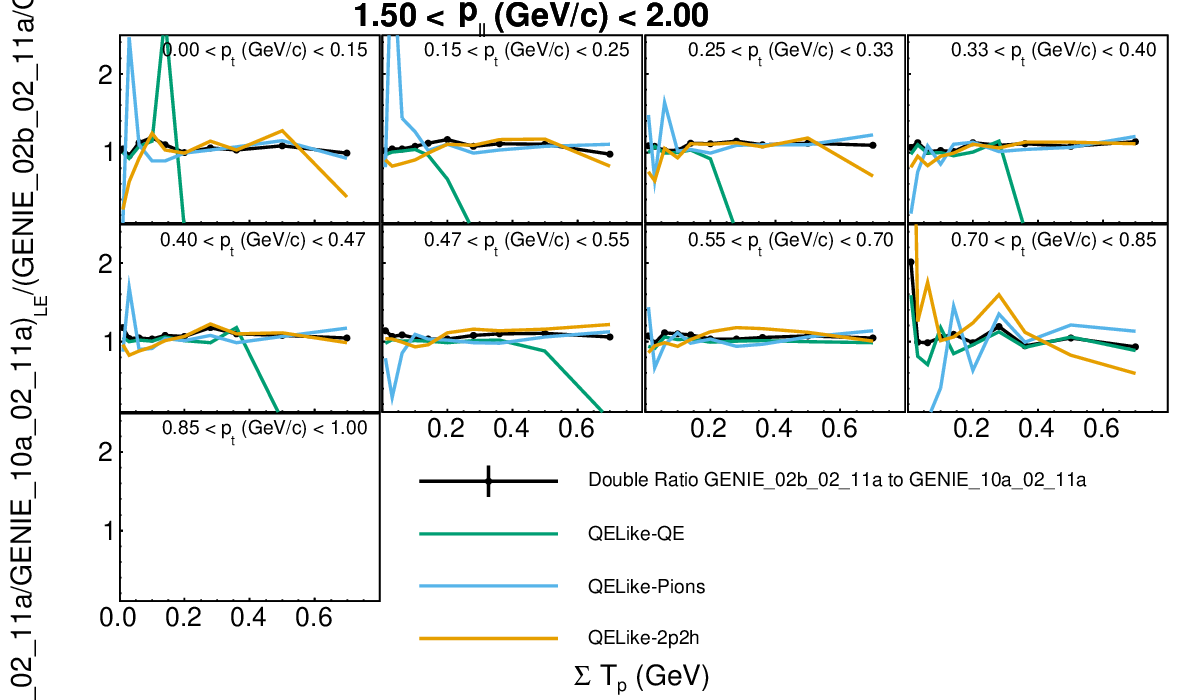}
    \includegraphics[width=\triplet]{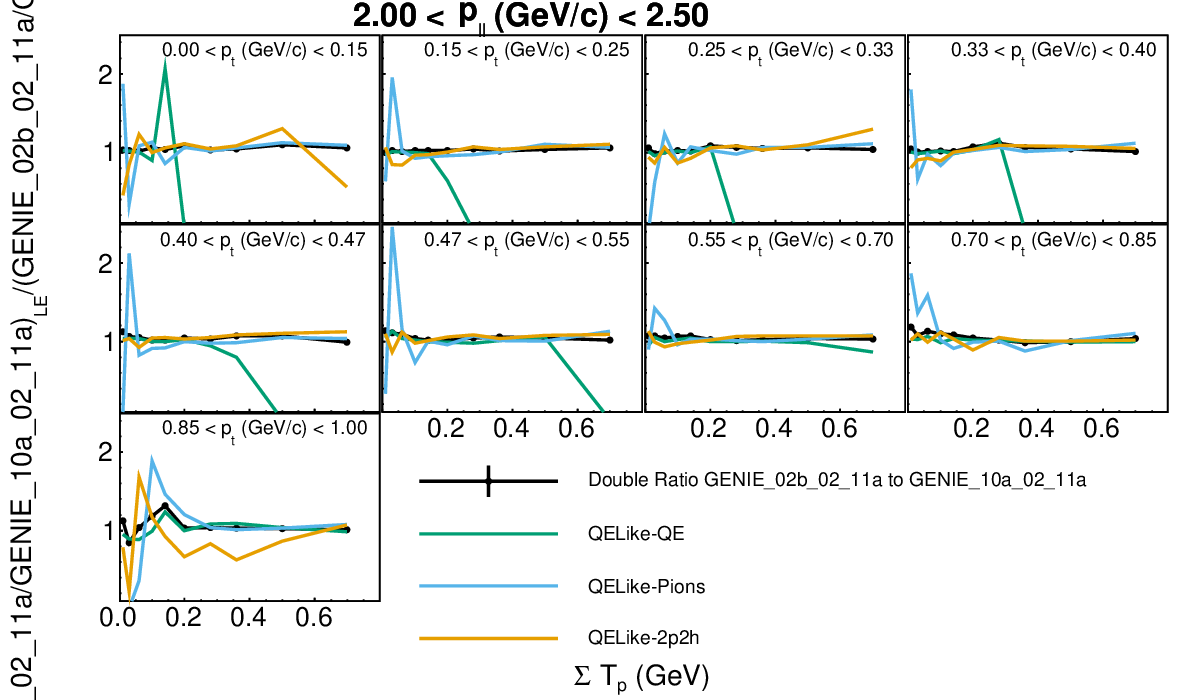}
        \includegraphics[width=\triplet]{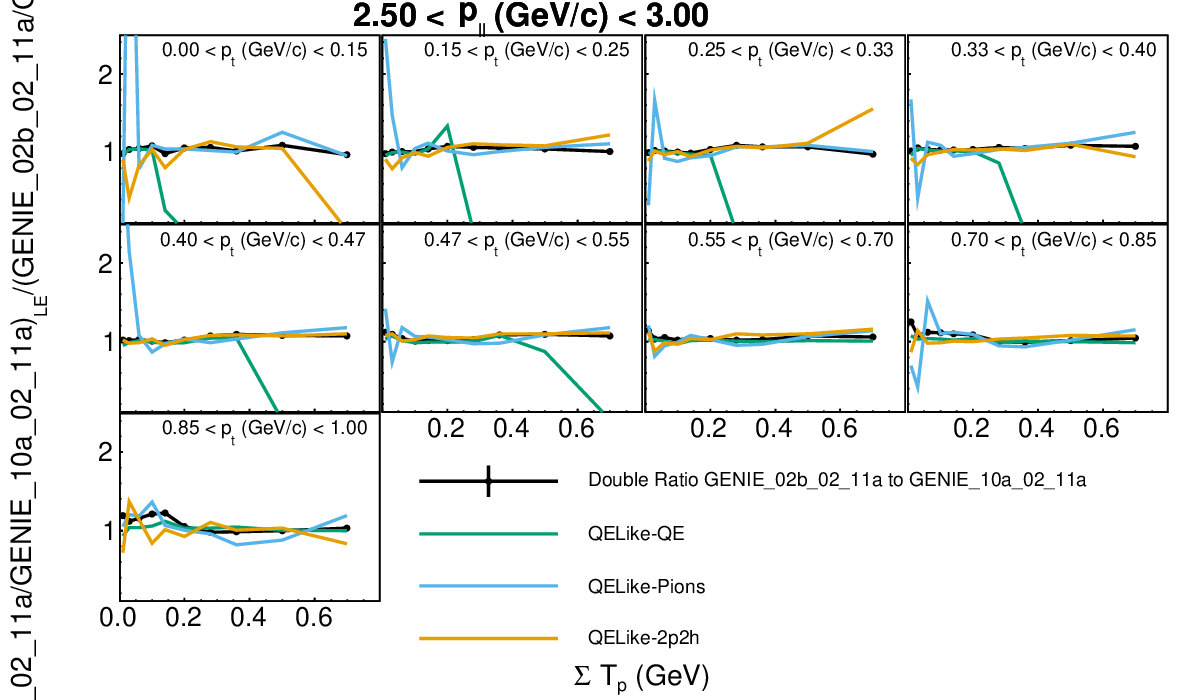}
    \caption{Pz bins 1 (top) and 2(bottom) comparing GENIE 02b against GENIE 10a.  The double ratios are also shown for individual sub-processes contributing to the signal, quasielastic (QE), multinucleon knockout (2p2h), and inelastic production of new particles (Pions).}
    \label{fig:ptpzsumtp_bin123_otherModels_G2b}
\end{figure*}

\begin{figure*}[tp]
    \centering    
    \includegraphics[width=\triplet]{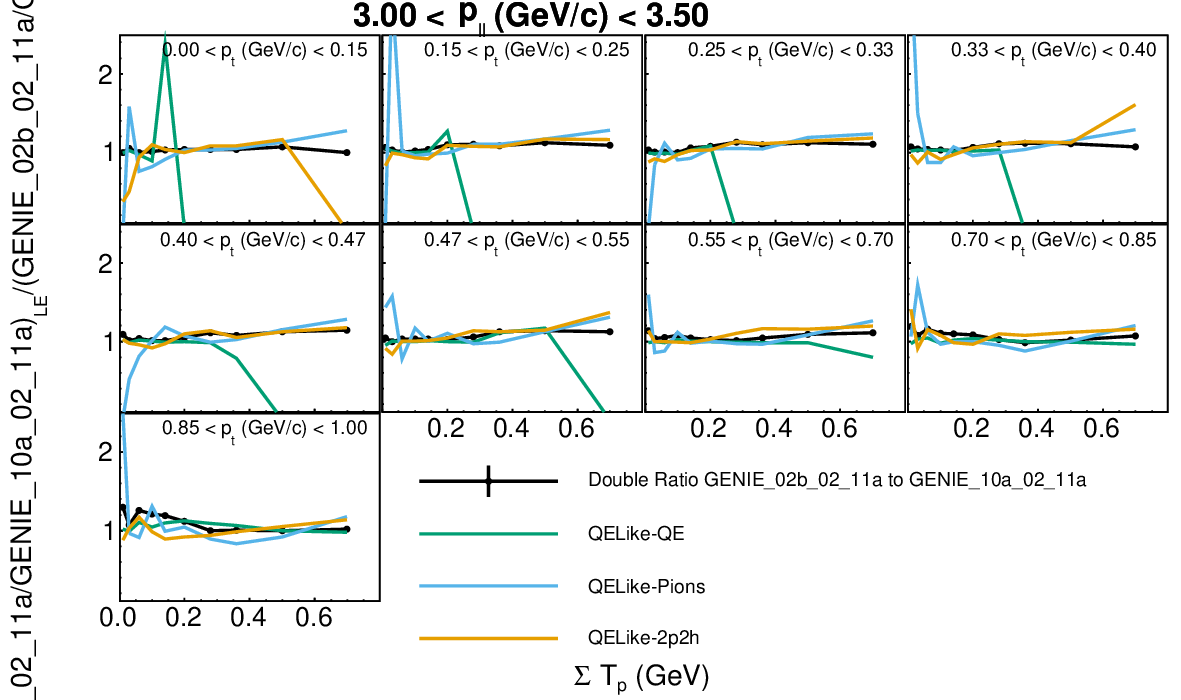}
    \includegraphics[width=\triplet]{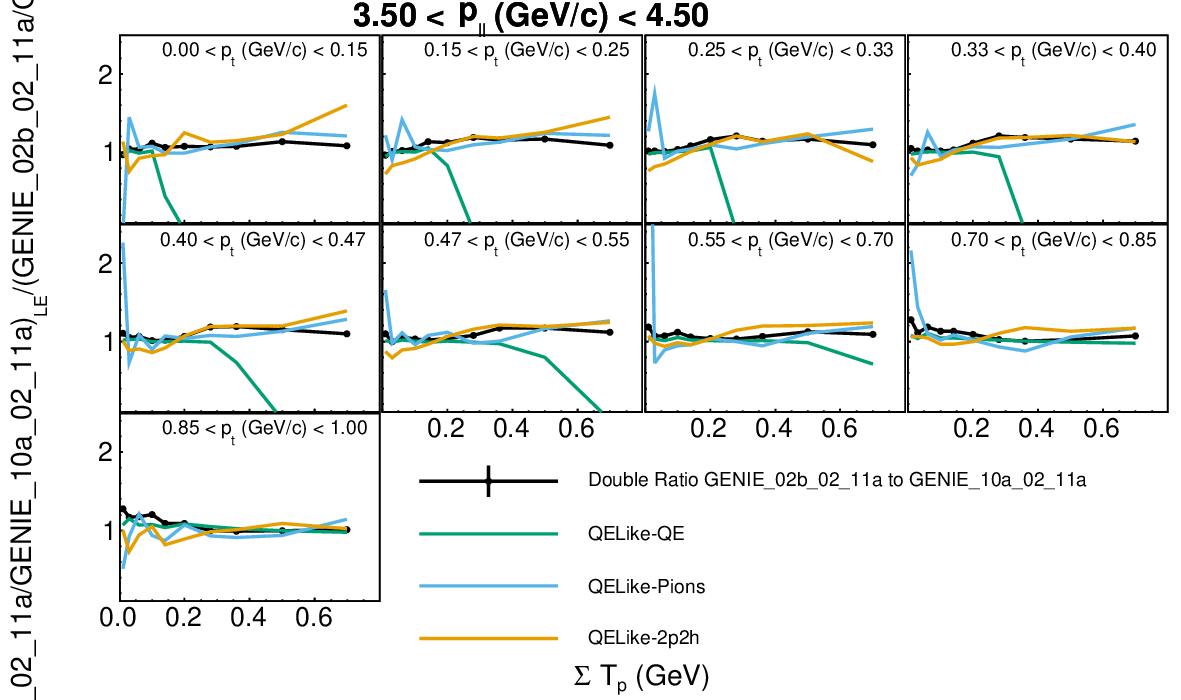}
    \caption{Pz bins 4 (top) and 5(bottom) comparing GENIE 02b against GENIE 10a.  The double ratios are also shown for individual sub-processes contributing to the signal, quasielastic (QE), multinucleon knockout (2p2h), and inelastic production of new particles (Pions).}
    \label{fig:ptpzsumtp_bin45_otherModels_G2b}
\end{figure*}

\subsection{GENIE 10b}
\begin{figure*}[tp]
    \centering    
    \includegraphics[width=\triplet]{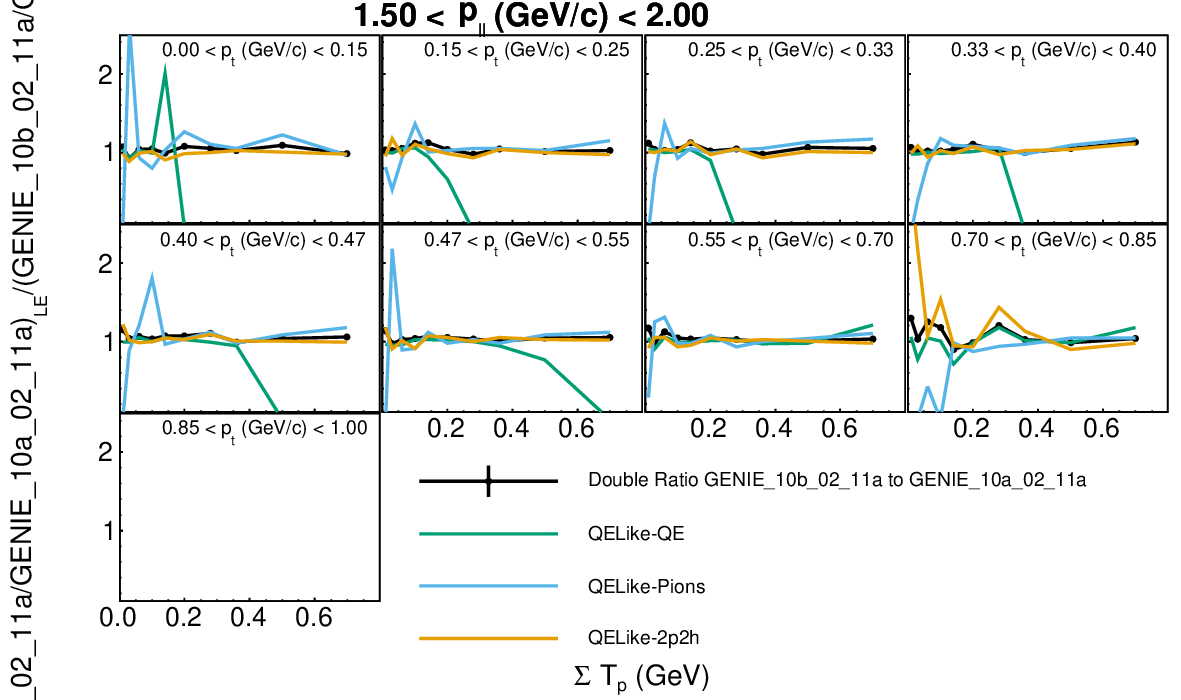}
    \includegraphics[width=\triplet]{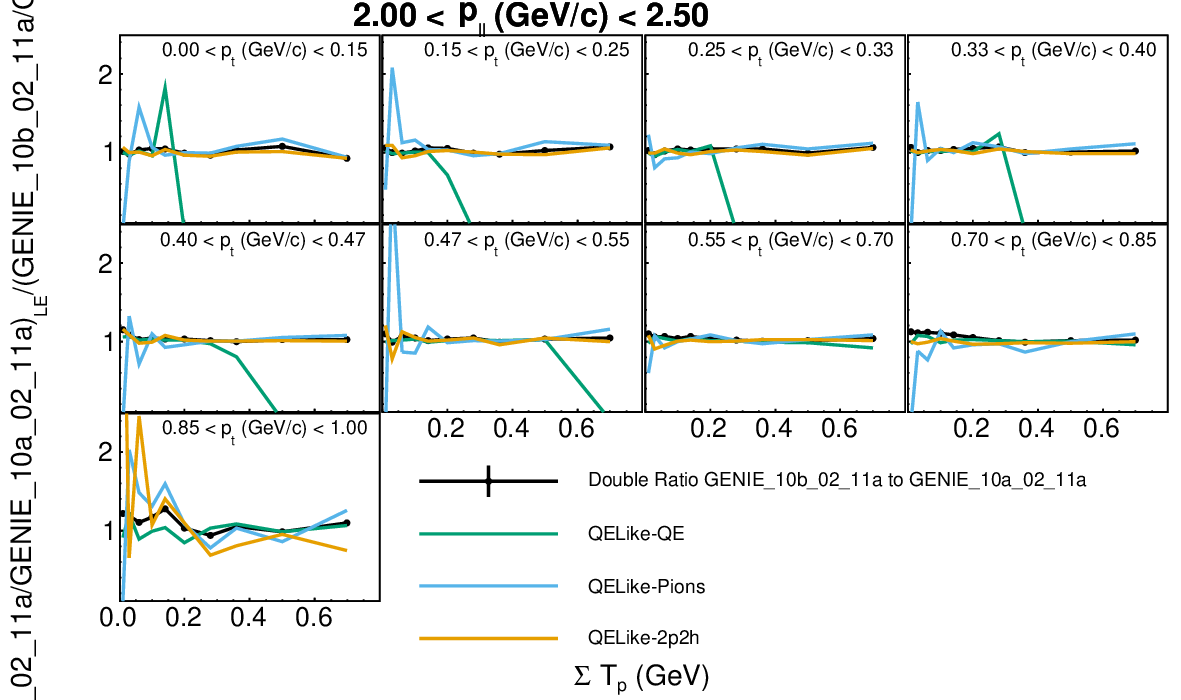}
        \includegraphics[width=\triplet]{NewPlots/DoubleRatio_ExternalModels_Models/GENIE_10b/LE_ME_RATIO_2.eps}
    \caption{Pz bins 1 (top) and 2(middle) and 3 (bottom) comparing GENIE 10b against GENIE 10a.  The double ratios are also shown for individual sub-processes contributing to the signal, quasielastic (QE), multinucleon knockout (2p2h), and inelastic production of new particles (Pions).}
    \label{fig:ptpzsumtp_bin523_otherModels_G10b}
\end{figure*}

\begin{figure*}[tp]
    \centering    
    \includegraphics[width=\triplet]{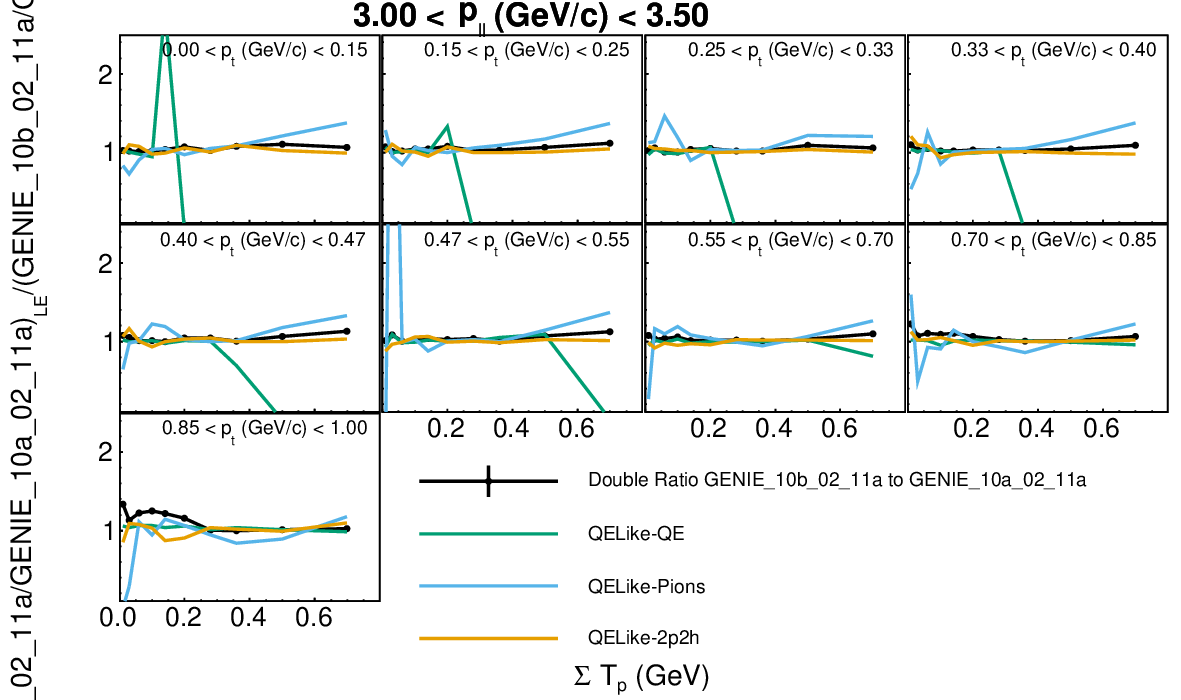}
    \includegraphics[width=\triplet]{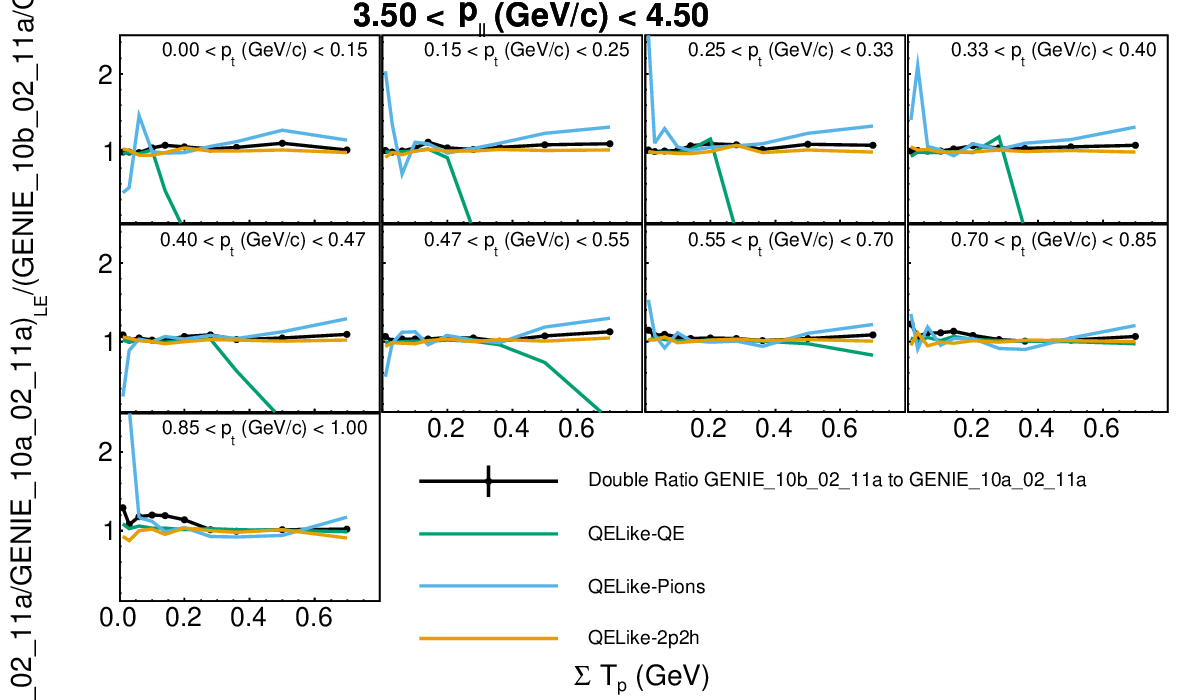}
    \caption{Pz bins 4 (top) and 5(bottom) comparing GENIE 10b against GENIE 10a.  The double ratios are also shown for individual sub-processes contributing to the signal, quasielastic (QE), multinucleon knockout (2p2h), and inelastic production of new particles (Pions).}
    \label{fig:ptpzsumtp_bin45_otherModels_G10b}
\end{figure*}

\subsection{GENIE AR23}
\begin{figure*}[tp]
    \centering    
    \includegraphics[width=\triplet]{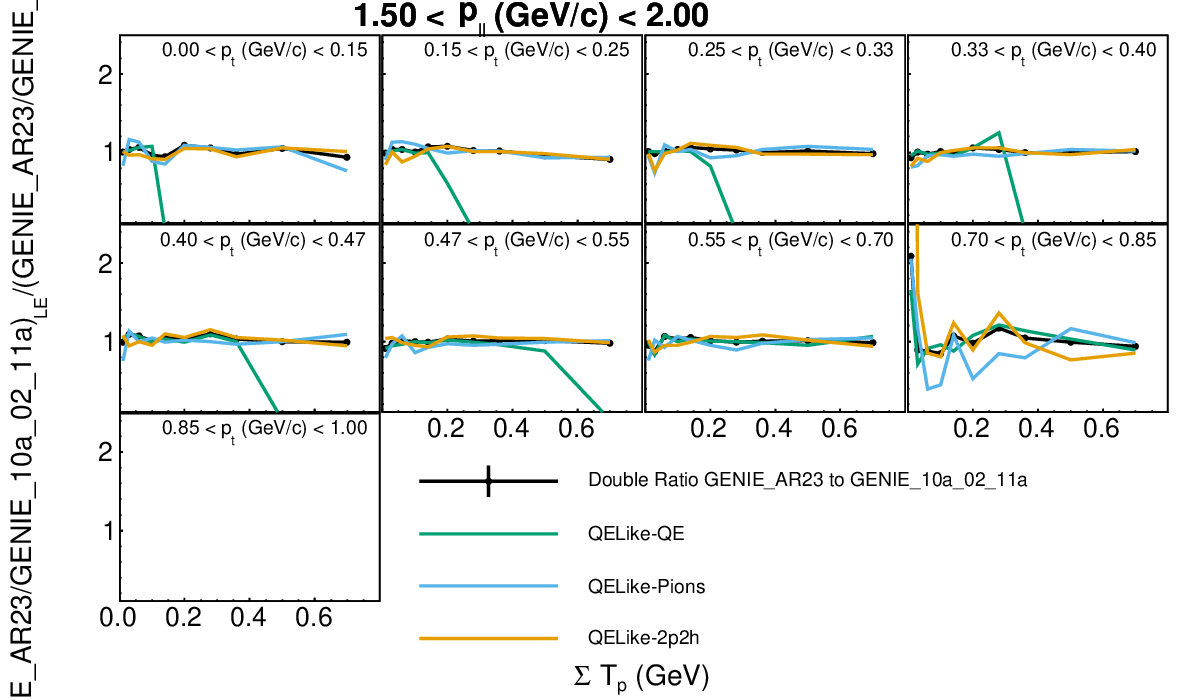}
    \includegraphics[width=\triplet]{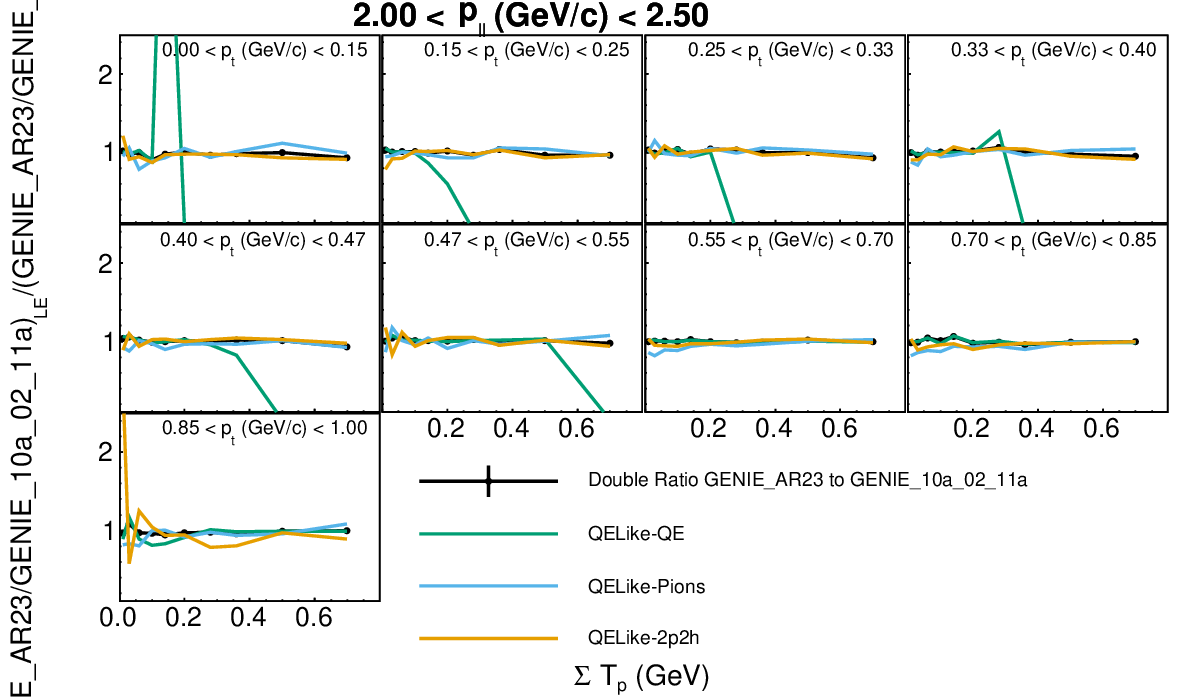}
        \includegraphics[width=\triplet]{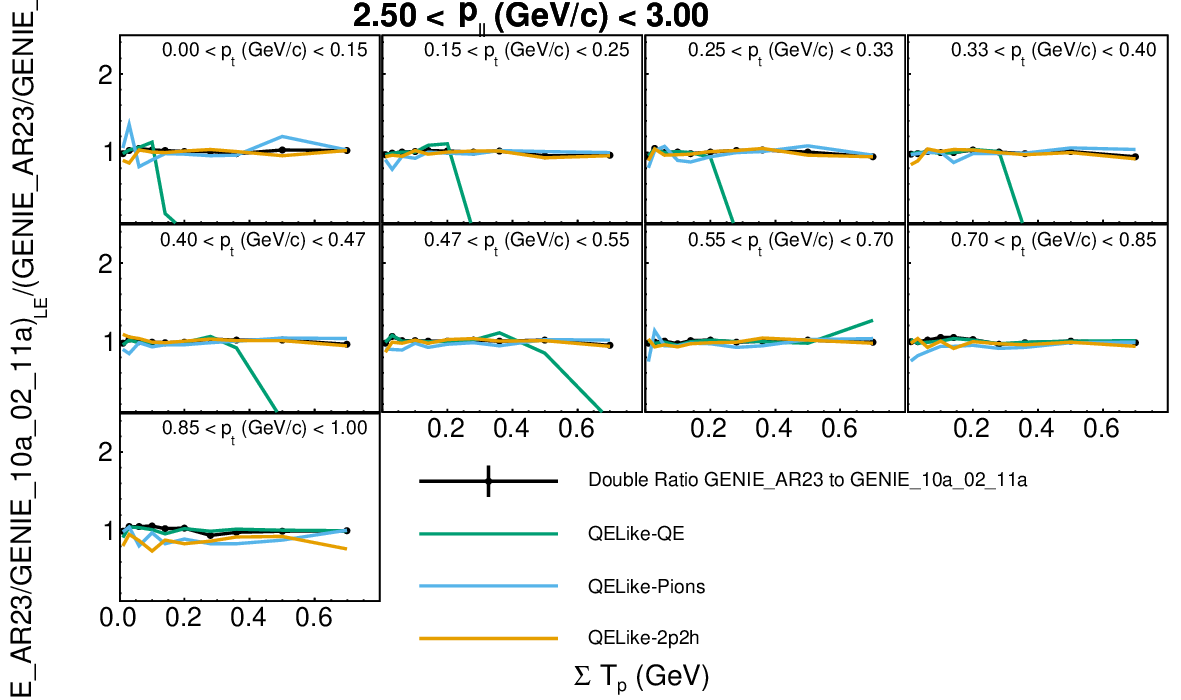}

    \caption{Pz bins 1 (top) 2(middle) and 3 (bottom) comparing GENIE AR23 against GENIE 10a.  The double ratios are also shown for individual sub-processes contributing to the signal, quasielastic (QE), multinucleon knockout (2p2h), and inelastic production of new particles (Pions).}
    \label{fig:ptpzsumtp_bin123_otherModels_GAR23}
\end{figure*}

\begin{figure*}[tp]
    \centering    
    \includegraphics[width=\triplet]{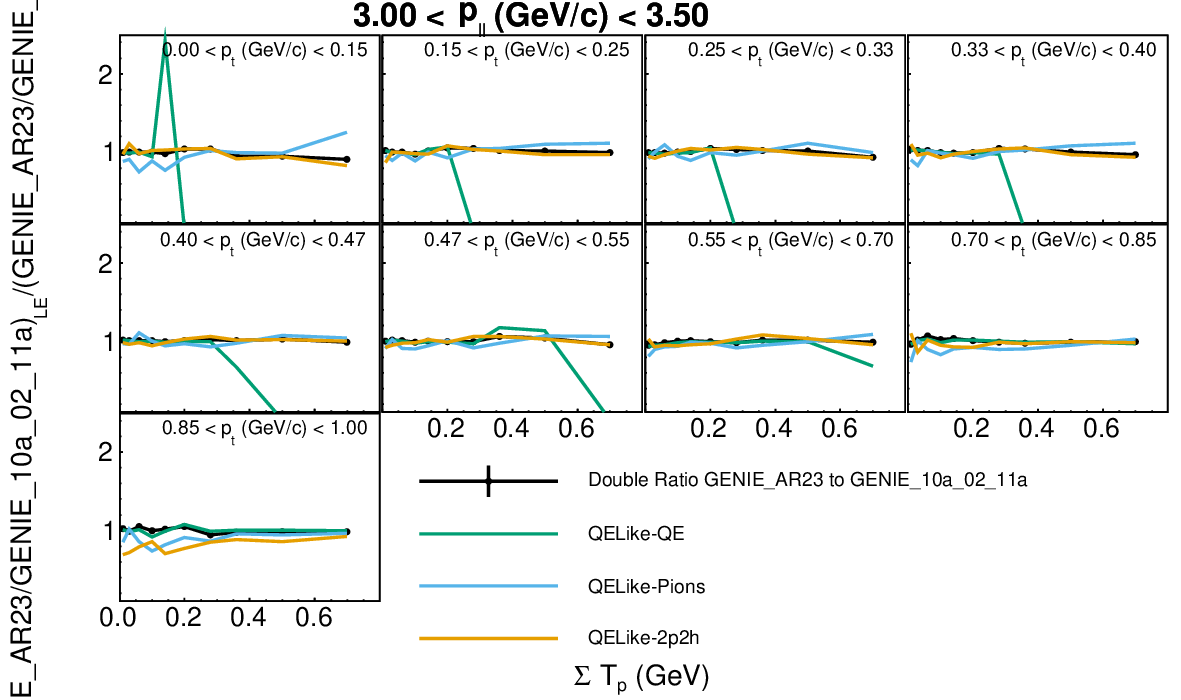}
    \includegraphics[width=\triplet]{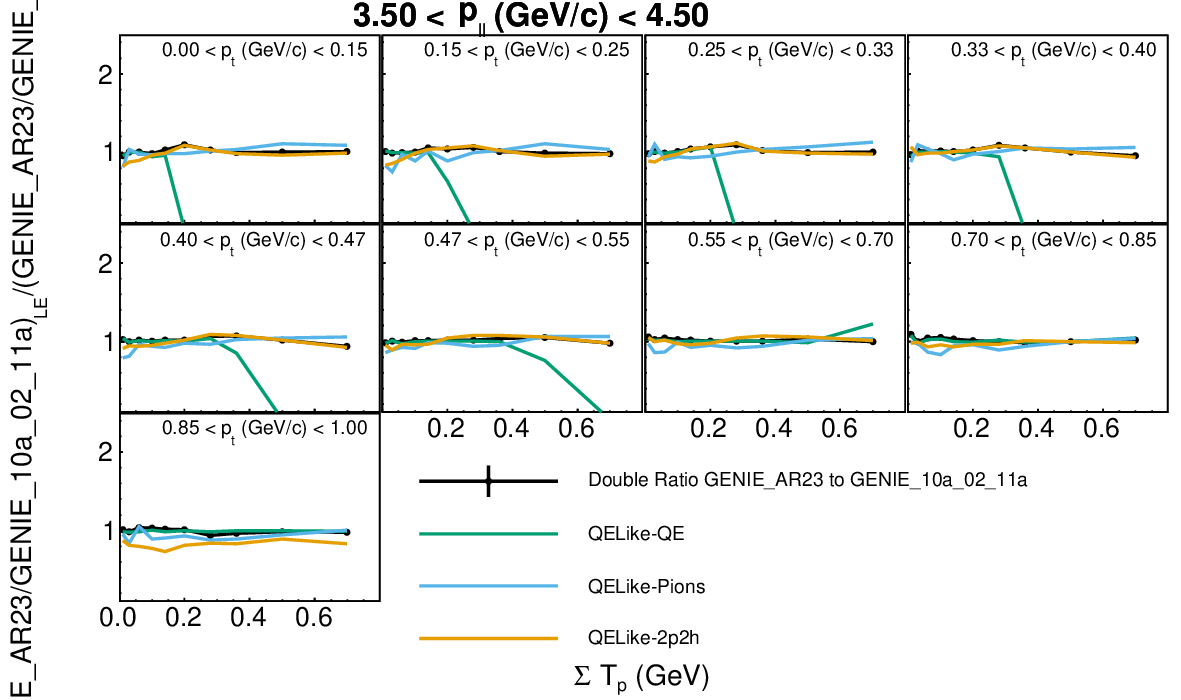}
    \caption{Pz bins 4 (top) and 5(bottom) comparing GENIE AR23 against GENIE 10a.  The double ratios are also shown for individual sub-processes contributing to the signal, quasielastic (QE), multinucleon knockout (2p2h), and inelastic production of new particles (Pions).}
    \label{fig:ptpzsumtp_bin45_otherModels_GAR23}
\end{figure*}

\subsection{NEUT SF}
\begin{figure*}[tp]
    \centering    
    \includegraphics[width=\triplet]{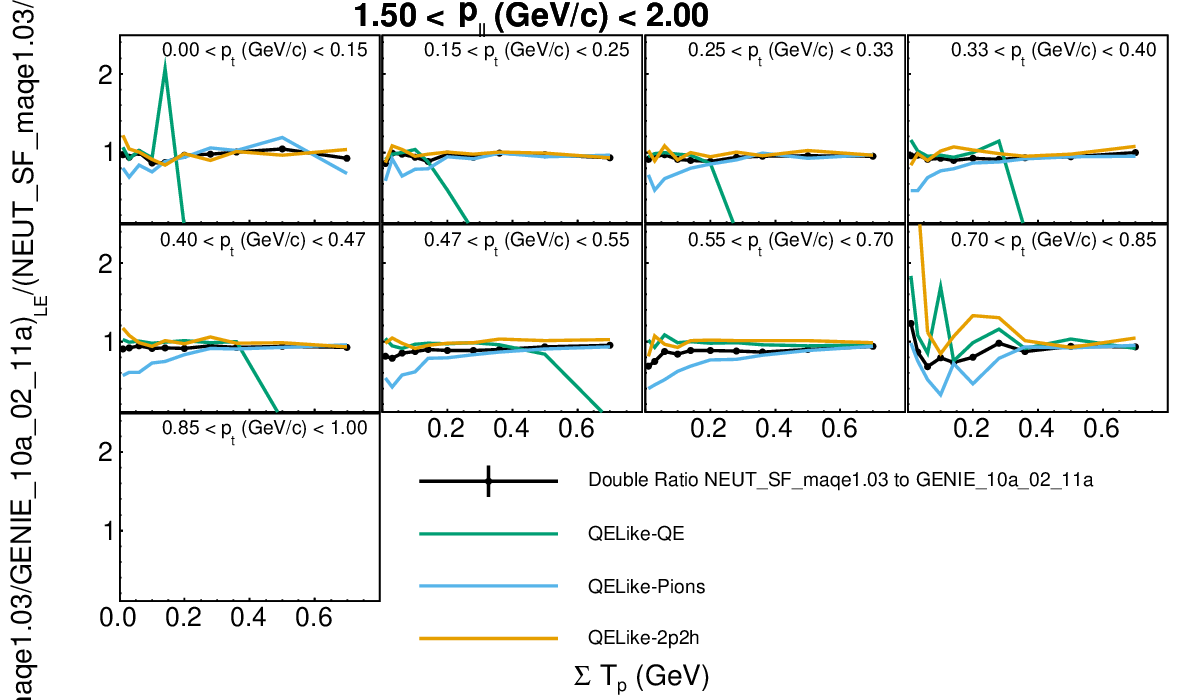}
    \includegraphics[width=\triplet]{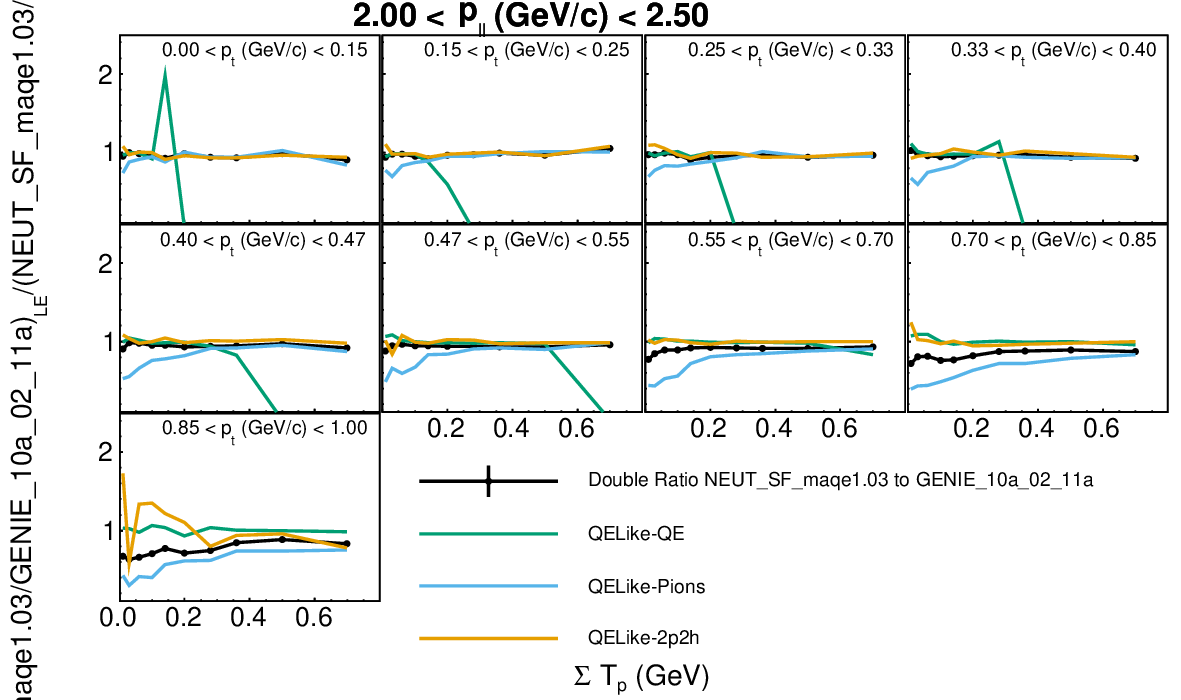}
        \includegraphics[width=\triplet]{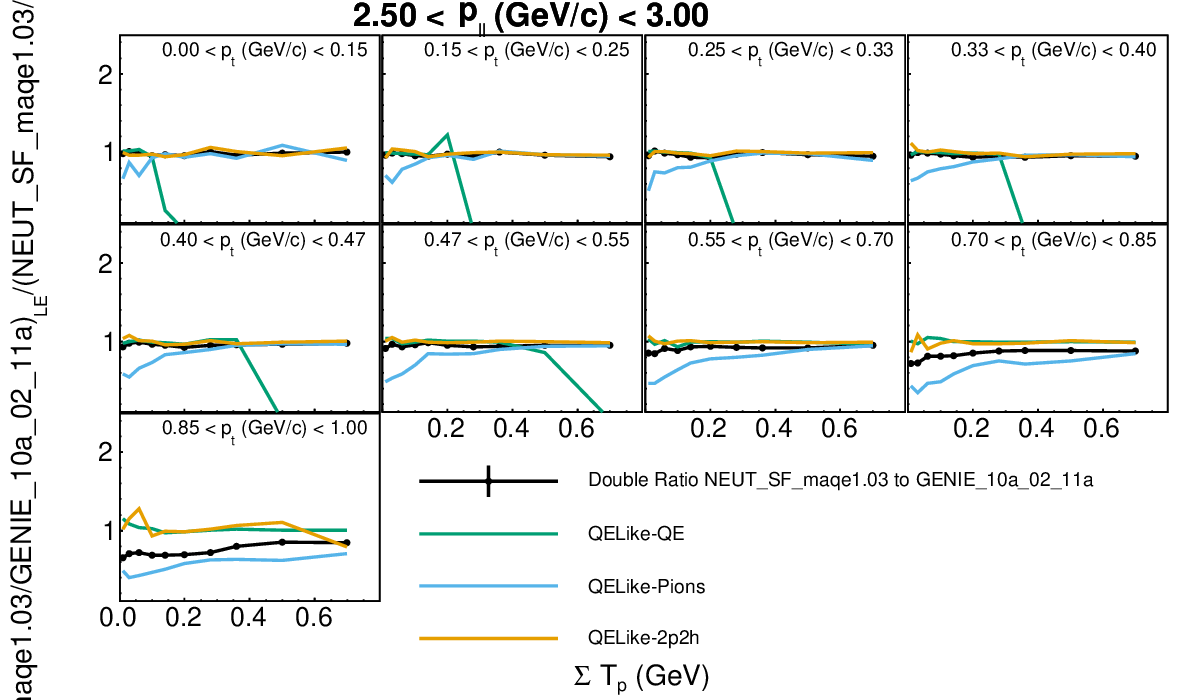}
    \caption{Pz bins 1 (top), 2(middle) and 3 (bottom) comparing NEUT SF against GENIE 10a.  The double ratios are also shown for individual sub-processes contributing to the signal, quasielastic (QE), multinucleon knockout (2p2h), and inelastic production of new particles (Pions).}
    \label{fig:ptpzsumtp_bin123_otherModels_NEUTSF}
\end{figure*}

\begin{figure*}[tp]
    \centering    
    \includegraphics[width=\triplet]{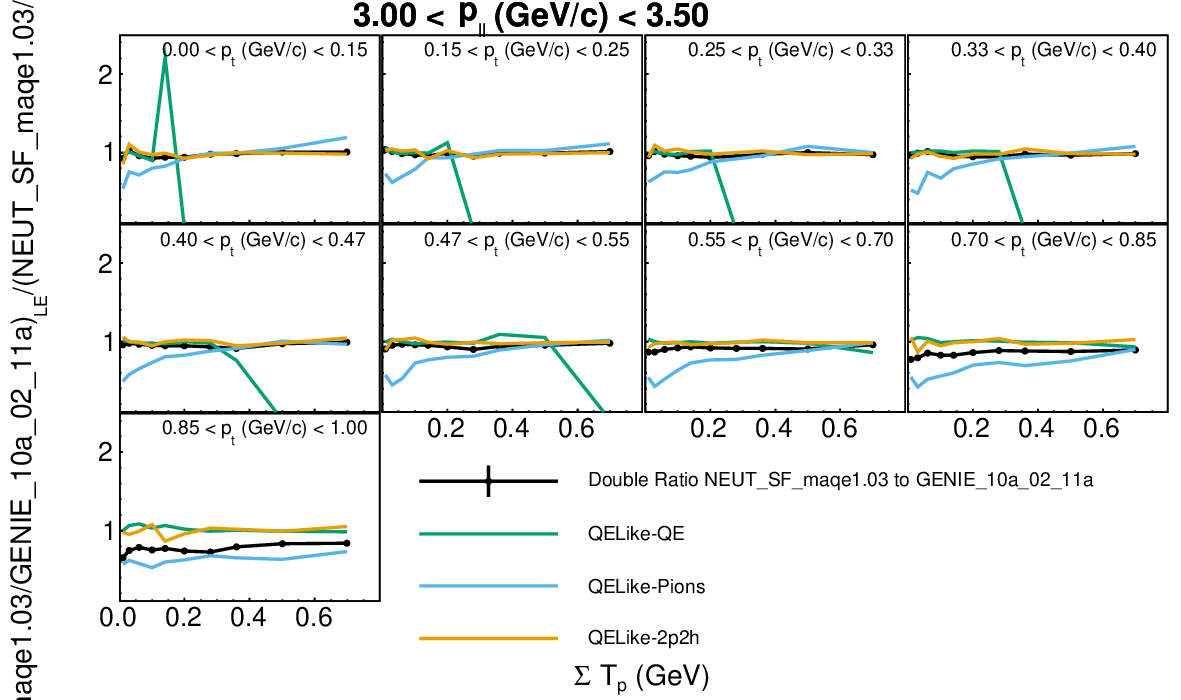}
    \includegraphics[width=\triplet]{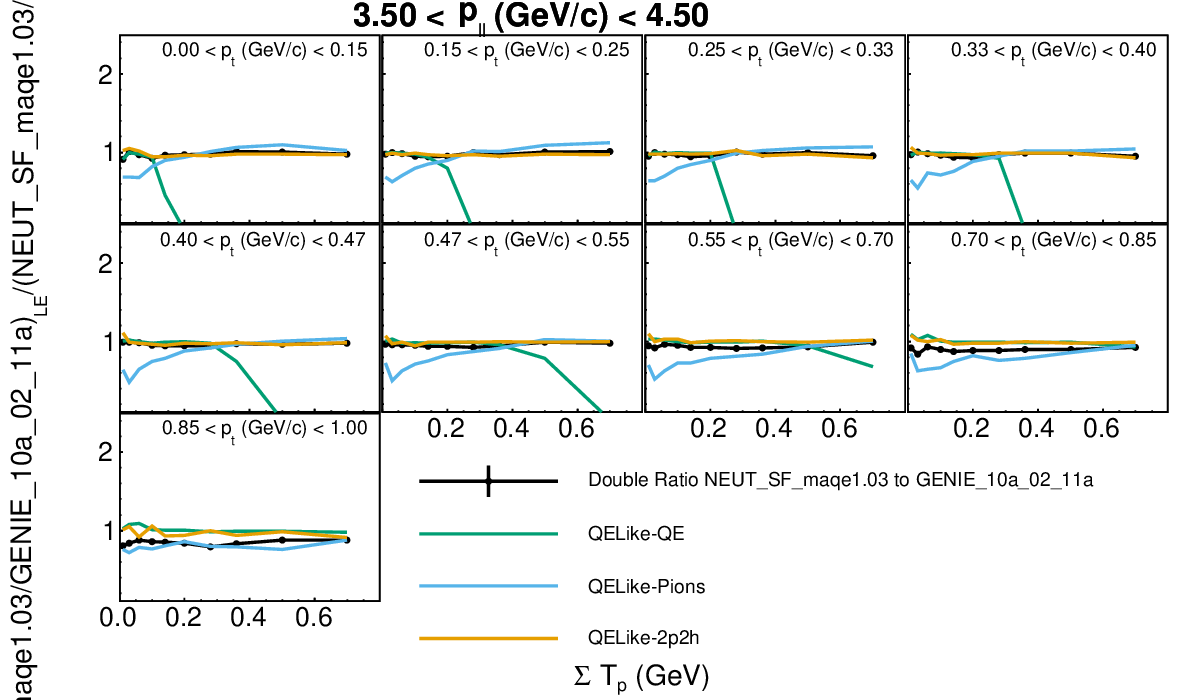}
    \caption{Pz bins 4 (top) and 5(bottom) comparing NEUT SF against GENIE 10a.  The double ratios are also shown for individual sub-processes contributing to the signal, quasielastic (QE), multinucleon knockout (2p2h), and inelastic production of new particles (Pions).}
    \label{fig:ptpzsumtp_bin45_otherModels_NEUTSF}
\end{figure*}

\subsection{NEUT LFG}
\begin{figure*}[tp]
    \centering    
    \includegraphics[width=\triplet]{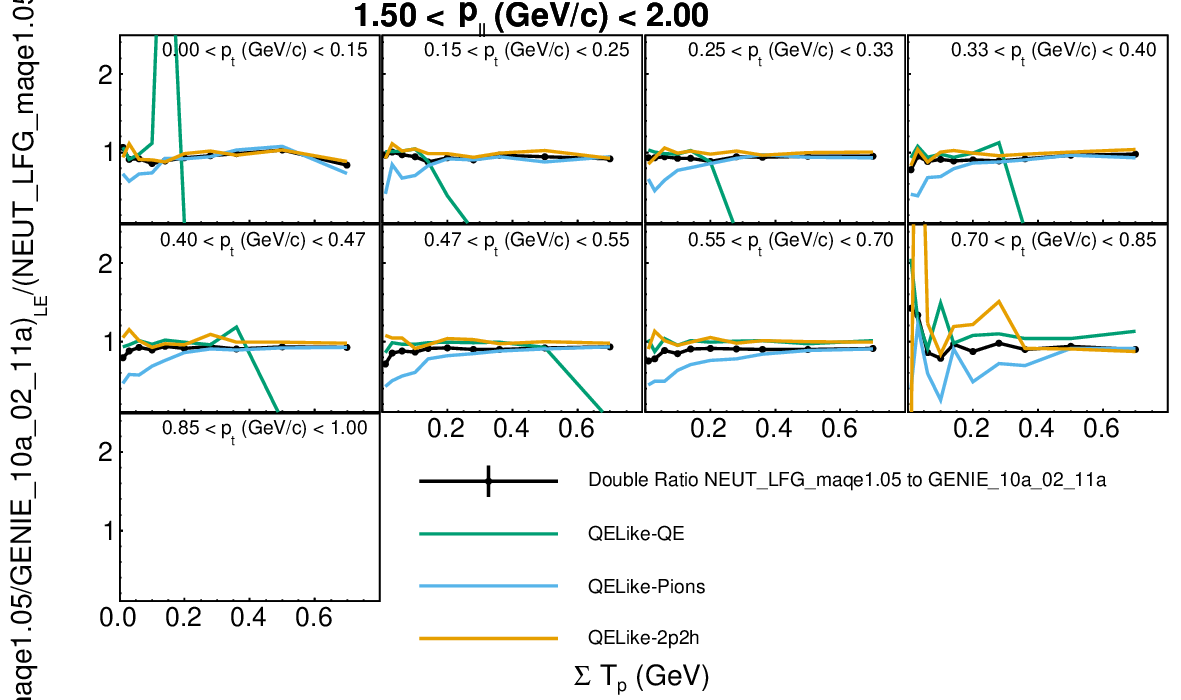}
    \includegraphics[width=\triplet]{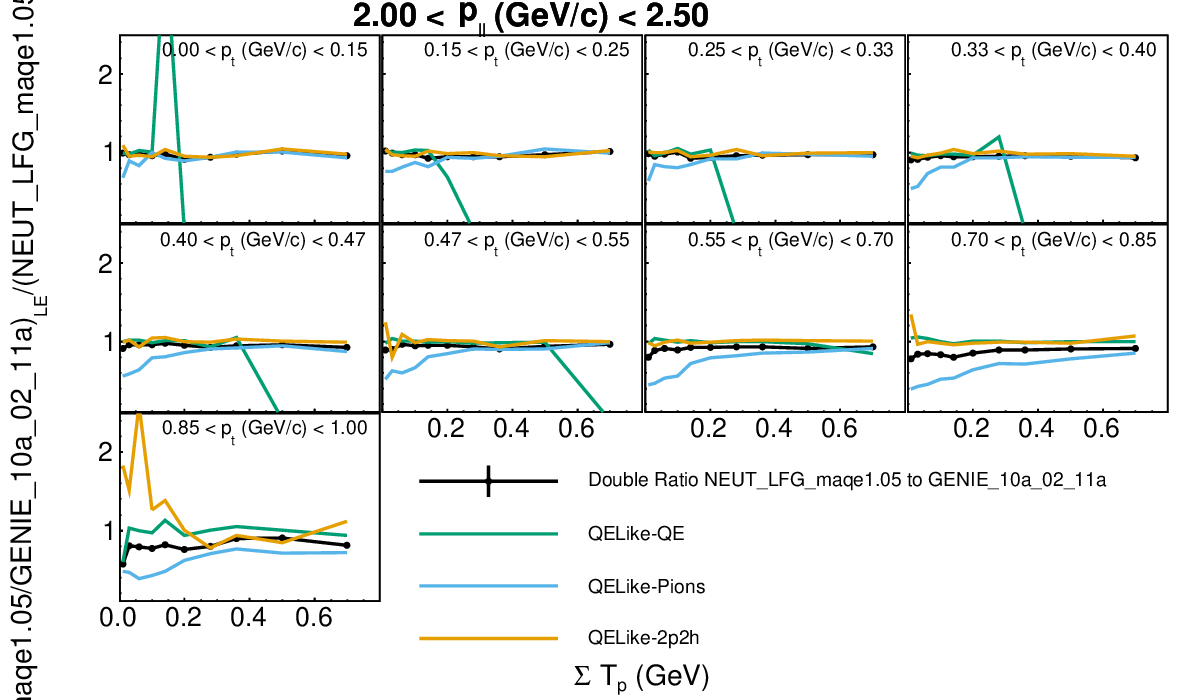}
        \includegraphics[width=\triplet]{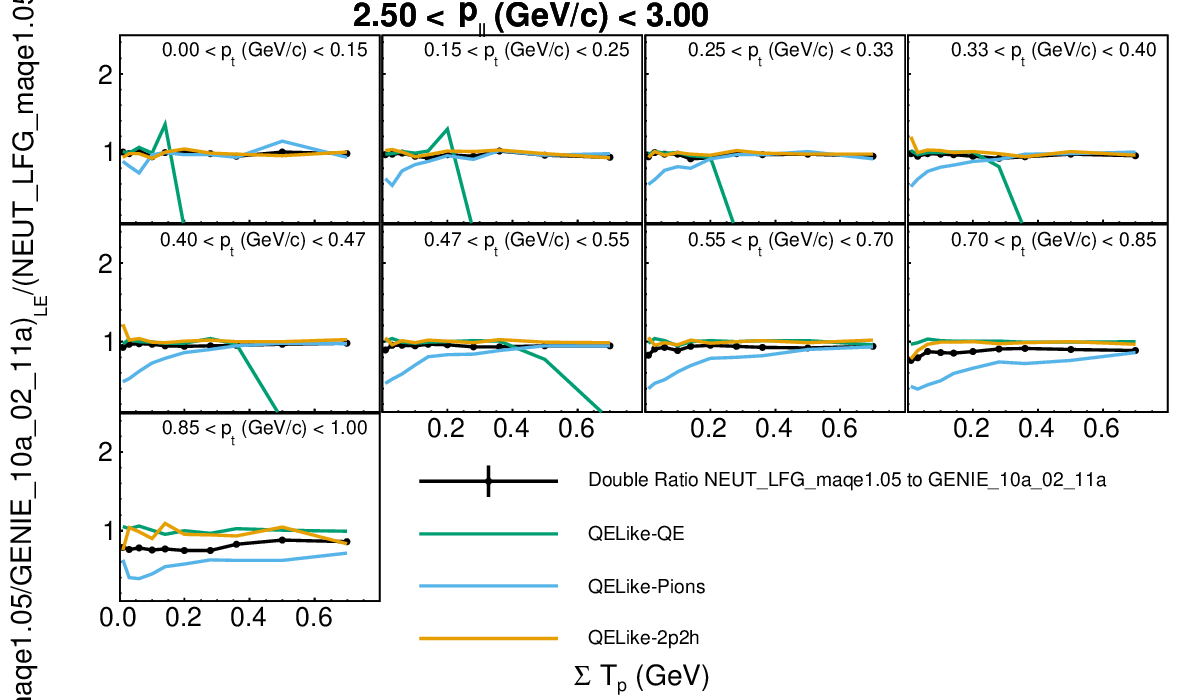}
    \caption{Pz bins 1 (top), 2(bottom) and 3 (bottom) comparing NEUT LFG against GENIE 10a.  The double ratios are also shown for individual sub-processes contributing to the signal, quasielastic (QE), multinucleon knockout (2p2h), and inelastic production of new particles (Pions).}
    \label{fig:ptpzsumtp_bin123_otherModels_NEUTLFG}
\end{figure*}

\begin{figure*}[tp]
    \centering    
    \includegraphics[width=\triplet]{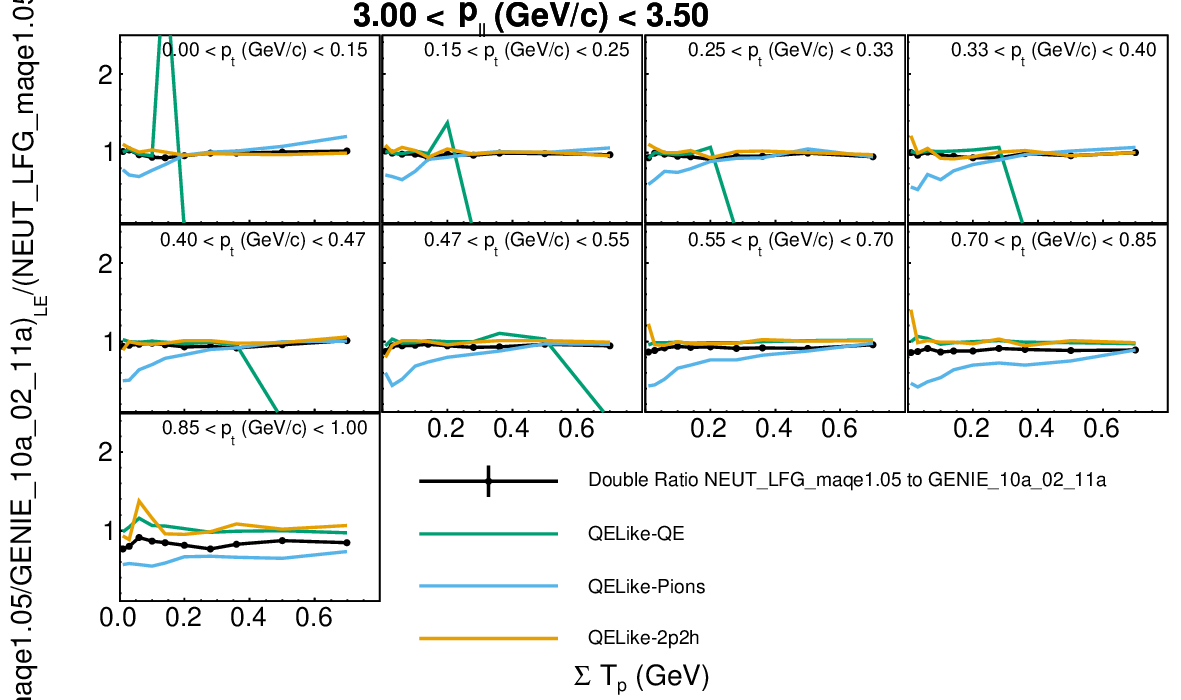}
    \includegraphics[width=\triplet]{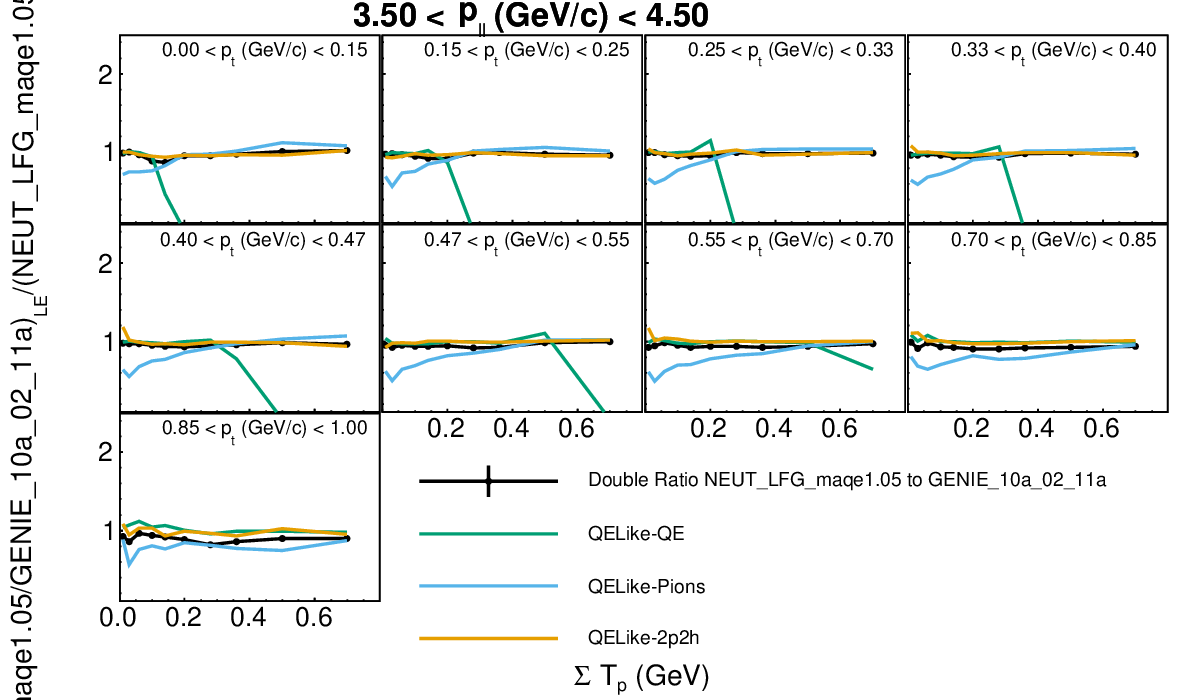}
    \caption{Pz bins 4 (top) and 5(bottom) comparing NEUT LFG against GENIE 10a.  The double ratios are also shown for individual sub-processes contributing to the signal, quasielastic (QE), multinucleon knockout (2p2h), and inelastic production of new particles (Pions).}
    \label{fig:ptpzsumtp_bin45_otherModels_NEUTLFG}
\end{figure*}

\subsection{NuWro LFG}
\begin{figure*}[tp]
    \centering    
    \includegraphics[width=\triplet]{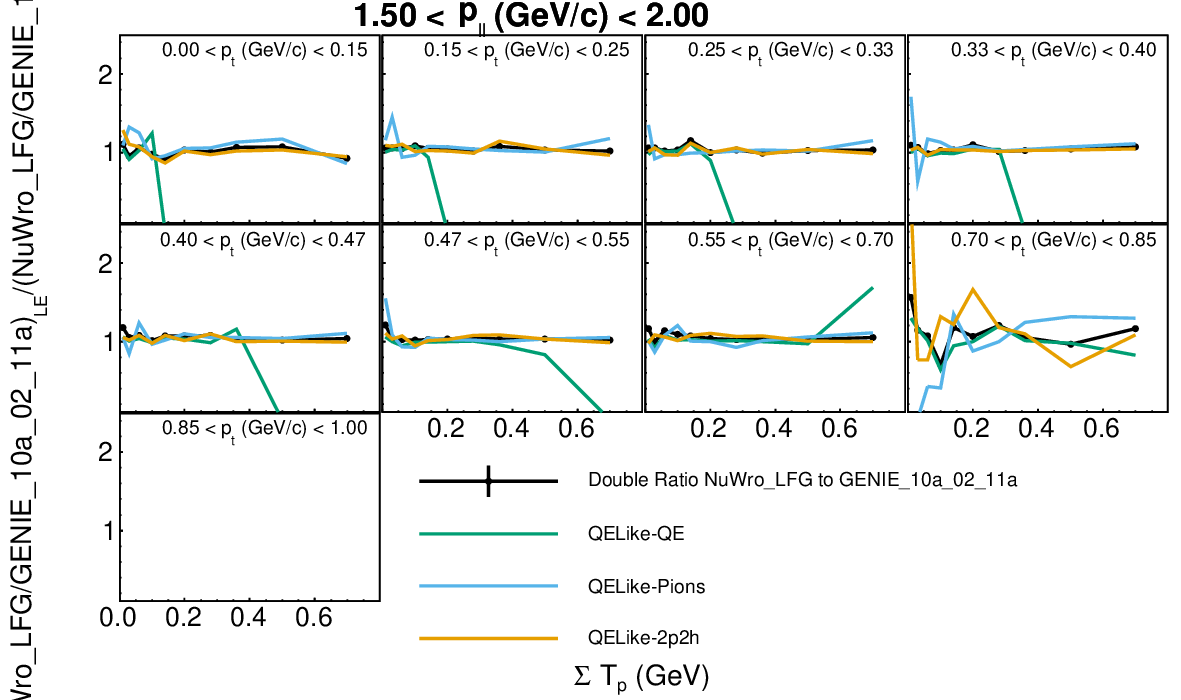}
    \includegraphics[width=\triplet]{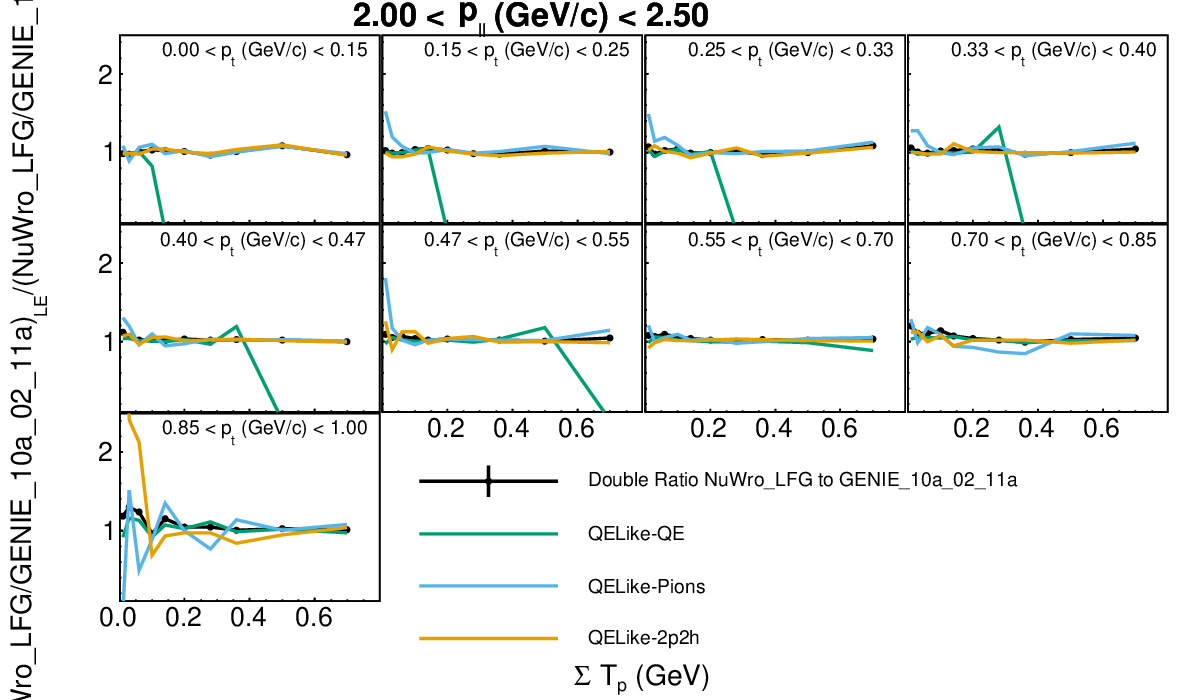}
        \includegraphics[width=\triplet]{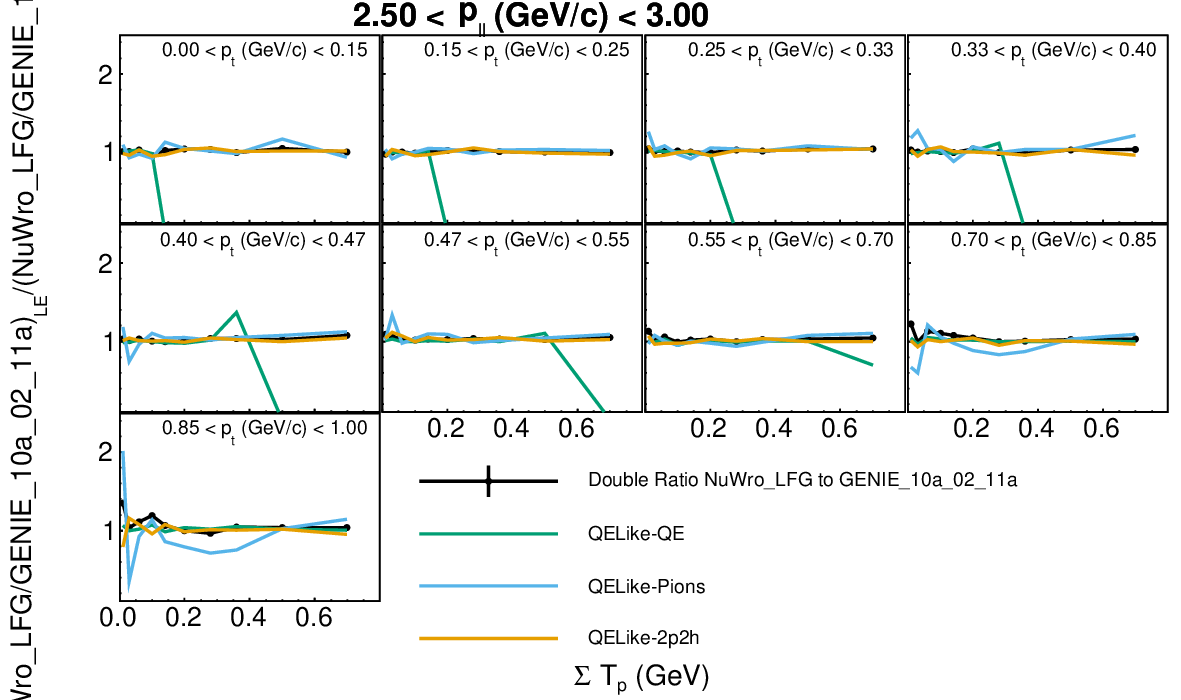}
    \caption{Pz bins 1 (top), 2(middle) and 3 (bottom) comparing NuWro LFG against GENIE 10a.  The double ratios are also shown for individual sub-processes contributing to the signal, quasielastic (QE), multinucleon knockout (2p2h), and inelastic production of new particles (Pions).}
    \label{fig:ptpzsumtp_bin123_otherModels_NuWroLFG}
\end{figure*}

\begin{figure*}[tp]
    \centering    
    \includegraphics[width=\triplet]{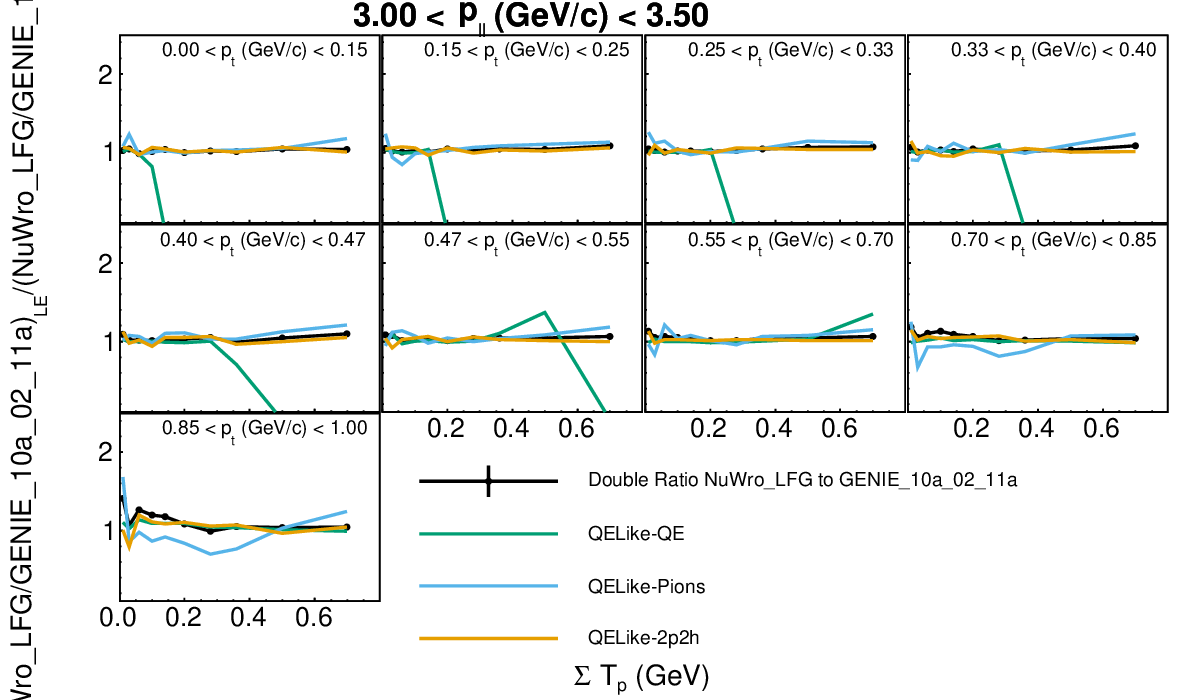}
    \includegraphics[width=\triplet]{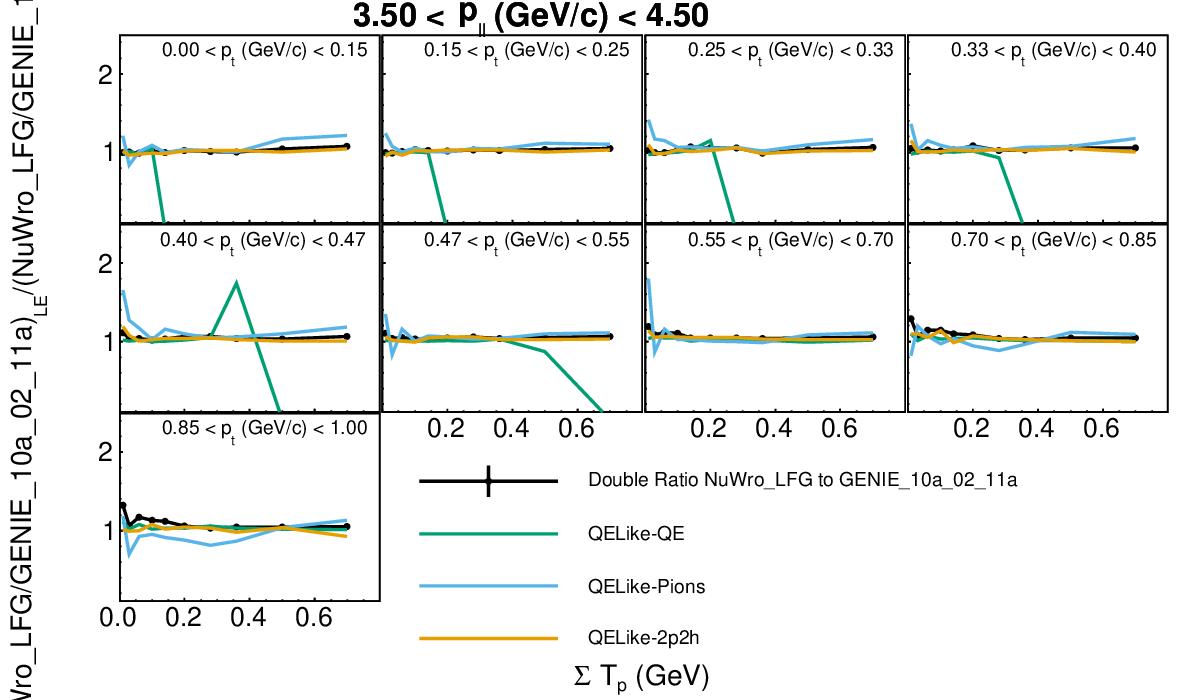}
    \caption{Pz bins 4 (top) and 5(bottom) comparing NuWro LFG against GENIE 10a.  The double ratios are also shown for individual sub-processes contributing to the signal, quasielastic (QE), multinucleon knockout (2p2h), and inelastic production of new particles (Pions).}
    \label{fig:ptpzsumtp_bin45_otherModels_NuWroLFG}
\end{figure*}

\subsection{NuWro SF}
\begin{figure*}[tp]
    \centering    
    \includegraphics[width=\triplet]{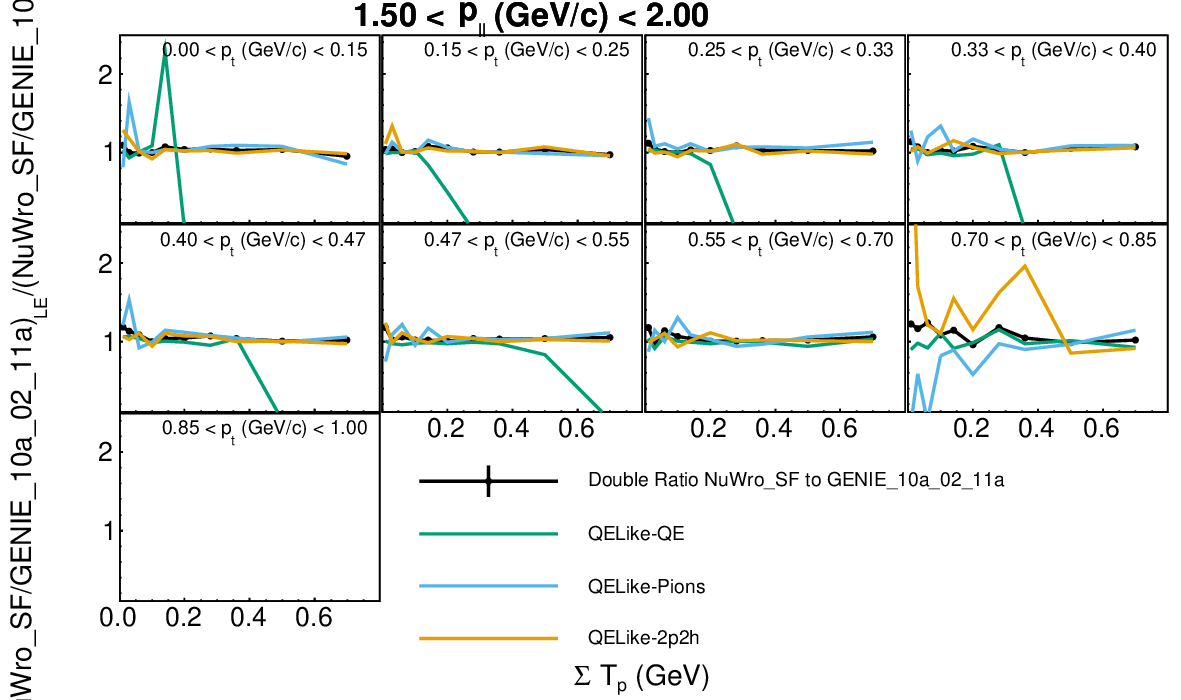}
    \includegraphics[width=\triplet]{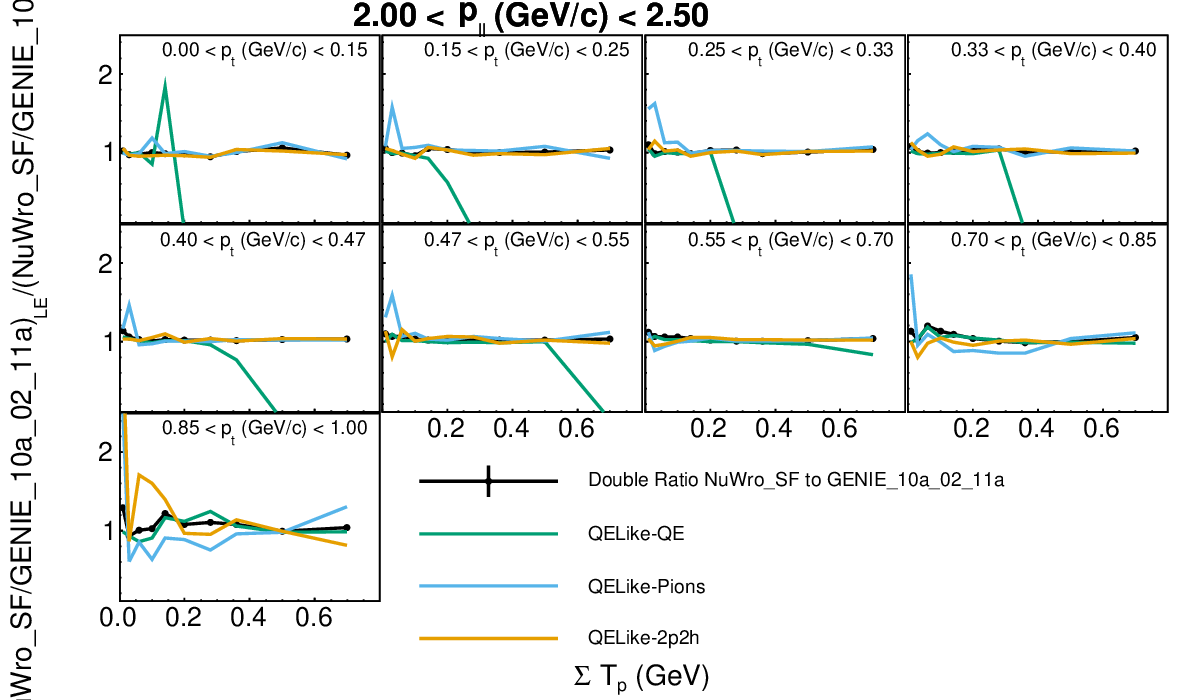}
        \includegraphics[width=\triplet]{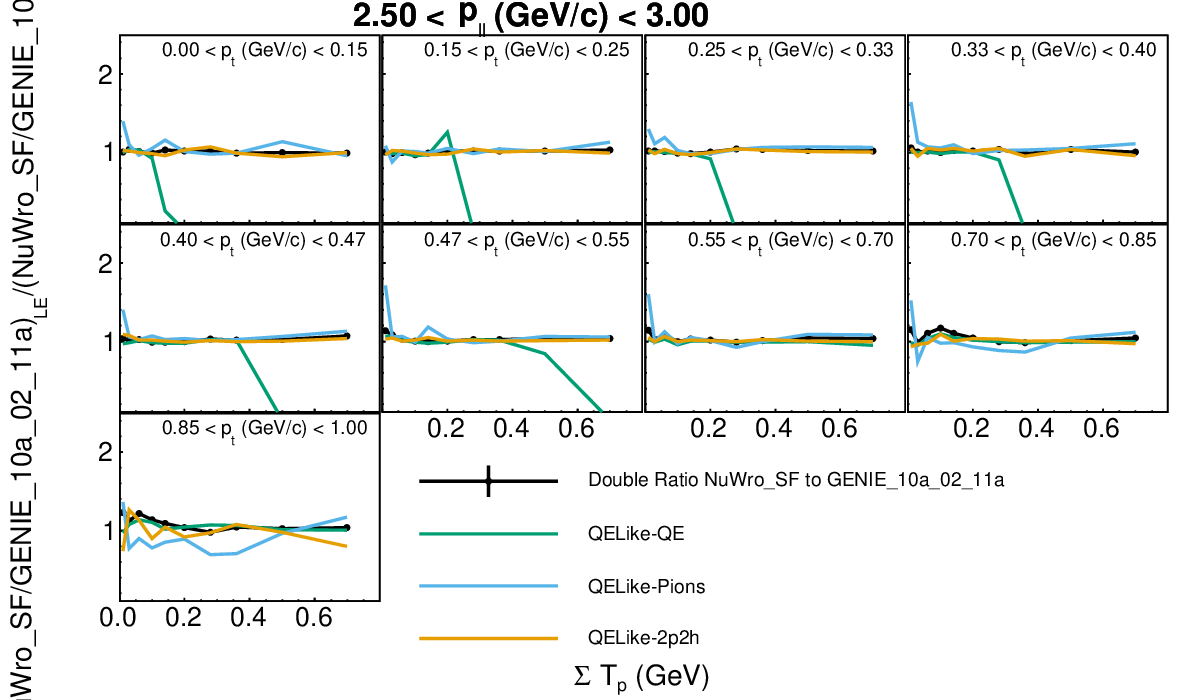}
    \caption{Pz bins 1 (top), 2(middle) and 3 (bottom) comparing NuWro SF against GENIE 10a.  The double ratios are also shown for individual sub-processes contributing to the signal, quasielastic (QE), multinucleon knockout (2p2h), and inelastic production of new particles (Pions).}
    \label{fig:ptpzsumtp_bin123_otherModels_NuWroSF}
\end{figure*}

\begin{figure*}[tp]
    \centering    
    \includegraphics[width=\triplet]{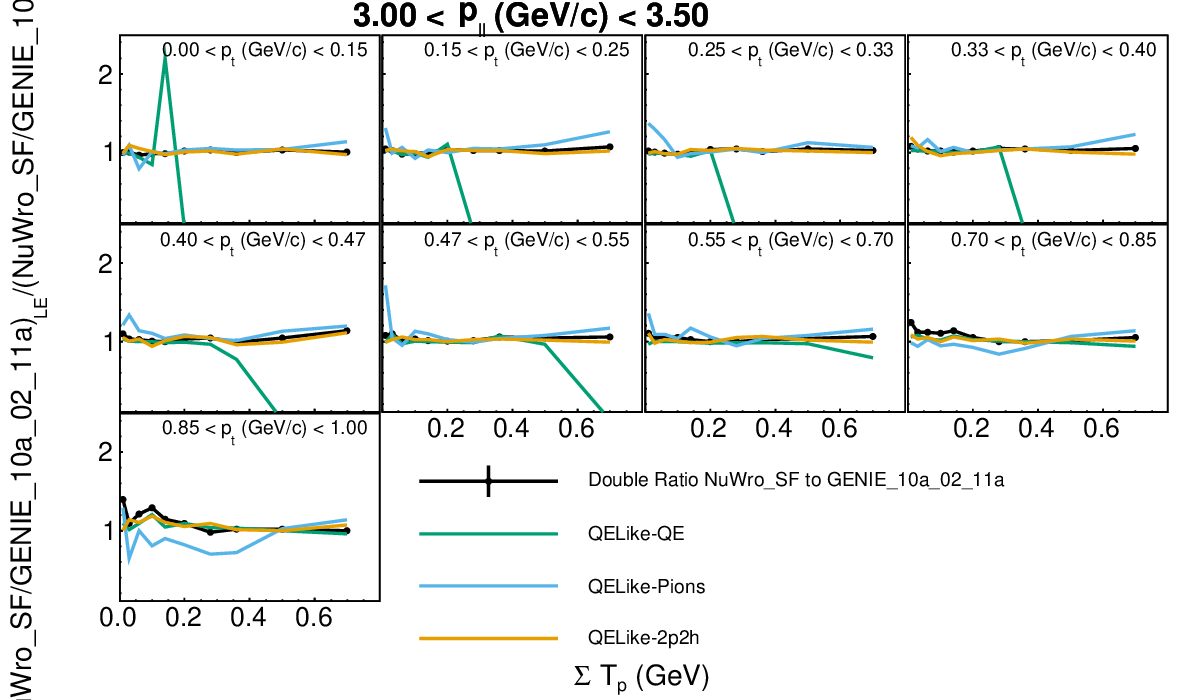}
    \includegraphics[width=\triplet]{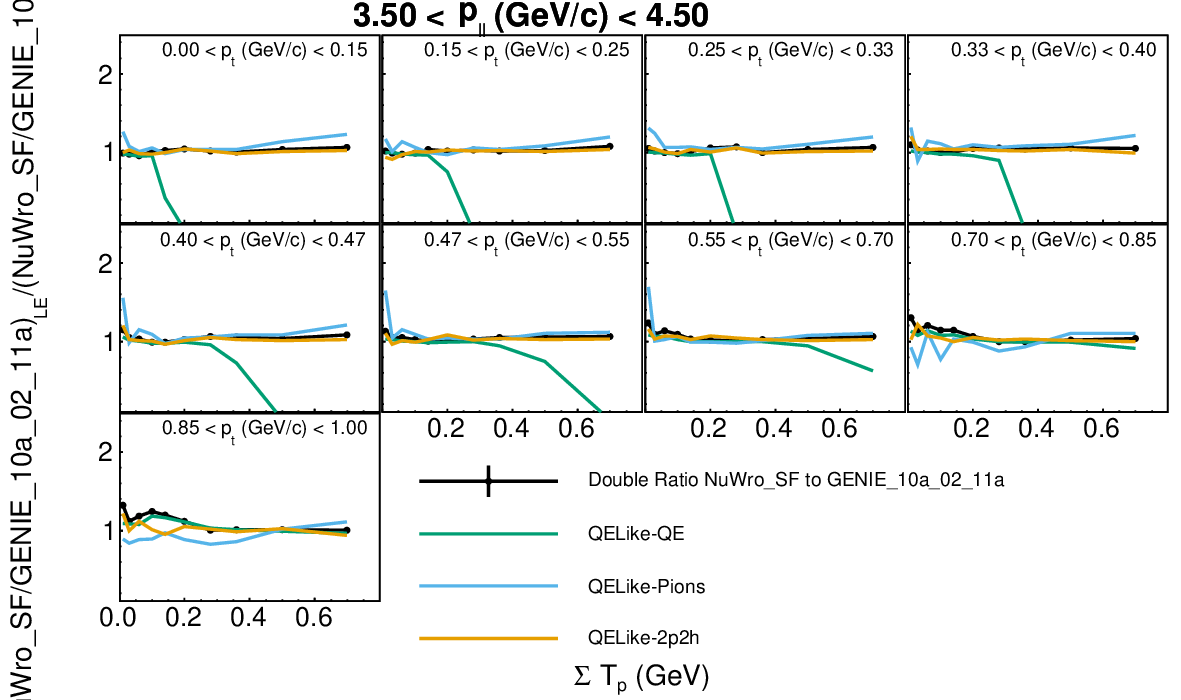}
    \caption{Pz bins 4 (top) and 5(bottom) comparing NuWro SF against GENIE 10a.  The double ratios are also shown for individual sub-processes contributing to the signal, quasielastic (QE), multinucleon knockout (2p2h), and inelastic production of new particles (Pions).}
    \label{fig:ptpzsumtp_bin45_otherModels_NuWroSF}
\end{figure*}

%
%

\FloatBarrier
\newpage 
\subsection{Event Distributions and predictions for all momentum bins}
\begin{figure*}[hp]
    \centering
    \includegraphics[width=.48\linewidth]{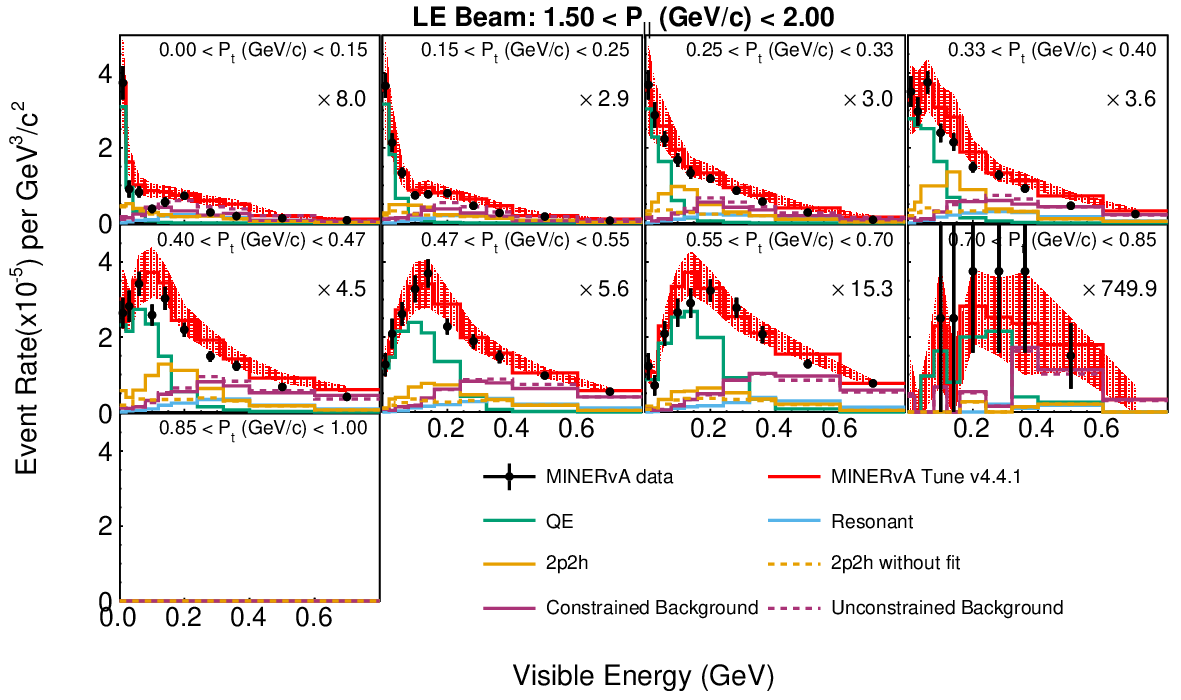}
    \includegraphics[width=.48\linewidth]{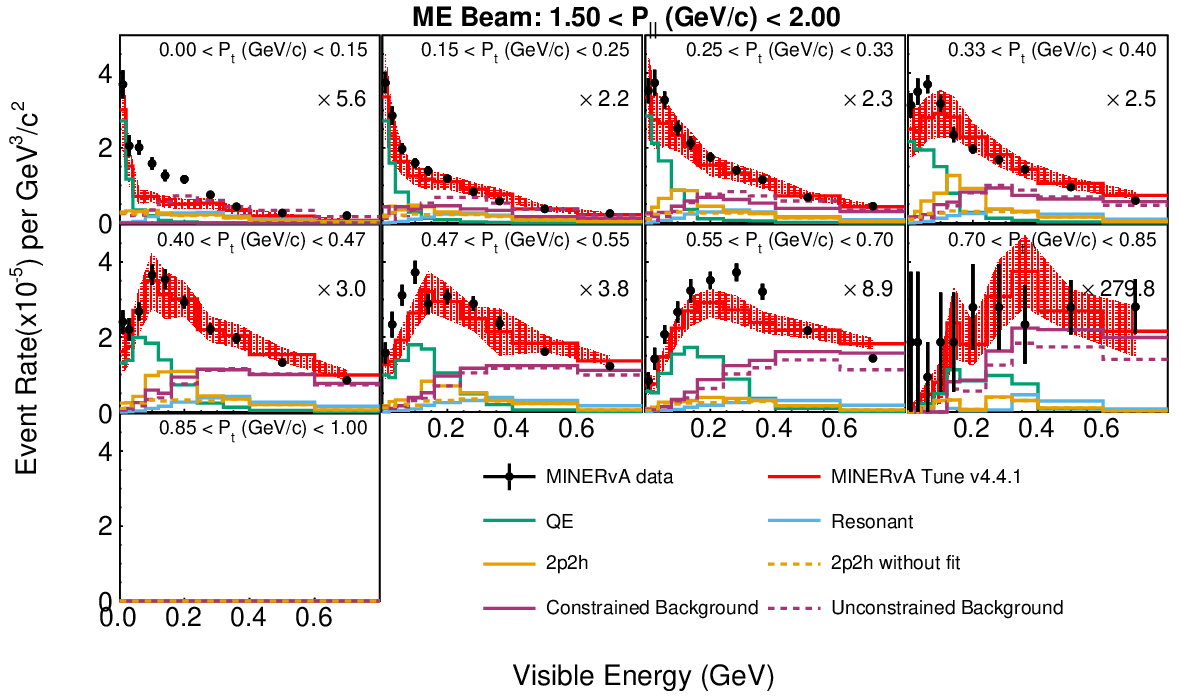}
    \caption{Event distributions in data and prediction after the background fits described in the text, for both the Low (top) and Medium (bottom) energy data, for $1.5 <\pz (GeV/c)< 2.0$ as a function of visible energy and \pt.}
    \label{fig:bckgd1}
\end{figure*}

\begin{figure*}[hp]
    \centering
    \includegraphics[width=.48\linewidth]{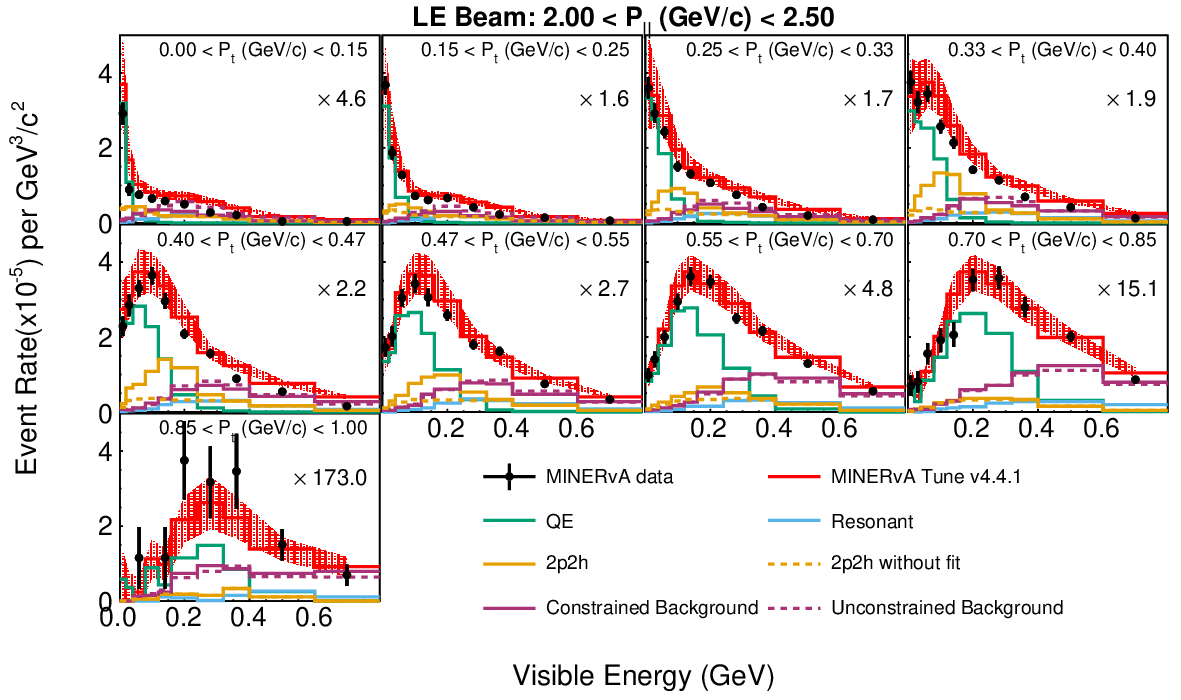}
    \includegraphics[width=.48\linewidth]{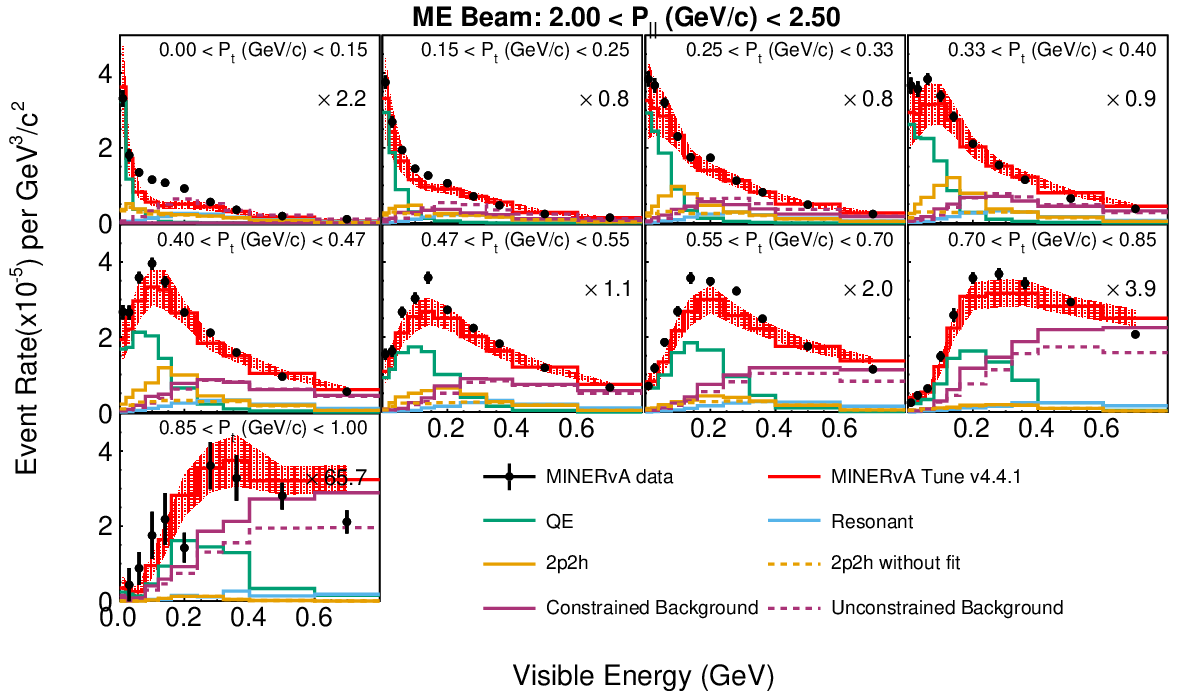}
    \caption{Event distributions in data and prediction after the background fits described in the text, for both the Low (top) and Medium (bottom) energy data, for $2.0 <\pz (GeV/c)< 2.5$ as a function of visible energy and \pt.}
    \label{fig:bckgd2}
\end{figure*}

\begin{figure*}[hp]
    \centering
    \includegraphics[width=.48\linewidth]{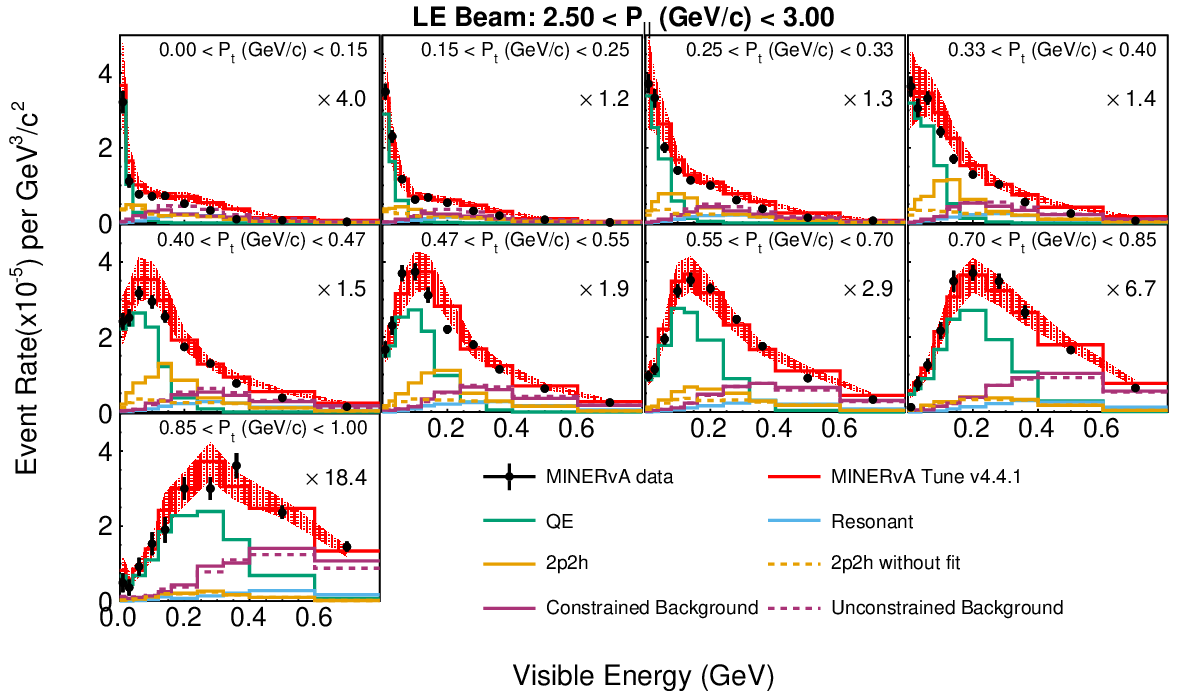}
    \includegraphics[width=.48\linewidth]{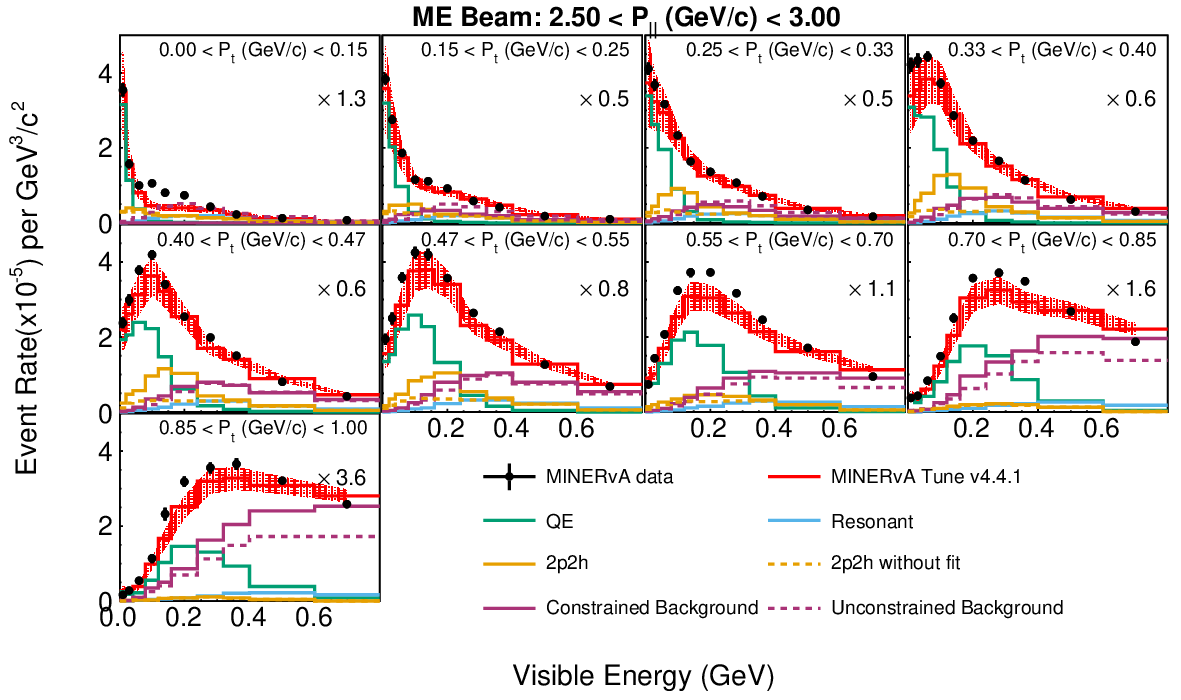}
    \caption{Event distributions in data and prediction after the background fits described in the text, for both the Low (top) and Medium (bottom) energy data, for $2.5 <\pz (GeV/c)< 3.0$ as a function of visible energy and \pt.}
    \label{fig:bckgd3}
\end{figure*}

\begin{figure*}[hp]
    \includegraphics[width=.48\linewidth]{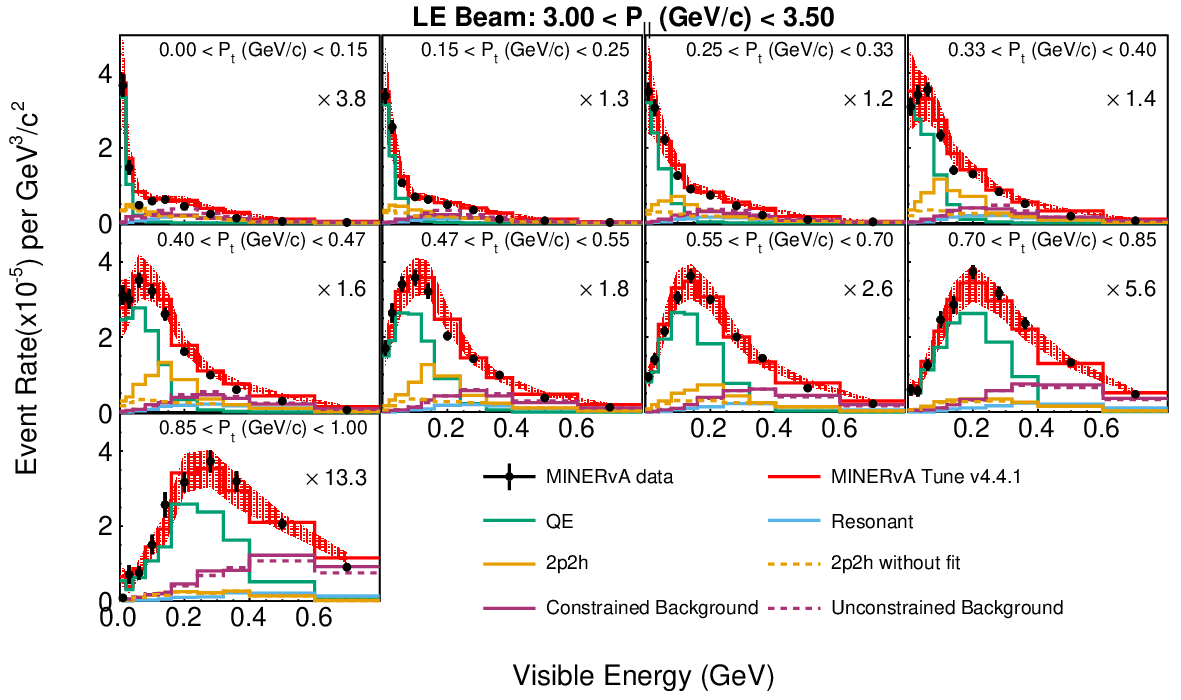}
    \includegraphics[width=.48\linewidth]{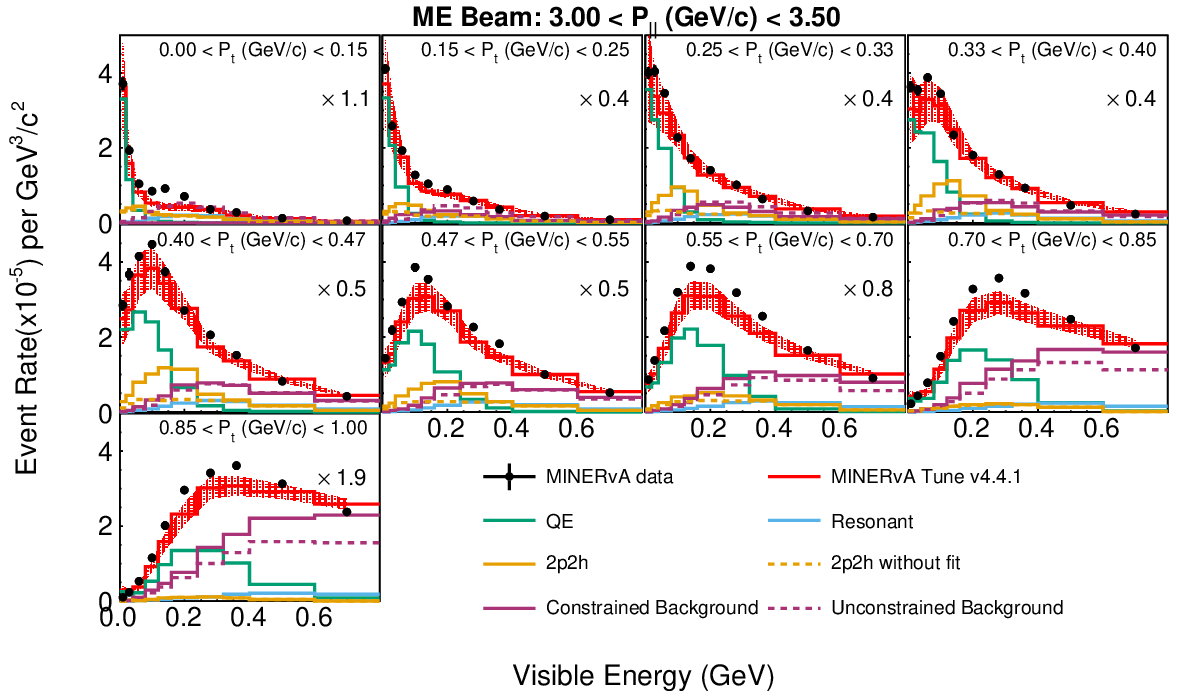}
    \caption{Event distributions in data and prediction after the background fits described in the text, for both the Low (top) and Medium (bottom) energy data, for $3.0 <\pz (GeV/c)< 3.5$ as a function of visible energy and \pt.}
    \label{fig:bckgd4}
\end{figure*}

\begin{figure*}[hp]
    \includegraphics[width=.48\linewidth]{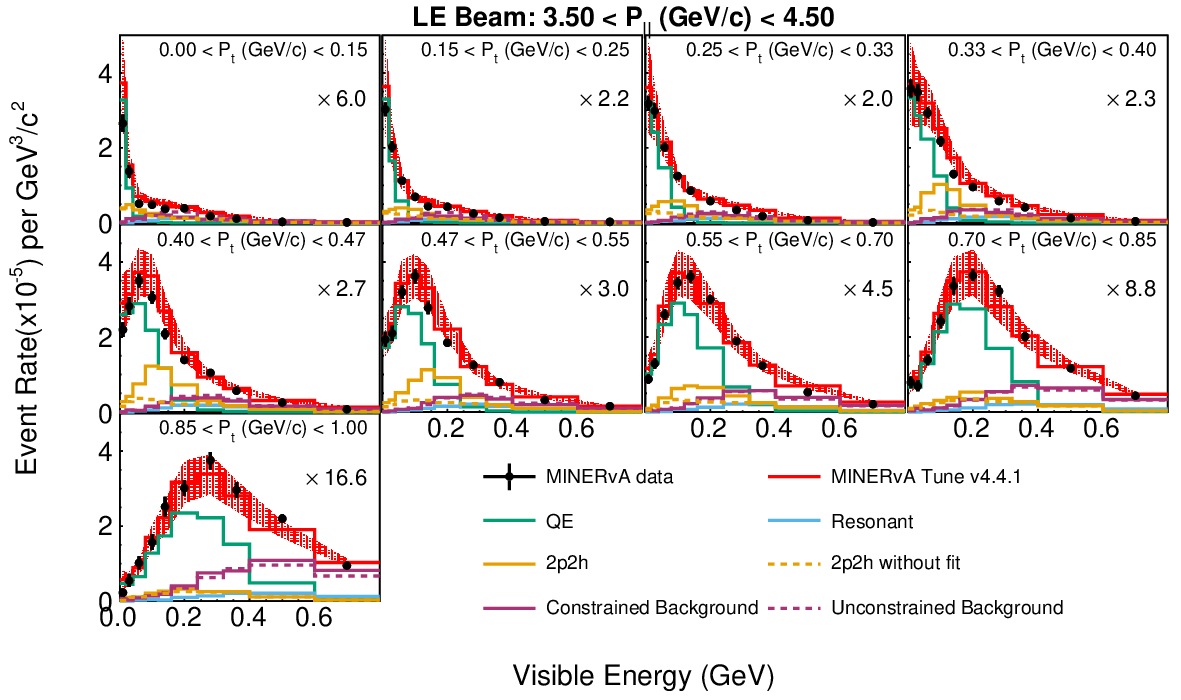}
    \includegraphics[width=.48\linewidth]{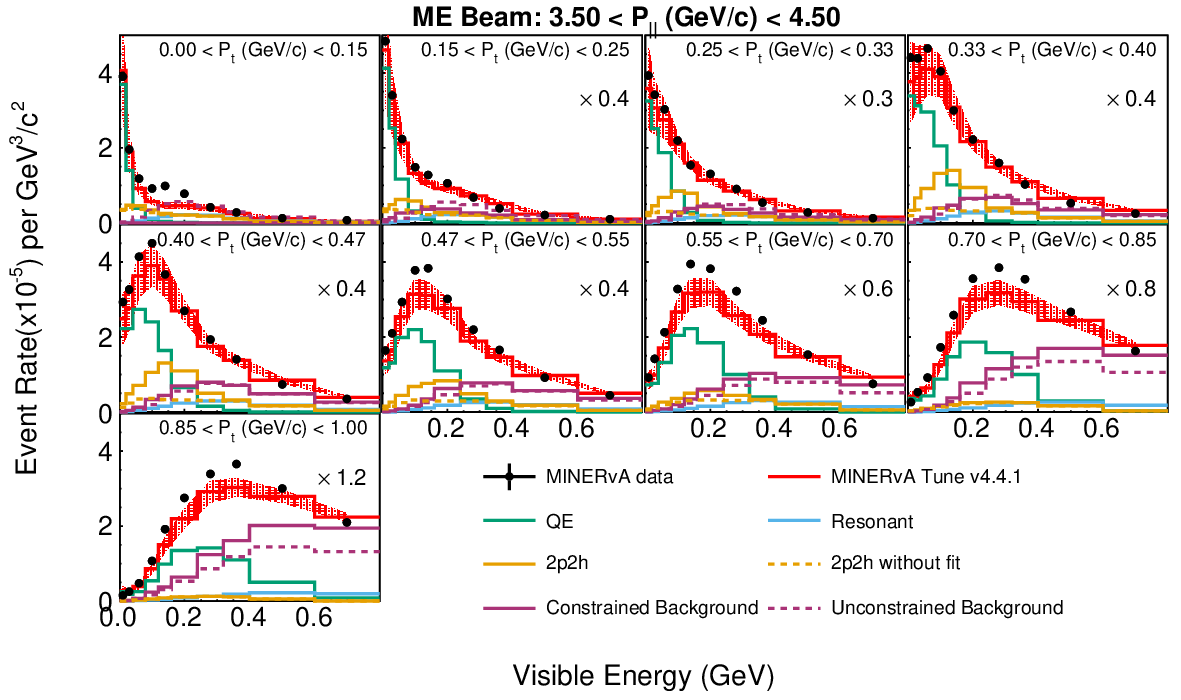}
    \caption{Event distributions in data and prediction after the background fits described in the text, for both the Low (top) and Medium (bottom) energy data, for $3.5 <\pz (GeV/c)< 4.5$ as a function of visible energy and \pt.}
    \label{fig:bckgd5}
\end{figure*}

\FloatBarrier
\newpage 
\subsection{Event Ratios for all \pz\ bins}
\begin{figure*}[hp]
    \centering
    \includegraphics[width=.48\linewidth]{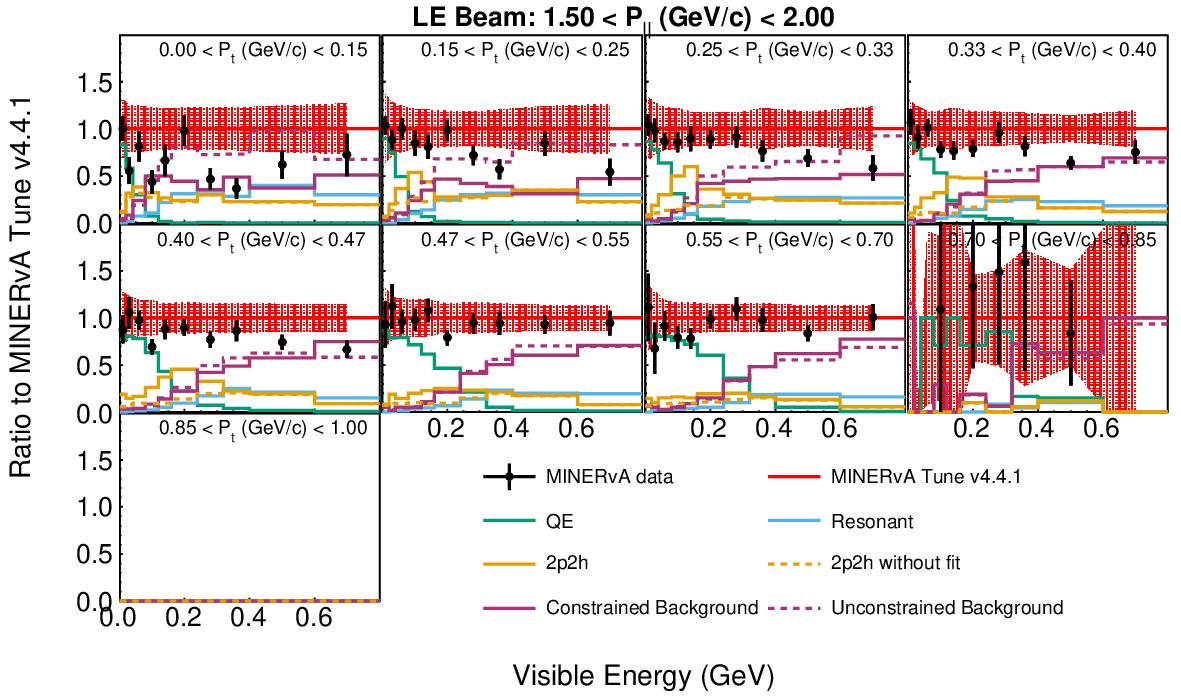}
    \includegraphics[width=.48\linewidth]{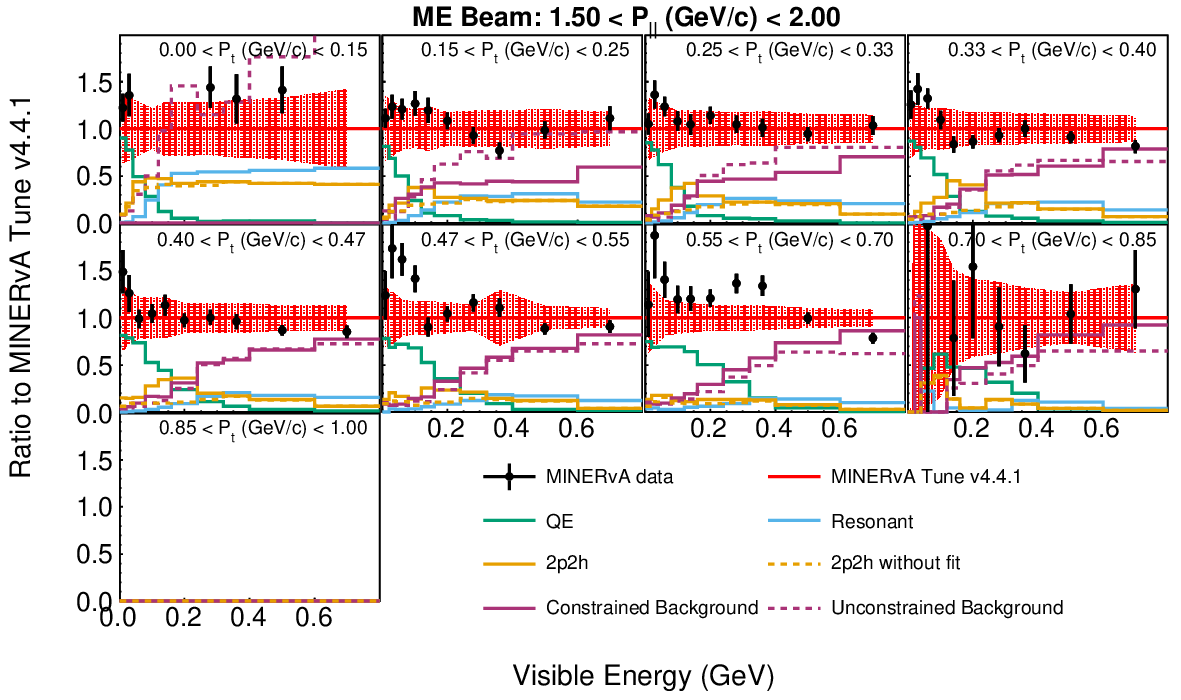}
    \caption{Ratios of measured to predicted event distributions after the background fits described in the text, for both the Low (top) and Medium (bottom) energy data, for $1.5 <\pz (GeV/c)< 2.0$ as a function of visible energy and \pt.}
    \label{fig:bckgdrat1}
\end{figure*}

\begin{figure*}[hp]
    \centering
    \includegraphics[width=.48\linewidth]{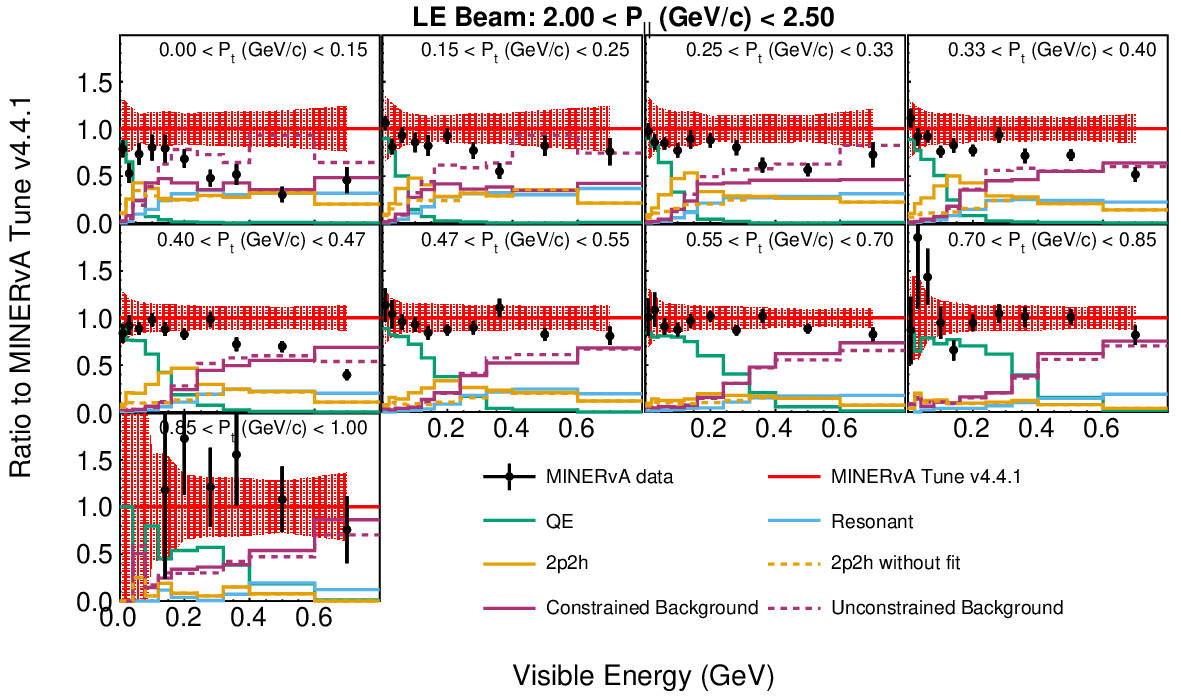}
    \includegraphics[width=.48\linewidth]{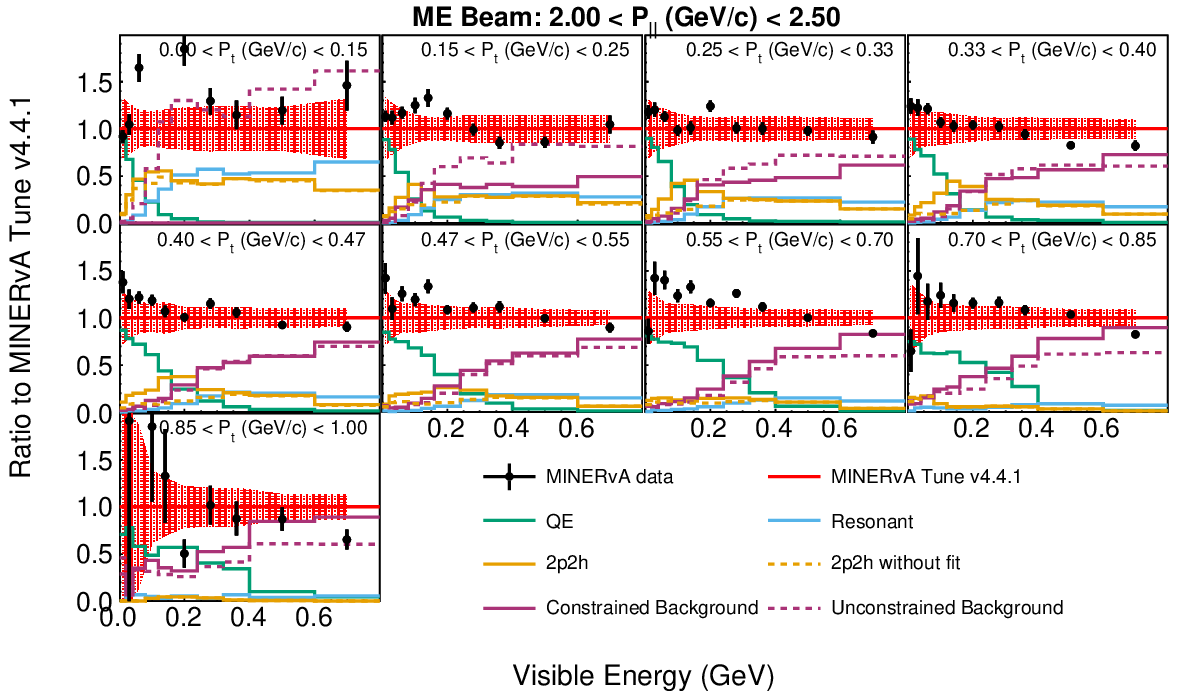}
    \caption{Ratios of measured to predicted event distributions after the background fits described in the text, for both the Low (top) and Medium (bottom) energy data, for $2.0 <\pz (GeV/c)< 2.5$ as a function of visible energy and \pt. }
    \label{fig:bckgdrat2}
\end{figure*}

\begin{figure*}[hp]
    \centering
    \includegraphics[width=.48\linewidth]{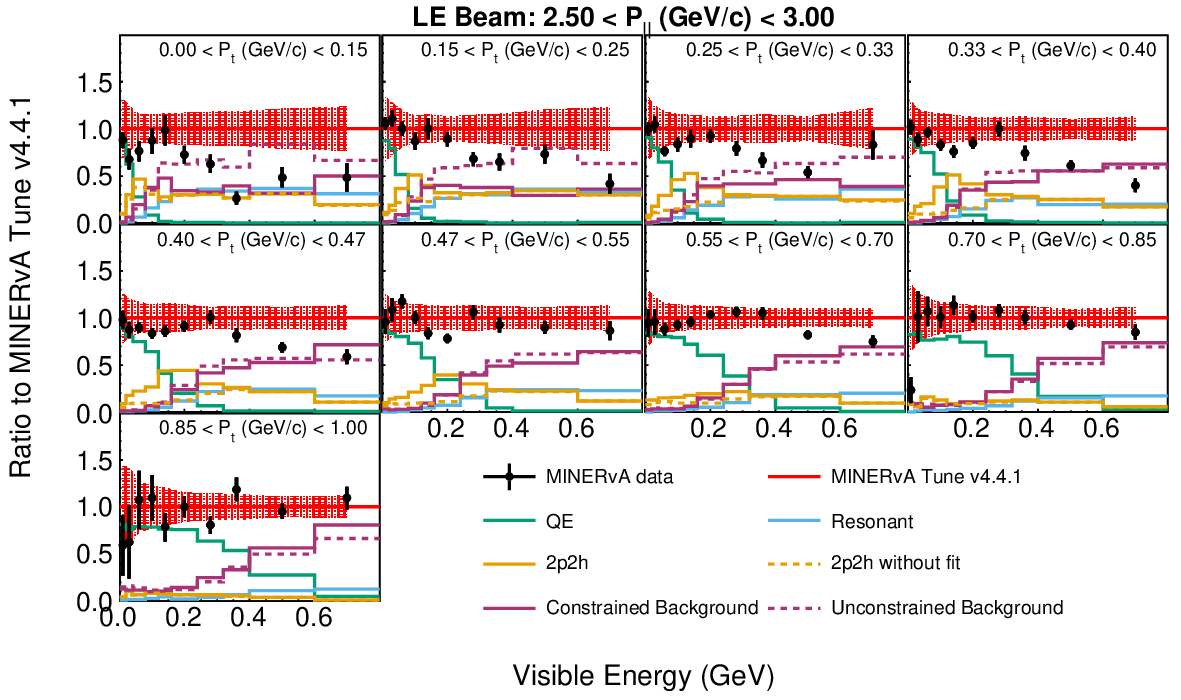}
    \includegraphics[width=.48\linewidth]{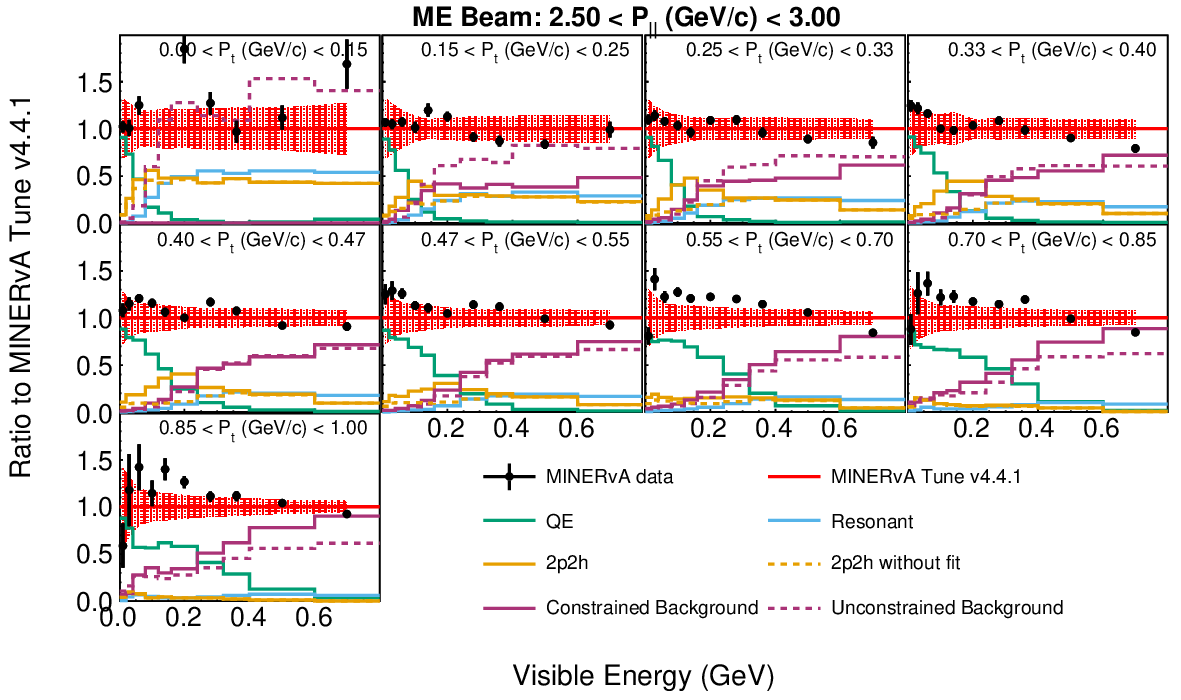}
    \caption{Ratios of measured to predicted event distributions after the background fits described in the text, for both the Low (top) and Medium (bottom) energy data, for $3.0 <\pz (GeV/c)< 3.5$ as a function of visible energy and \pt.}
    \label{fig:bckgdrat3}
\end{figure*}

\begin{figure*}[hp]
    \includegraphics[width=.48\linewidth]{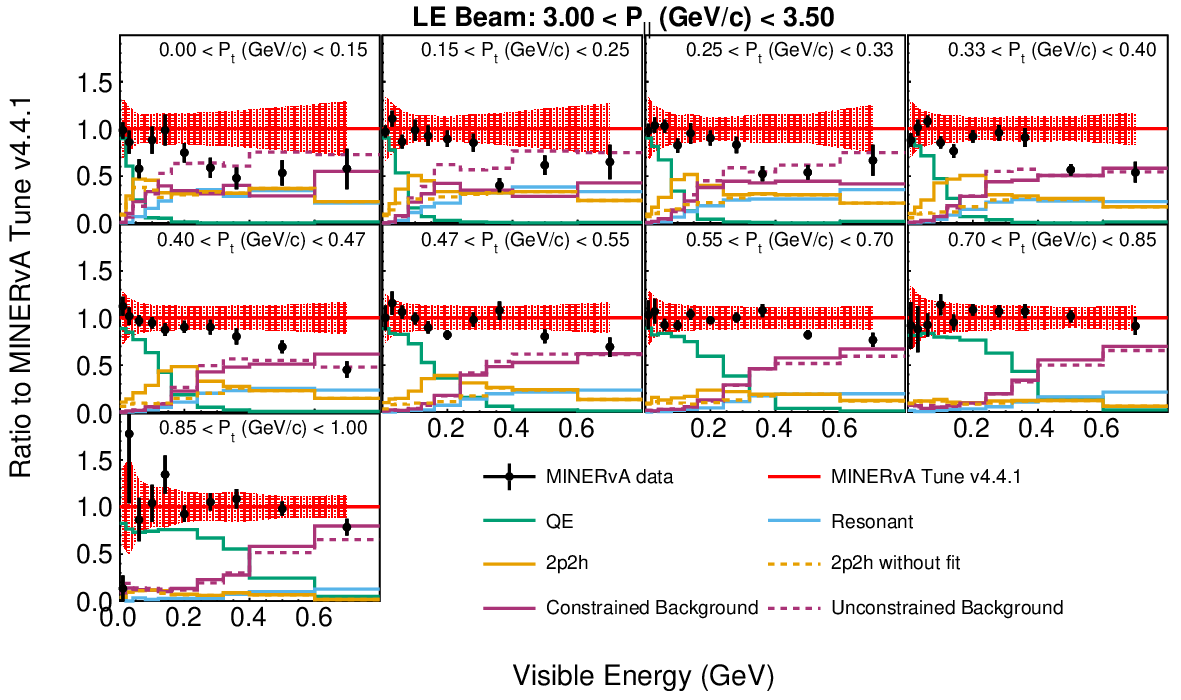}
    \includegraphics[width=.48\linewidth]{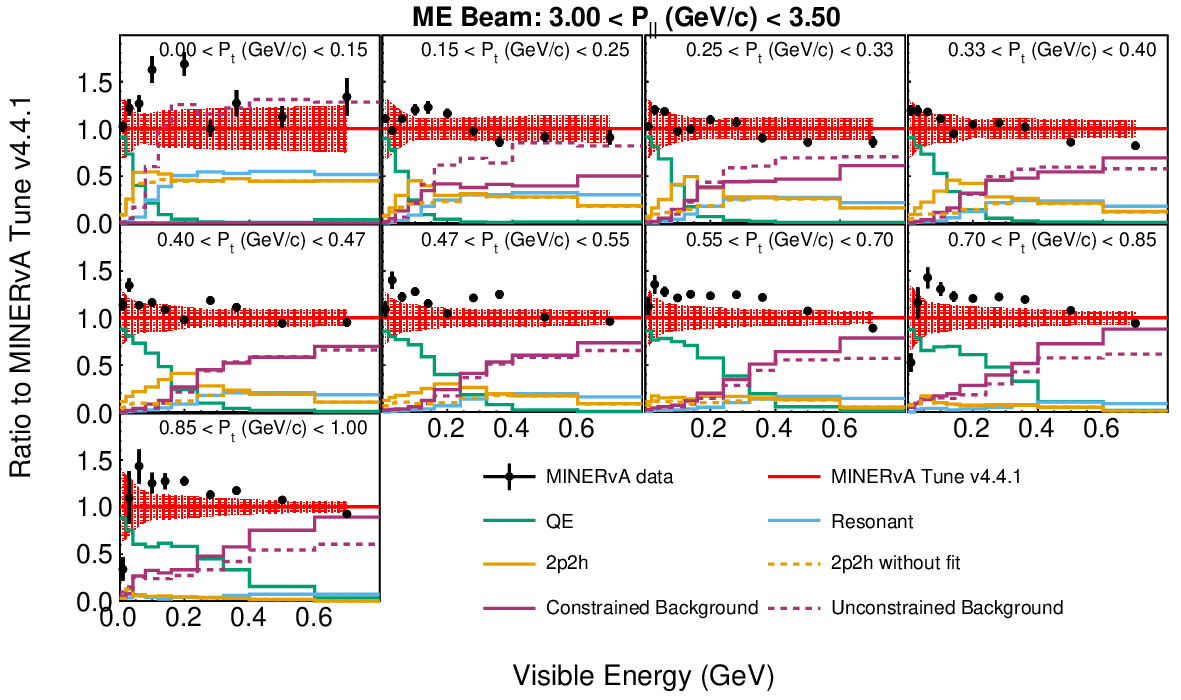}
    \caption{Ratios of measured to predicted event distributions after the background fits described in the text, for both the Low (top) and Medium (bottom) energy data, for $3.0 <\pz (GeV/c)< 3.5$ as a function of visible energy and \pt. }
    \label{fig:bckgdrat4}
\end{figure*}

\begin{figure*}[hp]
    \includegraphics[width=.48\linewidth]{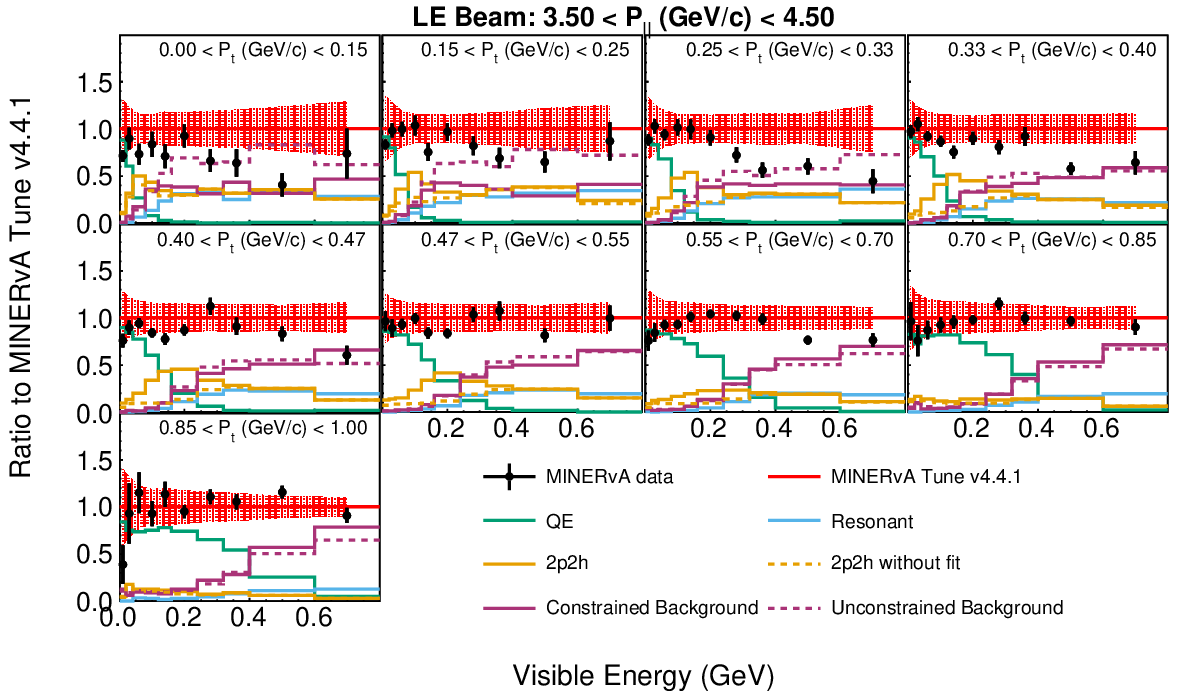}
    \includegraphics[width=.48\linewidth]{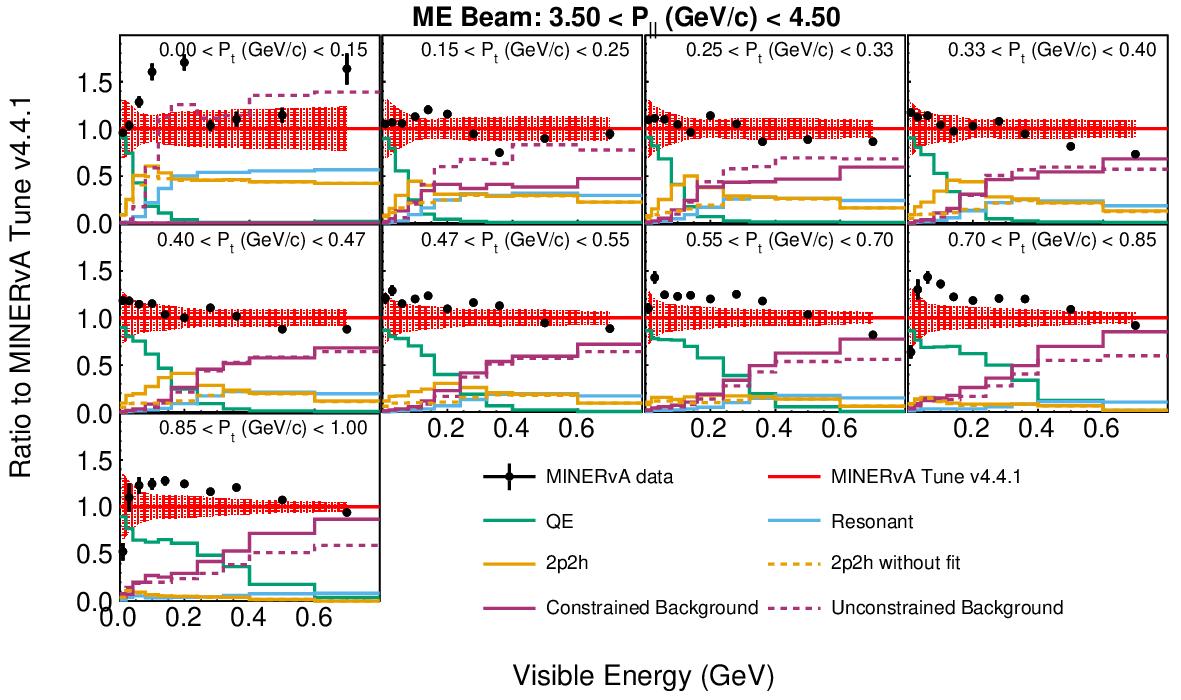}
    \caption{Ratios of measured to predicted event distributions after the background fits described in the text, for both the Low (top) and Medium (bottom) energy data, for $3.5 <\pz (GeV/c)< 4.5$ as a function of visible energy and \pt. }
    \label{fig:bckgdrat5}
\end{figure*}

\FloatBarrier
\newpage 

\subsection{Double Ratio Comparisons for all \pz\ bins}

\begin{figure*}
\includegraphics[width=.48\linewidth]{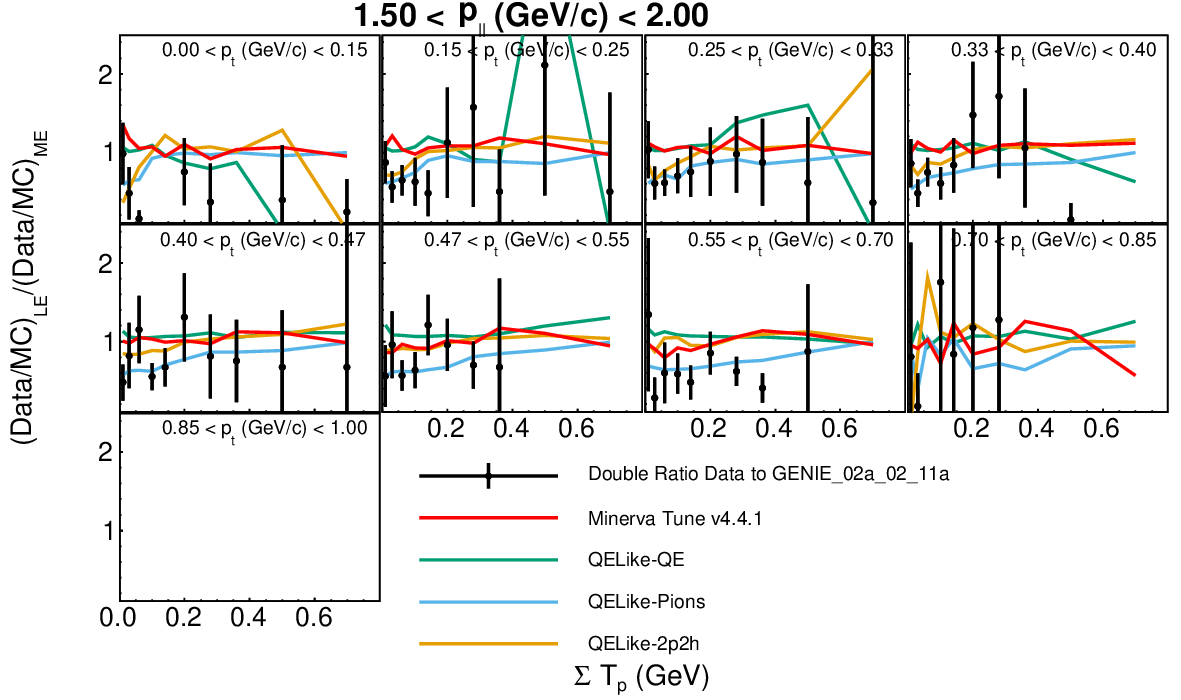}
\includegraphics[width=.48\linewidth]{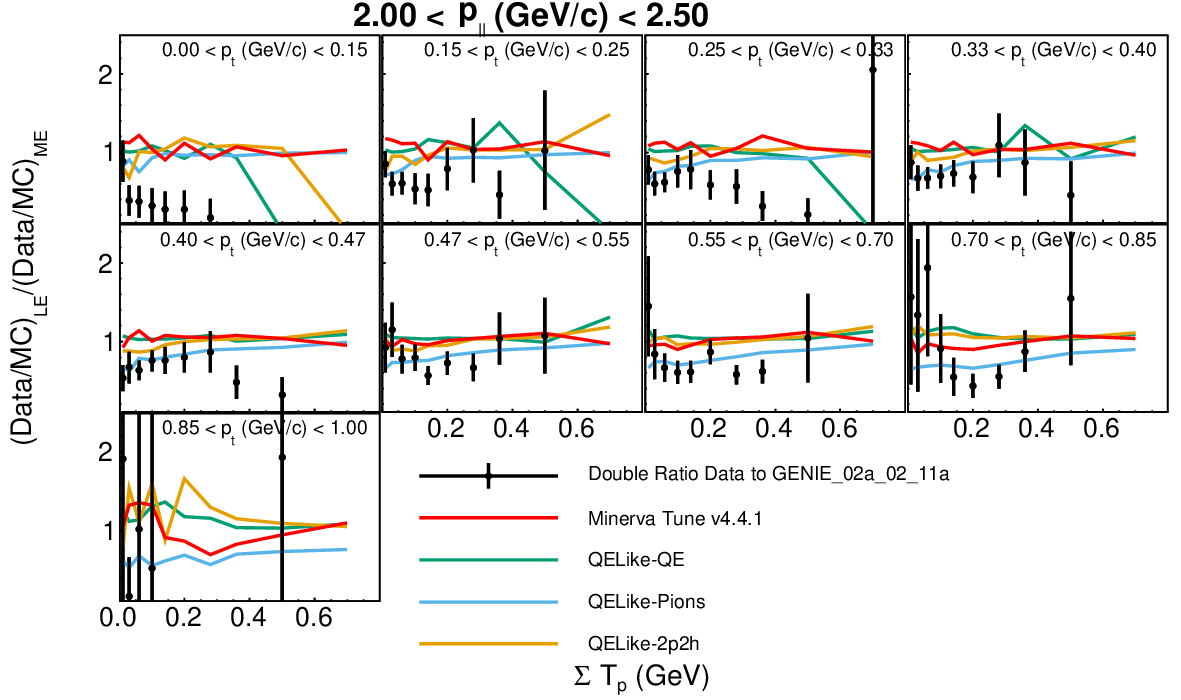}
\includegraphics[width=.48\linewidth]{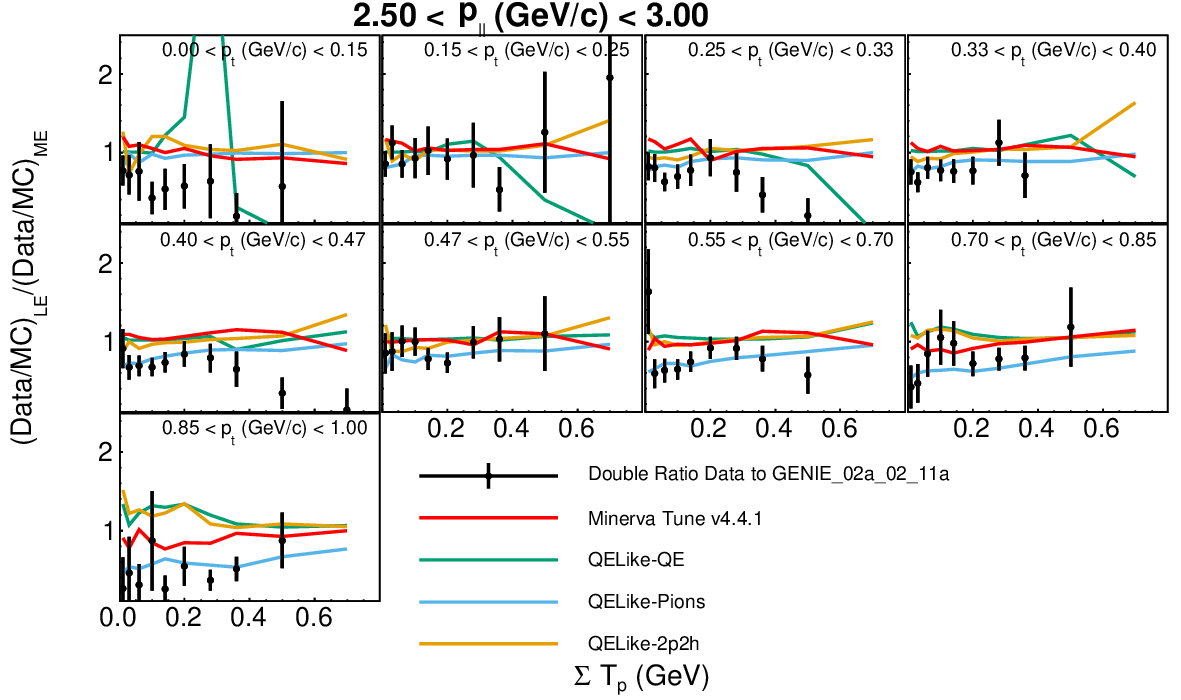}
\includegraphics[width=.48\linewidth]{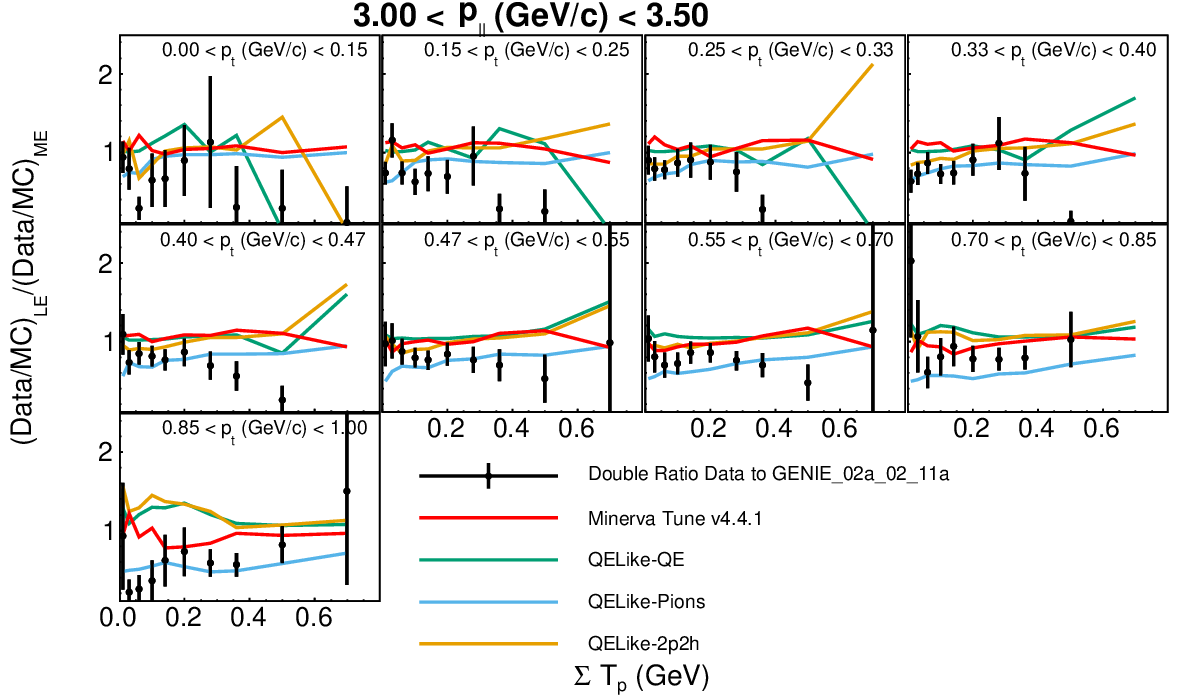}
\includegraphics[width=.48\linewidth]{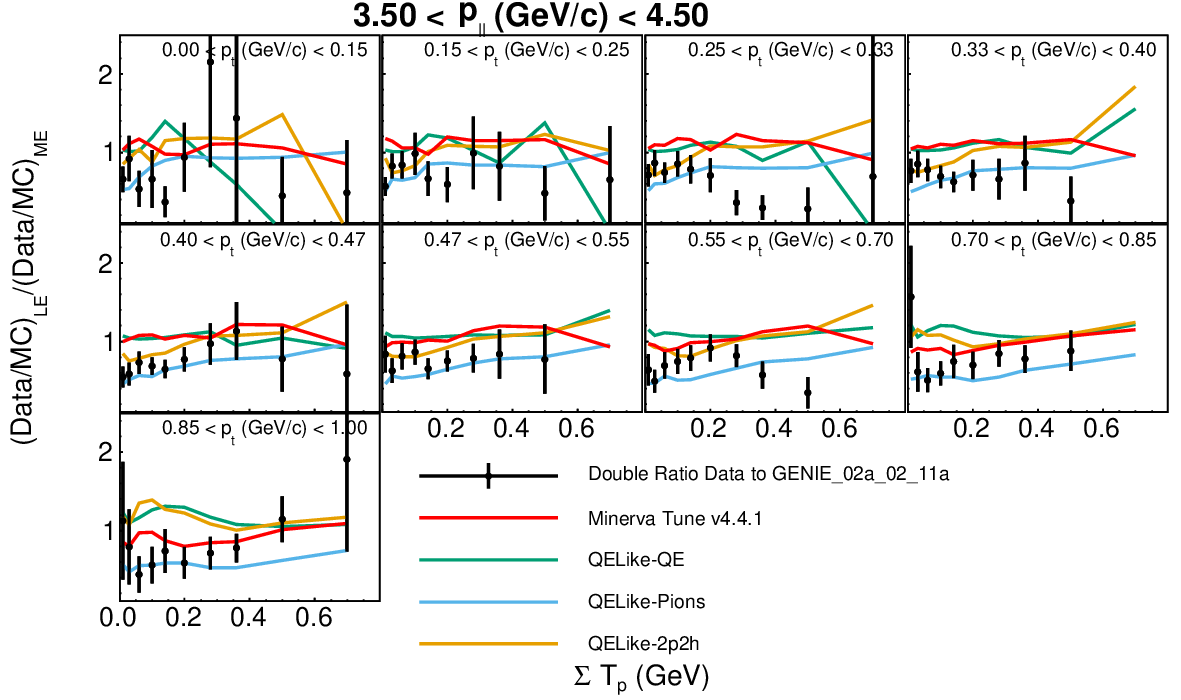}
\caption{Double Ratios to GENIE 02a for all momentum bins}
\label{fig:doubleratio_GENIE02a}
\end{figure*}
\begin{figure*}
\includegraphics[width=.48\linewidth]{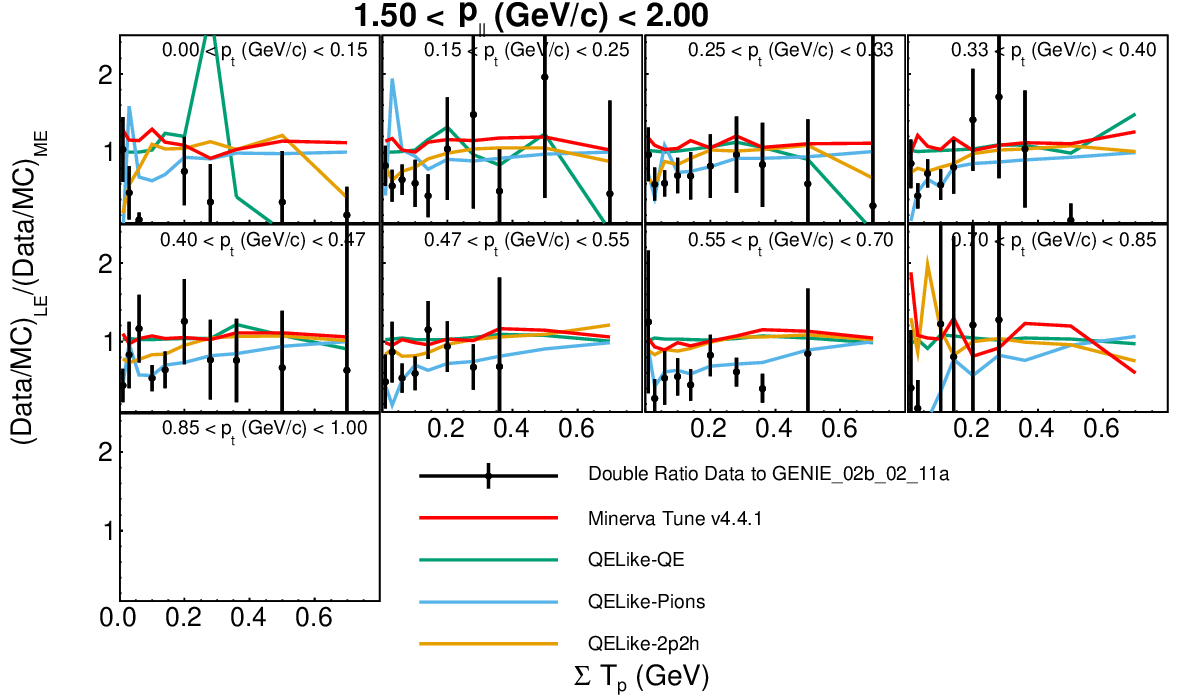}
\includegraphics[width=.48\linewidth]{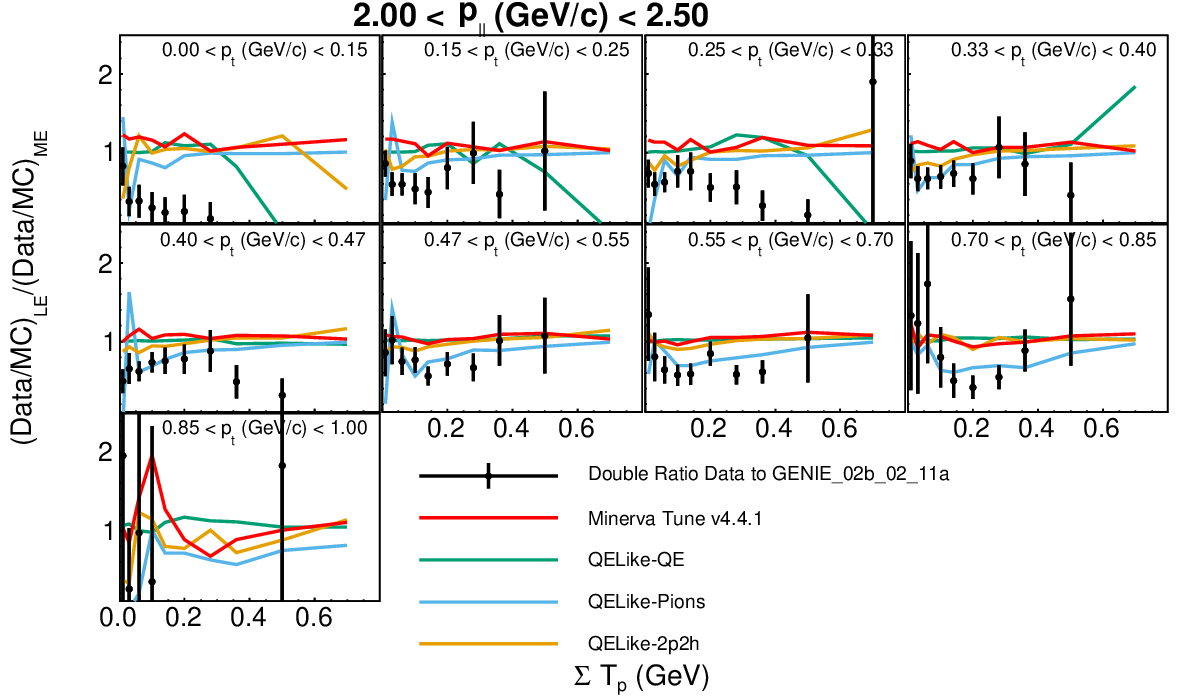}
\includegraphics[width=.48\linewidth]{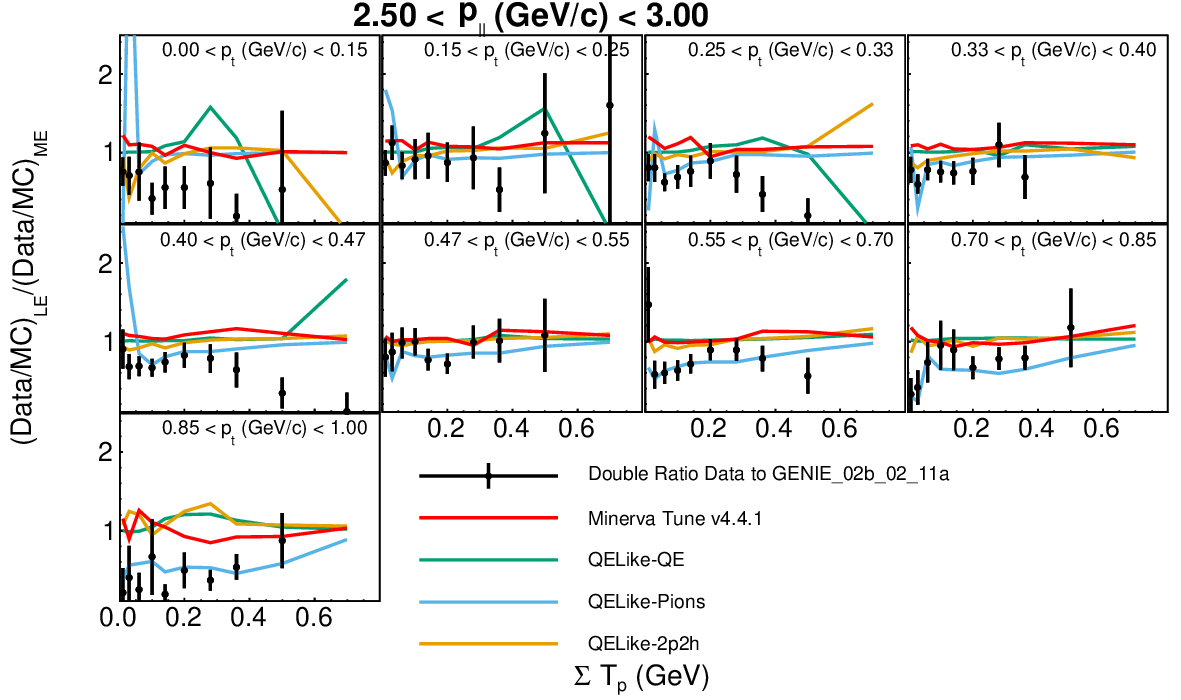}
\includegraphics[width=.48\linewidth]{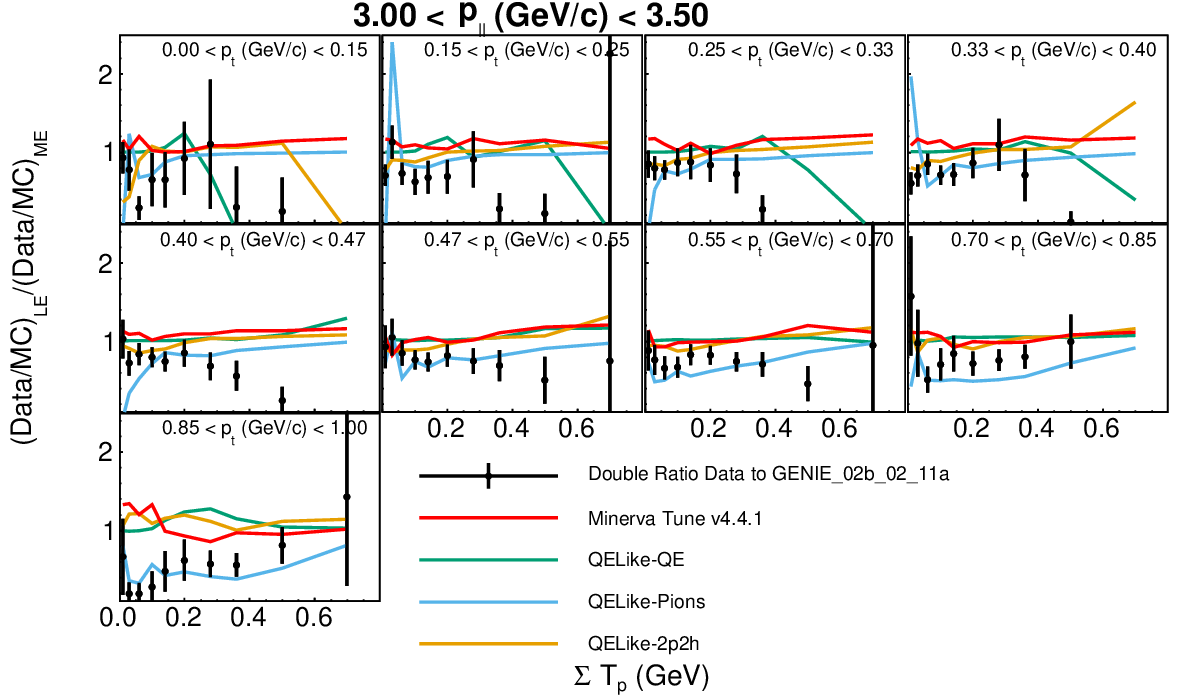}
\includegraphics[width=.48\linewidth]{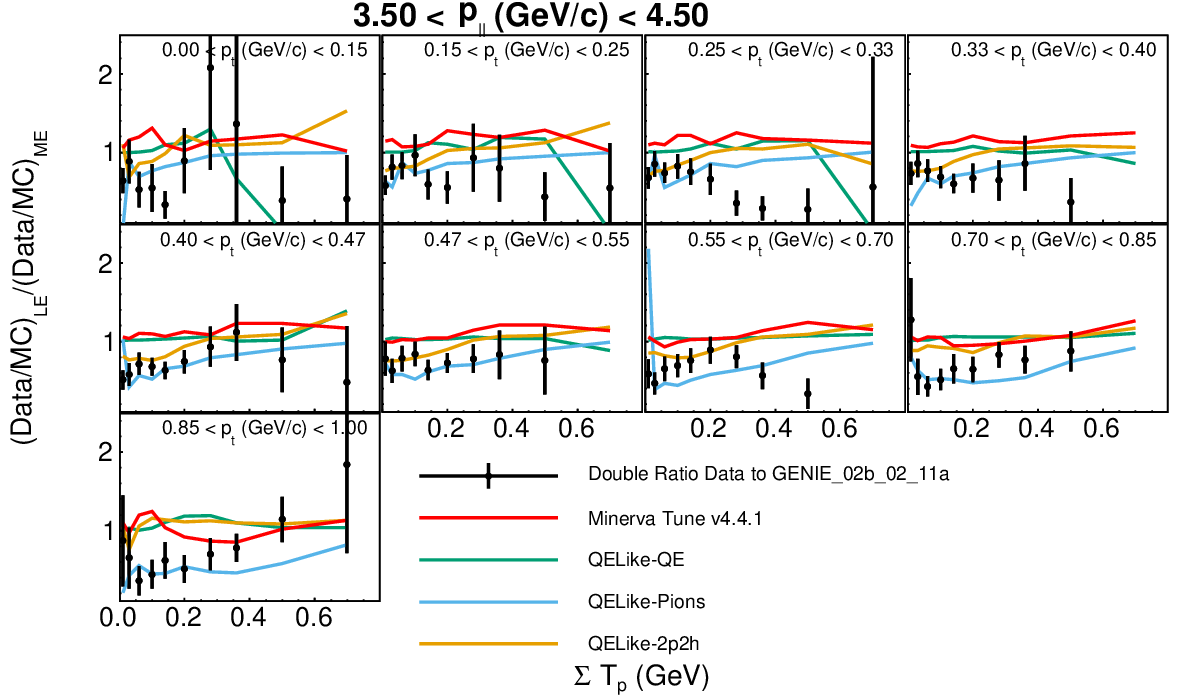}
\caption{Double Ratios to GENIE 02b for all momentum bins}
\label{fig:doubleratio_GENIE02b}
\end{figure*}
\begin{figure*}
\includegraphics[width=.48\linewidth]{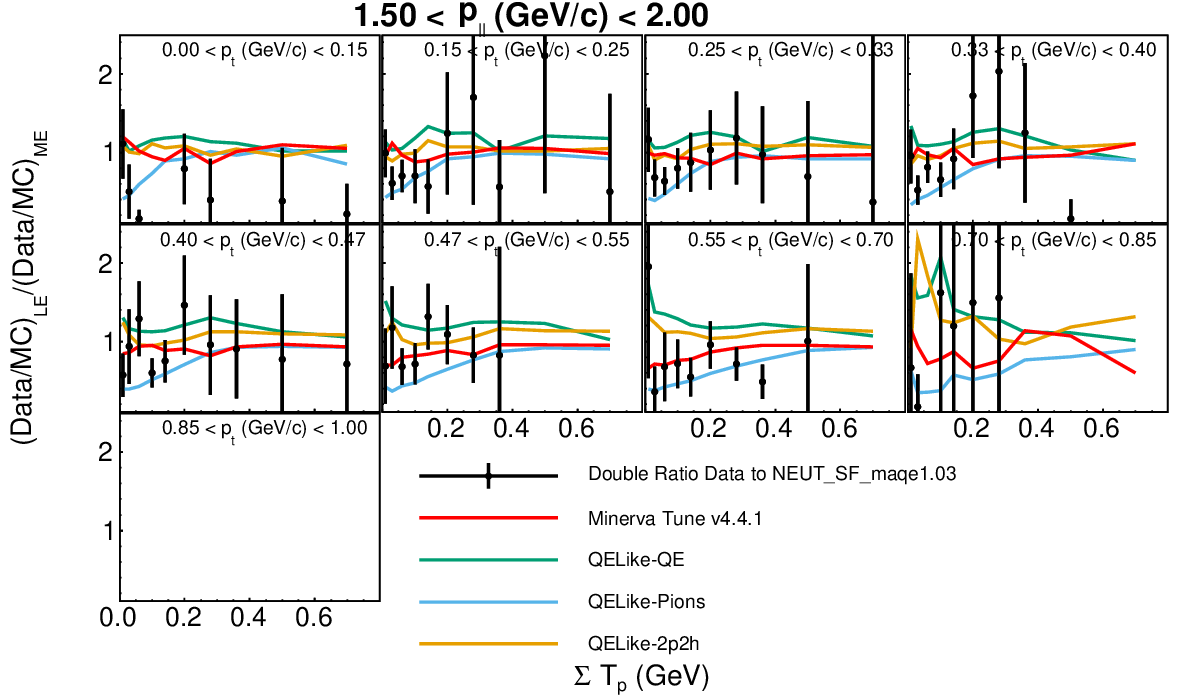}
\includegraphics[width=.48\linewidth]{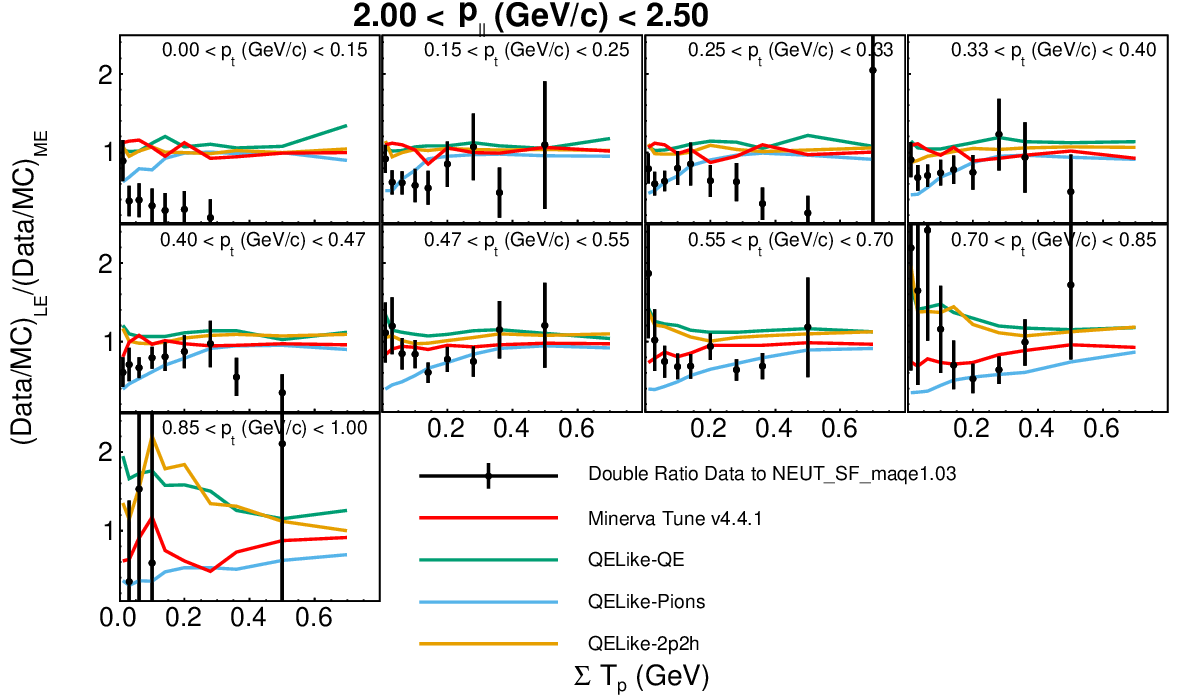}
\includegraphics[width=.48\linewidth]{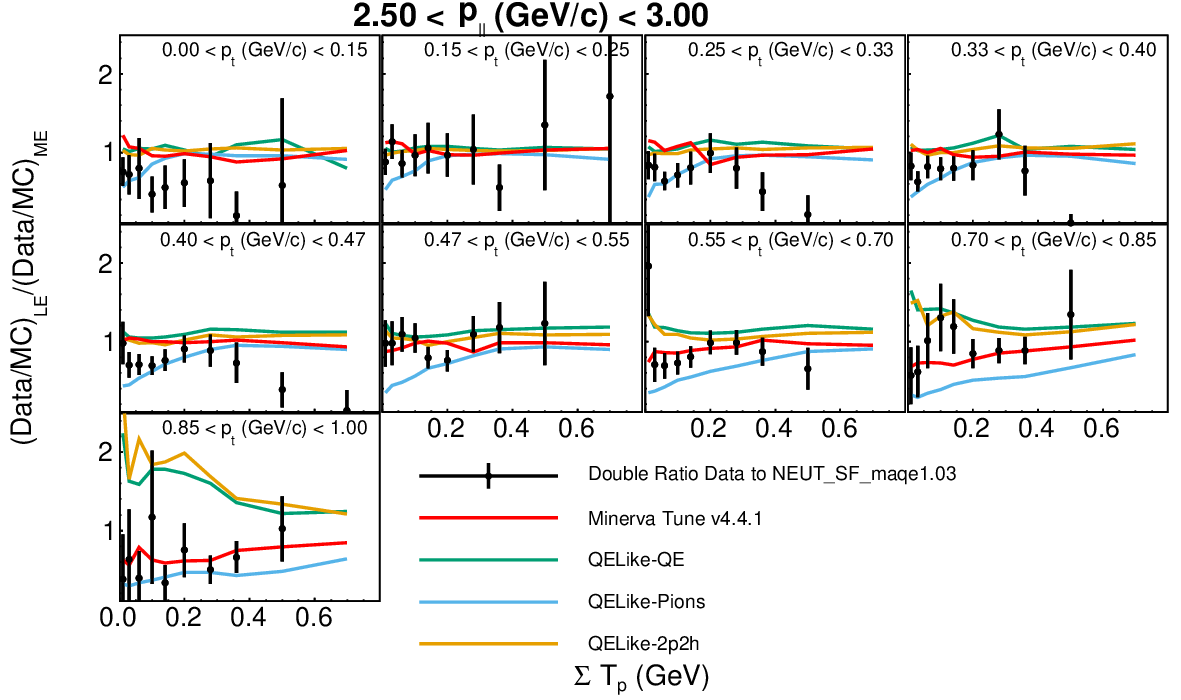}
\includegraphics[width=.48\linewidth]{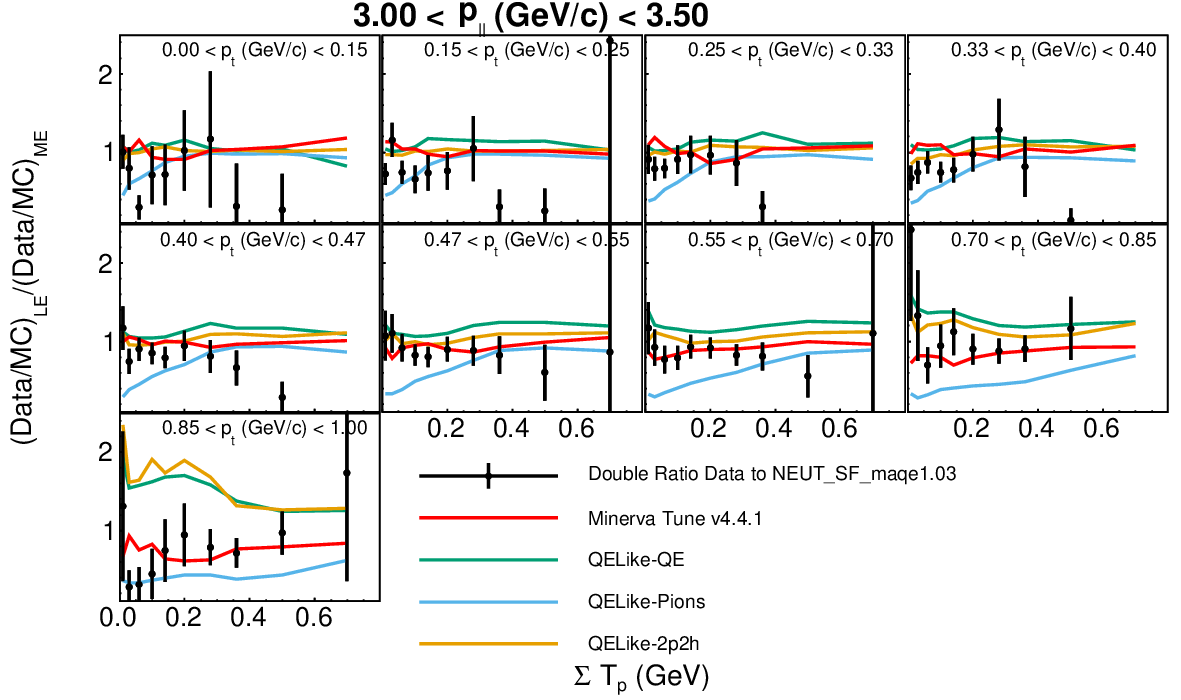}
\includegraphics[width=.48\linewidth]{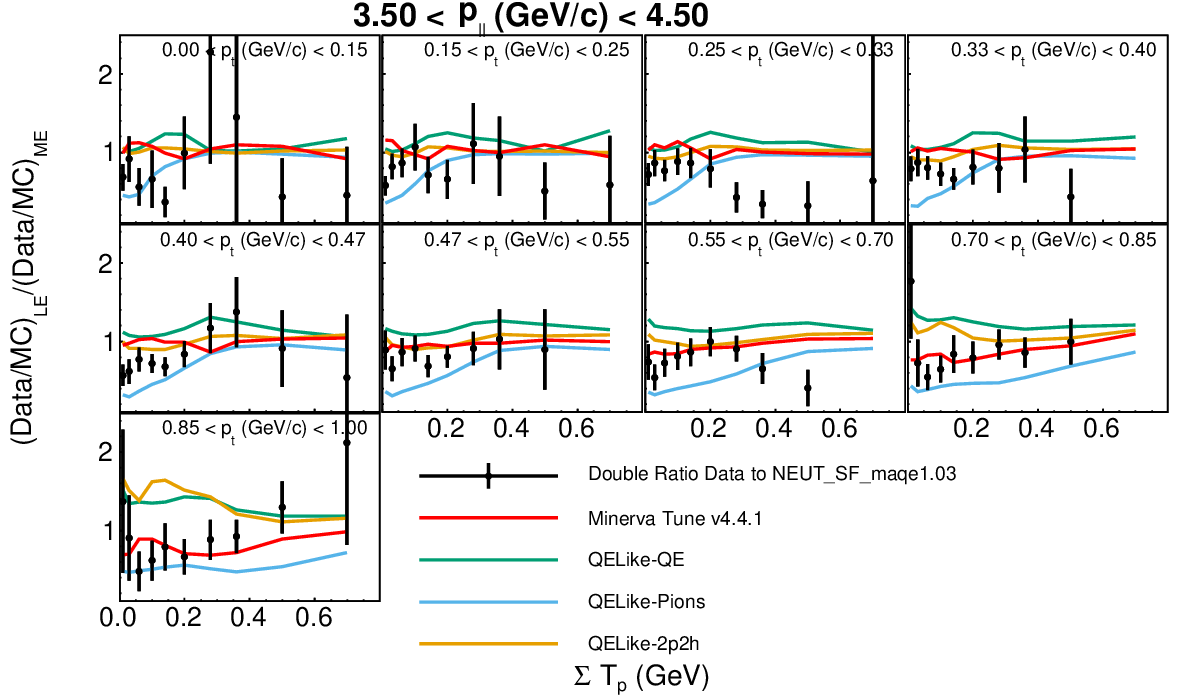}
\caption{Double Ratios to NEUT with a Spectral Function Model for all momentum bins}
\label{fig:doubleratio_NEUTSF}
\end{figure*}
\begin{figure*}
\includegraphics[width=.48\linewidth]{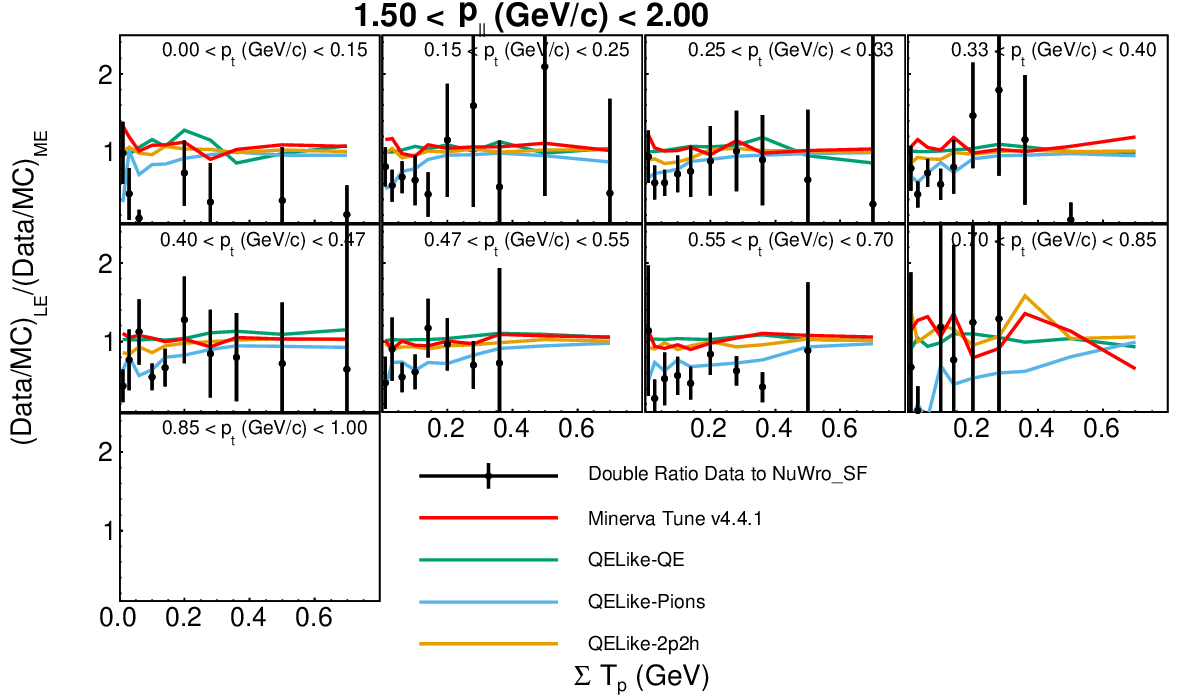}
\includegraphics[width=.48\linewidth]{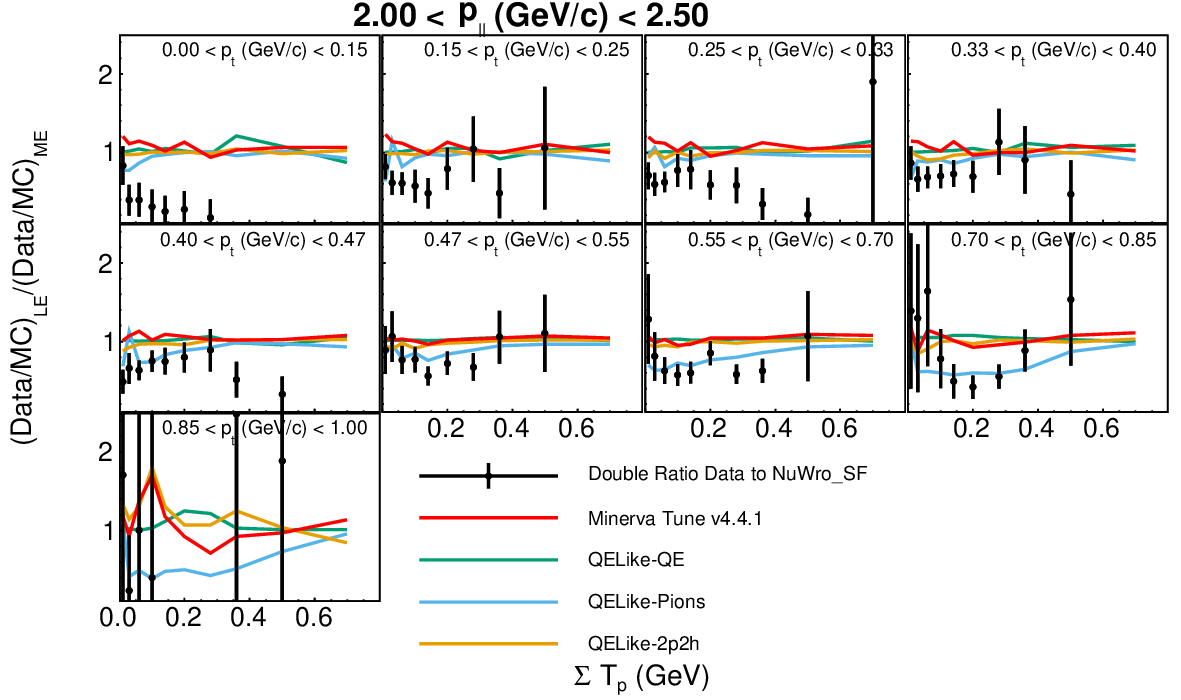}
\includegraphics[width=.48\linewidth]{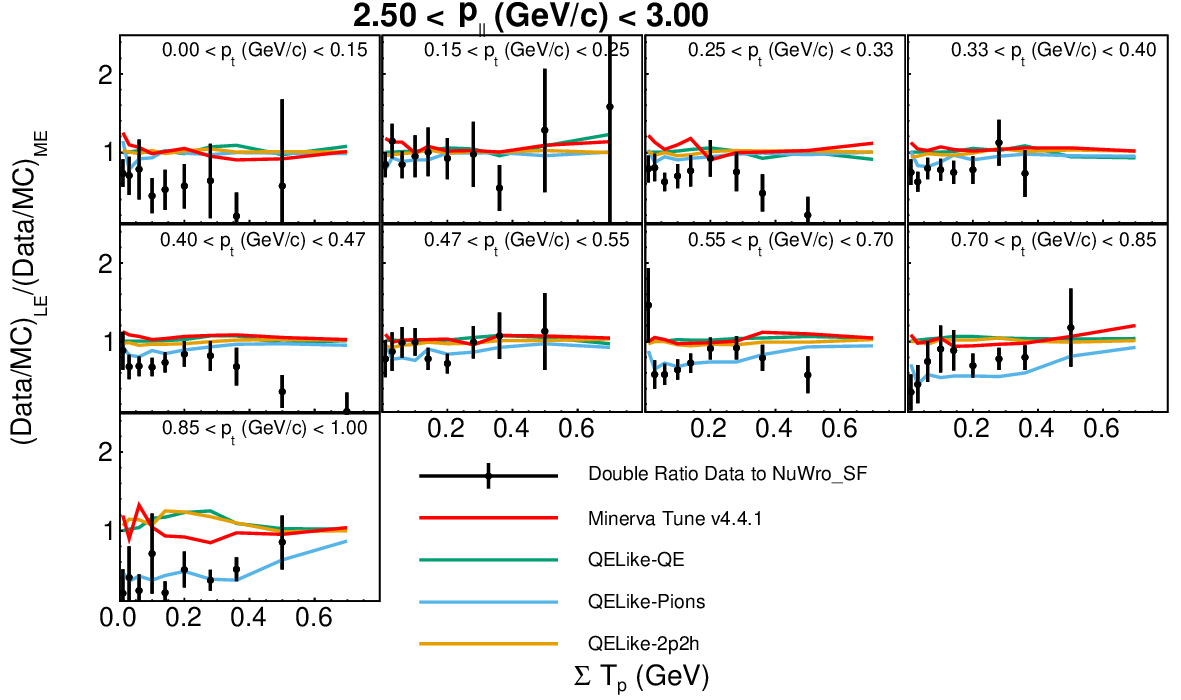}
\includegraphics[width=.48\linewidth]{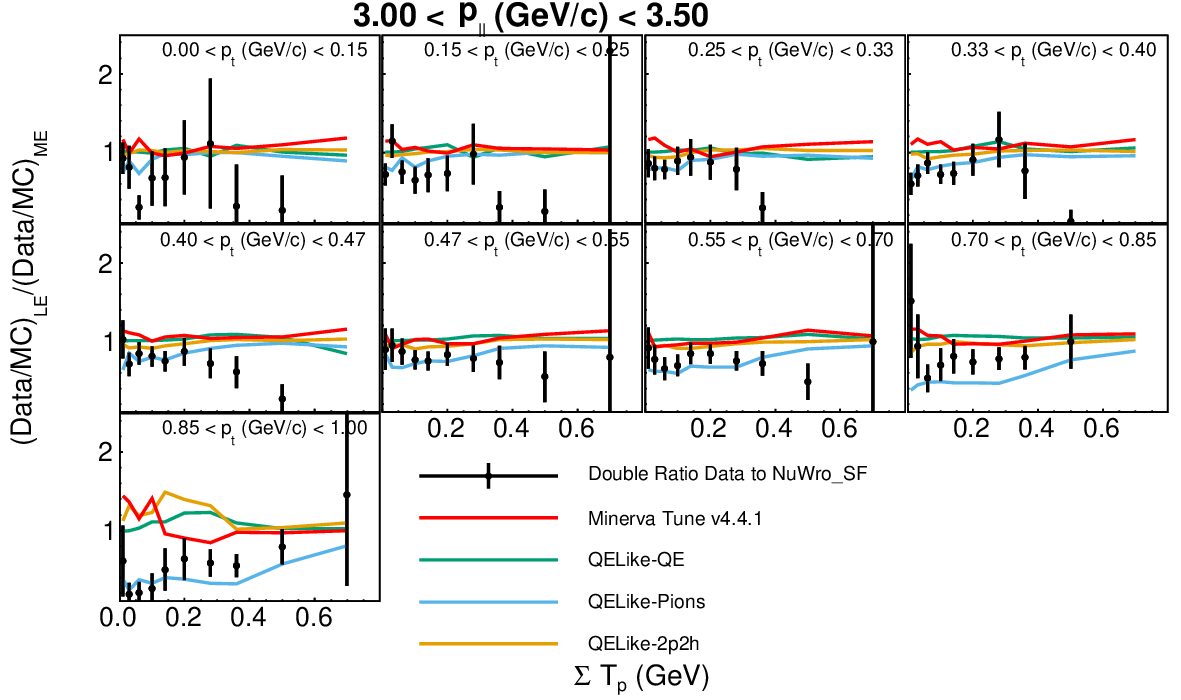}
\includegraphics[width=.48\linewidth]{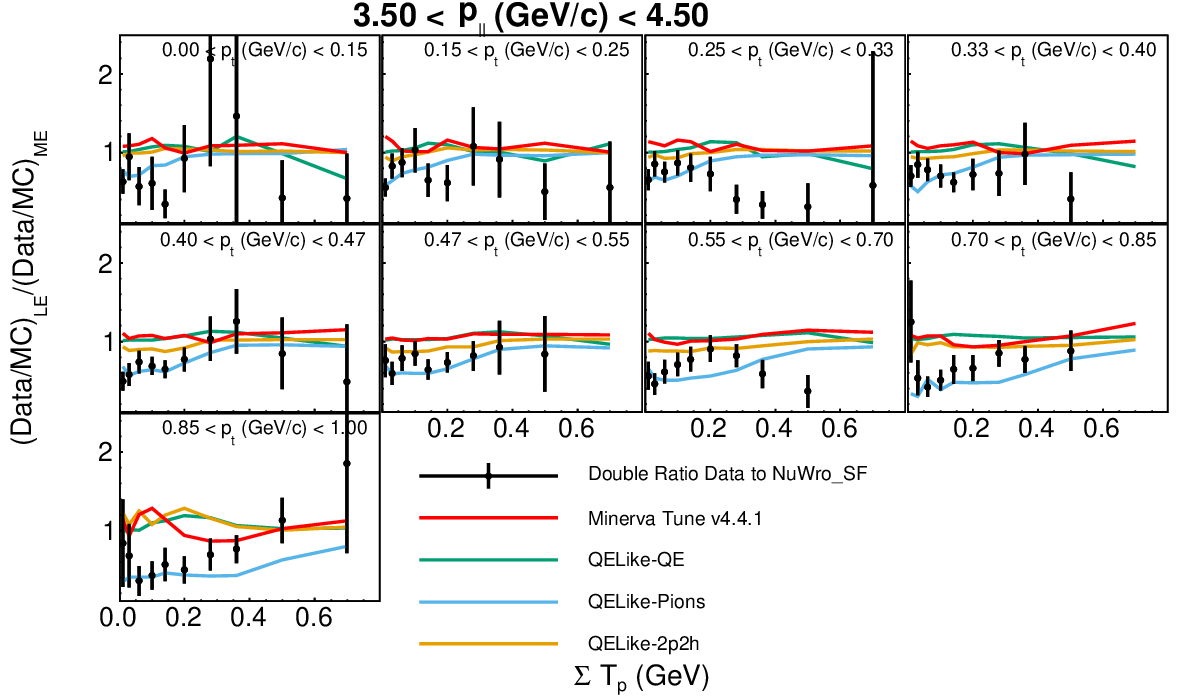}
\caption{Double Ratios to NuWro with a Spectral Function Model for all momentum bins}
\label{fig:doubleratio_NUWROSF}
\end{figure*}

\end{document}